\documentclass[3p,times,11pt]{elsarticle}

\newif\ifcomment

\usepackage[colorlinks=true, linktoc=page, linkcolor=blue, citecolor=blue, urlcolor=blue]{hyperref}

\usepackage[english]{babel}
\usepackage[T1]{fontenc}
\usepackage{graphics}
\usepackage{graphicx}
\usepackage{tabularx}
\usepackage{fancyhdr}
\usepackage{amsbsy}
\usepackage{pifont}
\usepackage{float}
\usepackage{delarray}
\usepackage{rotating}
\usepackage{makeidx}
\usepackage[normalem]{ulem}
\usepackage{color}
\usepackage[multiple]{footmisc}
\usepackage{xspace}
\usepackage{subfig}
\usepackage{wrapfig}
\usepackage{tikz}
\usetikzlibrary{patterns}
\usepackage{pgfplots}
\pgfplotsset{compat=1.12}

\usepackage{comment}
\usepackage{array}
\usepackage{longtable}
\usepackage{calc}
\usepackage{multirow}
\usepackage{hhline}
\usepackage{ifthen}
\usepackage{sectsty}

\usepackage[utf8]{inputenc}
\usepackage{natbib}
\biboptions{sort&compress}

\usepackage{amsmath}
\usepackage{amsfonts}
\usepackage{amssymb}
\usepackage{mathrsfs}
\usepackage{dcolumn}%
\usepackage{bm}%
\usepackage{amsfonts}
\usepackage{grffile} %
\usepackage{lineno}
\newcommand{\sqrts}{\ensuremath{\sqrt{s}}}
\newcommand{\sqrtsnn}{\ensuremath{\sqrt{s_{_{NN}}}}}
\newcommand{\alphas}{\ensuremath{\alpha_{\rm s}}}
\newcommand{\pT}{\ensuremath{P_T}\xspace}
\newcommand{\kT}{\ensuremath{k_T}\xspace}
\newcommand{\qT}{\ensuremath{q_T}\xspace}
\newcommand{\xF}{\ensuremath{x_{\rm F}}\xspace}

\newcommand{\ccbar}{\ensuremath{c\overline{c}}\xspace}
\newcommand{\bbbar}{\ensuremath{b\overline{b}}\xspace}
\newcommand{\jpsi}{\ensuremath{J/\psi}\xspace}
\newcommand{\psip}{\ensuremath{\psi(2S)}\xspace}
\newcommand{\ups}{\ensuremath{\Upsilon}\xspace}
\newcommand{\upsp}{\ensuremath{\Upsilon(2S)}\xspace}
\newcommand{\upspp}{\ensuremath{\Upsilon(3S)}\xspace}
\newcommand{\chic}{\ensuremath{\chi_c}\xspace}

\newcommand{\chico}{\ensuremath{\chi_{c1}}\xspace}
\newcommand{\chict}{\ensuremath{\chi_{c2}}\xspace}
\newcommand{\etac}{\ensuremath{\eta_c}\xspace}
\newcommand{\etacp}{\ensuremath{\eta_c(2S)}\xspace}
\newcommand{\chib}{\ensuremath{\chi_b}\xspace}
\newcommand{\etab}{\ensuremath{\eta_b}\xspace}

\newcommand{\pp}{\ensuremath{pp}\xspace}

\newcommand{\ppbar}{\ensuremath{p\bar{p}}\xspace}
\newcommand{\pA}{{\ensuremath{pA}}\xspace}
\newcommand{\Ap}{{\ensuremath{Ap}}\xspace}
\newcommand{\pN}{\ensuremath{pN}\xspace}
\newcommand{\pPb}{\ensuremath{p\rm{Pb}}\xspace}
\newcommand{\AaAa}{\ensuremath{AA}\xspace}
\newcommand{\PbPb}{\rm{PbPb}\xspace}
\newcommand{\PbXe}{\rm{PbXe}\xspace}
\newcommand{\PbAr}{\rm{PbAr}\xspace}

\newcommand{\Lumi}{\ensuremath{\mathcal L}\xspace}
\newcommand{\Lint}{\ensuremath{\int \mathcal L}\xspace}
\newcommand{\invmub}{\ensuremath{\mu\textrm{b}^{-1}}\xspace}
\newcommand{\invnb}{\ensuremath{\textrm{nb}^{-1}}\xspace}
\newcommand{\invpb}{\ensuremath{\textrm{pb}^{-1}}\xspace}
\newcommand{\invfb}{\ensuremath{\textrm{fb}^{-1}}\xspace}
\newcommand{\pythia}{{\sc pythia}\xspace}
\newcommand{\herwig}{{\sc herwig++}\xspace}
\newcommand{\mcfm}{{\sc mcfm}\xspace}

\newcommand*{\eg}{{e.g.}\xspace}
\newcommand*{\ie}{{i.e.}\xspace}
\newcommand*{\cm}{{\rm c.m.s.}\xspace}

\DeclareMathAlphabet{\pazocal}{OMS}{zplm}{m}{n}
\newcommand{\Q}{\pazocal{Q}}

\DeclareFontFamily{OT1}{pzc}{}
\DeclareFontShape{OT1}{pzc}{m}{it}{<-> s * [1.10] pzcmi7t}{}
\DeclareMathAlphabet{\mathpzc}{OT1}{pzc}{m}{it}

\newcommand{\red}[1]{\textcolor[rgb]{1,0,0}{#1}}

\newcommand{\pink}[1]{{\textcolor[rgb]{1,0,0.5}{#1}}}

\newcommand{\Remark}[2]{{\small \pink{Remark (by {#1}):\ {#2}}}}

\newcommand{\ce}[1]{Eq.~(\ref{#1})}
\newcommand{\cf}[1]{{Fig.~\ref{#1}}}
\newcommand{\ct}[1]{{Table~\ref{#1}}}

\def\RpA     {\ensuremath{R_{pA}}\xspace}
\def\RAA     {\ensuremath{R_{AA}}\xspace}

\def\RpPb    {\ensuremath{R_{p\rm Pb}}\xspace}

\def\pPb  {$p\mathrm{Pb}$}

\usepackage[shortcuts]{extdash} %

\begin{document}
\vspace*{-1cm}
\begin{frontmatter}
\title{Prospects for quarkonium studies at the high-luminosity LHC}

\date{\today}

\author[2]{\'{E}milien~Chapon\fnref{SE}}
\author[8]{David~d'Enterria\fnref{SE}}
\author[104]{Bertrand~Ducloué\fnref{SE}}
\author[5]{Miguel~G.~Echevarria\fnref{SE}}
\author[6]{Pol-Bernard~Gossiaux\fnref{SE}}
\author[7]{Vato~Kartvelishvili\fnref{SE}}
\author[Mainz]{Tomas~Kasemets\fnref{SE}}
\author[1]{Jean-Philippe~Lansberg\fnref{Editor}}
\author[9]{Ronan~McNulty\fnref{SE}}
\author[10]{Darren~D.~Price\fnref{SE}}
\author[11]{Hua-Sheng~Shao\fnref{SE}}
\author[9]{Charlotte~Van~Hulse\fnref{SE}}
\author[16]{Michael~Winn\fnref{SE}}
\fntext[SE]{Section editor}
\fntext[Editor]{Editor}

\author[37]{Jaroslav~Adam}
\author[22]{Liupan~An}
\author[6]{Denys~Yen~Arrebato~Villar}
\author[20]{Shohini~Bhattacharya}
\author[3,4,ECTstar,FBK]{Francesco~G.~Celiberto}
\author[35]{Cvetan~Cheshkov}
\author[14]{Umberto~D'Alesio}
\author[17]{Cesar~da~Silva}
\author[29]{Elena~G.~Ferreiro}
\author[15,322]{Chris~A.~Flett}
\author[1]{Carlo~Flore}
\author[26,22,31]{Maria~Vittoria~Garzelli}
\author[13,10]{Jonathan~Gaunt}
\author[32]{Jibo~He}
\author[4]{Yiannis~Makris}
\author[103]{Cyrille~Marquet}
\author[1]{Laure~Massacrier}
\author[18]{Thomas~Mehen}
\author[16]{Cédric Mezrag}
\author[200]{Luca Micheletti}
\author[30]{Riccardo~Nagar}
\author[101]{Maxim~A.~Nefedov}
\author[1]{Melih~A.~Ozcelik}
\author[14]{Biswarup~Paul}
\author[14]{Cristian~Pisano}
\author[25]{Jian-Wei~Qiu}
\author[14]{Sangem~Rajesh}
\author[21]{Matteo~Rinaldi}
\author[1,36]{Florent~Scarpa}
\author[7]{Maddie~Smith}
\author[103]{Pieter~Taels}
\author[7]{Amy~Tee}
\author[28]{Oleg~Teryaev}
\author[17]{Ivan~Vitev}
\author[25]{Kazuhiro~Watanabe}
\author[Kennesaw,RIKEN]{Nodoka~Yamanaka}
\author[19]{Xiaojun~Yao}
\author[8,33]{Yanxi~Zhang}

\address[2]{Institute of High Energy Physics, Beijing, China}
\address[8]{CERN, EP Department, CH-1211 Geneva 23, Switzerland}
\address[104]{Higgs Centre for Theoretical Physics, University of Edinburgh, Peter Guthrie Tait Road, Edinburgh EH9 3FD, Scotland}
\address[5]{Dpto.\ de F\'isica y Matem\'aticas, Universidad de Alcal\'a, 28805 Alcal\'a de Henares (Madrid), Spain}
\address[6]{SUBATECH, IMT Atlantique, CNRS/IN2P3, Universit\'e de Nantes, 4 rue Alfred Kastler, 44307 Nantes, France}
\address[7]{Department of Physics, Lancaster University, Lancaster, LA1 4YB, UK}
\address[Mainz]{PRISMA$^+$ Cluster of Excellence \& MITP, Johannes Gutenberg University, 55099 Mainz, Germany }
\address[1]{Universit\'e Paris-Saclay, CNRS/IN2P3, IJCLab, 91405 Orsay, France}
\address[9]{University College Dublin, Belfield, Dublin 4, Ireland}
\address[10]{Department of Physics \& Astronomy, University of Manchester, Manchester M13 9PL, UK}
\address[11]{Laboratoire de Physique Th\'eorique et Hautes Energies (LPTHE), UMR 7589,\\ Sorbonne Universit\'e et CNRS, 4 place Jussieu, 75252 Paris Cedex 05, France}
\address[16]{Irfu, CEA, Universit\'e Paris-Saclay, F-91191 Gif-Sur-Yvette, France}

\address[37]{Creighton University, Omaha, USA}
\address[22]{INFN, Sezione di Firenze, Firenze, Italy}
\address[27]{Department of Physics, University of Oslo, Norway}
\address[20]{Department of Physics, SERC, Temple University, Philadelphia, PA 19122, USA}
\address[3]{Dipartimento di Fisica, Universit\`a degli Studi di Pavia, via Bassi 6, I-27100 Pavia, Italy}
\address[4]{INFN, Sezione di Pavia, via Bassi 6, I-27100 Pavia, Italy}
\address[ECTstar]{European Centre for Theoretical Studies in Nuclear Physics and Related Areas (ECT*), I-38123 Villazzano, Trento, Italy}
\address[FBK]{Fondazione Bruno Kessler (FBK), I-38123 Povo, Trento, Italy}
\address[35]{Université de Lyon, Universite Lyon 1, CNRS/IN2P3, IP2I-Lyon, Villeurbanne, Lyon, France}
\address[14]{Dipartimento di Fisica, Universit\`a di Cagliari, and INFN Sezione di Cagliari, Cittadella Univ., I-09042 Monserrato (CA), Italy}
\address[17]{Los Alamos National Laboratory, Physics \& Theoretical Divisions, Mail Stops MS-H846 \& B283, Los Alamos, NM 87545, USA}
\address[29]{Dept. F{\'\i}sica de Part{\'\i}culas and IGFAE, Universidade de Santiago de Compostela, 15782 Santiago de Compostela, Spain}
\address[15]{Department of Mathematical Sciences, University of Liverpool, Liverpool, L69 3BX, UK}
\address[322]{Department of Physics, University of Jyv\"{a}skyl\"{a}, P.O. Box 35, 40014 University of Jyv\"{a}skyl\"{a}, Finland}
\address[26]{Dipartimento di Fisica e Astronomia, Universit\`a degli Studi di Firenze, Firenze, Italy}
\address[31]{Hamburg Universit\"at, II Institut f\"ur Theoretische Physik, D-22761 Hamburg, Germany}
\address[13]{CERN, TH Department, CH-1211 Geneva 23, Switzerland}
\address[32]{University of Chinese Academy of Sciences, Beijing, China}
\address[103]{Centre de Physique Th\'eorique, \'Ecole polytechnique, CNRS, I.P.\ Paris, F-91128 Palaiseau, France}
\address[18]{Department of Physics, Duke University, Durham, NC 27705 USA}
\address[200]{INFN, Sezione di Torino, Turin, Italy}
\address[30]{Dipartimento di Fisica, Universit\`a degli Studi di Milano-Bicocca, and INFN Sezione di Milano-Bicocca, I-20126 Milan, Italy}
\address[101]{Samara National Research University, Moskovskoe Shosse 34, 443086 Samara, Russia}
\address[25]{Theory Center, Jefferson  Laboratory, Newport News, VA 23606, USA}
\address[21]{Dipartimento di Fisica. Universit\`a degli studi di Perugia, and INFN Sezioni di Perugia. Via A. Pascoli, Perugia, 06123, Italy}
\address[36]{Van Swinderen Institute for Particle Physics and Gravity, University of Groningen, %
9747 AG Groningen, The Netherlands}
\address[28]{Joint Institute for Nuclear Research, 141980, Dubna, Russia}
\address[Kennesaw]{Department of Physics, Kennesaw State University, Kennesaw, GA 30144, USA}
\address[RIKEN]{Nishina Center for Accelerator-Based Science, RIKEN, Wako 351-0198, Japan}

\address[19]{Center for Theoretical Physics, Massachusetts Institute of Technology, Cambridge, MA 02139 USA}
\address[33]{Peking University, Beijing, China}

\vspace*{2cm}
\begin{abstract}
Prospects for quarkonium-production studies
accessible during the upcoming high-luminosity phases of the CERN Large Hadron Collider operation after 2021 are reviewed. 
Current experimental and theoretical open issues in the field are assessed together with the potential for future studies in quarkonium-related physics.
This will be possible through the exploitation of the huge data samples to be collected in proton-proton, 
proton-nucleus and nucleus-nucleus collisions, both in the collider and fixed-target modes. Such investigations include, among others, those of: (i) $J/\psi$ and $\Upsilon$ produced in association with other hard particles; (ii) $\chi_{c,b}$ and $\eta_{c,b}$ down to small transverse momenta; (iii) the constraints brought in by quarkonia on gluon PDFs, nuclear PDFs, TMDs, GPDs and GTMDs, as well as on the low-$x$ parton dynamics; (iv) the gluon Sivers effect in polarised-nucleon collisions; (v) the properties of the quark-gluon plasma produced in ultra-relativistic heavy-ion collisions and of collective partonic 
effects in general; and (vi) double and triple parton scatterings.
\end{abstract}
\end{frontmatter}

\tableofcontents

\section{Introduction}

The Large Hadron Collider (LHC) accelerator and detector systems are being upgraded to enable their optimal exploitation after 2027 with a ten–fold increase in their instantaneous luminosity in the proton-proton (\pp) running mode with respect to the nominal design values~\cite{Apollinari:2017cqg}. The ultimate high-luminosity phase of the collider (referred to as HL-LHC) will lead to the collection of huge data samples of \pp collisions from total integrated luminosities reaching $\Lumi = 3$~ab$^{-1}$ at ATLAS and CMS, and around $\Lumi = 0.3$~ab$^{-1}$ at LHCb, by the end of the LHC operation in 2035. In addition, integrated luminosities of about 13\,\invnb, 13\,\invnb, and 2\,\invnb of \PbPb data 
as well as 1.2\,\invpb, 0.6\,\invpb, and 0.6\,\invpb of \pPb\ data are expected to be obtained by each of the ATLAS/CMS, ALICE, and LHCb experiments until 2030, respectively~\cite{Citron:2018lsq}. Such unprecedented data sets will open up new exciting physics opportunities in the study of the Standard Model~\cite{Azzi:2019yne}, and in particular its heavy-flavour sector~\cite{Cerri:2018ypt}. This phase will also offer the possibility to collect data using the LHC proton and ion beams on Fixed Targets (FT). The corresponding physics programme~\cite{Brodsky:2012vg,Hadjidakis:2018ifr} of the LHC in the FT mode (referred to as FT-LHC) relies on extremely high {\it yearly} luminosities.
The FT mode is under study for ALICE and LHCb, where up to 10\,\invfb in \pp, 300\,\invpb in proton-nucleus (\pA), and 30\,\invnb in lead-nucleus ($Pb$A) collisions are expected.

The aim of this review is to highlight the impact that the upcoming operations of the LHC, in particular the HL-LHC and FT-LHC, will have on various sectors of quarkonium-production studies in \pp, \pA, and nucleus-nucleus (\AaAa) collisions. 
Not only is the mechanism underlying the inclusive production of quarkonia ($\Q$) still an outstanding problem in hadroproduction~\cite{Lansberg:2019adr}, but quarkonia can also serve as tools for the study of many other aspects of Quantum Chromodynamics (QCD). 
To name a few, charmonia and bottomonia can be used to probe the proton gluon content in terms of various parton densities such as parton distribution functions (PDFs, see \eg~\cite{Halzen:1984rq,Martin:1987ww,Martin:1987vw,Jung:1992uj,Lansberg:2020ejc}), transverse-momentum densities (TMDs, see \eg~\cite{Boer:2012bt,Dunnen:2014eta,Boer:2016bfj,Lansberg:2017dzg,Lansberg:2017tlc,Lansberg:2018fwx,Bacchetta:2018ivt,DAlesio:2019qpk,Kishore:2019fzb,Scarpa:2019fol,Boer:2020bbd}), and generalised parton densities (GPDs, see \eg~\cite{Diehl:2003ny,Ivanov:2004vd,Jones:2015nna,Flett:2019pux}); the gluon content of heavy nuclei through nuclear PDFs (see \eg~\cite{Ferreiro:2011xy,Lansberg:2016deg,Albacete:2016veq,Albacete:2017qng,Kusina:2017gkz}).
More generally, they allow the study of the initial stages of ultra-relativistic heavy-ion collisions (see \eg~\cite{Andronic:2015wma}) and at the same time, they offer new ways to investigate the dynamics of hard multi-parton interactions (see \eg~\cite{Kom:2011bd,Lansberg:2014swa,dEnterria:2016ids,Shao:2019qob}) or to measure the properties of the quark-gluon plasma (see \eg~\cite{Rapp:2008tf,Andronic:2015wma,Rothkopf:2019ipj}). 

The interested reader will find it useful to consult the following reviews~\cite{Kramer:2001hh,Brambilla:2004wf,Lansberg:2006dh,Brambilla:2010cs,ConesadelValle:2011fw} addressing HERA and Tevatron results, and more recent ones~\cite{Andronic:2015wma,Lansberg:2019adr} concerning recent advances in the field with the RHIC and LHC data. 
With regards to existing experimental results, the reader is guided to the HEPData database (\href{https://www.hepdata.net/}{\tt https://www.hepdata.net/}), to a dedicated repository of quarkonium measurements up to 2012 (\href{http://hepdata.cedar.ac.uk/review/quarkonii/}{\tt http://hepdata.cedar.ac.uk/review/quarkonii/}) documented in~\cite{Andronic:2013zim} and to a recent review  on RHIC results~\cite{Tang:2020ame}.

The document is organised as follows. Section~\ref{sec:pp} focuses on the studies accessible in \pp\ collisions. Section~\ref{sec:excl_diff} discusses quarkonium production in exclusive and diffractive interactions, while Section~\ref{sec:spin} is devoted to the impact on studies involving transverse-momentum-dependent observables. Quarkonium studies in \pA\ and \AaAa\ collisions are respectively covered in Sections~\ref{sec:pa} and~\ref{sec:aa}. Finally, Section~\ref{sec:dps} addresses quarkonium production in double parton scattering (DPS) and triple parton scattering (TPS). Following the structure of similar previous prospective quarkonium studies at the LHC, such as~\cite{Lansberg:2008zm}, each chapter starts with a short summary of the current state-of-the-art and open experimental and theoretical issues, followed by a succinct list of HL-LHC studies that should further improve the understanding of all quarkonium-related physics. Anywhere relevant we have indicated when higher luminosities are needed or when similar luminosities to those already collected during the previous runs will be sufficient with dedicated triggers for these new studies.

\section{Proton-proton collisions\protect\footnote{
Section editors: Darren Price, Hua-Sheng Shao.
}}
\label{sec:pp}

\subsection{Introduction: status and prospects}

The production of quarkonium in high-energy particle collisions has been linked to longstanding challenges in our understanding of quark confinement in QCD. The study of quarkonium production provides not only valuable information on non-perturbative QCD physics, but also crucial and often novel signatures for the exploration of new phenomena, multi-quark spectroscopy, probes of proton structure, and double parton scattering interactions, amongst other subjects. 

Various theoretical treatments tackling the production of quark-antiquark pairs and their subsequent formation of a quarkonium bound state have been proposed and confronted with experimental data. 
The most notable and used of these are the Colour Evaporation Model (CEM)~\cite{Fritzsch:1977ay,Gluck:1977zm,Barger:1979js,Amundson:1995em,Amundson:1996qr}, the Colour-Singlet Model (CSM)~\cite{Chang:1979nn,Berger:1980ni,Baier:1981uk}, and non-relativistic QCD (NRQCD)~\cite{Bodwin:1994jh} factorisation, which extends the CSM by the introduction of the Colour Octet (CO) mechanism. 
As regards their application to hadroproduction, in particular at the LHC, these are often employed within the framework of collinear factorisation, High-Energy (HE) factorisation\footnote{also referred to as \kT factorisation.} or even Colour Glass Condensates (CGC). Even though most of the discussion in this section focuses on NRQCD used with collinear factorisation, the proposed measurements are also relevant to test most of the theoretical models discussed in the literature. 

No single approach has been able to simultaneously describe all experimental observables collected to date, but the NRQCD approach is the most rigorous and thus has had greatest success. 
However, the predictive power of NRQCD relies heavily on the universality of the non-perturbative long-distance matrix elements (LDMEs) ensured by the factorisation approach. 
These LDMEs characterise the transition rates for various colour-spin states of a produced heavy-quark pair to become a physical quarkonium and should be process-independent. 
Despite the success of NRQCD in many phenomenological applications (see, \eg,~\cite{Brambilla:2010cs}), there are still challenges to understand the single-inclusive quarkonium production mechanism at colliders, notably a unified description of their cross section in different production modes, the polarisation of the vector states, the constraints set by the production of pseudoscalar mesons via NRQCD Heavy-Quark-Spin Symmetry (HQSS), not to mention further puzzles in associated production. 
Differences between various sets of LDMEs extracted from the accumulated data~\cite{Butenschoen:2011yh,Chao:2012iv,Gong:2012ug,Bodwin:2014gia,Shao:2014yta,Bodwin:2015iua,Han:2014kxa,Feng:2015wka} persist, and there is a long way to go to confirm the universality of these LDMEs~\cite{Lansberg:2019adr,Chung:2018lyq}. 
Hence, a coherent physical picture to interpret quarkonium production data is still missing today. 

This impasse in our understanding of quarkonium production, and its subsequent limitations on the use of quarkonium data as a tool for other physics processes, has motivated the critical need to establish novel observables that can serve to advance both goals. The study of various new final states containing quarkonia, and measurements of novel quarkonium observables in hadron-hadron collisions, $e^+e^-$, lepton-hadron, photon-hadron, photon-photon and nuclear collisions, can provide complementary sensitivity to different combinations of LDMEs (see~\cite{Lansberg:2019adr} and references therein) as well as insight into a wide range of phenomena.

The large \pp\ collision data sets expected to be collected at the HL-LHC for inclusive quarkonium production will provide a compelling setting for these investigations. In the following sections, we outline the potential that the HL-LHC experiments have to explore these topics, and highlight priorities for study. In Section~\ref{sec:quarkonium-pt-pp}, we begin by reviewing how measurements of the \jpsi and \ups transverse-momentum ($\pT$) distributions and polarisations is central to our understanding of their production and outline expectations for the HL-LHC period. Section~\ref{sec:oniacharacterisation} explains how studies of the surrounding hadronic activity in collision events containing quarkonia provide new insights to various QCD topics. In Section~\ref{sec:oniaUnconventional}, we address physics opportunities opened up through the study of unconventional quarkonium states, \eg\ the $C$-even $\eta_Q$ and $\chi_Q$ states. Section~\ref{sec:oniumassociate} examines how data on associated production of quarkonium with other heavy states provides a rich opportunity to explore topics as diverse as searches for new phenomena and multi-parton interactions at the HL-LHC. Finally, in Section~\ref{sec:pdfofpp} we outline how quarkonium data in the HL-LHC era can be a compelling tool for precision PDF determinations both at low $x$ and low scale, and at high $x$.
\subsection{\jpsi and \ups conventional measurements: \pT spectra and polarisation}

\subsubsection{\pT spectra: going higher}
\label{sec:quarkonium-pt-pp}
The study of $\pT$ distributions of heavy-quarkonium states produced at hadron colliders has played a critical role in the development of our understanding of the underlying production mechanism. 
The drastically different $\pT$ dependence observed between the Tevatron~\cite{Abe:1997jz,Abe:1997yz} data and the leading order (LO) theoretical predictions for $\jpsi$ and $\psip$ production at mid and high $\pT$ led to tremendous improvements in our understanding on how a produced heavy-quark pair at a large $\pT$ transmutes itself into a physical quarkonium state, and to the development of the NRQCD-factorisation approach for the production of heavy quarkonia~\cite{Bodwin:1994jh,Cho:1995vh}.

For a given heavy-quark mass, $m_Q$, heavy-quarkonium production in hadron-hadron collisions can be divided into three kinematic regimes: $\pT^2 \gg m_Q^2$, $\pT^2 \sim m_Q^2$, and $\pT^2 \ll m_Q^2$, which provide tools sensitive to very different, but often complementary, physics issues.  
For heavy-quarkonium production, QCD factorisation connects the colliding hadron(s) to the underlying quark-gluon scattering that produces a heavy-quark pair while NRQCD factorisation matches the pair, produced with various colour-spin states, to a physical quarkonium through the corresponding LDMEs. 
For both QCD and NRQCD factorisations, calculating quarkonium production in these three kinematic regimes requires different approximations and treatments. 

When $\pT^2 \gg m_Q^2$, quarkonium production is ideal to isolate the non-perturbative hadronisation part. Since the produced heavy-quark pair at high $\pT$ is so separated in phase space from the colliding hadron beams, one can thereby pin down the uncertainty associated to the LDMEs. However, the production involves two very different momentum scales, and resummation of large $\log(\pT^2/m_Q^2)$ terms is necessary~\cite{Nayak:2005rt,Kang:2014tta,Kang:2014pya}. 

Although reliable factorisation formalisms have been derived for, at least, the first two regimes where $\pT^2 \gg m_Q^2$ and $\pT^2 \sim m_Q^2$, a smooth matching between these two regimes is needed to be able to compare theoretical calculations with experimental data.  For example, when $\pT^2 \gg m_Q^2$, theoretical calculations are organised in terms of QCD collinear factorisation of the leading power and next-to-leading power contributions in the $1/\pT^2$ expansion~\cite{Nayak:2005rt,Kang:2014tta,Kang:2014pya}.  Since no QCD collinear factorisation is valid beyond the first subleading power contribution~\cite{Qiu:1990xy}, QCD factorisation formalisms in this regime can only include the leading $1/\pT^4$ and the first subleading $1/\pT^6$ factorised partonic hard parts.  On the other hand, for the regime where $\pT^2 \sim m_Q^2$, the leading term in the power of the strong coupling constant, $\alpha_s$, and heavy-quark relative velocity, $v$, in the NRQCD factorisation approach contains $1/\pT^6$ or $1/\pT^8$ terms depending on the colour-spin states of the pair. Clearly, the two factorisation approaches lead to different $\pT$ dependencies and a consistent matching between these two regimes is clearly needed, especially for fitting the LDMEs~\cite{Kang:2008zzd,LQSW:2020}.

On the experimental side, the Run 1 \& 2 LHC data already allow one to push further the reachable \pT domain for \jpsi and \ups production compared to the Tevatron range. The latter already reached $\pT \simeq 25$ GeV for $\psi$ but this was barely sufficient to assume $\pT^2 \gg m_Q^2$. Moreover, the Tevatron data sample was clearly insufficient for the $\ups$ states. Thanks to the extended LHC \pT reach  and to the advent of NLO NRQCD computations~\cite{Butenschoen:2011yh,Chao:2012iv,Gong:2012ug,Bodwin:2014gia,Shao:2014yta,Bodwin:2015iua,Han:2014kxa,Feng:2015wka}, one could set novel constraints on the NRQCD LDMEs needed to fit the $\psi$ data since high-\pT data prefer a dominance of {$^1S_0^{[8]}$ rather than the  $^3S_1^{[8]}$} state that was compatible with mid-\pT data at the Tevatron using LO estimates. Yet, there is still a debate about whether large $\log(\pT^2/m_Q^2)$ could affect the determination of the LDMEs. In this context, data at even higher \pT  will not be superfluous, especially for \psip for which  constraints from other colliding systems are very limited. Such data will also be useful in confirming the inability of the CEM to account for this high-\pT regime~\cite{Lansberg:2016rcx,Lansberg:2020rft}. The reader is guided to a recent review~\cite{Lansberg:2019adr} where the impact of the current LHC data on the phenomenology is explained in details. For the three \ups states, data at higher \pT, probably above 100~GeV, are certainly also welcome to be sure the fragmentation limit has been reached. These are certainly within the reach of HL-LHC.

\subsubsection{Polarisation: going further}
\label{sec:quarkonium-pol-pp}
The most studied quarkonia are the vector states, $\psi(nS)$ and $\ups(nS)$, because they are easily produced in $e^+e^-$ annihilation but also because they are easily detectable via their di-lepton decay channels. 
This also offers the possibility to directly measure their polarisation, also referred to as spin alignment, via the analysis of the angular distribution of their decay products, which can be parameterised as:
\begin{equation}
\frac{d^{2}N}{d\cos\theta d\phi} \propto 1+\lambda_\theta \cos^2\theta + \\ \lambda_\phi \sin^2\theta \cos2\phi + \lambda_{\theta\phi}\sin2\theta \cos\phi \,,
\label{eq:angularDistribution}
\end{equation}
where $\theta$ is the polar angle between the positively charged lepton momentum in the quarkonium rest frame, $p_{\ell^+}$,  and the spin-quantisation axis ($z$ axis) and $\phi$ is the azimuthal angle between the projection of $p_{\ell^+}$ on the $x-y$ plane (thus orthogonal to the spin-quantisation axis) and the $x$ axis. 
The decay angular coefficients, $\lambda_\theta$ (also known as $\alpha$), $\lambda_\phi$ and $\lambda_{\theta\phi}$ are related to specific elements of the spin density matrix but are frame dependent. They obviously depend on the choice of the  spin-quantisation axis.  The reader is guided to~\cite{Lansberg:2019adr} for an up-to-date discussion of the predicted values of these parameters in different production models and to~\cite{Andronic:2015wma} for an exhaustive
list of the existing measurements up to 2015.

It had been hoped that such polarisation measurements could verify  the smoking-gun prediction of NRQCD according to which vector quarkonia are produced transversely polarised~\cite{Cho:1994ih,Braaten:1994xb}, \ie\ $\lambda_\theta=+1$, in the helicity frame\footnote{In the helicity frame, the quantisation axis is the $\Q$ momentum in the \cm\ frame and the $xz$ plane contains the latter and the momenta of the colliding particles.} at large \pT. However, when Tevatron data (see \eg\ \cite{Aaltonen:2009dm}) became precise enough, 
it became clear that this prediction was wrong.  At the same time,  NRQCD results at NLO were found to deviate from this seemingly fundamental NRQCD result obtained at LO\footnote{For the record, the first NLO polarisation computations were performed in the CSM back in 2008~\cite{Gong:2008sn,Artoisenet:2008fc} showing a longitudinal yield at NLO instead of the transverse LO yield. The NLO NRQCD studies date back to 2012 by the Hamburg~\cite{Butenschoen:2012px}, PKU~\cite{Chao:2012iv,Shao:2014yta} and IHEP~\cite{Gong:2012ug} groups and their interpretations differ much. The Hamburg and IHEP NLO fits show increasingly 
transverse $\psi$ yields with increasing \pT in the helicity frame (at variance with Tevatron and LHC data). The PKU NLO fit --including Tevatron polarisation data at the beginning \cite{Chao:2012iv} but excluding them later \cite{Shao:2014yta}-- shows a quasi unpolarised yield at high \pT.}. From a smoking gun in the 1990's, the polarisation turned, in the 2010's, into a mere constraint via NRQCD fits. Indeed, the complex interplay between the different CO contributions at NLO renders NRQCD predictions for polarisation sensitive to tiny details of the fits. 

In this context, we would like to provide several recommendations:
\begin{itemize}
\item In the currently studied kinematic region, clear-cut theory predictions should not be expected. Leaving aside the feed-down effects, which however constitute a serious source of complications, the polarisation in fact essentially depends on a linear combination of LDMEs. New types of data, such as those discussed in the following sections are needed. In fact, even at extremely high \pT and for the 
feed-down-free $\psip$, it is not clear that a simple picture will emerge. Yet, it remains important to consolidate the current measurements especially for the excited states, which admittedly remain very limited. 
\item Precise polarisation measurements at low \pT in the collider mode certainly remain useful, especially in the central rapidity region, to get a more global view including RHIC and Tevatron measurements.  Along the same lines, measurements in the FT mode in the 100~GeV range, as can be realised at the FT-LHC, will be critical to complete the picture. 
\item It is essential to measure the three angular coefficients in order to avoid relying on theoretical and/or experimental assumptions. This also allows one to compute frame-invariant quantities~\cite{Faccioli:2010ej,Palestini:2010xu,Shao:2012fs,Martens:2017cvj,Peng:2018tty,Gavrilova:2019jea}, which can serve, by using determination of these invariants in multiple frames, as consistency checks of the experimental procedure. For further checks, it would also be interesting to measure the other angular distribution coefficients beyond \ce{eq:angularDistribution} like the $\lambda_{\phi}^\perp\sin^2{\theta}\sin{2\phi}$ and $\lambda^{\perp}_{\theta\phi}\sin{2\theta}\sin{\phi}$ terms, which are predicted to be exactly zero from parity invariance. 
\item Beside the extraction of these invariants from combinations of the angular coefficients, it is possible to extract them directly as functions of the lepton momenta~\cite{Teryaev:2011zza,Martens:2017cvj,Gavrilova:2019jea}. We encourage attempts in this direction.
\item We encourage theorists to compute the 3 angular coefficients and the related invariants.  We note that the first calculation of $\lambda_{\theta \phi}$ at NLO in NRQCD was only computed in 2018~\cite{Feng:2018ukp} for the \jpsi meson.
\item Polarisation measurements also remain very important in quantifying the acceptance corrections to be applied to pass from experimental cross-section measurements performed in a fiducial region to the inclusive ones. Even though the currently available results show no significant polarisation, it should be kept in mind that, in specific kinematic conditions, the polarisation could drastically change. 
Ideally, experiments should publish fiducial cross-section results, which would free their results from this additional source of uncertainty. However, it should be clear that advancing the theoretical predictions to higher precision  (we are not even yet at NNLO in \alphas) and, at the same time, providing predictions for the di-lepton angular distribution in designated fiducial regions, may not be possible due to the computational complexity. This certainly calls for a concerted effort between both experimental and theoretical communities.
\end{itemize}

\subsection{Characterisation of $\Q$ events}
\label{sec:oniacharacterisation}

\subsubsection{$\Q$ in jets}
\label{sec:oniainjets}

In the past few years, quarkonium production within jets has been attracting increasing attention as a probe of heavy-quark hadronisation and quarkonium production mechanisms. At the LHC, such a process has been measured by the LHCb~\cite{Aaij:2017fak} and CMS~\cite{Diab:2019she} collaborations. Both collaborations have observed striking deviations of data from Monte Carlo simulations, which use LO NRQCD complemented with subsequent parton showering. The observable measured in both experiments is the transverse momentum fraction, $z = \pT^{\psi}/\pT^{\text{jet}}$, carried by the quarkonium state $\jpsi$ inside the corresponding jet. This observable is indicative of the fragmenting-parton momentum carried by the quarkonium state. Theoretical predictions for the LHCb data have been provided in~\cite{Bain:2017wvk} by using the fragmentation-jet functions (FJF) and the Gluon Fragmentation Improved \pythia (GFIP), which correspond to a modified parton shower approach, where the quarkonium fragmentation is only implemented at the end of the shower

As shown in \cf{fig:in-jet}, the predictions reproduce many important features of the data. One however notes that the agreement depends both on the values of LDMEs (compare the bands between the 3 plots) and the fragmentation modelling (compare the red and grey bands in each plot). In principle, to improve the discriminating power of the data to the LDMEs, it is important not to integrate over  $\pT^{\text{jet}}$ and to look at how the probability to produce a \jpsi at fixed $z$ varies with $\pT^{\text{jet}}$. 
Even though these exploratory studies have shown that these new observables can provide deeper insights into quarkonium production at large \pT\,, more detailed studies are necessary to obtain a more comprehensive picture. The reader is guided to the review~\cite{Lansberg:2019adr} for a discussion of the theoretical caveats, and in particular that all the current
predictions rely on LO fragmentation functions (FF). As such, NLO corrections may be very large and diminish sensitivity to the LDMEs.

\begin{figure}[h!]
  \centerline{\includegraphics[width = \textwidth]{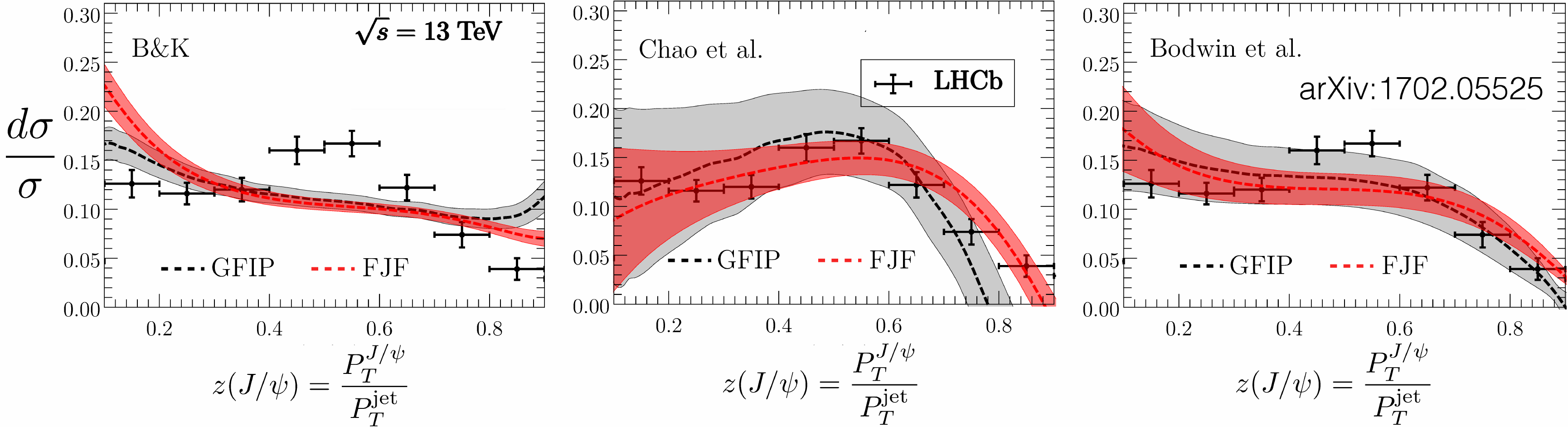}}
  \caption{Comparisons of the $z(\jpsi)$ LHCb measurements to the predicted $z(\jpsi)$ distribution using FJF (red) and GFIP (gray) for three choices of LDMEs (left,  centre, right).[Plots adapted from~\cite{Bain:2017wvk}]}
  \label{fig:in-jet}
\end{figure}

There are significant opportunities to expand these existing studies of \jpsi (and other quarkonia) in jets. Quarkonia in jets can be selected by data collected via leptonic triggers from the decay of the quarkonium or through the use of high-\pT hadronic triggers to select the jet candidate, potentially providing data spanning hundreds of GeV in jet transverse momenta. As events of interest will necessarily be characterised by leptons surrounded by significant hadronic activity, care will be needed to ensure such events are not vetoed by standard online or offline lepton- or jet-reconstruction algorithms optimised for HL-LHC conditions.
At the HL-LHC, we identify the following major subsequent studies with significant phenomenological impact:
\begin{itemize}
\item The transverse momentum fraction, $z$, should be measured with finer binning and for a wider range of jet transverse momentum. The transverse momentum of the jet, that sets the hard scale of the process, controls the size of evolution. This will allow the test of various aspects of the quarkonium fragmentation mechanisms, and their relative contributions. In addition, a detailed study of the regime $1-z\ll 1$ will give insights into the non-perturbative aspects of quarkonium hadronisation where the process is particularly sensitive to soft radiation.
\item To decouple modelling of the jet transverse momentum dependence from modelling of the fractional momentum associated with quarkonium states, we encourage experiments to perform measurements such as those illustrated in~\cf{fig:in-jet} in narrow intervals of jet transverse momentum, or ultimately as multi-differential measurements in jet \pT and $z$. As well as providing additional information to explore the observed discrepancies, this will allow for more detailed comparisons and combinations of data between experiments accessing complementary \pT ranges at HL-LHC.
\item  Jet substructure observables, such as thrust and other angularities, will give access to details of the radiation surrounding the quarkonium within the jet. The distribution of radiation in the jet is sensitive to the production mechanisms of quarkonium and could offer an additional handle on the numerical values of the LDMEs. A theoretical investigation of observables of this type has been performed in~\cite{Bain:2016clc}.
\item Multi-differential measurements of quarkonium energy fractions and transverse momentum with respect to the jet axis~\cite{Bain:2016rrv, Makris:2017hjk} can provide a three-dimensional picture of quarkonium fragmentation. 
\item Studies should be expanded to include measurement of other quarkonia, such as $\psip$ in jets or $\ups(nS)$ in jets for the first time. Such measurements are critical to provide a complementary way to constrain all LDMEs used in production modelling.
\end{itemize}

Compared to light hadrons, quarkonium production is anticipated to be less affected by background contributions from multi-parton interactions and underlying event activity. However, at the HL-LHC, the use of jet-grooming techniques might also be helpful in further removing contributions of soft radiation, and allow for an improved convergence of experimental studies and theoretical calculations. In particular, it is important to keep under control the impact of DPS, whereby a quarkonium is produced in one scattering simultaneously with a dijet from another, with one of these jets being so close to the quarkonium that the quarkonium is considered to belong to this jet.

\subsubsection{$\Q$ as a function of the particle multiplicity}
\label{sec:oniaparticlemultiplicity}

\begin{figure}[h!]
\centering
\includegraphics[width =  0.4\textwidth]{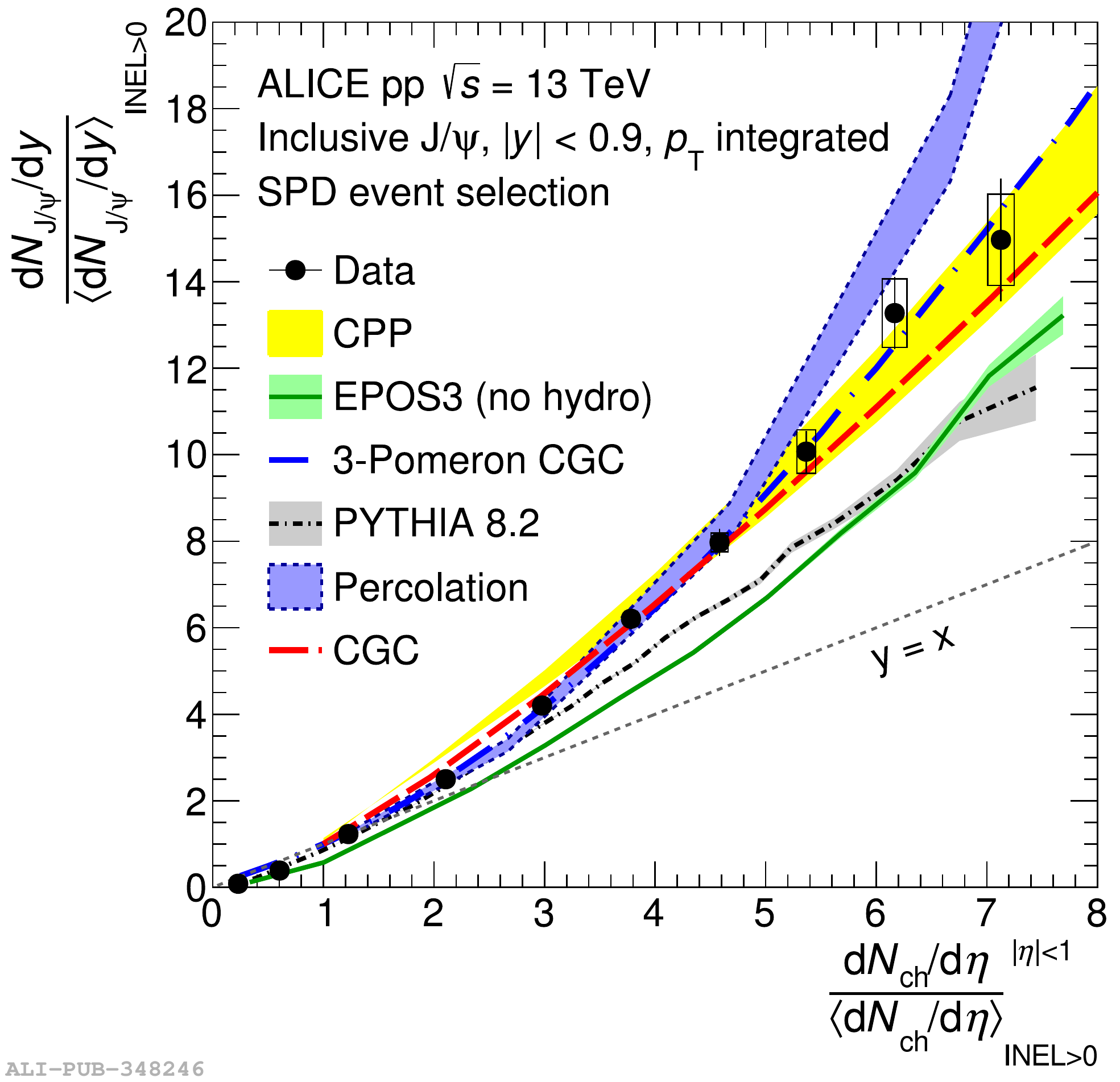}
\hspace{0.5cm}
\includegraphics[width =  0.4\textwidth]{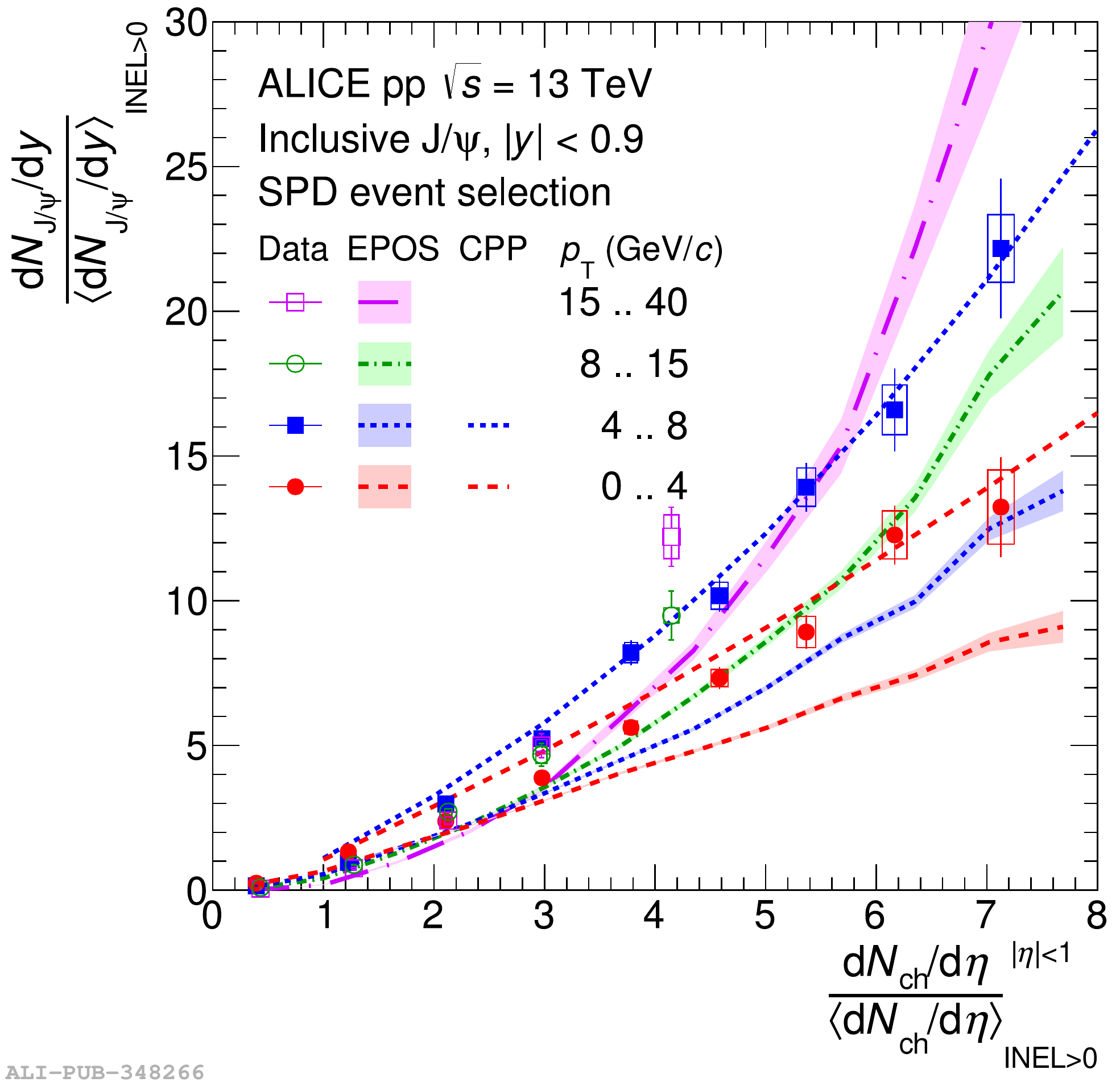}
  \caption{Normalised inclusive $\jpsi$ yield at mid-rapidity as a function of the normalised charged-particle pseudorapidity density at mid-rapidity for the $\pT$-integrated (left) and $\pT$ differential (right) cases. The data~\cite{Acharya:2020pit} are compared to theoretical predictions from EPOS-3~\cite{Werner:2013tya}, the coherent particle production model (CPP)~\cite{Kopeliovich:2013yfa}, the percolation model~\cite{Ferreiro:2012fb}, the CGC model~\cite{Ma:2018bax}, the 3-Pomeron CGC model~\cite{Siddikov:2019xvf}, and \pythia~8.2~\cite{Sjostrand:2014zea}. [Plots taken from~\cite{Acharya:2020pit}]
  } 
  \label{yields_vs_quarkonia_mid} 
\end{figure}

\begin{figure}[h!]
\centering
\includegraphics[width=0.45\textwidth]{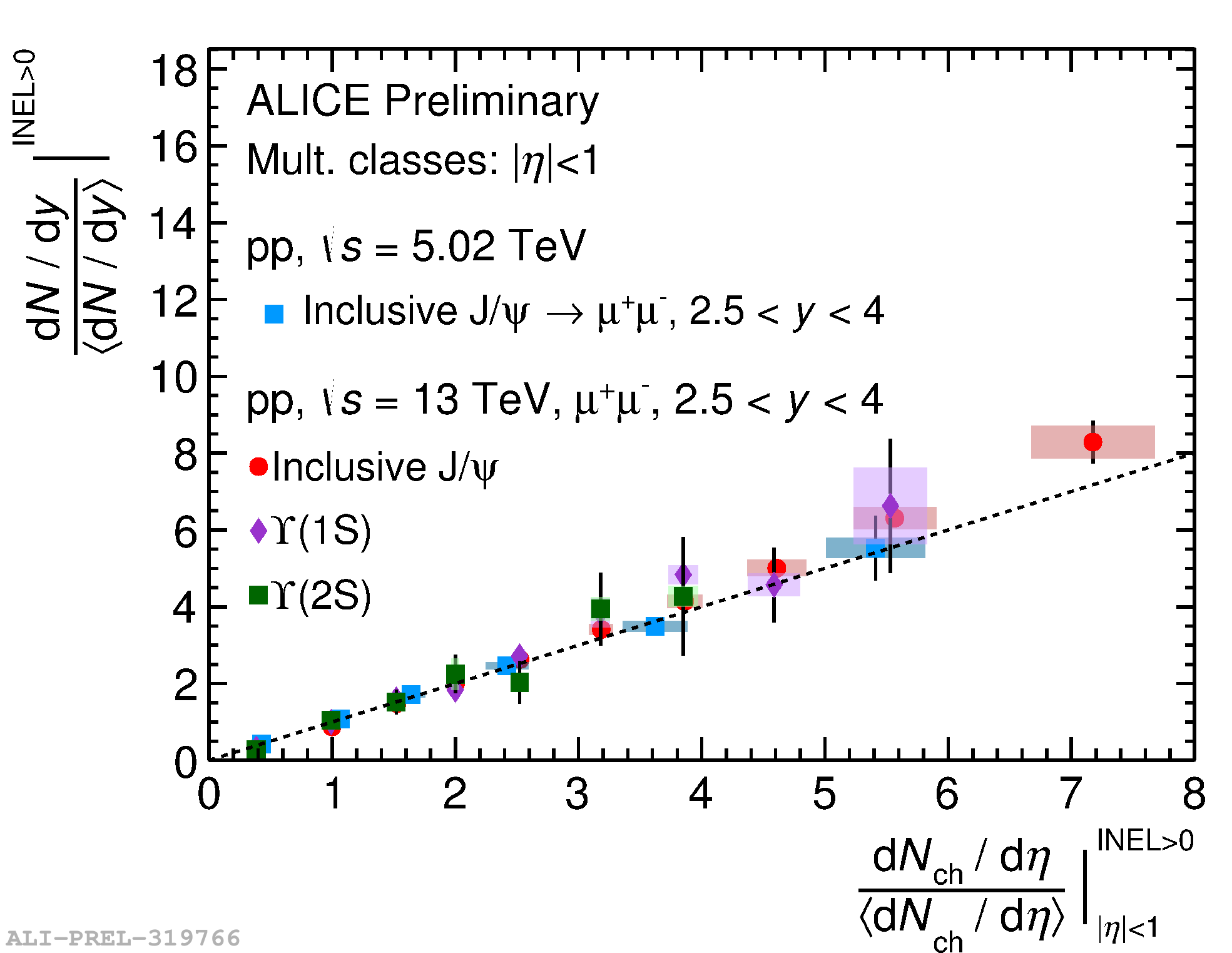}
\hspace{0.1cm}
\includegraphics[width=0.4\textwidth]{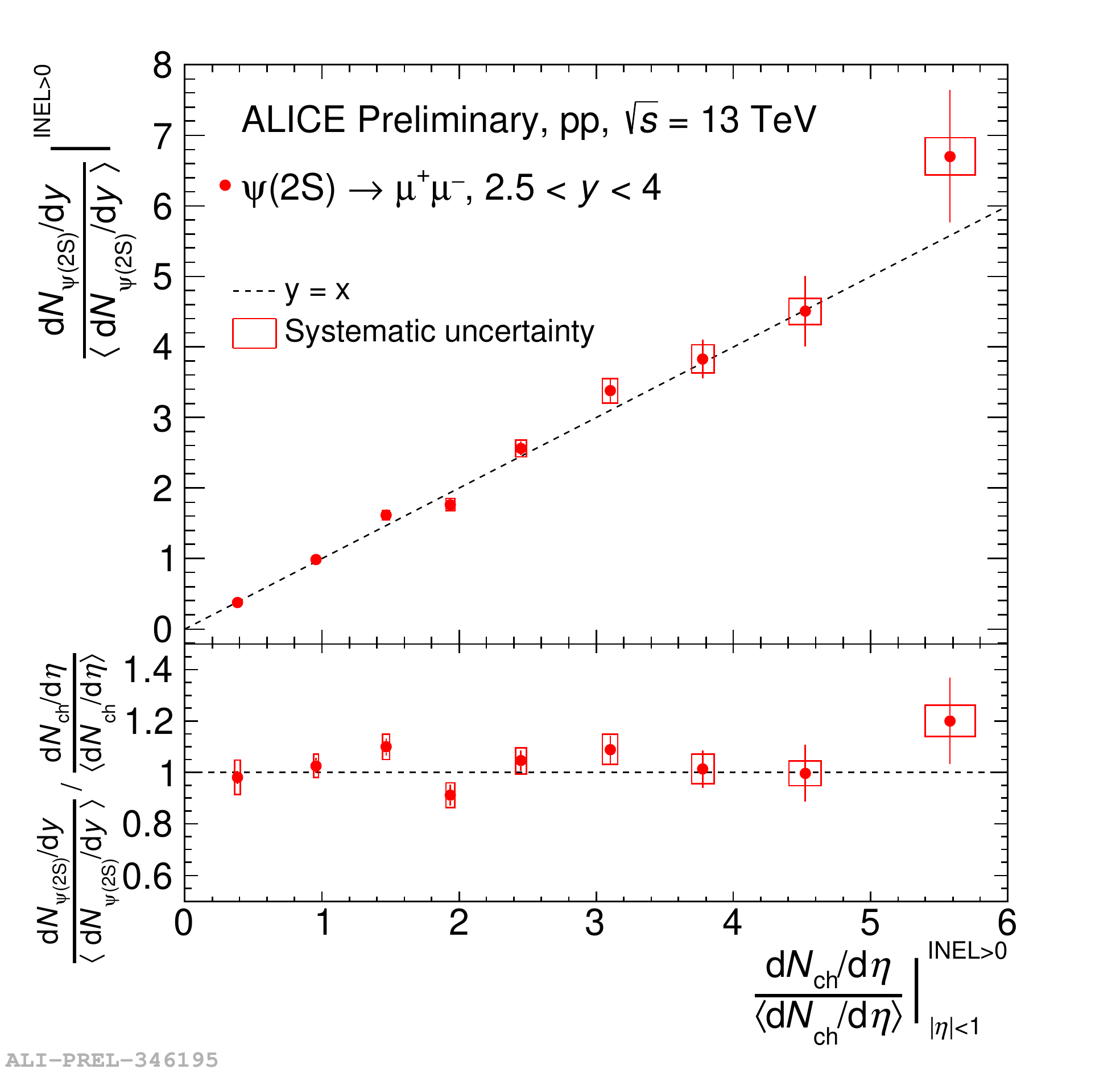}
\caption{Normalised yields of inclusive quarkonia at forward rapidity as a function of the charged-particle multiplicity at midrapidity d$N_{\rm ch}$/d$\eta$ in \pp\ collisions at $\sqrts$ = 5.02 (left) and 13 TeV (right). Both quantities are normalised by the corresponding value for minimum bias \pp\ collisions (d$N_{Q\bar{Q}}$/d$y$, d$N_{\rm ch}$/d$\eta$). The error bars represent the statistical uncertainty on the relative quarkonium yields, while the point-to-point systematic uncertainties on the relative quarkonium yield are depicted as boxes. The dotted linear line ($y=x$) is drawn to visualise the deviation from linearity of the data points. [Plots taken from~\cite{Acharya:2020pit}]
\label{yields_vs_quarkonia}
}
\end{figure}

Multiplicity-dependent studies of quarkonium production give insights into the correlations between hard (heavy-quark production) and soft (charged-particle multiplicity) processes, and improve our understanding of multi-parton interactions (MPI) and initial-state effects. Figures~\ref{yields_vs_quarkonia_mid} and~\ref{yields_vs_quarkonia} show the normalised yields of quarkonium production at mid- and forward-rapidity as a function of the charged-particle multiplicity, measured at mid-rapidity d$N_{\rm ch}$/d$\eta$ for \pp\ collisions at $\sqrts$ = 5.02 and 13 TeV, respectively~\cite{Acharya:2020pit}. The results at forward rapidity (Fig.~\ref{yields_vs_quarkonia}) include $\ups$ states as well as $\psip$ and $\jpsi$, whereas the central rapidity ones (Fig.~\ref{yields_vs_quarkonia_mid}) show only inclusive $\jpsi$.

The quarkonium normalised yield increases linearly as a function of the multiplicity at forward rapidity, while at mid-rapidity a faster-than-linear increase is observed both for the $\pT$-integrated and $\pT$-differential $\jpsi$ cases. Several theoretical models predict a correlation of the $\jpsi$ normalised yield with the normalised event multiplicity that is stronger than linear. These include a coherent particle production model~\cite{Kopeliovich:2013yfa}, the percolation model~\cite{Ferreiro:2012fb}, the EPOS3 event generator, a CGC+NRQCD model~\cite{Ma:2018bax}, the \pythia\,8.2 event generator~\cite{Sjostrand:2014zea}, and the 3-Pomeron CGC model~\cite{Siddikov:2019xvf}. In all these models, the predicted correlation is the result of a ($N_{ch}$-dependent) reduction of the charged-particle multiplicity. This can be an effect due to the colour string reconnection mechanism as implemented in \pythia, but the initial-state effects as implemented in the percolation or CGC models lead to a similar reduction in the particle multiplicity. Concerning the excited-over-ground yield ratios, recent results by the ALICE collaboration on charmonium production at forward rapidity are consistent with unity within the systematic uncertainties, while pointing at a possible reduction for an increasing normalised multiplicity~\cite{Gromada:2020knj}. Moreover, the preliminary results on the $\ups$ and $\upsp$ normalised yields versus the normalised multiplicity~\cite{Ding:2020stx} point at a stronger departure from linearity for the excited state when compared to the ground state, which would lead to a reduction of the ratio $\ups/\upsp$ at high multiplicities.

The measurements outlined here are currently limited by systematic uncertainties that will be reduced with the upcoming HL-LHC quarkonium data. With the expected much larger data samples and enlarged rapidity coverage, LHC experiments have good prospects to perform multi-differential studies among different kinematic variables, also at forward rapidity. This will enable focused studies of the dependence in specific regions of the phase space and precision analyses of the higher-excited quarkonium states, such as the $\upspp$.

\subsection{Production of unconventional states}
\label{sec:oniaUnconventional}

\subsubsection{Production of \etac and \etacp states\label{sec:etac}}

Studies of the \etac and \etacp states, as spin partners of the \jpsi and \psip, provide independent constraints on the LDMEs of the spin-triplet family based on HQSS, following the velocity-scaling rule of NRQCD. Next-to-leading order (NLO) NRQCD calculations~\cite{Butenschoen:2014dra,Han:2014jya,Zhang:2014ybe} show that there are only two relevant channels, a CO channel and the leading-$v^2$ CS channel, while the feed-down contribution is negligible. This greatly simplifies the corresponding theoretical analysis. On the other hand, all $S$- and $P$-wave CO channels, as well as significant feed-down contributions in the $\jpsi$ case, compete with each other in the $\psi$ hadroproduction processes. 

At the LHC, two LHCb measurements of  \etac exist~\cite{Aaij:2014bga,Aaij:2019gsn}, via the hadronic decay channel $\etac\to p\bar{p}$~\cite{Barsuk:2012ic} with a branching fraction of about $1.45\cdot 10^{-3}$~\cite{Tanabashi:2018oca}. The current trigger at the LHCb detector allows only the measurement of the $\pT$ spectrum of \etac with $\pT>6.5$ GeV. Nevertheless, such measurements have already presented surprises that indicate that the CS channel alone is already sufficient to account for the experimental data within the range $6.5<\pT<14$~GeV. Therefore, the data substantially constrain one CO LDME of the $\jpsi$ by using HQSS, essentially ruling out most of the LDME sets from the world data fits. With the much larger data samples anticipated at the HL-LHC, the extensions of the $\pT$ range in the measurements will be beneficial at, at least, two levels. On the one hand, the low-$\pT$ data will be very useful to extract the low-$x$ gluon PDF in the proton as discussed in Sec.~\ref{sec:pdfofpp}, given the dominance of the CS channel in this regime. The same measurement in the fixed-target mode~\cite{Hadjidakis:2018ifr} at the HL-LHC would allow a probe of the high-$x$ regime of the gluon density~\cite{Feng:2019zmn}. On the other hand, the higher-$\pT$ data will improve the sensitivity to the CO LDME, thanks to the harder $\pT$ spectrum of the CO compared to the CS channel.

\begin{figure}[h!]
\centering
\includegraphics[width=0.32\textwidth]{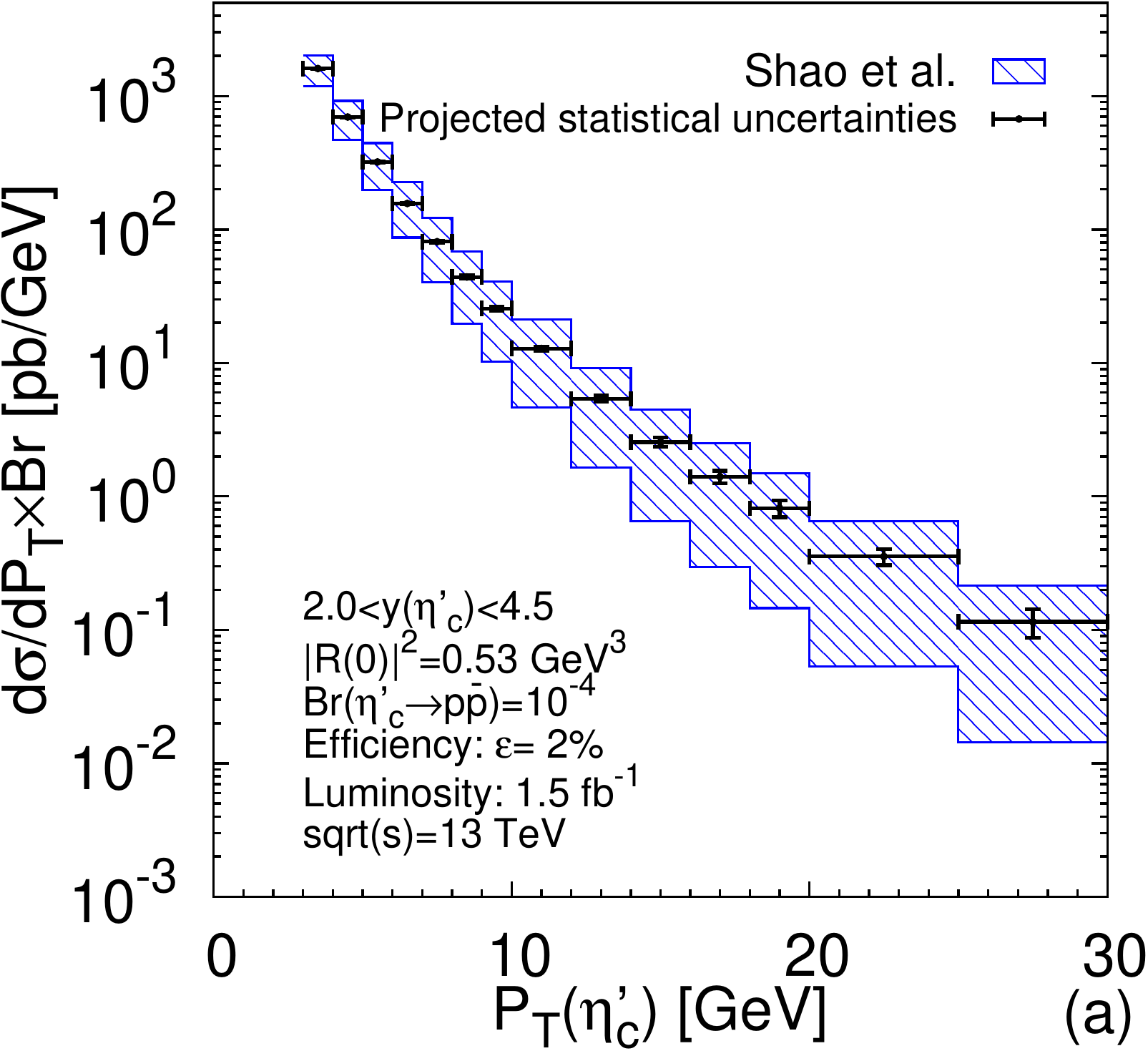}
\includegraphics[width=0.32\textwidth]{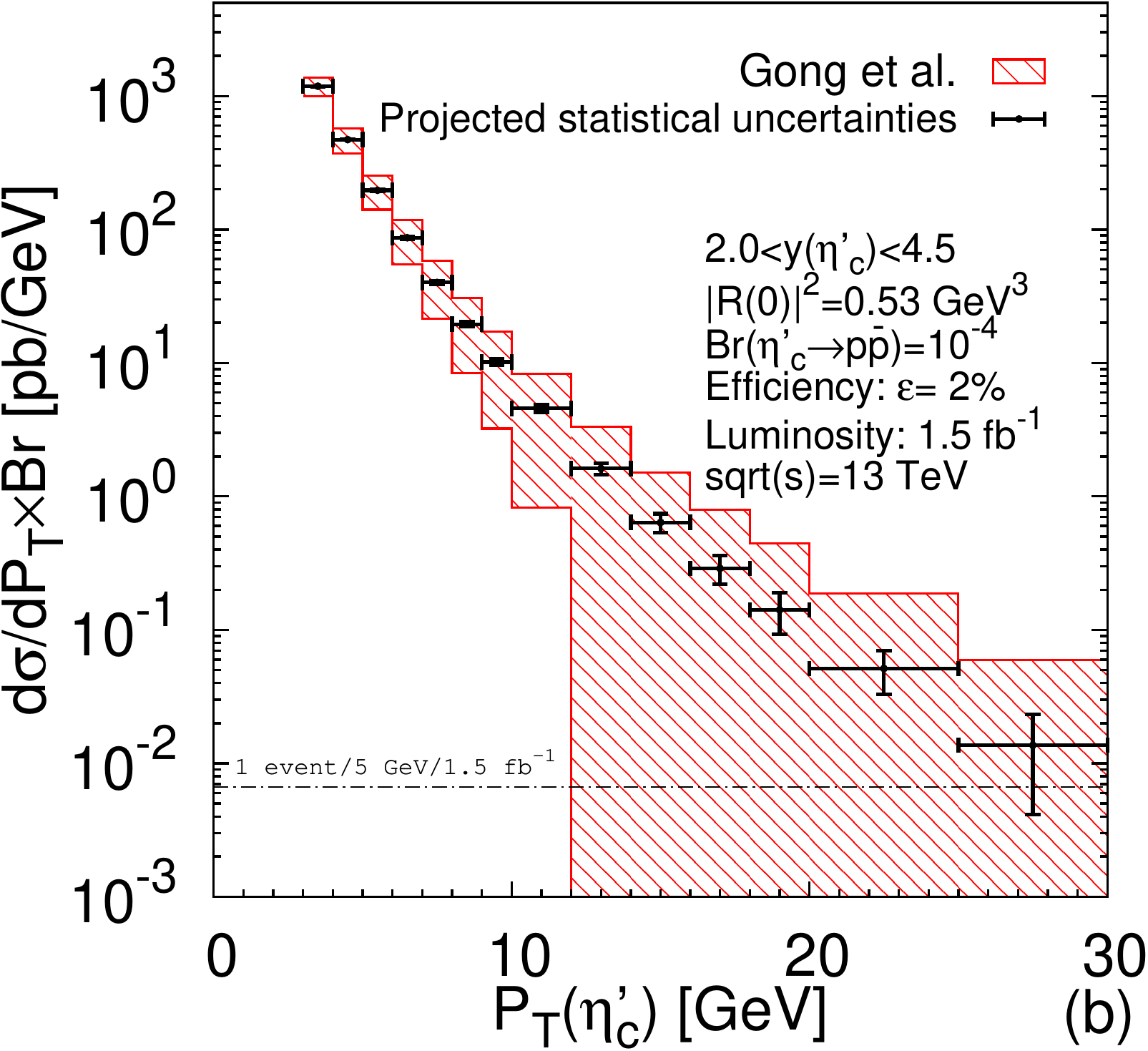}
\includegraphics[width=0.32\textwidth]{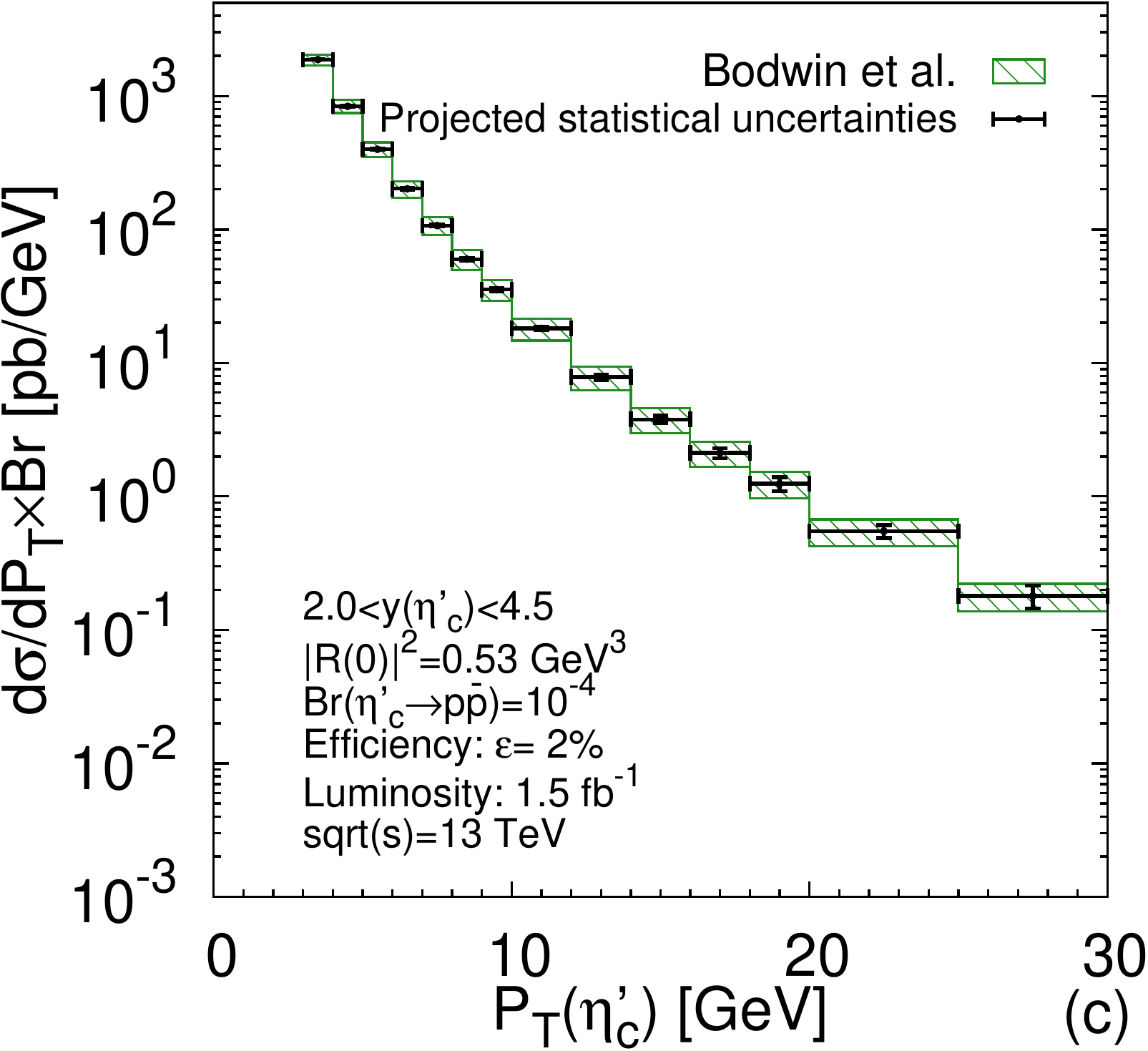}
  \caption{Differential-$\pT$ cross section for \etacp production times ${\cal B}(\etacp \to p \bar p)$ for the three CO LDME sets~\cite{Shao:2014yta,Gong:2012ug,Bodwin:2015iua} along the projected statistical uncertainties using the central theoretical values in each case, with an assumed efficiency of 2\% and a luminosity of 1.5~fb$^{-1}$ by the LHCb detector at $\sqrts=13$~TeV. [Plots taken from~\cite{Lansberg:2017ozx}]} 
  \label{fig:etac2S} 
\end{figure}

For the same reasons, the study of the \etacp state is interesting to understand the $\psi(2S)$ production mechanism based on HQSS. Its feasibility at the LHC has been explored in~\cite{Lansberg:2017ozx} via several decay channels. Figure~\ref{fig:etac2S} shows that the $\pT$-differential cross section strongly depends on the choice of the CO LDME set. A measurement, with a dedicated trigger, of \etacp at the HL-LHC will impact the final theoretical interpretation of the charmonium production data. 
Equivalent measurements with bottomonium are also feasible:
prospects for $\eta_b$ studies at the LHC have recently been discussed in~\cite{Lansberg:2020ejc}.

\subsubsection{Polarisations of $\chico$ and $\chict$ states\label{sec:chicpol}}

It has been advocated in~~\cite{Shao:2014fca,Faccioli:2018uik} that the measurement of the polarisations of the $\chico$ and $\chict$ mesons at the LHC would uncover how the $P$-wave states are produced at hadron colliders. These states contribute to around $30\%$ of the prompt $\jpsi$ yields via the radiative decays $\chi_{c}\to \jpsi+\gamma$ (see~\cite{Lansberg:2019adr} for an up-to-date discussion on the feed-down component). Such studies are also motivated by the simplicity of these states from the NRQCD perspective with the HQSS relation. Indeed, only a single CO LDME needs to be determined from the experimental data compared to three for the $S$-wave states to reach a comparable precision, and thereby NRQCD has a stronger predictive power for the $P$-wave states.

A first measurement by the CMS collaboration with 19.1~fb$^{-1}$ of data in \pp\ collisions at $\sqrts=8$ TeV appeared recently~\cite{Sirunyan:2019apc}. Such a measurement is very challenging at both ATLAS and CMS because the photons from these radiative transitions have to be measured through their conversions to $e^+e^-$~\cite{ATLAS:2014ala,Khachatryan:2014ofa} to achieve a high precision measurement. The analysis is thus limited by the large systematic uncertainty associated to muon and photon detection efficiencies. Consequently, only the difference between the $\chico$ and $\chict$ polarisations, from the angular dependence of the $\chict/\chico$ yield ratio, is available. In other words, the $\chico$ and $\chict$ polarisations are not yet known separately. 

\begin{figure}[h!]
\centering
\includegraphics[width=0.8\textwidth]{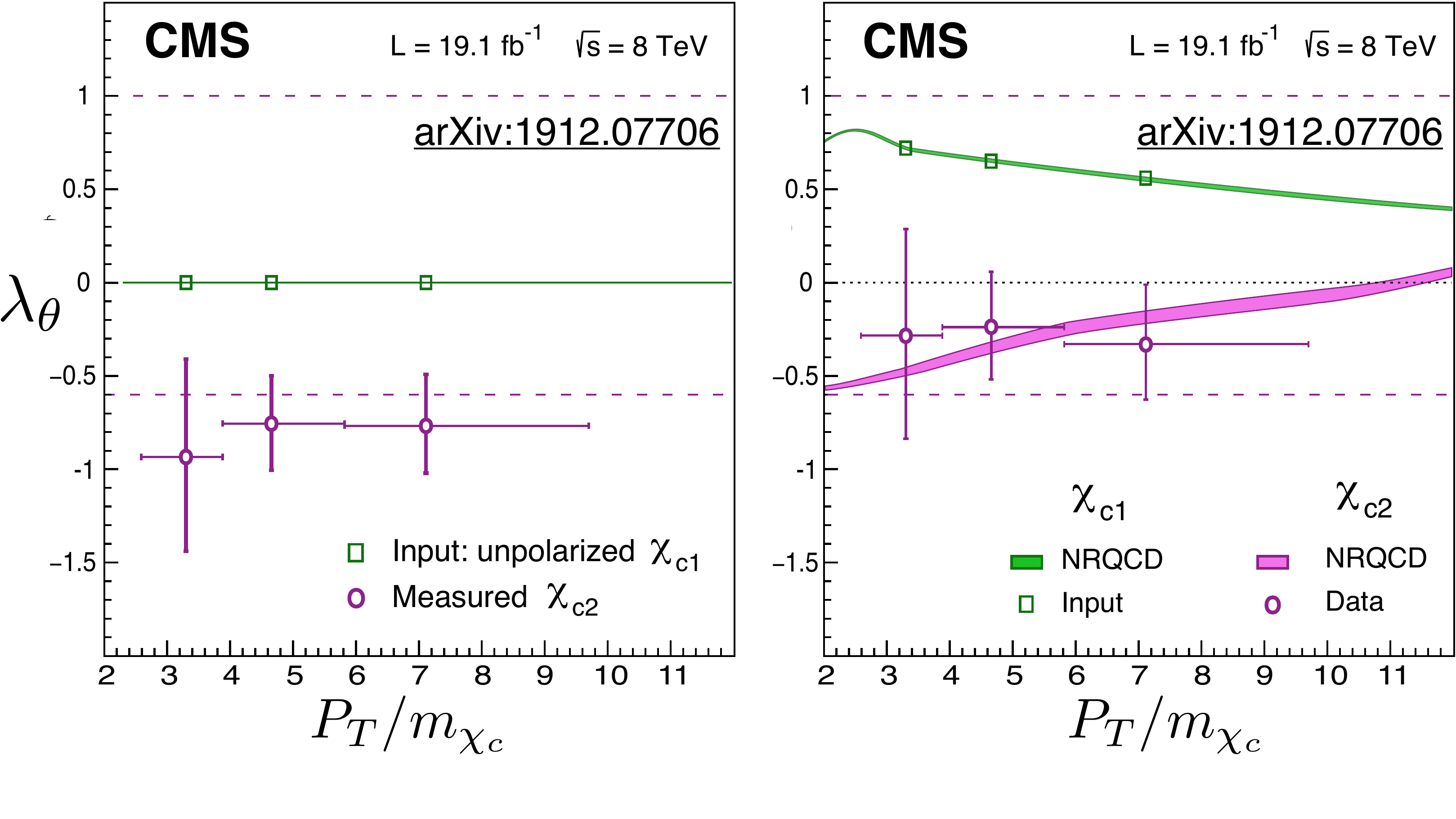}
  \caption{Measurements of the polarisation parameter $\lambda_{\theta}$ versus $\pT/m_{\chic}$ for $\chico$ and $\chict$ production in \pp\ collisions at $\sqrts=8$ TeV measured by the CMS collaboration compared to NRQCD predictions. [Plots adapted from~\cite{Sirunyan:2019apc}]} 
  \label{fig:chicpol} 
\end{figure}

Figure~\ref{fig:chicpol} displays the coefficient $\lambda_{\theta}$ %
in the decay chain $\chi_{c}\to \jpsi\gamma\to \mu^+\mu^-\gamma$, which follows the form $1+\lambda_{\theta}\cos^2{\theta}$~\cite{Faccioli:2010ji,Faccioli:2011be,Shao:2012fs}, where $\theta$ is the polar angle of $\mu^+$ in the rest frame of the $\jpsi$ meson. The left panel shows the polarisation pattern of $\chict$ when assuming unpolarised $\chico$.
The right panel compares data with a NLO NRQCD prediction~\cite{Shao:2014fca,Faccioli:2018uik} for $\chict$ polarisation after fixing the $\chico$ polarisation to the correponding NRQCD values. 

At the HL-LHC, the limitation of measuring only the $\chico$ and $\chict$ polarisation difference can be lifted, with the much larger data samples and a better control of systematic uncertainties. This will require a commitment to collect high-statistics calibration data at low $\pT$ for the determination of muon and photon conversion efficiencies. It is thus very desirable to measure the independent polarisation for each $P$-wave state in the future. The equivalent measurements for the $\chib$ states will also be useful for understanding the corresponding bottomonium sector. In fact, these measurements are also necessary for a proper interpretation of the \jpsi and \ups polarisation measurements to properly account for the effect of the $\chi_Q$ feed-down component. 

\subsubsection{Production of the exotic $X$, $Y$ and $Z$ states}
\label{sec:pp-xyz}

Many $X$, $Y$ and $Z$ states have been found following the $X(3872)$ observation~\cite{Choi:2003ue}, but their underlying (multi-quark or hadron-molecular) nature is still unclear. Studying the production of $X$, $Y$ and $Z$ states, both theoretically and experimentally, can provide important information to understand the formation and properties of multi-quark states.

Let us take as an example the $X(3872)$, now also referred to as $\chi_{c1}(3872)$. Its prompt production rate has been estimated within the NRQCD factorisation approach~\cite{Artoisenet:2009wk} assuming that it is a pure charm-meson molecule. In order to be able to adequately describe measurements of its production rate from the CDF collaboration, the charm-meson rescattering mechanism was introduced in~\cite{Artoisenet:2009wk}. However, the LO calculation with the non-perturbative matrix element determined from the CDF data leads to much bigger yields than experimentally measured by the CMS collaboration~\cite{Chatrchyan:2013cld}, as shown in~\cf{fig:X3872HadroProd} (left). On the contrary, the authors of~\cite{Meng:2013gga} suggested that the $X(3872)$ is a mixture of $\chico(2P)$ and $D^0\bar{D}^{0*}$ states, and the hadroproduction proceeds dominantly through its $\chico(2P)$ component. The cross section through the charm-meson molecular component has been assumed to be negligible in~\cite{Meng:2013gga}, and the fraction of the charmonium component in $X(3872)$ has been tuned to the CMS data.

\begin{figure}[h!]
\centering
\raisebox{15pt}{\includegraphics[width=0.49\textwidth]{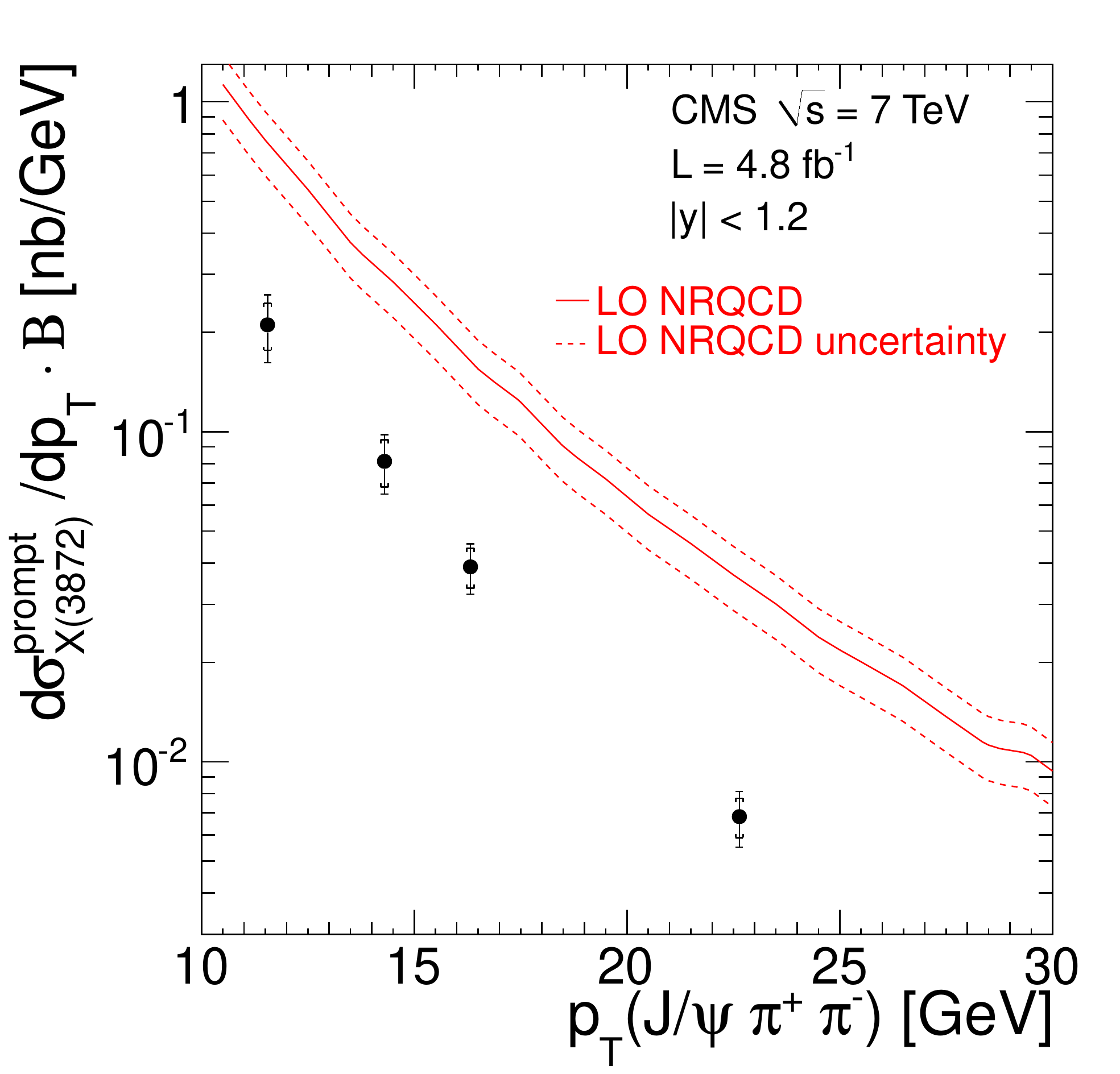}}
\includegraphics[width=0.45\textwidth]{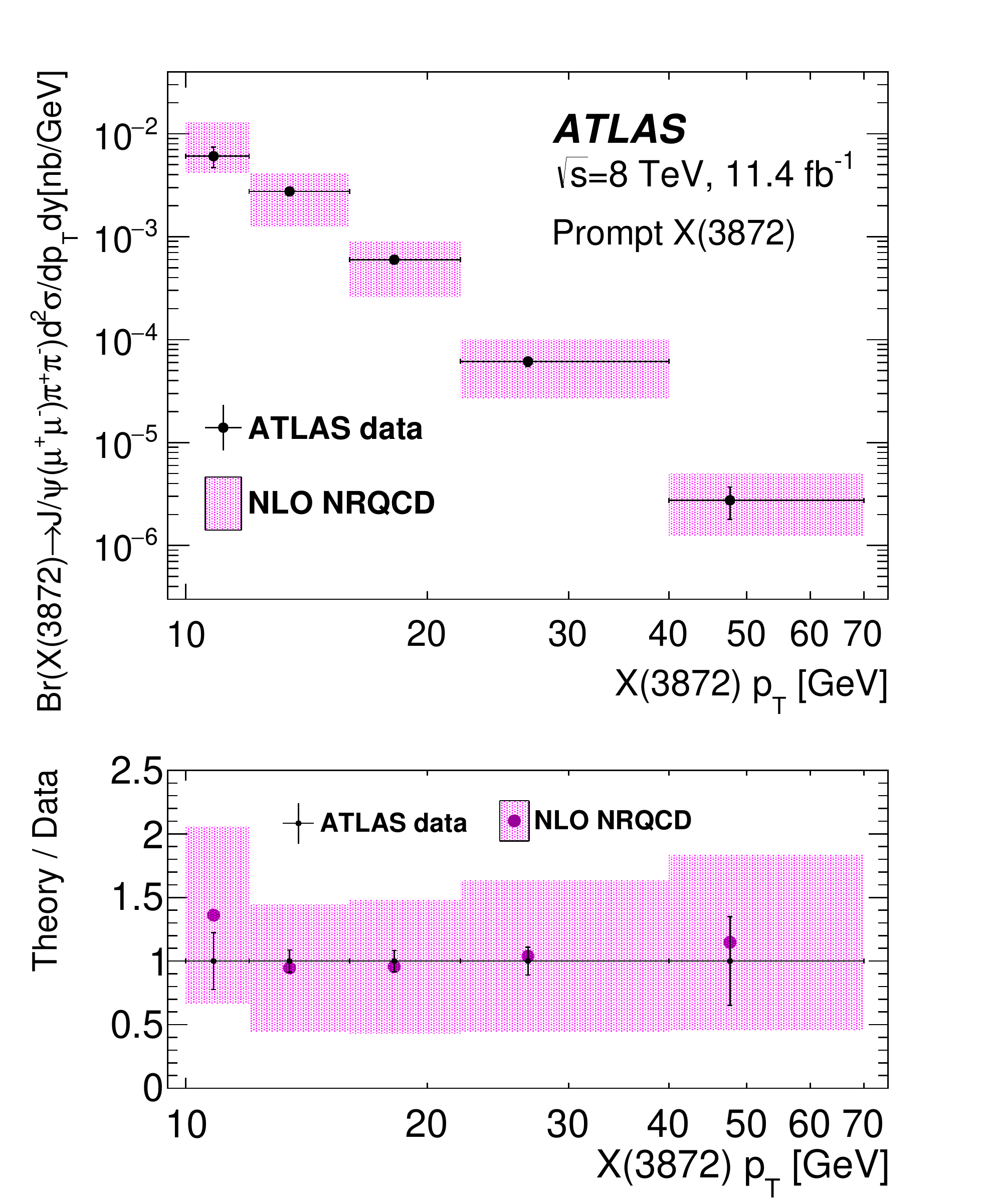}
  \caption{\pT\ differential cross section times branching fraction as a function of $\pT$ for the prompt production of the $X(3872)$ state, measured by the CMS collaboration~\cite{Chatrchyan:2013cld} (left), and by the ATLAS collaboration~\cite{Aaboud:2016vzw} (right), compared to the theoretical predictions from~\cite{Artoisenet:2009wk} and~\cite{Meng:2013gga}, respectively. [Plots taken from~\cite{Chatrchyan:2013cld,Aaboud:2016vzw}]}
  \label{fig:X3872HadroProd} 
\end{figure}

Once the overall normalisation of the cross section from CMS observations was fit by fixing this charmonium fraction, the NRQCD approach was found to describe the data across a much larger range of \pT as observed by ATLAS~\cite{Aaboud:2016vzw} (Fig.~\ref{fig:X3872HadroProd}, right).  It is worth noting that the NRQCD calculations plotted in both panels of~\cf{fig:X3872HadroProd} have very different underlying assumptions. The LO NRQCD curve~\cite{Artoisenet:2009wk} in the left (CMS) plot assumes $X(3872)$ to be a pure loosely bound charm-meson molecular state and the model includes the charm-meson rescattering mechanism, whereas the NLO NRQCD curve in the right (ATLAS) panel, takes the $X(3872)$ as a mixed state of charmonium and charm-meson molecule and assumes its production via the charmonium component is dominant. On the other hand, the non-prompt $X(3872)$ production rate measured by ATLAS~\cite{Aaboud:2016vzw} was found to be poorly described by fixed-order next-to-leading-log (FONLL) predictions~\cite{Cacciari:2012ny} in contrast to the good agreement observed for the charmonium $\psip$ case. Analysis of the non-prompt $X(3872)$ lifetime distribution indicated the presence of an anomalously large short-lifetime component consistent with decays via the $B_c$, which has yet to be fully understood.  Given the unclear nature of the $X(3872)$, and different model/theoretical assumptions in various existing cross-section calculations, it is not yet conclusive which picture can successfully reproduce the world data on the $X(3872)$.

\begin{figure}[h!]
\centerline{\includegraphics[width=0.49\linewidth]{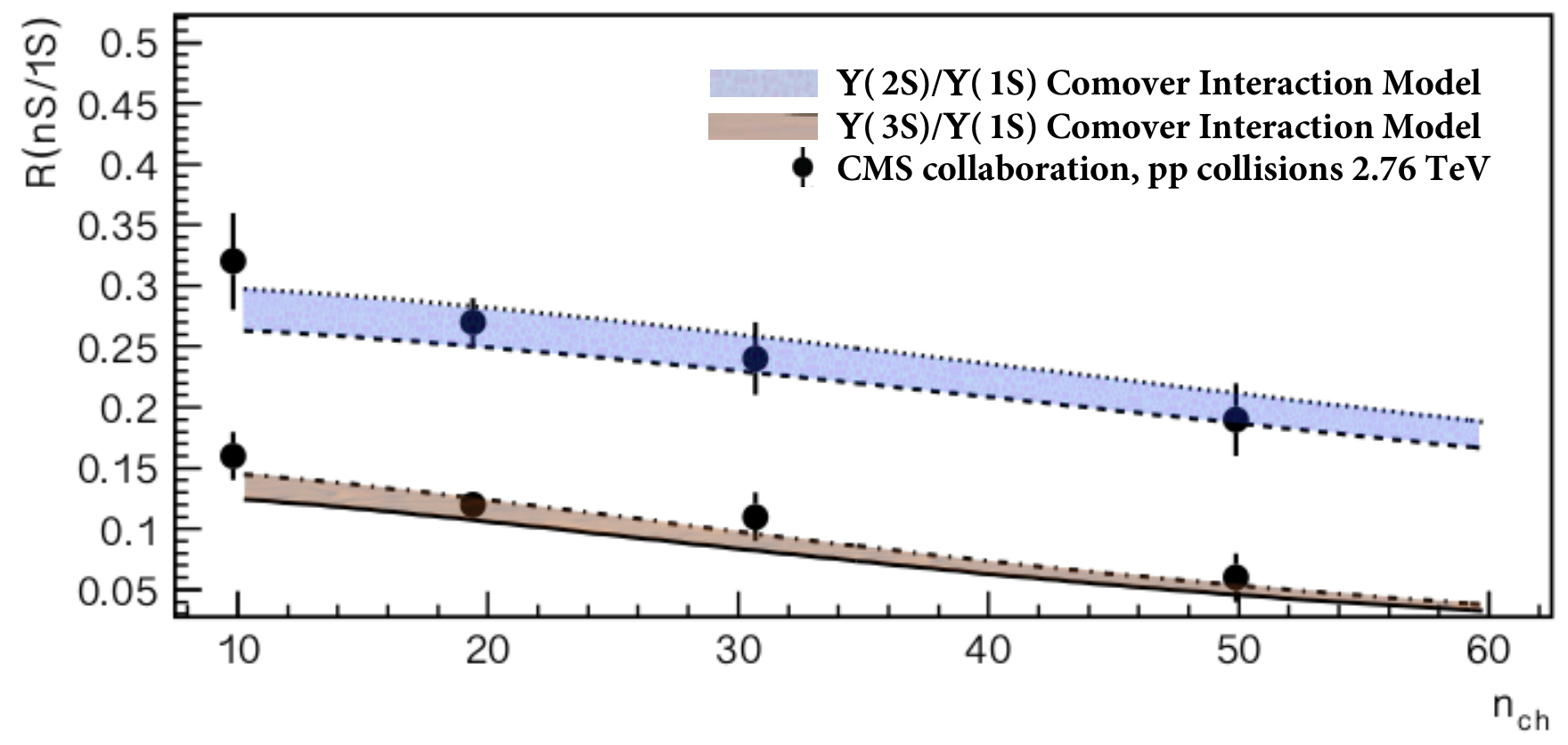}
\includegraphics[width=0.49\linewidth]{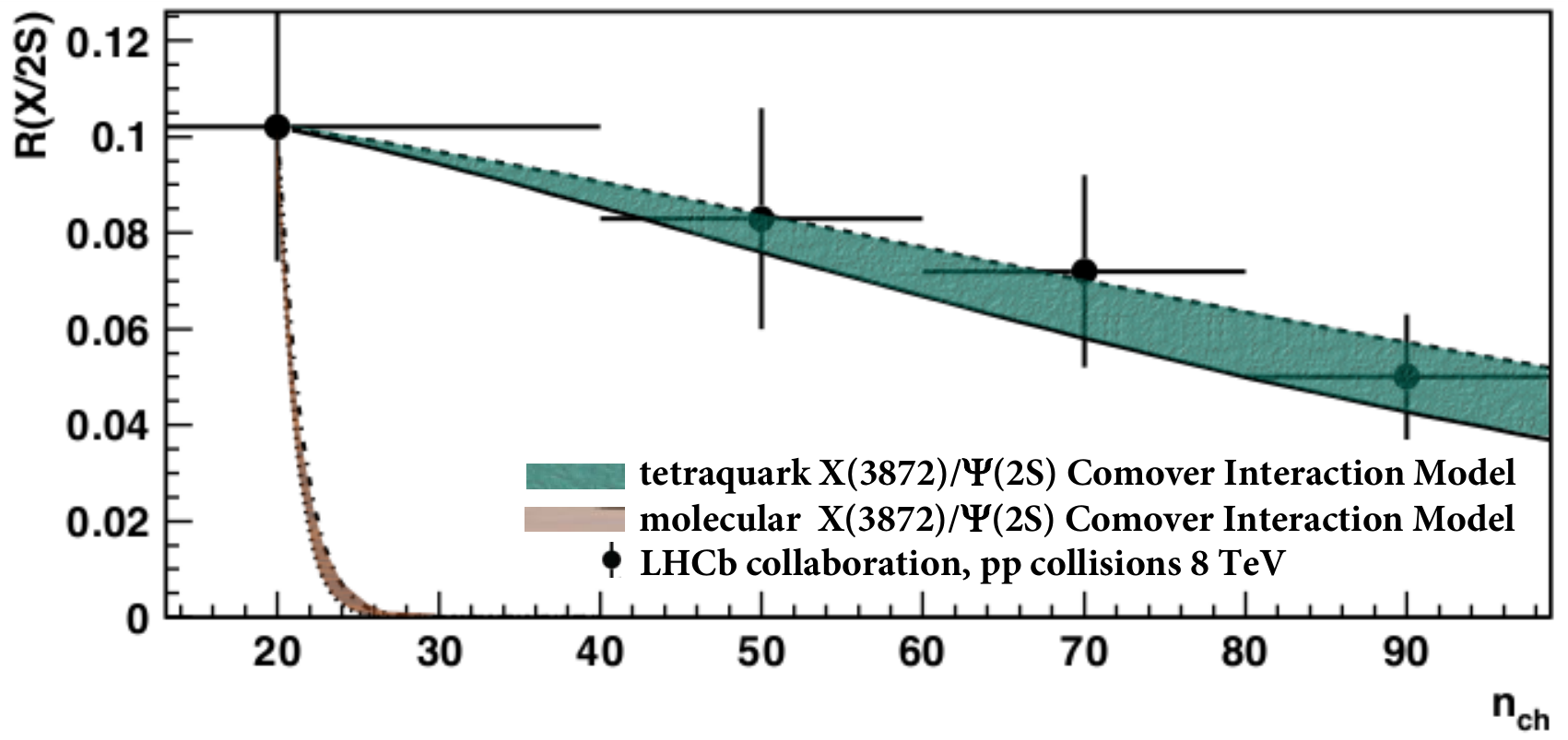}}
\caption{\label{fig:figppUpsilonX}
(Left) Relative yields of excited-to-ground-state $\ups$ mesons as a function of event multiplicity in \pp\ collisions at $\sqrts = 2.76$~TeV at central rapidities as measured by the CMS collaboration~\cite{Chatrchyan:2013nza}. (Right) Relative yields of $X(3872)$ over $\psi(2S)$ as a function of charged-particle multiplicity at $\sqrts = 8$~TeV in the forward region as measured by the LHCb collaboration~\cite{Aaij:2020hpf}. The bands represent the CIM results~\cite{Ferreiro:2018wbd,Esposito:2020ywk}, while the points refer to the experimental data. In the right panel, the brown band assumes $X(3872)$ is a molecular state with a radius of 5~fm, while the green band assumes it is a compact tetraquark state with a radius of 0.65~fm.%
}
\end{figure}

The study of the relative production of $X(3872)$ over conventional quarkonium states as a function of particle multiplicity, as performed by the LHCb collaboration~\cite{Aaij:2020hpf} in \pp\ collisions at 8 TeV, can help to discriminate the nature of this exotic state. The preliminary LHCb data on the relative yields of the exotic $X(3872)$ over $\psi(2S)$ meson show a similar behaviour to the relative yields of excited-over-ground-state $\ups$ mesons reported by the CMS collaboration~\cite{Chatrchyan:2013nza} in \pp\ collisions at 2.76 TeV, pointing to a possible common origin.
Both data sets have been studied in the framework of the comover interaction model (CIM), \ie\ including final-state interactions with the comoving medium~\cite{Ferreiro:2018wbd,Esposito:2020ywk}. The results are shown in~\cf{fig:figppUpsilonX}. Within this approach, the radius of the $X(3872)$ is about twice that of the $\psip$. 
This finding supports the $X(3872)$ being a tetraquark state and disfavours the molecular interpretation that would need a much larger radius, close to 5 fm.

With the data delivered by the HL-LHC, many more rare $X$, $Y$ and $Z$ states are likely to be collected. These future datasets offer the opportunity to measure their double differential production cross section as functions of transverse momentum and rapidity, or set much more stringent upper limits on their hadroproduction cross sections. In addition, the polarisations of some $X$, $Y$ and $Z$ states in \pp\ collisions, \eg\ as proposed in~\cite{Butenschoen:2019npa} for the $X(3872)$, can also be measured, as done recently by the CMS collaboration for the $\chi_{c1,2}$ mesons~\cite{Sirunyan:2019apc} (see also the discussion in Sec.~\ref{sec:chicpol}).

\subsection{$\Q$-associated-production processes}
\label{sec:oniumassociate}
\subsubsection{Associated production of $\Q$ and vector bosons }
\label{sec:onium_bosons}

The study of associated-production processes, such as the combined production of quarkonia with a vector boson, provides a new tool to study perturbative and non-perturbative QCD, novel approaches to searches for new phenomena for both light and heavy states~\cite{Clarke:2013aya,Aaltonen:2014rda}, and an additional probe of multiple-parton-scattering interactions complementary to $W+\mathrm{jet}$,  $WW$, and di-quarkonium production processes, which are discussed in Section~\ref{sec:dps}.

Associated $\jpsi+Z$ and $\jpsi+W$ production have both been observed~\cite{Aad:2014rua,Aad:2014kba,Aaboud:2019wfr} by the ATLAS collaboration. These are extremely rare processes, with only approximately one in every $10^{6}$ $W$ or $Z$ boson-production events also producing a $\jpsi$ in the fiducial volume of the ATLAS detector. Yet they can provide rich physics opportunities well-suited to precision studies with the large data-sets at the HL-LHC.  The presence of a vector boson allows for more efficient event triggering than that would be possible for inclusive quarkonium processes. The resulting relatively high-$\pT$ multi-lepton signatures mean that selections are resilient to the expected high instantaneous luminosities (and correspondingly large numbers of multiple simultaneous \pp\ pileup interactions) anticipated at the HL-LHC. Based on existing selections, this means that the ultimate HL-LHC data-sets of 8\,500 prompt $\jpsi+Z$ events and  30\,000 prompt $\jpsi+W$ events, and double as many non-prompt events, can be expected to be recorded for study by each of the general purpose detectors. Similar measurements of $\jpsi+\gamma$ associated production, as well as equivalent processes with bottomonium production and with excited quarkonium states, together provide a rich laboratory for future exploration.

\cf{fig:zjpsi_atlas} shows an example of the measured differential $\jpsi+Z$ rates for prompt $\jpsi$ production compared with CO and CS NLO NRQCD predictions and data-driven estimates of the DPS contribution. Existing measurements point to discrepancies that can in part be explained by: enhanced DPS rates inconsistent with measurements from inclusive vector boson and hadronic jet processes; non-trivial correlations in DPS interactions; or enhanced contributions to single parton scattering (SPS) rates that become particularly important at large transverse momenta.

The limited data currently available implies that the existing DPS extractions are approximate (see detailed discussions in Section~\ref{sec:dps}). Data from the HL-LHC will enable more detailed studies of DPS dynamics necessary to decouple DPS from SPS interactions at low momentum transfer. Current $\jpsi$ measurements are limited to differential rates versus $\pT$, but measurements of other observables and two-dimensional differential distributions, such as the difference in the azimuthal angle between the boson and the quarkonium, $\Delta\phi(Z, \jpsi)$, versus $\pT(\jpsi)$, are recommended in the future to decouple SPS and DPS dynamics and allow precision studies and reinterpretation of these data. First studies in this direction have been produced by the ATLAS collaboration, that
observed~\cite{Wjpsi2020aux} no strong $\pT(\jpsi)$ dependence on the distribution of $\Delta\phi(W, \jpsi)$.

\begin{figure}[h!]
\centering
\includegraphics[width=0.47\textwidth]{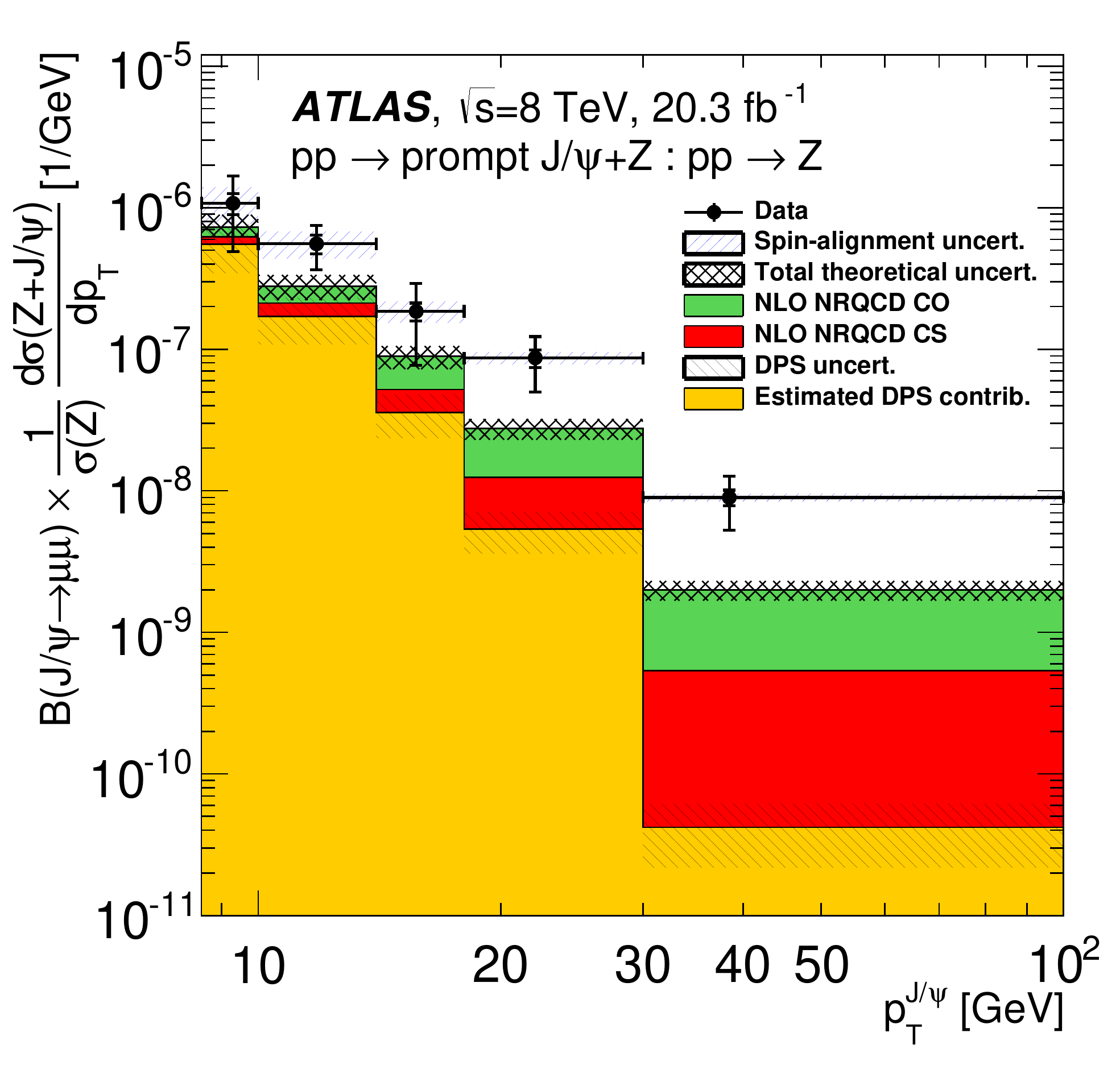}
\includegraphics[width=0.47\textwidth]{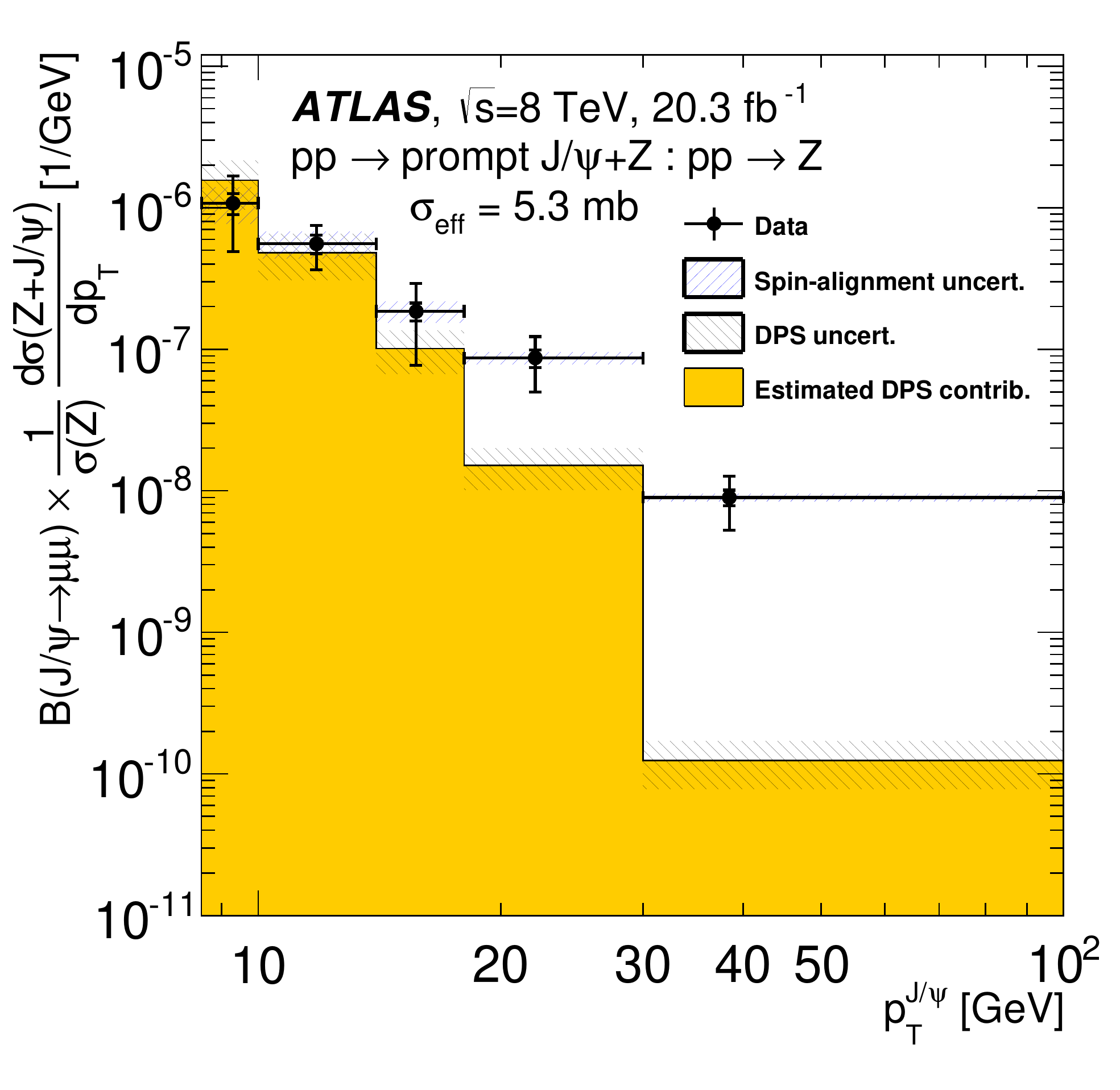}
  \caption{Measured differential prompt $\jpsi+Z$ rates as a function of $\jpsi$ $\pT$ compared to SPS predictions and a DPS rate normalised to $\sigma_\mathrm{eff}=15$~mb (left), where
  the normalisation is fit to data at low $\Delta\phi(Z, \jpsi)$ resulting in a preferred $\sigma_\mathrm{eff}=5.3$~mb (right). [Plots taken from~\cite{Aad:2014kba}.]} 
  \label{fig:zjpsi_atlas} 
\end{figure}

At high \pT, SPS can be expected to dominate over DPS processes. As $\jpsi$ produced in association with a vector boson have been observed to be produced with harder $\pT$ spectra than inclusive $\jpsi$ (a two- to three-orders of magnitude drop from 10---100~GeV in the former compared to a six-order of magnitude drop~\cite{Aad:2015duc,Khachatryan:2015rra} in the latter), data at the HL-LHC are expected to provide a comparable high-$\pT$ reach to current inclusive measurements for detailed testing of perturbative QCD calculations. 

\cf{fig:wjpsi_atlas}~(Left) illustrates how rates of $\jpsi+W$ production can be described by NLO CEM predictions together with a DPS contribution with an effective cross section of $\sigma_\mathrm{eff} = 6.1^{+3.3}_{-1.9}{}^{+0.1}_{-0.3}$~mb~\cite{Lansberg:2017chq}, compatible with the minimum of $6.3\pm 1.9$~mb determined by ATLAS~\cite{Aaboud:2019wfr} and with the lower (68\% C.L.) limit of $5.3$~mb determined in the $\jpsi+Z$ process~\cite{Aad:2014kba}. However, new data, shown in \cf{fig:wjpsi_atlas}~(right), illustrate that challenges remain in describing associated production in the high-$\pT$ regime, where perturbative calculations underestimate the data by an order of magnitude. 

\begin{figure}[h!]
\centering
\raisebox{12pt}{\includegraphics[width=0.47\textwidth]{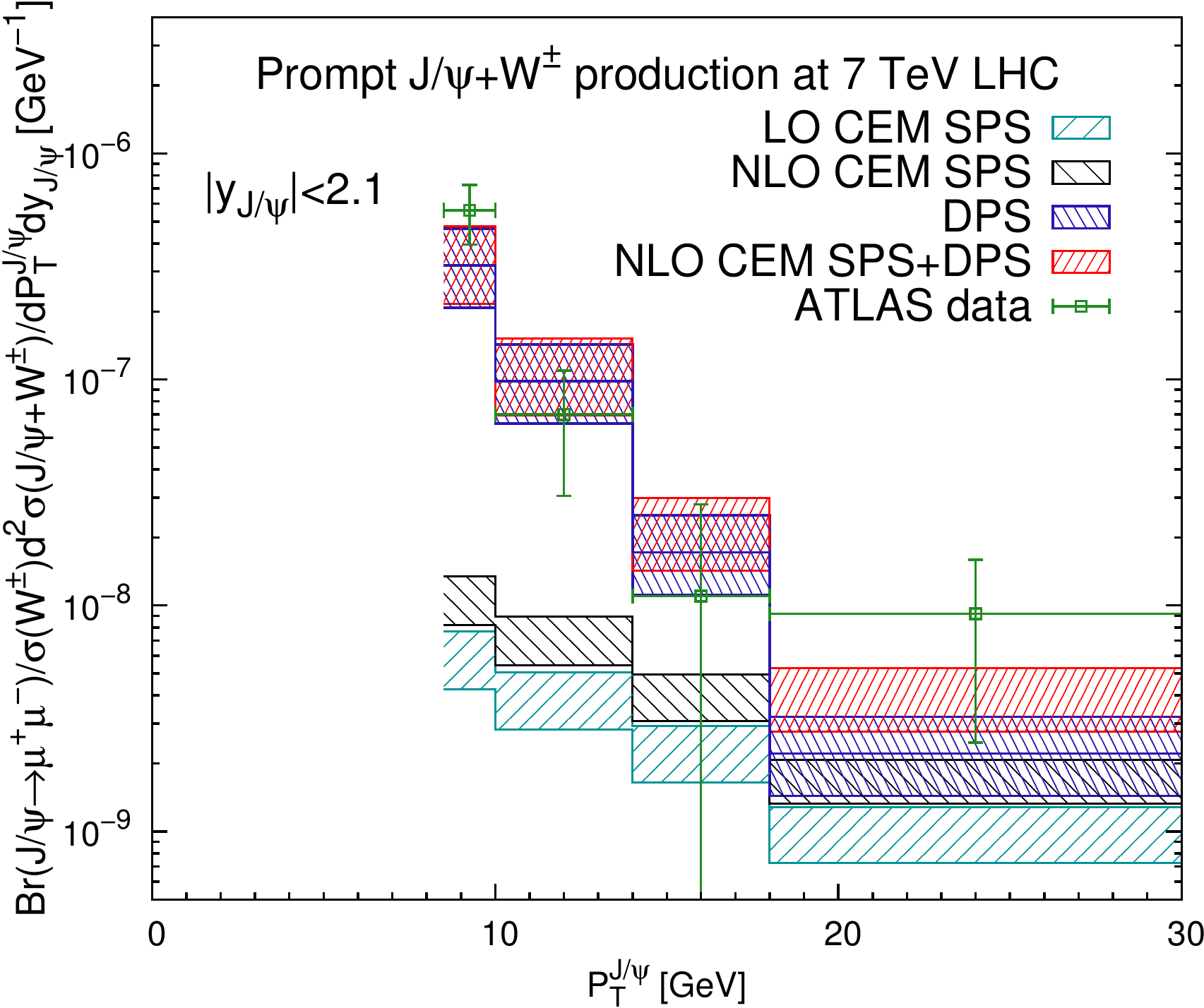}}
\hspace{0.5cm}
\includegraphics[width=0.45\textwidth]{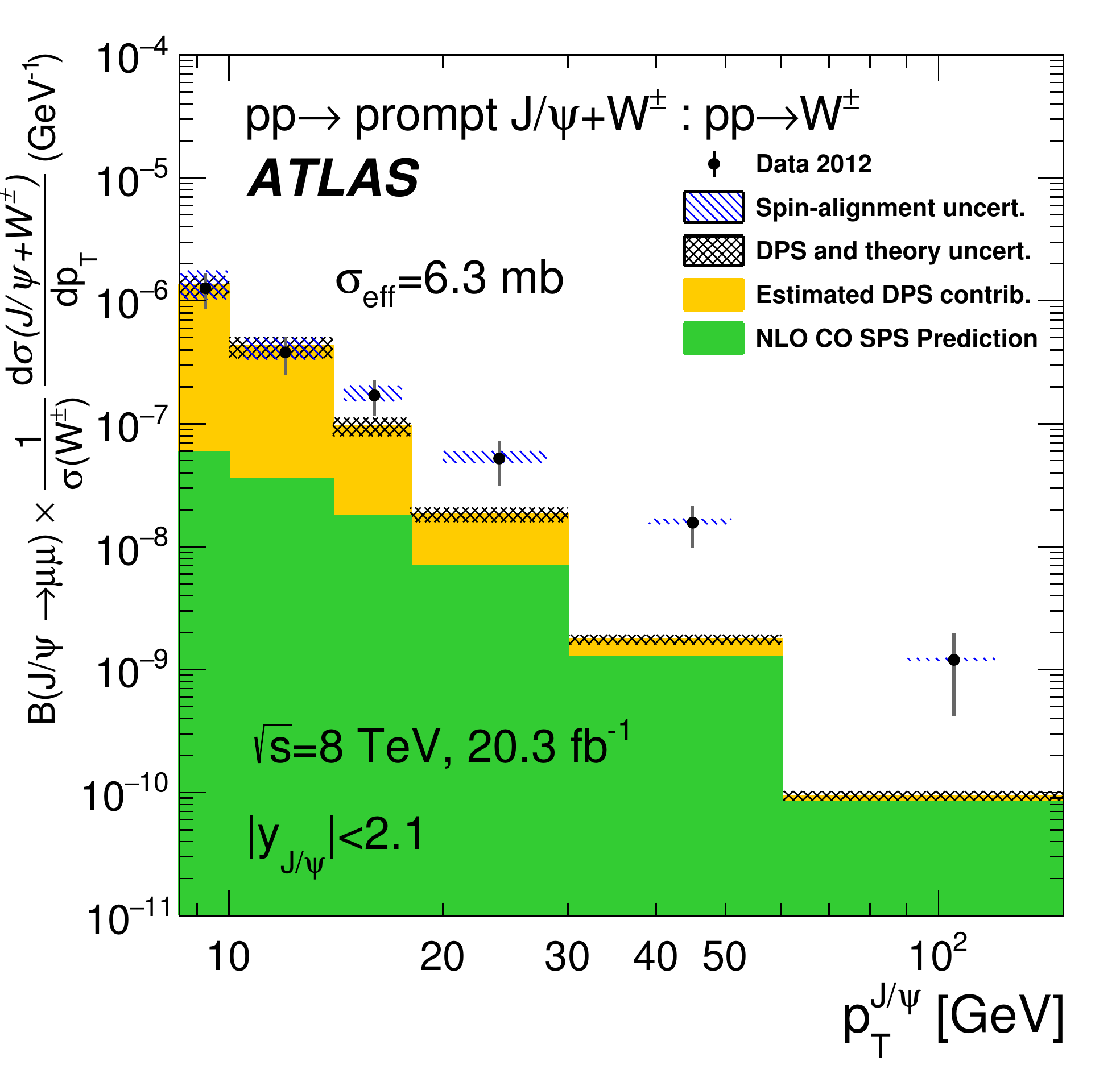}
  \caption{(Left) Measured differential prompt $\jpsi+W$ rates as a function of $\jpsi$ $\pT$ in \pp\ collisions at 7~TeV and (Right) at 8~TeV extending to $\pT(\jpsi)>100$~GeV compared to CEM SPS and DPS predictions. The data--theory agreement is good on the left, but the data exceed the predictions in the right distribution.
  [Plots and data taken from~\cite{Lansberg:2017chq} and~\cite{Aad:2014rua,Aaboud:2019wfr}.]} 
  \label{fig:wjpsi_atlas} 
\end{figure}

Studies of other observables such as the spin alignment of quarkonia in associated production are only expected to become possible in the HL-LHC era, and will provide additional information on the underlying production mechanisms. The NLO CS contributions, which are expected to dominate at high $\pT$, predict the polar anisotropy of direct $\jpsi$ and $\ups$ produced in association with a $Z$ boson to become strongly longitudinal ($\lambda_\theta<0$)
as the $\pT$ of the quarkonium increases~\cite{Gong:2012ah}. This stands in contrast to observations of very weak spin alignment in inclusive-production modes~\cite{Chatrchyan:2013cla,Aaij:2013nlm} and so it is an area where even  measurements of limited precision can provide useful inputs.

Associated production of a photon and a quarkonium state offers an additional new probe of production dynamics. This process has yet to be observed, although these final 
states have been the subject of study for both $P$-wave quarkonium states~\cite{Aad:2011ih,LHCb:2012ac,Chatrchyan:2012ub,ATLAS:2014ala,Sirunyan:2018dff} and exclusive $H^0$ boson decays~\cite{Aaboud:2018txb,Sirunyan:2018fmm}. The non-resonant production of $\gamma+\Q$ is challenging to distinguish due to low photon-reconstruction efficiencies or large experimental backgrounds at low transverse momenta.  These processes are however predicted to have large SPS NLO cross-sections~\cite{Lansberg:2009db} that make them well-suited for future study, in particular to constrain gluon TMDs (see Section~\ref{sec:jpsigammaTMD}). Expected DPS rates are not well-known, a fact that can compound the interpretation of measurements at low $\pT$ in a similar fashion to the existing $\jpsi+W/Z$ measurements. The requirement of a high-$\pT$ photon in experimental measurements will likely suppress DPS contributions and enable excellent prospects for study at the HL-LHC if photons can be reliably associated to the quarkonium-production vertex.

Similarly, associated vector-boson processes involving bottomonium states have yet to be observed. This is in part due to lower expected production rates and the (slightly) larger combinatorial backgrounds (from leptonic $B$ meson decays) present in the $\ups(nS)\to\mu^+\mu^-$ invariant mass region.

Such processes may be sensitive to heavy-quark-gluon-fusion contributions, in addition to
the gluon-gluon  and light-quark-gluon processes present for the charmonium-production modes, and
provide further complementary tools with which to study SPS and DPS dynamics.

Associated $\Q+W/Z/\gamma$ production provides a new opportunity to study heavy-flavour production in association with a vector boson, which has to-date otherwise been predominantly tested in hadronic jet final states~\cite{Chatrchyan:2012vr,Aaboud:2017skj,Aad:2020gfi}. Quarkonium-production modes provide an opportunity to test theoretical predictions of $W/Z/\gamma+b(c)$ in a complementary regime at low transverse momentum and at small opening angles sensitive to gluon splitting contributions.
Open-heavy-flavour states have begun to be exploited for the study of charm~\cite{Aad:2014xca,Sirunyan:2018hde}, and $\jpsi+b$   final states have demonstrated their effectiveness for study of heavy-flavour modelling~\cite{Aaboud:2017vqt} in topologies that are challenging to study with hadronic jet final states.
 Existing measurements of non-prompt $\jpsi+Z$ production~\cite{Aad:2014kba} have established
these processes as having relatively high production rates and small DPS contributions. Non-prompt $\jpsi+Z$ production has been found~\cite{Lansberg:2016muq} to be a sensitive probe of $Z+b$ production which is complementary to $b$-jet identification approaches and that will, in particular, benefit from the enlarged acceptances and increased datasets at the HL-LHC.

Prompt-$\Q+V$ processes also represent a tool and an opportunity to study a variety of potentially new phenomena. Prompt-$\Q+V$ production has been proposed as a compelling prospect for the study of rare decays of the $H^0$ boson~\cite{Doroshenko:1987nj,Kartvelishvili:1988pu,Gonzalez-Alonso:2014rla}, or new heavy states~\cite{Diaz:1994pk,Davoudiasl:2012ag,Falkowski:2014ffa,Clarke:2013aya}. Such searches have begun to be explored experimentally~\cite{Aaltonen:2014rda,Aaboud:2018txb,Aad:2020hzm}, but the potential of such searches will only be fully realised in the HL-LHC era. In addition to searches for resonant phenomena decaying into $Q\bar{Q}+V$ final states, such measurements can be re-purposed in the search for new light-mass states produced in association with a vector boson. A study~\cite{Clarke:2013aya} using initial $\jpsi+W$-observation data~\cite{Aad:2014rua} was able to set competitive limits on the production of a light scalar near the $\jpsi$ mass, exceeding constraints both from dedicated low-mass di-lepton searches at the LHC, as well as from searches via radiative $\ups$ decays from $e^+e^-$ experimental data. A dedicated programme of searches for new phenomena  in $Q\bar{Q}+V$ final states has yet to be performed within the LHC collaborations but has fruitful prospects. The potential inclusion of $Z$-boson associated-production modes, and its extension to include higher di-lepton masses, up to and beyond the \ups, \upsp and \upspp resonances, together with the large HL-LHC datasets, offer the opportunity for the LHC to far surpass~\cite{Clarke:2013aya} the current bounds from LEP data.

\subsubsection{$\Q$-pair production}
\label{sec:onium_pair_pp}

The production of pairs of quarkonia also offers rich opportunities for the study of both SPS and DPS as well as of searches for rare decays and for new particles. Unlike for inclusive quarkonium production, which is presumably dominated by quarkonium plus jet(s) or minijet(s) final states (see, \eg,~\cite{Artoisenet:2008fc,Lansberg:2008zm,Shao:2018adj,Flore:2020jau}), the quarkonium pair production processes, including double charmonia, double bottomonia and charmonium+bottomonium, can in principle provide independent handles to investigate the SPS quarkonium production mechanism at LHC energies. Such measurements are, however, contaminated by sizeable DPS contributions. Conversely, the measurements of these processes are also strongly motivated~\cite{Kom:2011bd,Lansberg:2014swa} by their potential for study of DPS interactions in their own right (see Section~\ref{sec:dps}), and as a probe of the linearly polarised gluons inside the proton~\cite{Lansberg:2017dzg,Scarpa:2019fol} (see Section~\ref{sec:spin}), as well as for the search for new exotic states predicted in QCD~\cite{Chen:2018cqz,Aaij:2020fnh} and in beyond the Standard Model theories, and for searches for rare decay modes of $H^0$ and $Z$ bosons~\cite{Sirunyan:2019lhe}. 
The di-\jpsi final state has proven the potential of such searches with the large
datasets beginning to become available at the LHC with the recent
observation by the LHCb collaboration of a new state~\cite{Aaij:2020fnh} at a mass of 6.9~GeV, widely interpreted as a fully-charm tetraquark state.

Due to their rare but distinctive four-lepton signatures, di-quarkonia are excellent candidates for precision studies with the large datasets expected at the HL-LHC. The experimental challenge will be to ensure wide kinematic coverage and high event-selection efficiency in the complex HL-LHC environment. Searches for their production in the decay of $H^0$ bosons~\cite{Aaboud:2018fvk,Sirunyan:2019lhe},
or other high-mass states, will benefit from Lorentz boosts in systems with large invariant mass. However, for new particle searches below the $Z$ boson peak (and particularly below the $B\bar{B}$ threshold)~\cite{Aaij:2018zrb,Sirunyan:2020txn}, an effective use of the unprecedented luminosities delivered in the HL-LHC programme requires that the experiments (in particular, the general purpose detectors) maintain a high efficiency for reconstruction of low transverse-momentum leptons, $\mathcal{O}(2-4~\mathrm{GeV})$, in four-lepton signature events containing one or more quarkonium candidates.

The di-$\jpsi$ final states have been the focus of many theoretical studies~\cite{Qiao:2002rh,Li:2009ug,Qiao:2009kg,Ko:2010xy,Berezhnoy:2011xy,Kom:2011bd,Lansberg:2013qka,Li:2013csa,Sun:2014gca,Lansberg:2014swa,Lansberg:2015lva,He:2015qya,Baranov:2015cle,Likhoded:2016zmk,Borschensky:2016nkv,Lansberg:2017dzg,Scarpa:2019fol,Gridin:2019nhc,Lansberg:2019fgm,He:2019qqr,Lansberg:2020rft}, reflected in the concentration of measurements from the LHC so far into the same final state~\cite{Aaij:2011yc,Abazov:2014qba,Khachatryan:2014iia,Aaboud:2016fzt,Aaij:2016bqq}. 
The experimental picture has been recently broadened with measurements of di-$\ups$ production by CMS~\cite{Khachatryan:2016ydm,Sirunyan:2020txn}, and a study of
$\jpsi+\ups$~\cite{Abazov:2015fbl} production by D\O\ at the Tevatron. 

From the theoretical perspective, di-\ups production is more-or-less similar to the di-\jpsi process, while both are significantly different with respect to $\jpsi+\ups$. The bulk of SPS events can be accounted for by the leading-$v^2$ CS channel in the former ones, while, for the latter, the complete study of~\cite{Shao:2016wor} reveals that the contributions from CO plus the feed-down contribution from excited quarkonium states are larger than the CS channel in SPS production of $\jpsi+\ups$. Given this, $\jpsi+\ups$ would be a priority for study from the viewpoint of investigating the CO mechanism.
Although plagued by a significant fraction of events from DPS interactions, the CO contributions can potentially be determined at the HL-LHC through measurement of their invariant mass distribution as shown in~\cf{fig:psiYplot}. Such a process has never been observed experimentally, while the D\O\ collaboration only found $3\sigma$ evidence at the Tevatron for the inclusive process, which should be expected to be composed of SPS and DPS components. It is thus desirable to carry out an analysis at the LHC to establish observation and decouple the SPS component for further study.

\begin{figure}[h!]
\centering
\includegraphics[width=0.49\textwidth,draft=false]{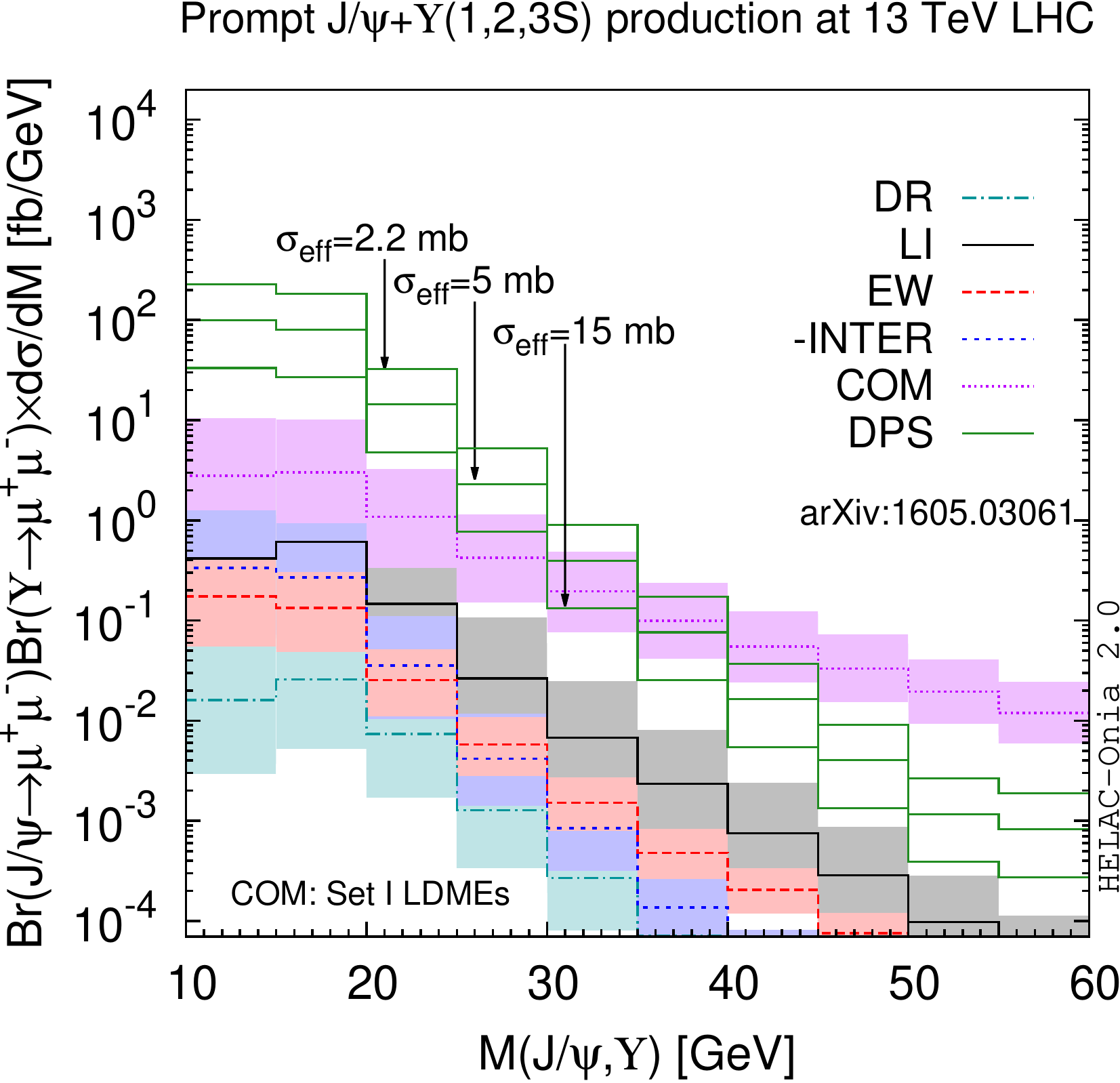}
\caption{Prediction for the invariant mass distribution of $\jpsi+\ups$ production in \pp\ collisions at $\sqrts = 13$~TeV within the fiducial volume $2<y_{\jpsi},\,y_{\ups}<4.5$. 
[Plot taken from~\cite{Shao:2016wor}]
\label{fig:psiYplot}}
\end{figure}

Besides the continuing measurements on $\jpsi+\jpsi$ and $\ups(1S)+\ups(1S)$, it would be difficult to achieve a coherent picture without a survey of all the excited states, and mapping the different production modes will require the large datasets in the HL-LHC era. Measurements of these excited states are eagerly anticipated: different combinations of quarkonia provide independent tests for the existence of new anticipated~\cite{Wu:2016vtq,Chen:2016jxd,Karliner:2016zzc} and unanticipated states, and will not only provide information on feed-down contributions to existing production measurements but also provide a direct probe of their own production mechanisms.

\subsubsection{Associated production of $\Q$  and jets}
\label{sec:onium_jets}
The production mechanisms for high-$\pT$ quarkonium-production result in the production of one or more hadronic jets accompanying the quarkonium state. The multiplicity of these jets and the radiation patterns relative to the produced state (or the similar $\jpsi$--hadron azimuthal correlation studies carried out \eg\ by STAR~\cite{Abelev:2009qaa}) provide valuable insights into the underlying production mechanisms of quarkonia, and are sensitive to CS/CO contributions as well as to higher-order quantum corrections.

{We} encourage the HL-LHC experimental collaborations to measure the production cross section of quarkonium states as functions of both jet multiplicity (for jets above some fixed transverse momentum scale, $\pT>Q_0$) and quarkonium transverse momentum. Ideally, these measurements should be matched to corresponding measurements of inclusive jet production in the same fiducial volume. As well as serving as an alternative tool for the study of single parton production, such datasets can find further application as an additional probe of DPS through measurements of angular correlations and $\pT$ balance observables, analogous to those performed in $V+\mathrm{jets}$~\cite{Aad:2013bjm,Chatrchyan:2013xxa} and other quarkonium final states (see Section~\ref{sec:DPSonia}).

Although initial studies can already be performed with the existing LHC data, high-$\pT$ $\Q$ + inclusive~jets signatures can be efficiently recorded at the HL-LHC, where the large datasets will enable access to events with high jet multiplicities as well as the hadronic activity accompanying very high-$\pT$ ($100$--$300$~GeV) quarkonium, a regime that inclusive production measurements are now starting to probe~\cite{ATLAS:2019ilf}.
Such studies would be complementary to measurements of quarkonium produced {\em in} jets 
(Section~\ref{sec:oniainjets}) as well as studies of quarkonium correlations with soft 
hadronic activity at low $\pT$ (Section~\ref{sec:oniaparticlemultiplicity}).

The high performance of jet flavour tagging at ATLAS/CMS offers the potential for novel studies of associated quarkonium and heavy-flavour production in final states with hadronic jets. Measurements of such processes have not yet been performed. The $\pT$ dependence of the production rates of $\jpsi+c$-jet or $\ups+b$-jet events is sensitive to the CS and CO contributions. Measurements of the topology of such events would provide valuable information, with CS transitions expected to dominate for quasi-collinear $\jpsi+c$ or $\ups+b$ production, while CO contributions would dominate in topologies with two heavy-flavour jets recoiling against the quarkonium state. Production rates~\cite{Artoisenet:2007xi} for these processes are sufficiently large at high $\pT^\Q>20$~GeV (needed to ensure jets can be adequately reconstructed and tagged), and thereby there are good prospects for study of these processes at the HL-LHC. Assuming a combined trigger and dimuon efficiency of approximately 50\%~\cite{Aad:2012dlq}
and low-$\pT$ $c$-tagging and $b$-tagging efficiencies of 25\% and 50\%, respectively~\cite{Aad:2015ydr,Sirunyan:2017ezt}, estimated yields of $\jpsi+c\bar{c}$ and $\ups+b\bar{b}$  (with at least one identified heavy-flavour jet) of 7\,500 and 150\,000 can be expected at both ATLAS and CMS detectors with the HL-LHC dataset, {\em if} the current efficiencies for triggering on quarkonium states down to $\pT\approx$~20~GeV can be maintained. %

\subsection{Constraining the gluon PDF in the proton using $\Q$\label{sec:pdfofpp}}

The progressive accumulation of large amounts of experimental data at the LHC, with associated increasingly reduced statistical uncertainties, requires a parallel effort to decrease the theoretical uncertainties of the corresponding predictions. Theoretical hadronic cross sections exhibit dependencies on intrinsic scales such as the factorisation and renormalisation scales. The inclusion of higher-order perturbative QCD corrections in calculations of partonic cross sections, mostly at NLO accuracy today, will lead to a reduction of such scale uncertainties. For many theoretical calculations, the PDF uncertainties are often the leading uncertainty today. In particular, the gluon density in the small-$x$ domain ($x\lesssim 10^{-4}$), extrapolated from larger-$x$ regions, is essentially unconstrained by experimental data, and affects quarkonium cross sections in the low-$\pT$ and/or forward regimes. Low-$x$ studies are also relevant in searches for phenomena beyond the DGLAP linear QCD evolution equations~\cite{Gribov:1972ri,Dokshitzer:1977sg,Altarelli:1977zs}, such as those driven by Balitsky-Fadin-Kuraev-Lipatov (BFKL)~\cite{Fadin:1975cb,Kuraev:1976ge,Kuraev:1977fs,Balitsky:1978ic} or non-linear parton saturation~\cite{Gribov:1984tu,Mueller:1985wy,Gelis:2010nm}. 

It is thus helpful to exploit new paths to perform PDF fits, by trying to incorporate new precise data already available, which are not traditionally considered in PDF global fits. A possibility in this direction is offered by the LHC data on hidden and open-heavy-flavour production, characterised by very high statistical precision. There are recent small-$x$ gluon constraints, though not yet in the global analyses, from inclusive open-charm production~\cite{Zenaiev:2015rfa,Gauld:2016kpd, Bertone:2018dse, Zenaiev:2019ktw} and from $\jpsi$ in exclusive reactions~\cite{Flett:2020duk} (see section~\ref{sec:dif_gluonpdf} for details). These data probe the very low $x$, down to $x\simeq 3\cdot 10^{-6}$, and low scale (a few GeV$^2$) domain. However, tensions arise between the open charm and the exclusive $\jpsi$ extractions. It would therefore be valuable to have new insights from additional independent determinations with similar inputs, such as from the inclusive quarkonium data. 

Given the lack of consensus on the production mechanisms, the inclusion of inclusive quarkonium production data in the PDF global fits, which was initially proposed in the 1980s~\cite{Martin:1987vw, Martin:1987ww}, has been abandoned. In addition, the $\psi$ and $\ups$ hadroproduction processes suffer from large LDME uncertainties from several competitive CO channels. These non-perturbative CO LDMEs could be largely determined from the corresponding experimental data for each PDF choice. The usage of the $\psi$ and $\ups$ data in a PDF fit, albeit still lacking the coherent picture, could be taken correlated with the corresponding LDME determinations, which is analogous to the role of the strong coupling $\alpha_s$ in other PDF analysis.

The situation could be radically improved if, for a given quarkonium observable, one is able to identify a single dominant channel or mechanism. One typical example is \etac hadroproduction at the LHC at $\pT \lesssim 12$~GeV (see also discussions in Section~\ref{sec:etac}), where it is understood that only the leading-$v^2$ CS channel is relevant. The remaining obstacle in using the quarkonium data to pin down the small-$x$ gluon density in the proton is the large intrinsic theoretical uncertainties in the cross section calculations, that are at NLO accuracy today. Similar to the open-charm case, these scale uncertainties can be largely mitigated by looking at ratios of (differential) cross sections (such as the ratios of two independent measurements at different centre-of-mass energies~\cite{Mangano:2012mh} or of cross sections in two different rapidity bins~\cite{Zenaiev:2015rfa}). These ratios have the extra advantage of cancellation of some of the systematic uncertainties, such as those related to the (single) LDME in theoretical calculations and the correlated systematical errors in experimental measurements. Exploiting the \etac LHCb Run-2 measurement~\cite{Aaij:2019gsn} is however not competitive as it is dominated by large statistical uncertainties. The HL-LHC is clearly able to significantly increase the precision of the measurement. All such inclusive quarkonium data are able to improve our knowledge of the proton PDFs lying in the low $x$ ($x<10^{-5}$) and low {scales} (a few GeV$^2$) regime, which are expected to be hard to constrain in general.

In addition, the collider mode, FT-LHC~\cite{Hadjidakis:2018ifr} will allow one to probe the high-$x$ range of the proton PDFs. In such a colliding configuration, the probed $x$ range of the parton (gluon, charm and valence quarks) densities can reach $x\simeq 0.5$, if not larger, by using various final states, including open-heavy-flavour hadrons and quarkonia.

For further aspects on the experimental side, the experimental collaborations should provide all information needed 
to include their data with the appropriate uncertainties in PDF fits. In particular, information~\cite{Abdallah:2020pec} on bin-by-bin correlations of systematic uncertainties, in the form of covariance error matrices in differential distributions, as well as those on correlations between different distributions, are essential to perform a fully meaningful statistical analysis and extraction of best-fit PDFs accompanied by reliable uncertainty estimates. Moreover, it is obvious that the quarkonium data used in a standalone way are not enough to perform PDF fits at all values of $x$ and {scale}. Therefore, it is ideal to use them in conjunction with all other data traditionally already used in PDF global fits, in particular those on inclusive and semi-inclusive Deep-Inelastic-Scattering (DIS), as a complementary tool to extend the ($x$, scale) coverage of the latter. 

The last considerations {of this section regard recent theory progresses}. It has been known for a long time that, for some choices of parameterisation, where gluon PDFs are rather flat, the open-charm and charmonium \pT-integrated cross sections at low scales can become negative~\cite{Schuler:1994hy,Mangano:1996kg,  Feng:2015cba,Accardi:2016ndt,Ozcelik:2019qze} {at high \sqrts}. Such pathological behaviours appear at (N)NLO for open charm and at NLO for charmonium. Hence, imposing the positivity of these cross sections, {assuming that} the missing higher order QCD corrections do not completely change the picture, {would} add additional constraints on the gluon PDF. {This would also go along the lines of a recent} exploratory study on the positivity of the $\overline{{\rm MS}}$ PDF itself~\cite{Candido:2020yat}. 

{However, it was recently found in~\cite{Lansberg:2020ejc} that the unphysical behaviour of the $\eta_c$ cross section at NLO for increasing \sqrts\  (but not necessarily for extremely large \sqrts) could efficiently be tempered by a specific factorisation-scale choice. The resulting cross section indeed then shows  a reduction of the renormalisation scale uncertainty while remaining very sensitive to the gluon PDF at low scales. If a similar scale choice can be used for $J/\psi$ for which numerous \pT-integrated cross sections have been measured, it could certainly be used in the future to fit the gluon PDF at NLO.}

\section{Exclusive and diffractive   production\protect\footnote{Section editors: Charlotte Van Hulse, Ronan McNulty.
}
}
\label{sec:excl_diff}

The diffractive production of quarkonia differs from inclusive production, discussed in the previous section, by the presence of colourless particle exchanges that lead to rapidity gaps, devoid of any hadronic activity, in the final state of the event. Diffractive processes are called exclusive if the final state, including the forward scattered protons, is fully determined. In hadron-hadron collisions, such events are generally characterised by two large rapidity gaps with a centrally produced object, which can consist of a single particle or a pair of particles.

Diffractive quarkonium production at hadron colliders offers a unique tool to study the nature of both $C$-even pomerons  and $C$-odd odderons, multi-gluon colourless systems exchanged in scatterings with hadrons, which are fundamental to the understanding of soft hadron interactions. In the perturbative regime, the pomeron and odderon can roughly be interpreted as consisting of two and three gluons, respectively, though in general these are non-perturbative objects.

Diffractive processes can provide an improved understanding of the production of quarkonium states. Different Feynman diagrams contribute in inclusive, diffractive, and exclusive quarkonium production, which can be accessed through a comparison of results; \eg\ in exclusive $\jpsi$ production, CO contributions are entirely absent.   
In addition, exclusive production presents a particularly clean experimental environment and, sometimes, a simpler theoretical domain, which may assist with the identification of exotic quarkonia. In the large \cm energy limit, diffractive processes serve as a special testing ground for the BFKL resummation of HE logarithms entering at all orders of the perturbative expansion.

One of the most fruitful applications of exclusive quarkonium production is their use as probes of the partonic structure of the colliding objects.
Exclusive measurements are the only way to probe the 3D distribution of partons as functions of their longitudinal momentum and transverse position (through single-particle production), and their 5D distribution in terms of transverse position, longitudinal, and transverse momentum (through the production of pairs of particles or jets). 
These 5D distributions are related to Generalised Transverse Momentum Distributions (GTMDs), which are Fourier transforms of Wigner distributions. They are known as the ``mother distributions", since they contain the most complete information on the nucleon structure. Integrating them over the parton transverse momentum gives the generalised parton distributions (GPDs) and taking the forward limit of the GPDs results in the PDFs.

Selected experimental results on exclusive and diffractive quarkonium measurements are presented in Section~\ref{sec:dif_exp}, with some discussion of open questions for theory and experiment. A measurement of diffractive quarkonium production for the study of BFKL resummation is introduced in Section~\ref{sec:dif_production}. In  Section~\ref{sec:excl_quark}, the physics accessible in  single vector-quarkonium production is discussed under three headings. Firstly, Section~\ref{sec:dif_upc} presents processes in hadron-hadron interactions useful for the extraction of GPDs: unlike DIS data, which have been extensively used to constrain GPDs, hadron-hadron collider data have not yet been exploited. Secondly, exclusively-produced quarkonia can, with certain approximations, provide information on PDFs, but until now such measurements have not been included in global PDF fits. Section~\ref{sec:dif_gluonpdf} discusses the theoretical framework and proposes a method for the extraction of PDFs from exclusive $\jpsi$ production.  Thirdly, with FT-LHC, a kinematic region complementary to that in the collider mode could be accessed, as discussed in Section~\ref{sec:dif_FT}. The exclusive production of pairs of quarkonia (and jets) is discussed in Section~\ref{sec:dif_wig}. Until recently, it was not known how to access GTMDs, but now it has been shown that they can be extracted from pairs of particles or jets both in DIS and photoproduction.  DIS measurements await the future Electron-Ion Collider (EIC)~\cite{Accardi:2016ndt}, but for photoproduction in hadron-hadron collisions, the LHC is the ideal machine. A discussion of some of the most favourable experimental channels at the HL-LHC is provided in Section~\ref{sec:dif_wig}. 

As already discussed, three main distinct modes of operation are foreseen for HL-LHC: \pp, \pA\ and \AaAa\ collisions\footnote{Collisions using O, Ar, Kr and Xe beams may also be envisioned.} where $A$ is an ion, usually lead. Compared to previous LHC running, data taken in HL-LHC \pp  collisions will be difficult to use for exclusive measurements because of the high number of \pp interactions per beam collision. Measurements in such an environment may still be possible using proton taggers that could allows identification of the separate \pp primary vertices, or through dedicated data collection with a lower number of interactions per beam collision.

Collisions  in the \pA mode offer several advantages for the study of the nucleon, and have been under-utilised to date, due to the low integrated luminosities taken thus far. For HL-LHC, a nearly tenfold increase of data is foreseen~\cite{Citron:2018lsq}. This will not result in many multiple interactions per beam crossing and will provide a more appropriate channel to perform photoproduction measurements, exploiting the enhanced photon flux from the nucleus (which goes approximately as $Z^2$). Compared to \pp\ collisions (in the absence of a proton tagger), \pA\ collisions also have the advantage of identifying the photon emitter. 
In addition, they might offer a handle on constraining nuclear distributions, through photoproduction on the nucleus in the nucleus-going direction. 
While the foreseen tenfold increase in luminosity is important for a wide range of physics, exclusive measurements would clearly benefit from even higher luminosities in order to exploit their full potential (\eg for $\Upsilon$ and charmonium-pair production, as discussed later).

Access to nuclear distributions (PDFs, GPDs, and Wigner distributions) is best provided through the study of ion-ion (\AaAa) collisions. A nearly tenfold increase of data collection in PbPb collisions is foreseen for the HL-LHC. For photoproduction processes, the ambiguity in the identity of the photon emitter can in part be lifted through the detection of neutrons emitted by one of the Pb ions, \eg\ in zero-degree calorimeters~\cite{Guzey:2013jaa}.

\subsection{Experimental results}

\subsubsection{Selected experimental results}

\label{sec:dif_exp}

Exclusive and diffractive production has been studied in 
lepton-hadron interactions, both at the FT experiments HERMES~\cite{Airapetian:2001yk, Airapetian:2006zr, Airapetian:2008aa, Airapetian:2009aa, Airapetian:2009bm, Airapetian:2011uq, Airapetian:2010aa, Airapetian:2010dh, Airapetian:2012mq, Airapetian:2012pg, Airapetian:2014gfp, Airapetian:2015jxa, Airapetian:2017vit}, COMPASS~\cite{Alexakhin:2007mw, Adolph:2012ht, Adolph:2013zaa, Adolph:2014lvj, Adolph:2014hba, Adolph:2016ehf,  Akhunzyanov:2018nut, Alexeev:2019qvd}, and the experiments at Jefferson Lab
~\cite{Camacho:2006qlk, Mazouz:2007aa, Collaboration:2010kna, Defurne:2015kxq, Hattawy:2018liu, HirlingerSaylor:2018bnu, Park:2017irz, Hattawy:2017woc, Bedlinskiy:2017yxe, Bosted:2016hwk, Bosted:2016spx, Girod:2007aa, Stepanyan:2001sm}, and at the collider experiments H1 (see, among others, ~\cite{H1:2020lzc, H1:2015bxa, Andreev:2015cwa, Andreev:2014yra, Alexa:2013xxa, Aaron:2012ad, Aaron:2010su, Aaron:2009xp, Aaron:2009ac, Aaron:2008ab, Aaron:2007ab, Aktas:2007bv, Aktas:2007hn, Aktas:2006up, Aktas:2006hy, Aktas:2006hx, Aktas:2006qe}) and ZEUS (see, among others,~\cite{ZEUS:2017nkv,Abramowicz:2015vnu,Aaron:2012hua,Abramowicz:2011pk,Abramowicz:2011fa, Chekanov:2009zz, Chekanov:2008vy, Chekanov:2008cw, Chekanov:2007zr, Chekanov:2007pm, Chekanov:2005cqa, Chekanov:2005vv, Chekanov:2004mw}). 
It has also been studied in hadron-hadron collisions at the Tevatron, RHIC, and LHC. While lepton-hadron interactions offer the advantage of high-precision measurements by using a {point probe} to study hadrons, hadron colliders can reach a higher \cm energy, hence providing access to lower values of the parton fractional momentum, $x$.

Future experiments are envisaged with expanded possibilities for exclusive and diffractive measurements. For the study of lepton-hadron interactions, the EIC construction is in development~\cite{accardi2012electron}. The EIC will allow the collection of large samples of data at variable \cm energies, thus making possible high-precision, multi-differential measurements with a vast kinematic coverage. Moreover, measurements with polarised nucleons and (unpolarised) nuclear-ion beams will allow one to probe the nucleon  spin structure and nuclear matter, respectively. Stepwise upgrades for the LHC and the LHC experiments are also ongoing and planned, on a time scale preceding the EIC, at \cm\ energies about 50 times larger than those accessible at the EIC. Such a programme has the potential to access rarer diffractive and exclusive processes. In addition, there exists the possibility to perform measurements with FT collisions at the LHC, covering \cm\ energies similar to those at the EIC. These provide access to the high-$x$ region. Also, ideas and studies for measurements with a polarised target at the LHC, sensitive to spin-related physics, are underway~\cite{Kikola:2017hnp,Hadjidakis:2018ifr}. Furthermore, planned data collection with a heavy-ion beam and nuclear targets provides the possibility to study the nuclear parton density. 

\begin{figure}[h!]
    \centering
    \includegraphics[scale=0.35]{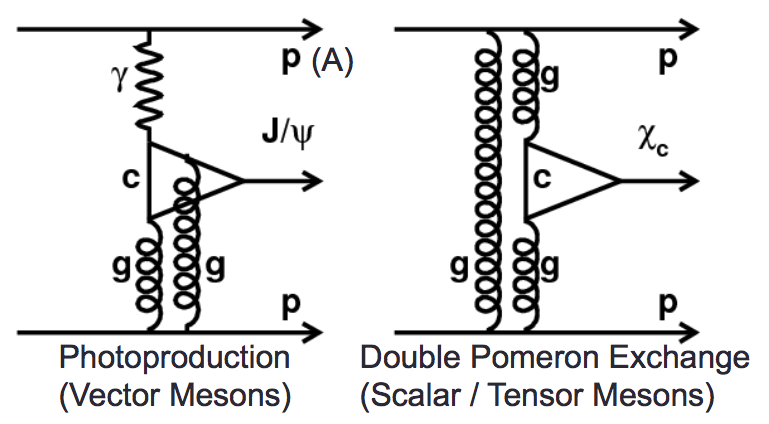}
    \caption{Photoproduction (Left) and double-pomeron exchange production (Right) of charmonium at hadron colliders.} 
    \label{fig:fd_exclusive}
\end{figure}

Exclusive quarkonium production at hadron colliders is commonly studied in ultra-peripheral collisions (UPCs)~\cite{Baltz:2007kq,dEnterria:2007zra,Contreras:2015dqa,Klein:2017nqo}. In such collisions, impact parameters are typically larger than the sum
of the nuclear radii, so strong interactions are suppressed, while electromagnetic interactions are favoured. Exclusive meson production has been studied at the Tevatron in $p\bar{p}$ collisions, at RHIC in AuAu and $p$Au collisions, and at the LHC in proton-proton, \pPb, and \PbPb collisions. The measurements cover light vector-meson production, such as single-$\rho$ production~\cite{Sirunyan:2019nog, Sirunyan:2020cmr, Adam:2020sap, Adamczyk:2017vfu, Agakishiev:2011me, Abelev:2008ew, Abelev:2007nb, Aaltonen:2015uva, Albrow:2010yb, Albrow:2014yma}, and heavier single and pair-produced quarkonia $\jpsi$, $\psi(2S)$, $\chic$ and $\ups$~\cite{Aaij:2014rms, Aaij:2014iea, Aaij:2018arx, Aaij:2015kea,Sirunyan:2018sav,Afanasiev:2009hy,Albrow:2014yma, LHCb:2011dra}. In heavy-quarkonium production, the charm and bottom quarks provide the hard scale that makes possible a theoretical perturbative expansion, and the interpretation of the results in terms of PDFs, GPDs, or Wigner distributions. The majority of the measurements so far were performed using unpolarised hadrons, but preliminary measurements with a transversely polarised proton have been performed at RHIC and these allow access to spin-dependent PDFs, GPDs, and Wigner distributions~\cite{jarda_talk}. Differential cross-sections have been measured as functions of quarkonium rapidity and the Mandelstam variable $t$, which can be approximated by the $\pT^2$ of the produced meson system.

Single vector-meson production involves a single-pomeron exchange (SPE), and is the most studied process so far in central exclusive production, while scalar and tensor quarkonium are produced through double-pomeron exchange (DPE); the different production mechanisms are shown in Fig.~\ref{fig:fd_exclusive}. 
Since only gluon propagators are present in DPE, central exclusive production is a fertile hunting ground for glueballs, tetraquarks, and quark-gluon hybrid states, with the potential advantage of a lower background contamination compared to non-exclusive measurements.
Such gluon-rich media are also a good environment to study the odderon, predicted in QCD but not unambiguously observed~\cite{Boussarie:2019vmk,McNulty:2020ktd}.
A very promising channel that would provide strong evidence for the existence of the odderon is exclusive photoproduction of C+ quarkonia, which can only be produced if the photon fuses with another C- propagator.
Searches for the exclusive production of scalar~\cite{Bartels:2004hb,Czyzewski:1996bv} or tensor quarkonia in pA or AA collisions are therefore of great interest and require high luminosity.
Exclusive, in comparison to inclusive, measurements can also give insight into the production mechanisms of charmonia and could indirectly help to distinguish between different frameworks used for inclusive production.

The HL-LHC operation warrants a future programme of work for experimentalists and theorists in which the different frameworks can be better disentangled through the comparison of suitably chosen high-precision observables in exclusive, diffractive, and inclusive reactions.  For example, in exclusive reactions, CO states are absent and, in non-exclusive processes, it is plausible that, as the produced quarkonium becomes more isolated, the CO contributions become more suppressed.

By virtue of the Landau-Yang theorem~\cite{Landau:1948kw,Yang:1950rg}, which states that a spin-1 particle cannot couple to two identical massless vector particles, the exclusive production of  $\chi_{c1}$  is expected to be heavily suppressed compared to its spin partners, the $\chi_{c0}$ and $\chi_{c2}$. In inclusive production, this suppression may not be as pronounced because of the CO contributions. If the initial gluons are allowed to be off-shell or if a third gluon is emitted, this suppression is lifted but the $\chi_{c1}$ rates remain partly suppressed~\cite{Khoze:2004yb,Pasechnik:2009bq} compared to the  $\chi_{c2}$. In the exclusive case, CO contributions are absent and any gluon emission is forbidden. Thus, 
measuring the yield ratio $\chi_{c2}/\chi_{c1}$ here would directly probe the degree of off-shellness of the gluons compared to the inclusive mode discussed in Section~\ref{sec:beyond_TMD}.

Such investigations can be further expanded by measuring quarkonium polarisation. The first study on polarisation of prompt $\chico$ and $\chict$ in inclusive production~\cite{Sirunyan:2019apc} uncovered a significant difference in polar anisotropy, $\lambda_\theta$, in agreement with NRQCD. Measurements that avoid the need to detect photon conversions are expected to improve the experimental resolution~\cite{Aaij:2017vck}. This may in fact be possible in the exclusive mode where the polarisation invariants (see Section~\ref{sec:quarkonium-pol-pp}) can reach their extremal values.

While photoproduction of quarkonia at heavy-ion colliders is typically studied in UPCs, there are now indications of contributions from photoproduced $\jpsi$ in peripheral collisions (with partial hadronic overlap), both at LHC by the ALICE experiment~\cite{Adam:2015gba}, and at RHIC by the STAR experiment~\cite{STAR:2019yox}. At low \pT, an excess of $\jpsi$ yields compared to that expected from hadroproduction is observed, which can be explained by contributions from photon-induced $\jpsi$ production. In this context, it would be extremely useful to study the polarisation of the $\jpsi$, since it is measured to be unpolarised in hadroproduction and transversely polarised in photoproduction, as inherited from the {parent (real) photon}. This study could be expanded to include $\psip$ and $\ups$.

Contradictory results have been found for the ratio of coherently photoproduced $\psip$ to $\jpsi$.  While the data at central rapidity in \PbPb\ collisions at $\sqrtsnn = 2.76$~TeV showed a ratio almost twice that in \pp\ collisions~\cite{Adam:2015sia}, new data at forward rapidity in \PbPb\ collisions at 5.02~TeV give a ratio consistent with the \pp\ results~\cite{Acharya:2019vlb}.

 Studying diffractive processes where the probed beam hadron breaks up provides some sensitivity to the non-uniformity of the gluon distribution in the transverse impact-parameter space (see~\cite{Cepila:2018zky,Mantysaari:2020axf} and references therein). Cross sections and cross-section ratios of coherent and incoherent $\jpsi$ and $\ups$ photoproduction are a valuable tool to study the nucleon shape~\cite{Cepila:2018zky, Mantysaari:2017dwh,Cepila:2016uku,Cepila:2017nef}.

The identification of diffractive processes in general and the separation of diffractive processes where the beam particles break up or stay intact are experimentally and theoretically challenging, especially in collider experiments. Some aspects involved in the identification of exclusive and non-exclusive diffractive processes are discussed next.

\subsubsection{Experimental identification of diffractive processes}

The experimental identification of diffractive events usually relies
on the identification of a large rapidity gap, found by
ordering all charged particles in pseudorapidity and noting the
largest difference, $\Delta \eta$, between adjacent particles.
There are at least two practical problems with this approach,
due to the fact that detectors are not hermetic.
Firstly, how big a gap is required to identify an event as diffractive?
Secondly, how does one deal with the dissociation of the projectiles, which
occurs within a few units of rapidity of the beams and often enters
uninstrumented regions?  Both these issues are briefly discussed below:
neither is fully resolved and each is difficult to approach
both theoretically and experimentally.

\begin{figure}[h!]
    \centering
    \includegraphics[scale=0.41]{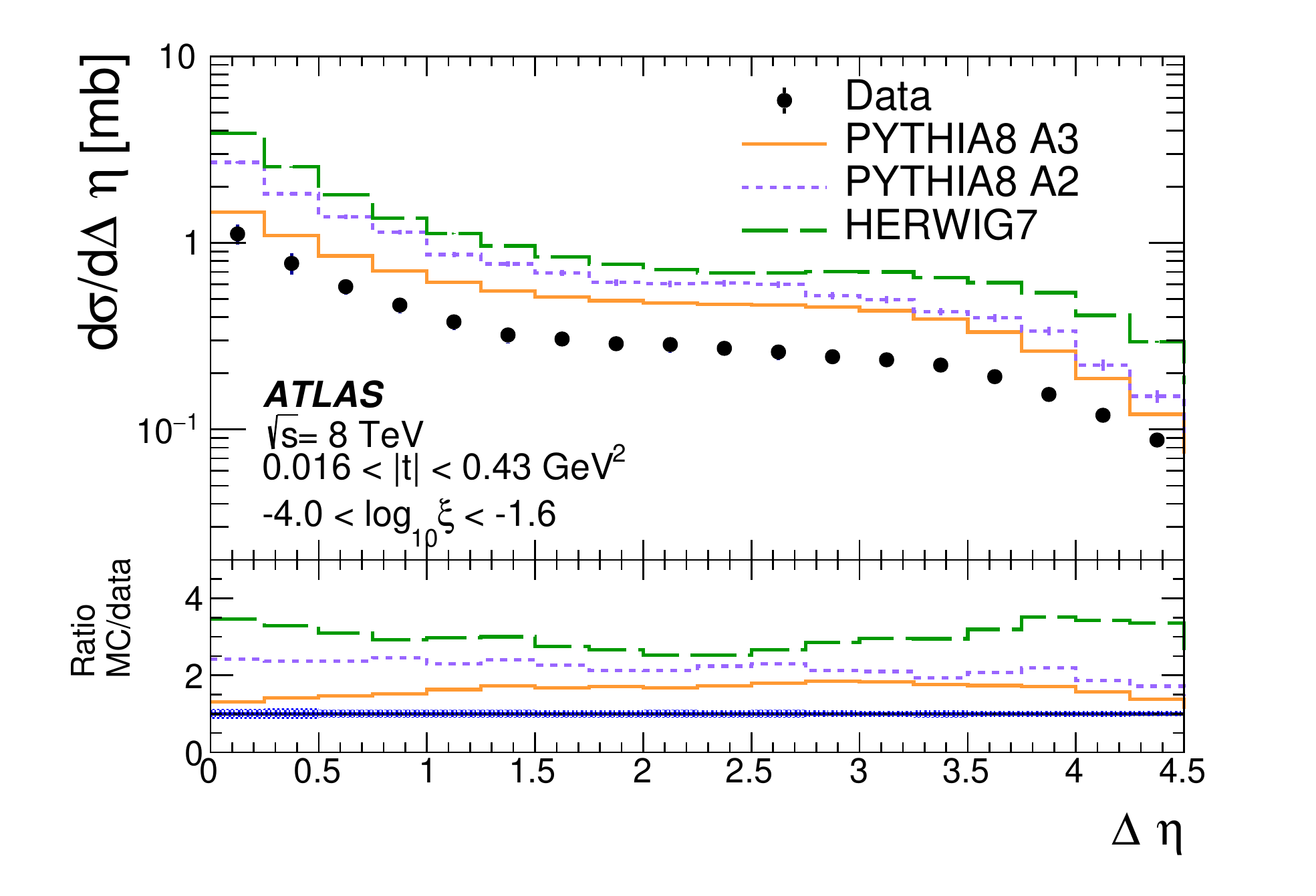}
    \raisebox{6pt}{\includegraphics[scale=0.40]{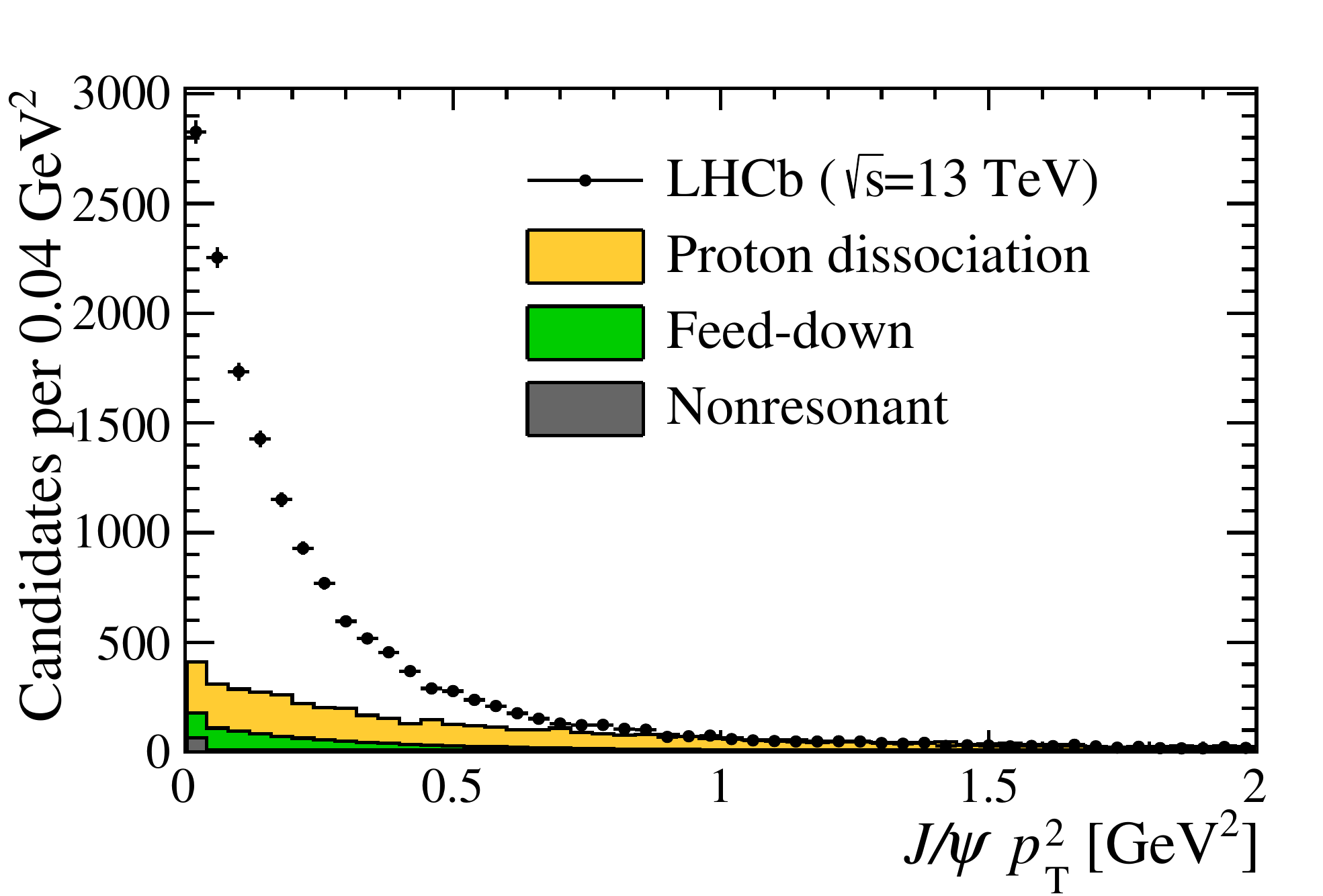}}%
\caption{Left:
Hadron-level differential single-diffractive cross section as a function of $\Delta\eta$,
comparing the measured data with \pythia and \herwig~7 predictions.
Right:
Transverse momentum squared distribution of $\jpsi$ candidates showing
estimated fractions of exclusive, feed-down, and proton-dissociation contributions. [Figures taken from~\cite{Aad:2019tek} and~\cite{Aaij:2018arx}.]}
\label{fig:iso}
\end{figure}

One way to investigate the size of the gap required to tag an
event as diffractive is to look at the (background) gap sizes in inclusive events.
However, modelling this in generators is difficult as the results depend
on non-perturbative effects: a single soft particle can destroy the gap.
It was shown in~\cite{Khoze:2010by} that, if the threshold for detecting
tracks is relatively high ($\pT>1$ GeV), similar results are obtained with
cluster hadronisation and string fragmentation models. However, order-of-magnitude
differences occur at lower transverse momenta: the probability of
$\Delta\eta>4$ in minimum bias \pp\ events at $\sqrts=7$ TeV was found
to be about 0.1 using a cluster hadronisation model~\cite{Winter:2003tt}
and 0.02 for string fragmentation~\cite{Sjostrand:2003wg}.
A recent measurement by ATLAS~\cite{Aad:2019tek},
comparing the largest gap in \pp\ events
at $\sqrts=8$ TeV with \pythia~\cite{Sjostrand:2007gs}
and \herwig~\cite{Bahr:2008pv} (Fig.~\ref{fig:iso}, Left),
shows that the data exhibit fewer large gaps than the models predict.

A related problem occurs in central exclusive production, where large
gaps should exist on either side of the central system (\eg\ $\pp\rightarrow p\oplus \jpsi\oplus p$ as discussed further in Section~\ref{sec:dif_gluonpdf}).
Various methodologies have been employed in such systems to determine
whether the candidate events are truly isolated
or whether the proton dissociates.
A simple approach, taken in the analysis of exclusive $\pi\pi$ production
by CMS~\cite{Sirunyan:2020cmr}, is to fit additional neutral energy deposits
in known non-exclusive events, and extrapolate to the signal region, assuming similar behaviour for like-sign and unlike-sign combinations.
Another approach, by LHCb, in the analysis of
exclusive $\jpsi$ production~\cite{Aaij:2018arx}, uses Regge theory to
fit the $\pT$ distribution in known non-exclusive events to model
the dissociative process and combines this
with the signal shape to determine the purity of a sample of candidate
exclusive events (Fig.~\ref{fig:iso}, Right).
A more complex approach was presented in a recent H1 analysis of the
photoproduction of $\rho$ mesons~\cite{Bolz:2019znd}.
Dissociative events are not well described either at generator level or in the 
detector simulation.  Therefore, a sophisticated re-weighting of the
DIFVM generator~\cite{List:1998jz} was employed, tuned using control samples from data.

An elegant solution to the problem of identifying exclusive events is found if the intact protons can be reconstructed.  
This requires dedicated detectors installed at very low angles to the beam, typically in Roman pots located several hundred metres from the interaction point.  
Both ATLAS (through the AFP spectrometer~\cite{Tasevsky:2015xya}) and CMS-TOTEM (using CT-PPS~\cite{Albrow:2015ois}) use such technology, which has the additional advantage of providing an independent measurement of the mass of the central system.  

\subsection{Forward \jpsi + backward jet production}
\label{sec:dif_production}

In Section \ref{sec:onium_jets}, experimental studies were motivated towards the measurement, in inclusive reactions,  of quarkonia associated with jets, following the first proposal of~\cite{Lansberg:2019adr}. Motivations for studying them in  diffractive reactions are now considered.

Diffractive reactions featuring a semi-hard scale hierarchy~\cite{Gribov:1984tu}, \ie\ $\sqrts \gg \{Q\} \gg \Lambda_{\rm QCD}$, with $\sqrts$ the centre-of-mass energy and $\{Q\}$ a (set of) characteristic hard scale(s), serve as a special testing ground for the dynamics of strong interactions in the High-Energy (HE) limit.
Here, a genuine Fixed-Order (FO) treatment based on collinear factorisation fails, since large energy logarithms enter the perturbative series in the strong coupling, $\alpha_s$, with a power that increases with the order. In particular, large final-state rapidities (or rapidity distances), typical of single forward emissions (or double forward/backward emissions) with colourless exchanges in the $t$-channel, directly enhance the weight of terms proportional to $\ln (s)$.
The HE factorisation based on the BFKL equation %
performs an all-order resummation of these large energy logarithms both in the leading-logarithmic approximation (LL), which means inclusion of all terms proportional to $\alpha_s^n \ln (s)^n$, and in the next-to-leading-logarithmic approximation (NLL), including all terms proportional to $\alpha_s^{n+1} \ln (s)^n$.

\begin{figure}[h!]
 \centering
 \includegraphics[scale=0.45]{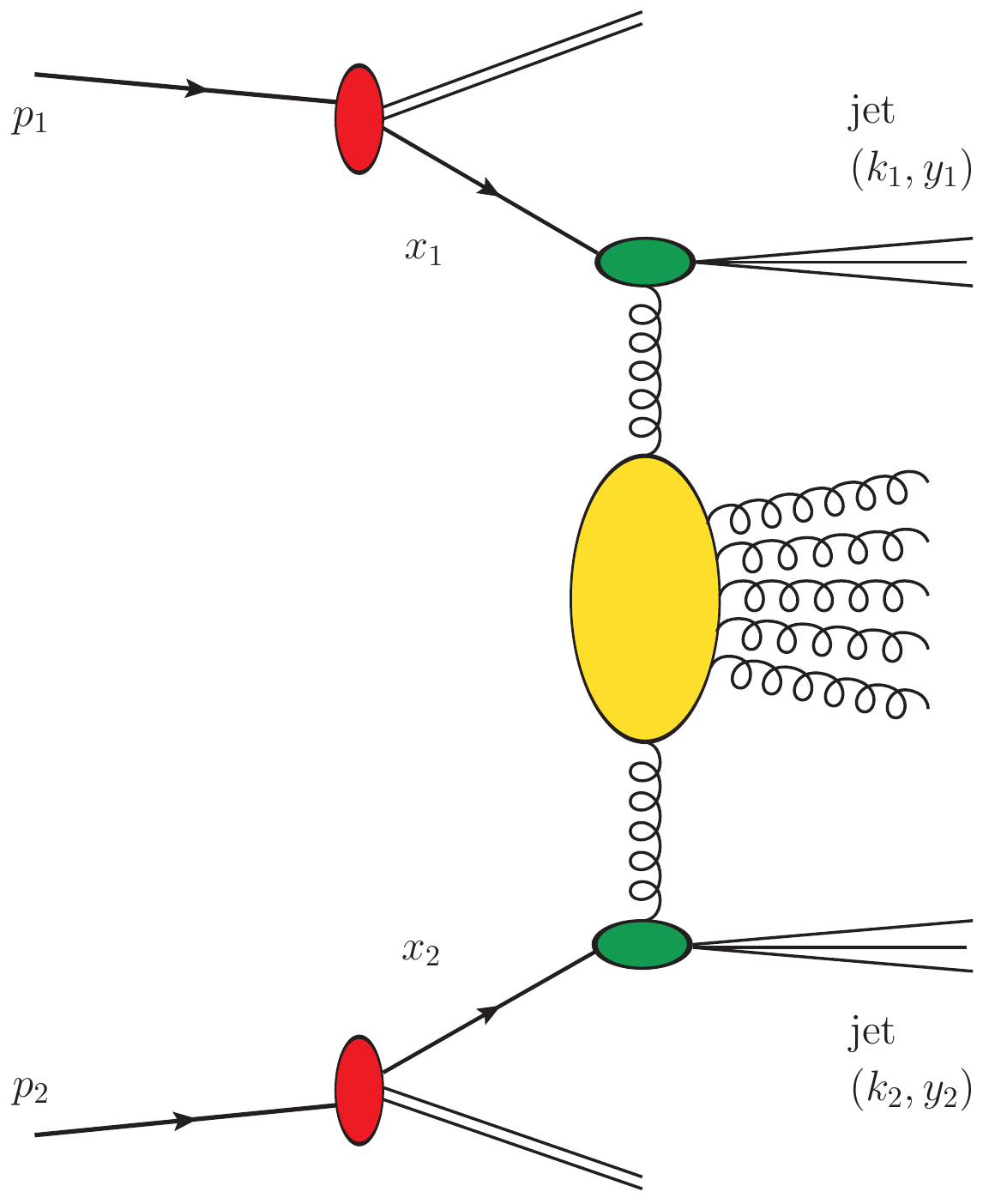}
 \hspace{1.5cm}
 \includegraphics[scale=0.45]{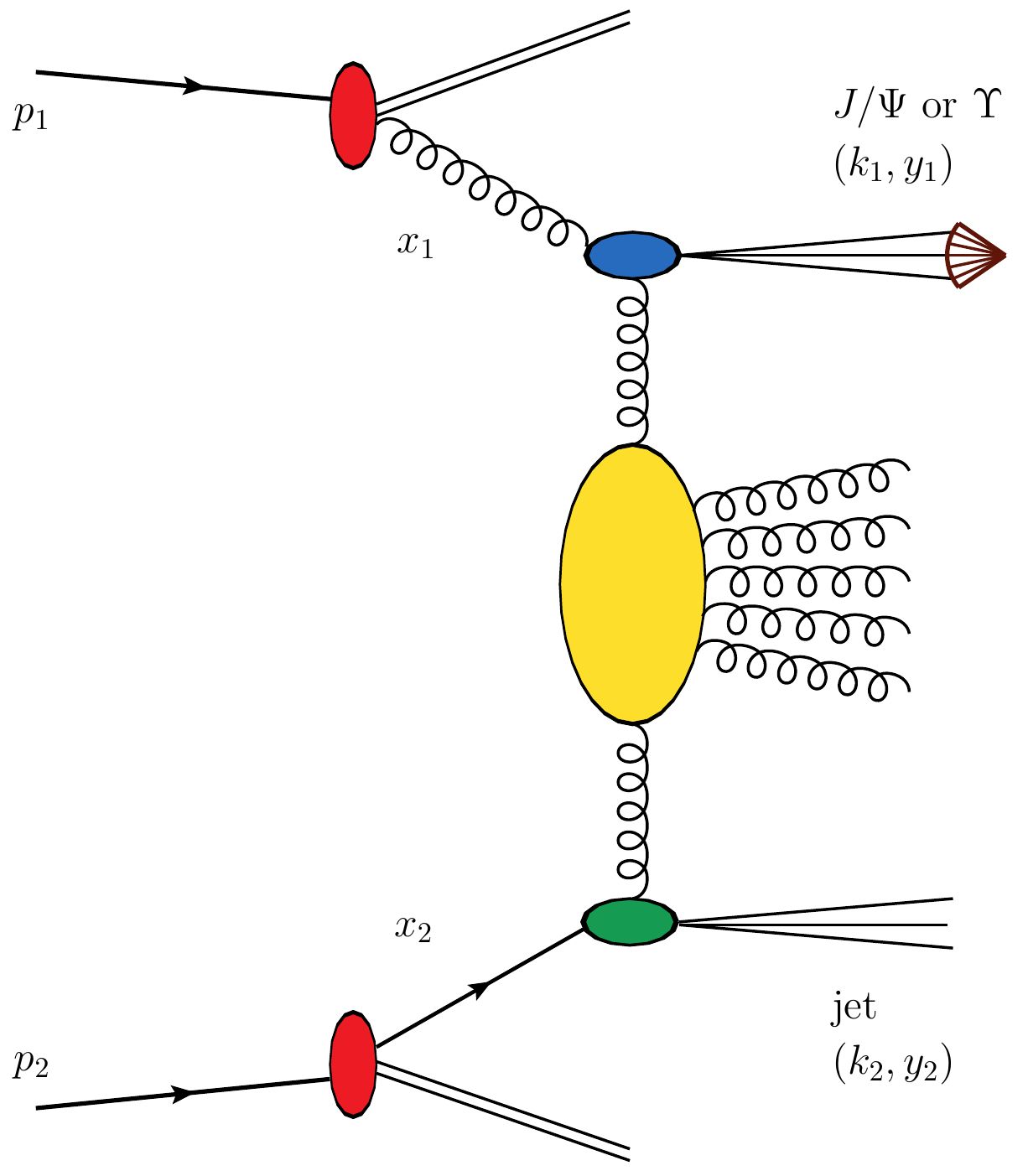}
 \\ \vspace{0.25cm}
 \hspace{1.25cm} 
 a) Mueller-Navelet jets
 \hspace{3.75cm}
 b) Forward quarkonium + backward jet
 \caption[]
 {Pictorial representation of two semi-hard reactions in the hybrid ``HE + collinear'' factorisation. Red blobs denote collinear PDFs, whereas green (blue) ones refer to the hard part of impact factors accounting for jet (quarkonium) emissions. They are connected to the BFKL gluon Green's function, schematically represented in yellow, via pomeron lines.}
 \label{fig:semi-hard}
\end{figure}

Over the last few years, predictions for observables in a wide range of semi-hard final states have been proposed~\cite{Caporale:2015int,Caporale:2016soq,Chachamis:2015crx,Caporale:2016xku,Caporale:2016zkc,Caporale:2015vya,Celiberto:2016hae,Celiberto:2016vhn,Celiberto:2017ptm,Celiberto:2020rxb,Bolognino:2018oth,Bolognino:2019yqj,Golec-Biernat:2018kem,Xiao:2018esv,Celiberto:2020tmb,Celiberto:2017nyx,Bolognino:2019ouc,Bolognino:2019yls}.
Among them, azimuthal correlations between two Mueller-Navelet jets~\cite{Mueller:1986ey} have been identified as favourable observables in the discrimination between BFKL- and FO-inspired calculations~\cite{Celiberto:2015yba,Celiberto:2015mpa,Celiberto:2020wpk}. This channel, depicted in~\cf{fig:semi-hard} (a), is characterised by hadroproduced jets with high transverse momenta, a large difference in rapidity, and a secondary undetected gluon system.\footnote{Although featuring secondary gluon emissions in the final state, this reaction can be classified as a diffractive one. Indeed, the imaginary part of its cross section, dominant with respect to the real part in the HE limit, can be directly linked to the forward elastic-scattering amplitude via the optical theorem.} 

\begin{figure}[h!]
	\centering
	\includegraphics[scale=1.25]{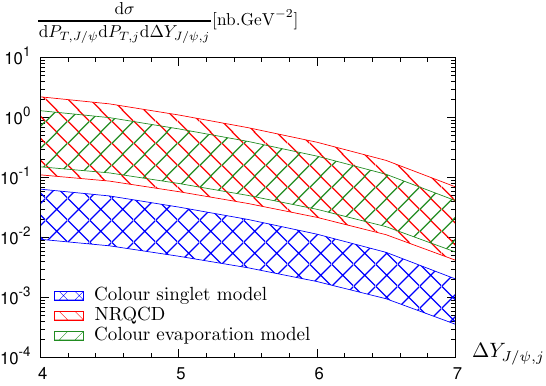}
	\includegraphics[scale=1.25]{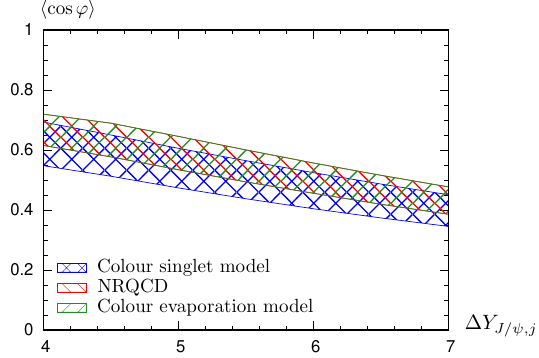}\\
	\small CMS ($-4.5 < y_j < 0 \, , \, 0 < y_{\jpsi} < 2.5 \, , \, P_{T, j} = P_{T, \jpsi}=10$ GeV)
	\\ \vspace{0.25cm}
	\includegraphics[scale=1.25]{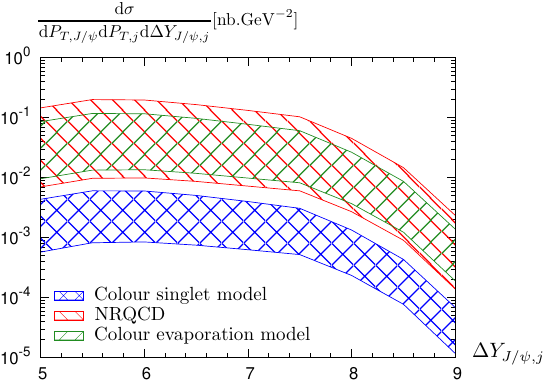}
	\includegraphics[scale=1.25]{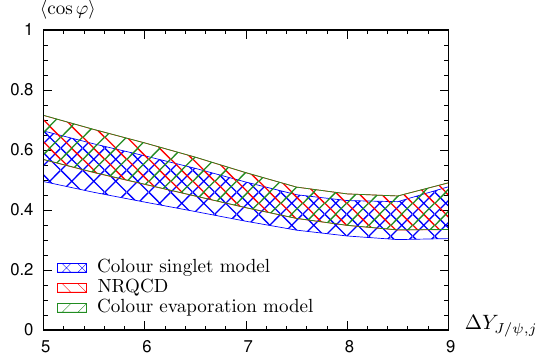}\\
	CMS+CASTOR ($-6.5 < y_j < -5 \, , \, 0 < y_{\jpsi} < 2.5 \, , \, P_{T, j} = P_{T, \jpsi}=10$ GeV)
	\caption[]
	{Cross section (Left) and $\langle \cos \varphi \rangle$ (Right) at $\sqrts = 13$ TeV as a function of the rapidity distance $\Delta Y_{\jpsi, j}$ between the $\jpsi$ and the jet as obtained in the BFKL approach in~\cite{Boussarie:2017oae}, for three different $\jpsi$ hadronisation models. [Figures adapted from~\cite{Boussarie:2017oae}.]}
	\label{fig:onium-jet}
\end{figure}

Several phenomenological studies have been conducted so far~\cite{Marquet:2007xx,Colferai:2010wu,Caporale:2012ih,Ducloue:2013bva,Ducloue:2013hia,Caporale:2013uva,Caporale:2014gpa,Colferai:2015zfa,Mueller:2015ael,Celiberto:2016ygs,Colferai:2016inu,Celiberto:2017ydk,Caporale:2018qnm} and have been found to be in fair agreement with data collected by the CMS collaboration~\cite{Khachatryan:2016udy}.
The theoretical description relies on the so-called \emph{hybrid factorisation}, where DGLAP ingredients are elegantly combined with the HE resummation. On the one hand, longitudinal momentum fractions of the two jets are assumed to be large enough so that the collinear factorisation applies, thus permitting a description of the incoming partons in terms of the usual PDFs. On the other hand, transverse momenta exchanged in the $t$-channel are not negligible due to the large rapidity interval in the final state, thus calling for a HE-factorised treatment, genuinely afforded by the BFKL approach. 

In line with these studies, the inclusive detection of a forward $\jpsi$ and a very backward jet in hadronic collisions at the LHC was recently proposed~\cite{Boussarie:2017oae} as a novel semi-hard channel (\cf{fig:semi-hard}, b). Here, at variance with most of the previous analyses, calculations are done in a hybrid HE + collinear factorisation with partial NLL BFKL accuracy, while different quarkonium  production mechanisms are at work. This study allows a probe of the dynamics of the HE resummation, its effect being emphasised by the smaller values of transverse momentum at which identified mesons can be tagged with respect to jets (thus heightening the weight of secondary, undetected gluons). At the same time, it offers an intriguing and complementary opportunity to probe different approaches for the description of the production of quarkonium states.

Predictions for the differential cross section and for the azimuthal correlation, $\langle \cos \varphi \rangle$, with $\varphi = \varphi_{\jpsi} - \varphi_j - \pi$, the difference of the azimuthal angles of both emitted objects, are presented in~\cf{fig:onium-jet} as a function of the rapidity interval, $\Delta Y_{\jpsi, j}$, between the $\jpsi$ and the jet at $\sqrts = 13$ TeV.
The meson is detected in the forward rapidity region of the CMS detector, $0 < y_{\jpsi} < 2.5$, while two possibilities are considered for the backward-jet emission: it can be tagged a) by CMS $-4.5 < y_j < 0$, or b) inside the ultra-backward CASTOR detector~\cite{CMS:2016ndp}, $-6.5 < y_j < -5$. Notably, case b) compensates %
for the smaller rapidities at which mesons can be detected (which represents a major drawback in the detection of a $\jpsi$ instead of a jet), thus restoring the rapidity intervals typical of the Mueller-Navelet jet production. Both the $\jpsi$ and jet \pT are required to be above 10~GeV.
The uncertainty bands combine the effect of the variation of the renormalisation and the factorisation scales, together with the running of the non-perturbative constants related to the hadronisation of the $\jpsi$. In particular, the CS LDME $\langle {\cal O}^{^3S^{[1]}_1}_{\jpsi} \rangle$ is varied between 1.16 and 1.32 GeV$^3$, as obtained in~\cite{Eichten:1995ch} and~\cite{Bodwin:2007fz} respectively. The CO LDME $\langle {\cal O}^{^3S^{[8]}_1}_{\jpsi} \rangle$ is varied between $0.224 \times 10^{-2}$ and $1.1 \times 10^{-2}$ GeV$^3$~\cite{Butenschoen:2011yh,Chao:2012iv,Bodwin:2014gia}.\footnote{Note that the NRQCD result presented in~\cite{Boussarie:2017oae} only takes into account the $3S^{[1]}_1$ CS and $^3S^{[8]}_1$ CO states and it should be kept in mind that the other CO states could also give a sizeable contribution.} In the CEM, the parameter $F_{J/\psi}$ represents the fraction of the $c\bar{c}$ pairs produced in the invariant mass range $[2m_c,2m_D]$ hadronising into $\jpsi$ mesons and it is varied between 0.02 and 0.04~\cite{Nelson:2012bc} (see also~\cite{Lansberg:2016rcx}).

The inspection of results in~\cf{fig:onium-jet} leads to significant cross sections that can be studied at the HL-LHC. In the NRQCD approach, the CO contribution prevails over the CS one~\cite{Boussarie:2017oae}, while the CEM exhibits a behaviour similar to the NRQCD (CS+CO) result. 
Azimuthal correlations show patterns very similar to the ones obtained for the Mueller-Navelet dijet and, in general, for all the semi-hard channels investigated so far: large rapidity intervals enhance the weight of undetected hard-gluon radiation, thus leading to a loss of correlation between the two final-state particles in the azimuthal plane.

Future studies will extend this work to: (\emph{i}) a full NLL BFKL analysis, (\emph{ii}) the integration of transverse momenta of the $\jpsi$ and the jet over kinematic ranges accessible at the LHC, (\emph{iii}) the evaluation of possible DPS effects~\cite{Ducloue:2015jba}.

\subsection{Single vector-$\Q$ exclusive photoproduction}
\label{sec:excl_quark}

Measurements of quarkonia in UPCs allow one to probe various parton distributions. In general, as discussed in Section~\ref{sec:dif_upc}, exclusive processes provide access to GPDs. At very low values of $x$ and $t$, the GPD can be related to the conventional integrated PDF, via the Shuvaev transform, as discussed in Section~\ref{sec:dif_gluonpdf}. While data collected at the LHC in the collider mode probes the low-$x$ region, data collected with a FT, as presented in Section~\ref{sec:dif_FT}, can constrain GPDs at high $x$. In both the low- and high-$x$ regions, measurements are scarce and hence the distributions currently suffer from large uncertainties. 

\begin{figure}[h!]
    \centering
    \includegraphics[scale=0.5]{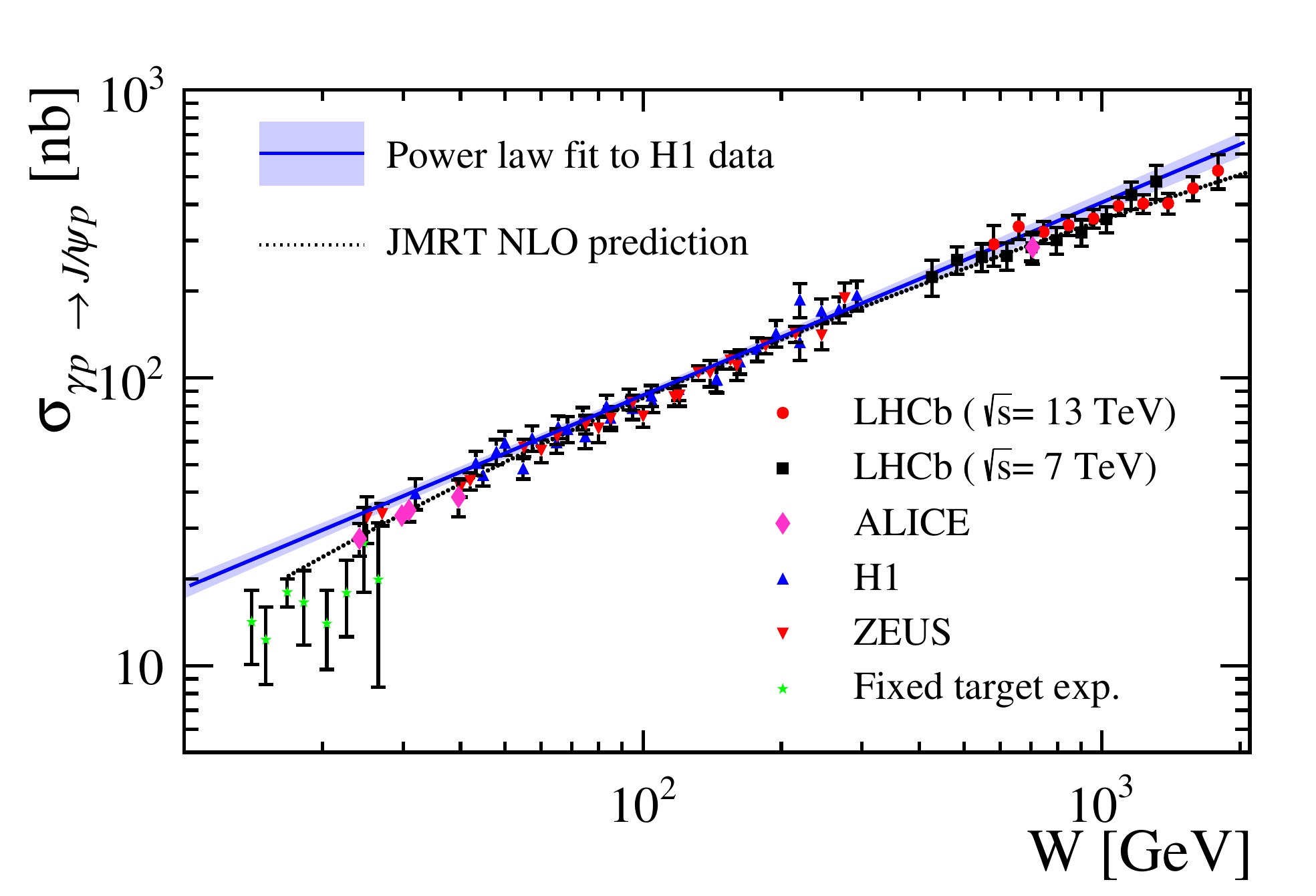}
\caption{Cross section for exclusive $\jpsi$ production as a function of 
the photon-proton \cm energy, $W$, extracted from data collected in proton-proton collisions by LHCb (red circles and black squares), proton-lead collisions by ALICE (magenta diamonds), $\ell p$ collisions by H1 and ZEUS (triangles), and FT experiments. [Figure taken from~\cite{Aaij:2018arx}.]}
\label{fig:ExclpgamJpsi}
\end{figure}

Exclusive production of $\jpsi$ and $\psip$ has been measured in \pp~\cite{Aaij:2013jxj,Aaij:2014iea,Aaij:2018arx}, \pPb~\cite{TheALICE:2014dwa,Acharya:2018jua}, and \PbPb~\cite{Abelev:2012ba,Abbas:2013oua,Adam:2015sia,Acharya:2019vlb,Khachatryan:2016qhq,LHCb-CONF-2018-003} collisions by the experiments at the LHC, in AuAu collisions at RHIC~\cite{Afanasiev:2009hy}, and in $p\bar{p}$ collisions at the Tevatron~\cite{Aaltonen:2009kg}. Exclusive production of $\ups$ has also been analysed in \pp~\cite{Aaij:2015kea} and in \pPb~\cite{Sirunyan:2018sav} collisions at the LHC. \cf{fig:ExclpgamJpsi} presents the $\gamma p$ cross section for exclusive $\jpsi$ production  as a function of the $\gamma p$ \cm energy, $W$,  extracted by the LHCb~\cite{Aaij:2018arx}, ALICE~\cite{TheALICE:2014dwa}, H1~\cite{Alexa:2013xxa}, ZEUS~\cite{Chekanov:2002xi}, and FT experiments. Good consistency over two orders of magnitude in energy is seen between photoproduction in diverse experimental conditions, which hints at the universality of the underlying physics. 

Exclusive production of vector quarkonia also has sensitivity to odderon production.  
In addition to the photon-pomeron fusion process shown in Fig.~\ref{fig:fd_exclusive}, vector quarkonia can be produced through odderon-pomeron fusion.  
It was shown in ~\cite{Bzdak:2007cz} that the odderon contribution may be significant at the LHC and that it may dominate at large transverse momenta. 
The two production mechanisms can therefore potentially be separated through the transverse momentum distribution. 
Although the precise shape of each spectrum is somewhat uncertain, an excess of events at high \pT could be evidence for the odderon.
One possibility for HL-LHC would be to measure the \pT spectrum precisely in pA collisions, where any odderon production is heavily suppressed with respect to photoproduction, and then to compare that to the spectrum obtained in pp collisions.  
The presence of proton-taggers would greatly assist this measurement as it would allow the major background in the high-\pT region due to proton dissociation (see Fig.~\ref{fig:iso}, Right)  to be heavily suppressed.

\subsubsection{Accessing GPDs from data collected in UPCs}
\label{sec:dif_upc}

Introduced more than 20 years ago~\cite{Ji:1996nm,Mueller:1998fv,Radyushkin:1997ki}, GPDs have been since studied both theoretically and experimentally. They provide access to the quark and gluon orbital angular momenta~\cite{Ji:1996ek}, the 3D distribution of quarks and gluons as a function of their longitudinal momentum and transverse position~\cite{Burkardt:2000za,Burkardt:2002hr}, and the distribution of pressure and shear forces inside the nucleon~\cite{Polyakov:2018zvc,Lorce:2018egm}. 

The channels to experimentally access GPDs are exclusive processes with a hard scale. Their extraction requires a measurement that is doubly differential in $x$ and $t$. So far GPDs have mainly been constrained in the high-to-medium $x$ region from measurements of deeply virtual Compton scattering (DVCS)~\cite{Airapetian:2001yk, Airapetian:2006zr, Airapetian:2008aa, Airapetian:2009aa, Airapetian:2009bm,  Airapetian:2011uq, Airapetian:2010aa, Airapetian:2012mq, Airapetian:2012pg, Akhunzyanov:2018nut, Camacho:2006qlk, Mazouz:2007aa, Defurne:2015kxq, Hattawy:2018liu, HirlingerSaylor:2018bnu, Hattawy:2017woc, Girod:2007aa, Stepanyan:2001sm, Aaron:2009ac, Aaron:2007ab, Chekanov:2008vy} and exclusive meson production in DIS~\cite{Airapetian:2010dh, Airapetian:2014gfp, Airapetian:2015jxa, Airapetian:2017vit, Alexakhin:2007mw, Adolph:2012ht, Adolph:2013zaa, Adolph:2014lvj, Adolph:2016ehf, Alexeev:2019qvd, Collaboration:2010kna, Park:2017irz, Bedlinskiy:2017yxe, Bosted:2016spx, Bosted:2016hwk, H1:2020lzc, H1:2015bxa, Aaron:2009xp, Abramowicz:2011pk, Chekanov:2007zr,Chekanov:2005cqa}, where the hard scale is provided by the large virtuality, $Q$, of the photon exchanged between the incoming lepton and nucleon.  Each of these processes provides complementary information, with a sensitivity to different types and flavour combinations of the GPDs. 

Instead of requiring a highly virtual incoming photon, a real photon can be used as a probe if the final-state particle is a heavy quarkonium (ideally $\Upsilon$), where now the hard scale is provided by the large mass of the quarkonium. Alternatively, GPDs can be probed in timelike Compton scattering (TCS), characterised by a real incoming photon and producing a highly virtual outgoing photon that provides the hard scale. The $ep$ collider experiments H1 and ZEUS measured the photoproduced heavy quarkonia, $\jpsi$ and $\Upsilon$~\cite{Alexa:2013xxa, Abramowicz:2011fa, Chekanov:2009zz, Chekanov:2004mw}, but did not have sufficiently large data samples to measure TCS. 

Hadron-hadron UPCs can also photoproduce quarkonia and TCS. The large \cm energy at the LHC offers the unique advantage of providing access to the very low-$x$ region, down to $x \approx10^{-6}$ (for photon virtualities of 1~GeV$^2$). In the case of heavy-ion UPCs, there is a further benefit compared to \pp\ or $\ell p$ collisions of an increased photon flux, since it is proportional to $Z^2$.  The cleanest extraction of GPDs at the HL-LHC would be obtained in \pA\ collisions, which necessitates a high luminosity for this collision mode due to the double-differential nature of the measurement.

Exclusive production of quarkonia (\cf{fig:fd_exclusive}, Left) is an ideal channel in UPCs to study gluon GPDs, since it is already sensitive to gluons at LO. In contrast, access to quark GPDs in UPCs is provided at LO by the TCS  process~\cite{Berger:2001xd,Boer:2015fwa}. At the same time, TCS shows some sensitivity to gluons due to NLO contributions, which are sizeable at the LHC~\cite{Moutarde:2013qs}. In the FT mode, with polarised and unpolarised targets, exclusive quarkonium production and TCS~\cite{Lansberg:2015kha} provide additional information to constrain gluon and quark GPDs, respectively.  Exclusive quarkonium measurements in FT collisions are discussed in more detail in Section~\ref{sec:dif_FT}.

In general, exclusive measurements in \pp\ collisions allow the study of nucleon GPDs, while the analysis of \AaAa\ collisions gives access to nuclear GPDs.
\pA\ collisions can access both nucleon and nuclear GPDs. Indeed, depending on the rapidity of the final-state particles, $\gamma p$ or $\gamma A$ interactions dominate~\cite{Guzey:2013taa}. Hence, with a  non-central detector, as for example LHCb, measurements in \pA\ and in \Ap\ collisions offer important complementary information.

Some caveats regarding the study of GPDs in UPCs should also be kept in mind. 
At present, there is still no all-order factorisation proof
of exclusive quarkonium production. In addition, higher-twist, higher-order, and mass corrections could play a sizeable role when evaluating the process amplitude.

\subsubsection{Probing the low-$x$ and low-scale gluon PDF with exclusive $\Q$ production}
\label{sec:dif_gluonpdf}

In~\cite{Flett:2019ept, Flett:2019pux}, the utility of the exclusive $\jpsi$ data, measured recently by the LHCb collaboration in the forward rapidity interval $2<y_\Q<4.5$, as a means of probing and ultimately determining the low-$x$ and low-$Q$ gluon PDF is discussed. To date, the exclusive data have not been included in global PDF analyses for two reasons. First, the underlying theory prediction within collinear factorisation at NLO for exclusive $\jpsi$ production suffered from a large scale uncertainty and exhibited poor perturbative stability. Second, 
one could not readily extract a PDF to compare to the $\overline{\text{MS}}$ collinear distributions determined in the global fits due to the off-forward kinematics and the description of the process via GPDs with the skew parameter $\xi$.
However, both of these problems have recently been overcome and the reader is pointed to~\cite{Jones:2015nna, Jones:2016ldq} and~\cite{Shuvaev:1999fm} for more details. 

At small $x$ and skewness $\xi$ values, one may relate the conventional collinear PDFs to the GPDs via the Shuvaev transform~\cite{Shuvaev:1999fm}. This approach exploits the observation that the evolution of the conformal moments of GPDs is similar to that of the Mellin moments of PDFs. The polynomiality in the series of $\xi$ of the conformal moments of the GPDs allows an identification of the leading term as the Mellin moments of the PDFs. In turn, one may then systematically construct all the conformal moments of the GPDs at small $\xi$ with an accuracy of $O(\xi^2)$ at LO. At NLO, the evolution becomes non-diagonal and the accuracy is lowered to $O(\xi)$. Still, for the diffractive processes of interest, this is more than adequate. 
Therefore, by virtue of the exclusive $\jpsi$ process sitting at low $x$ and at a low $Q$ scale, one can relate the underlying GPD inputs to the conventional PDFs.

After a systematic taming within the NLO result, amounting to a resummation of a class of large logarithms and implementation of a low $Q_0$ cut within the NLO coefficient function, the cross-section predictions utilising state-of-the-art NLO global parton fits describe the data well in the HERA region, yet produce vastly different results in the LHCb region, see~\cf{fig:chris_label} (Left) and~\cite{Flett:2019pux}. This large uncertainty {\it between} the global predictions is tantamount to the lack of data constraints for $x<10^{-3}$, where the global parton behaviour in the low $(x,Q^2)$ domain is based on extrapolating the input PDF Ansatz from larger $x$.  As shown in the right panel of~\cf{fig:chris_label}, the propagation of this, currently large, uncertainty at small $x$ to the exclusive $\jpsi$ cross section demonstrates the sizeable uncertainty for any given parton function set and provides support for the claim that the exclusive $\jpsi$ data are in a position to reliably constrain the low-$x$ gluon.

\begin{figure}[h]
    \centering
    \includegraphics[scale=0.450]{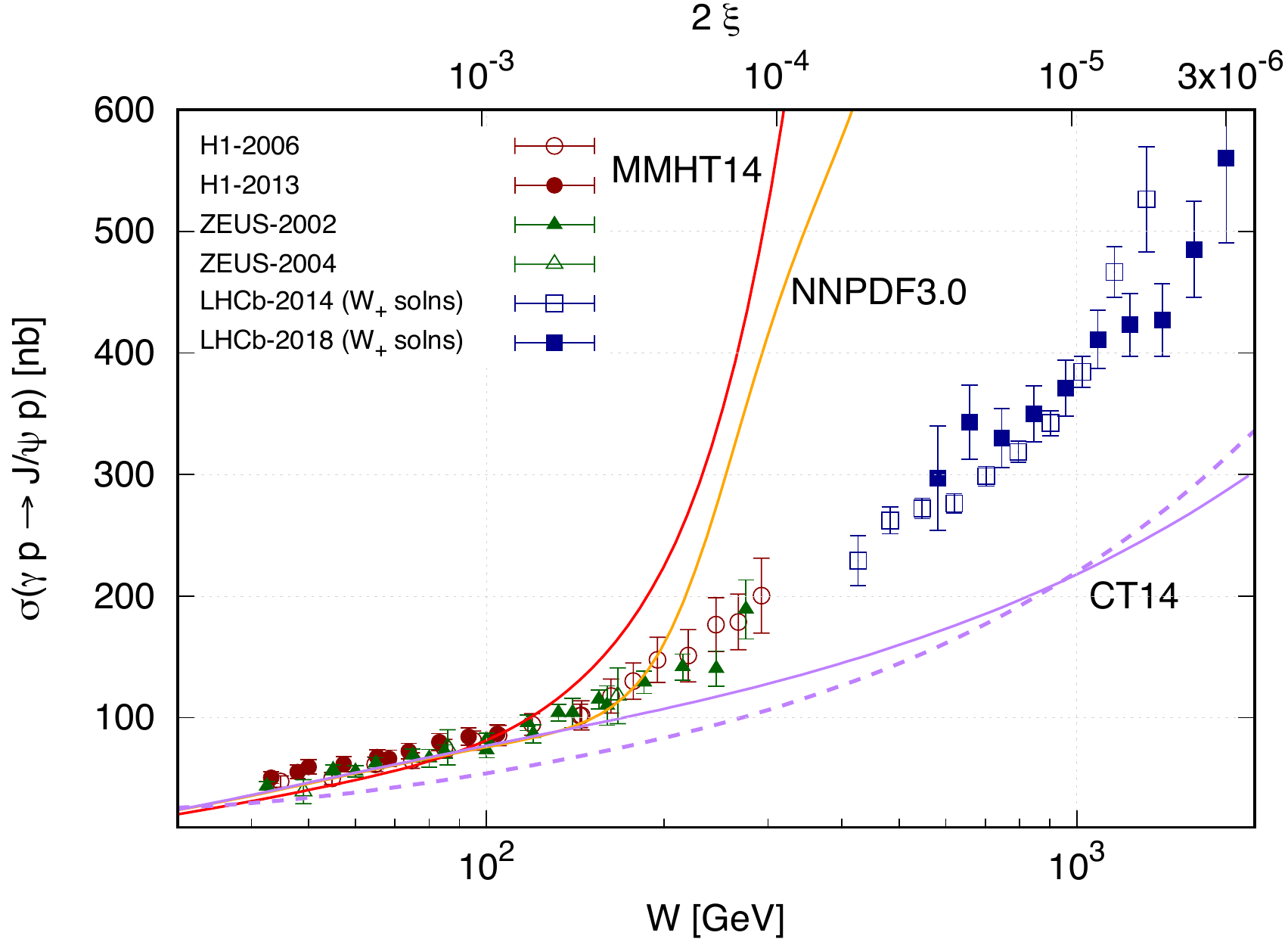}
    \qquad
    \includegraphics[scale=0.445]{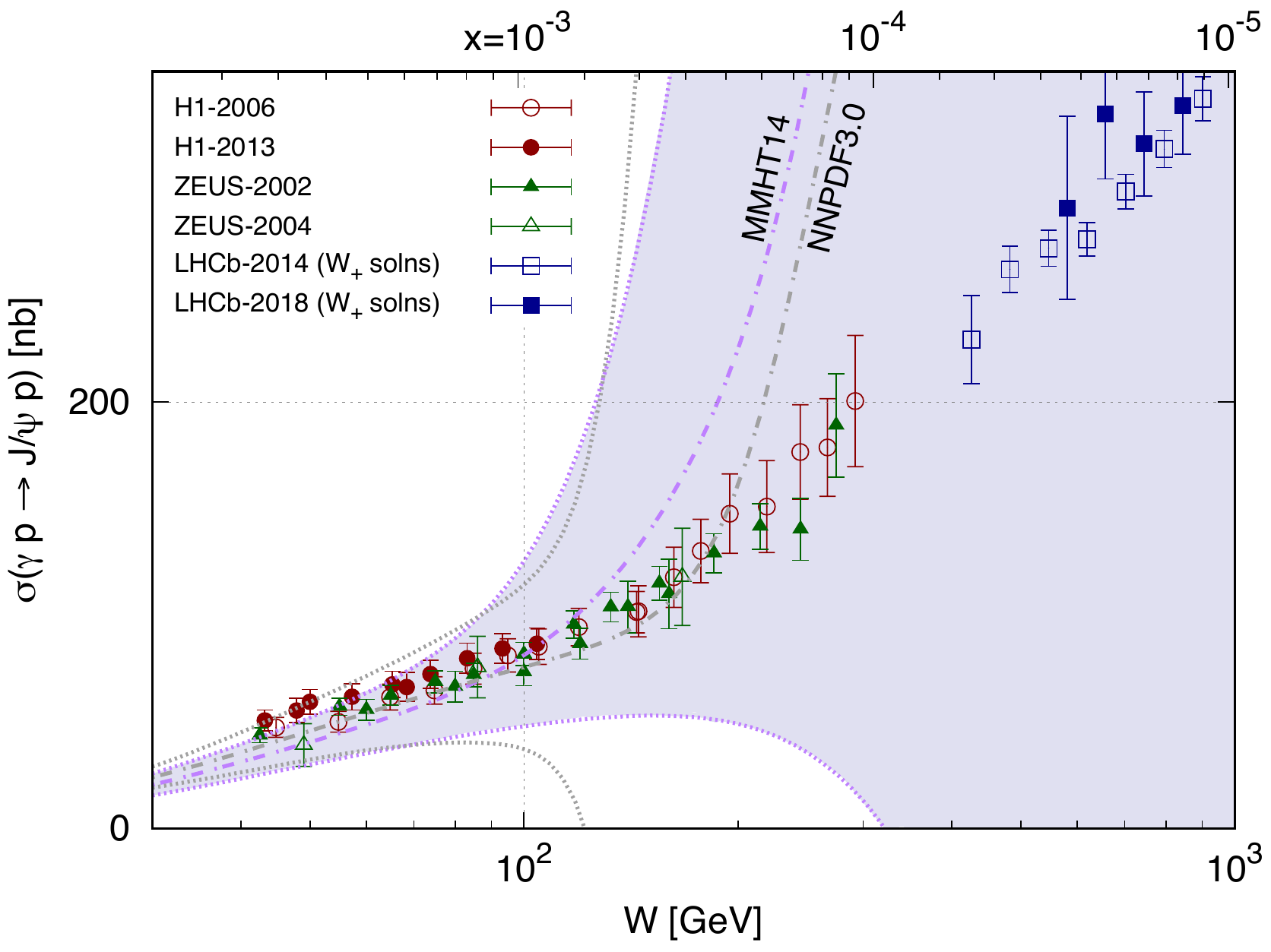}
    \caption{Cross sections for exclusive $\jpsi$ photoproduction in $ep$ and $pp$ collisions as a function of $\gamma p$ \cm\ energy. Left: Data compared to predictions using three distinct sets of global PDFs with scales $\mu_F^2 = \mu_R^2 = m_c^2$ (solid lines). Also shown for CT14 is the prediction with scales $\mu_F^2 = \mu_R^2 = 2m_c^2$ (dashed line), demonstrating the stability of the theory with respect to scale variations. Right: Data compared to two sets of global PDFs, showing the global PDF $1\sigma$ uncertainty, which greatly exceeds the experimental uncertainty. The data are from~\cite{Chekanov:2002xi,Chekanov:2004mw,Aktas:2005xu,Alexa:2013xxa} and the LHCb $W_+$ solutions are constructed from~\cite{Aaij:2014iea, Aaij:2018arx}. [Plots taken from~\cite{Flett:2019ept}.]}
    \label{fig:chris_label}
\end{figure}

For the future, there are three points of note regarding exclusive $\jpsi$ production via UPCs and the extraction of low-$x$ gluon PDFs.
Firstly, there is no indication of gluon density saturation down to $x=10^{-5}$ -- the data are compatible with a rising power law. With increasing data quality and statistics in the upcoming HL-LHC phase together with, in time, higher collision energies, some sensitivity to the effects of saturation might be seen.
Secondly, there is a need to reconcile the differing behaviour of the low-$x$ gluon PDF obtained from independent analyses using the inclusive ($D$-meson)~\cite{Gauld:2016kpd} and exclusive ($\jpsi)$~\cite{Flett:2020duk} sectors. It is unclear whether this is a question of data quality or whether the theory framework needs improving. Thirdly, measurements of the inclusive forward $C$-even charmonia ($\chi_{c0}$, $\chi_{c2}$, $\eta_c$) production, (and indeed bottomonia) integrated over $\pT$ are of high value. The NLO gluon can be probed down to approximately the same $x$ and $Q$ values as in exclusive $\jpsi$, but now in the conventional inclusive mode. From a phenomenological standpoint, it would be interesting to compare the low-$x$ gluons obtained from fits to scalar, vector, and tensor charmonia.

The same methodology applies to making NLO predictions for $\psip$ and $\ups$ production.  
For the latter, the scale dependencies are significantly reduced and the predictions are more robust (see \cite{Aaij:2015kea}).
However, the experimental precision for both is poorer due to lower statistics. This situation can be remedied at the HL-LHC.  
The ideal situation is to measure both these processes in high-luminosity \pA\ collisions.
At present, though, the anticipated integrated luminosity for this phase of running is probably not sufficient to make a competitive measurement.
Further studies are required in order to determine whether $\pp$ collisions could be used.  
The increase in luminosity at HL-LHC means an increased pileup of $\pp$ interactions, but it may still be possible to select exclusive $\psip$ and $\ups$ production through their characteristic signals of precisely two muons consistent with a primary interaction point and/or using forward proton tagging.

\subsubsection{FT measurements of $\Q$ photoproduction}
\label{sec:dif_FT}
UPCs are not only unique tools to study photoproduction processes with hadron beams in the collider mode, but also in the FT mode. LHC FT collisions can release a \cm energy of $\sqrts=115$ GeV with the LHC 7 TeV proton beam~\cite{Brodsky:2012vg}, giving access to the high-$x$ range of the parton distributions. 
FT collisions have already been achieved by the LHCb collaboration, thanks to its System for Measuring Overlap with Gas (SMOG)~\cite{Aaij:2018ogq}. In the upcoming Run~3 of the LHC, a new system (SMOG2) will be installed, consisting of a target cell, in which gas is injected in the centre and pumped out at both ends~\cite{smog2_a,Bursche:2649878}. The HL-LHC plans to have an upgraded polarised SMOG system. It is worth noting that the ALICE collaboration is also studying the feasibility to conduct a FT programme after Run~3 using a solid target coupled to a bent crystal to deflect the beam halo~\cite{Galluccio:2671944}. 

FT measurements with an unpolarised target in general access spin-independent (TMD) PDFs, GPDs, and Wigner distributions, while polarised targets access different spin-dependent objects. Exclusive quarkonium measurements in polarised FT collisions are discussed in~\cite{Lansberg:2018fsy}.  With transversely polarised protons, the measurement of exclusive photoproduction of vector quarkonia is sensitive to the gluon GPD, $E_{g}$, which in turn allows, in principle, a determination of the gluon orbital angular momentum, $L_{g}$. Both $L_{g}$ and $E_{g}$ are currently essentially unknown. 
 
 With an integrated yearly luminosity of only 150~pb$^{-1}$ foreseen for Run~3 at the LHC, one could expect to produce about 3000 $\jpsi$ in the LHCb acceptance to perform preliminary studies of the multi-dimensional gluon content of the proton.   Similar studies can in principle be conducted with a Pb beam on a H gas target but would only produce a few tens of $\jpsi$~\cite{Lansberg:2018fsy}.

\cf{fig:FixedTarget} shows projections for HL-LHC, for an LHCb-like detector.  These assume an integrated luminosity of 10~fb$^{-1}$, corresponding to the maximum luminosity that can be obtained in a year of running at the LHC in $p$H collisions using a storage cell gas target with a longitudinal dimension of 1~m. The left panel shows the differential cross section as a function of $y_\psi$ in the laboratory frame before and after applying kinematic cuts for the LHCb region, covering a rapidity between 2 and 5. The top $x$ axis shows the photon-proton \cm\ energy, $W_{(\gamma p)}$, while the right $y$ axis shows the yearly yield per 0.1 rapidity unit. A yearly yield of about $2\times10^5$ in the di-muon decay channel is expected.  The right panel of \cf{fig:FixedTarget} shows the projection of the single transverse-spin asymmetry (STSA) ($A_{N}$) of the $\jpsi$  as a function of Feynman $x_{\rm F}$, for two ranges in \pT and one year of FT-LHC data taking in $p$H collisions. The asymmetry can be measured with an absolute precision ranging from 1 to 4$\%$, making possible a first measurement sensitive to $E_{g}$ in FT mode at the LHC with a polarised target, likely before the EIC.

Projections also exist for the ALICE detector operated in the FT mode for HL-LHC, assuming the usage of a polarised gas system with a storage cell technology.%
The maximum yearly integrated luminosities considered are mainly limited by the detector rate capabilities and amount to 260~pb$^{-1}$ in $p$H collisions (polarised or unpolarised), leading to a photoproduced $\jpsi$ yield of about 1300 $\jpsi$ in the ALICE muon spectrometer. The statistics are even more scarce in the ALICE central barrel, which is covering the very backward rapidity region, at the edge of the $\jpsi$ phase space.

\begin{figure}[h!]
    \centering
    \includegraphics[scale = 0.3]{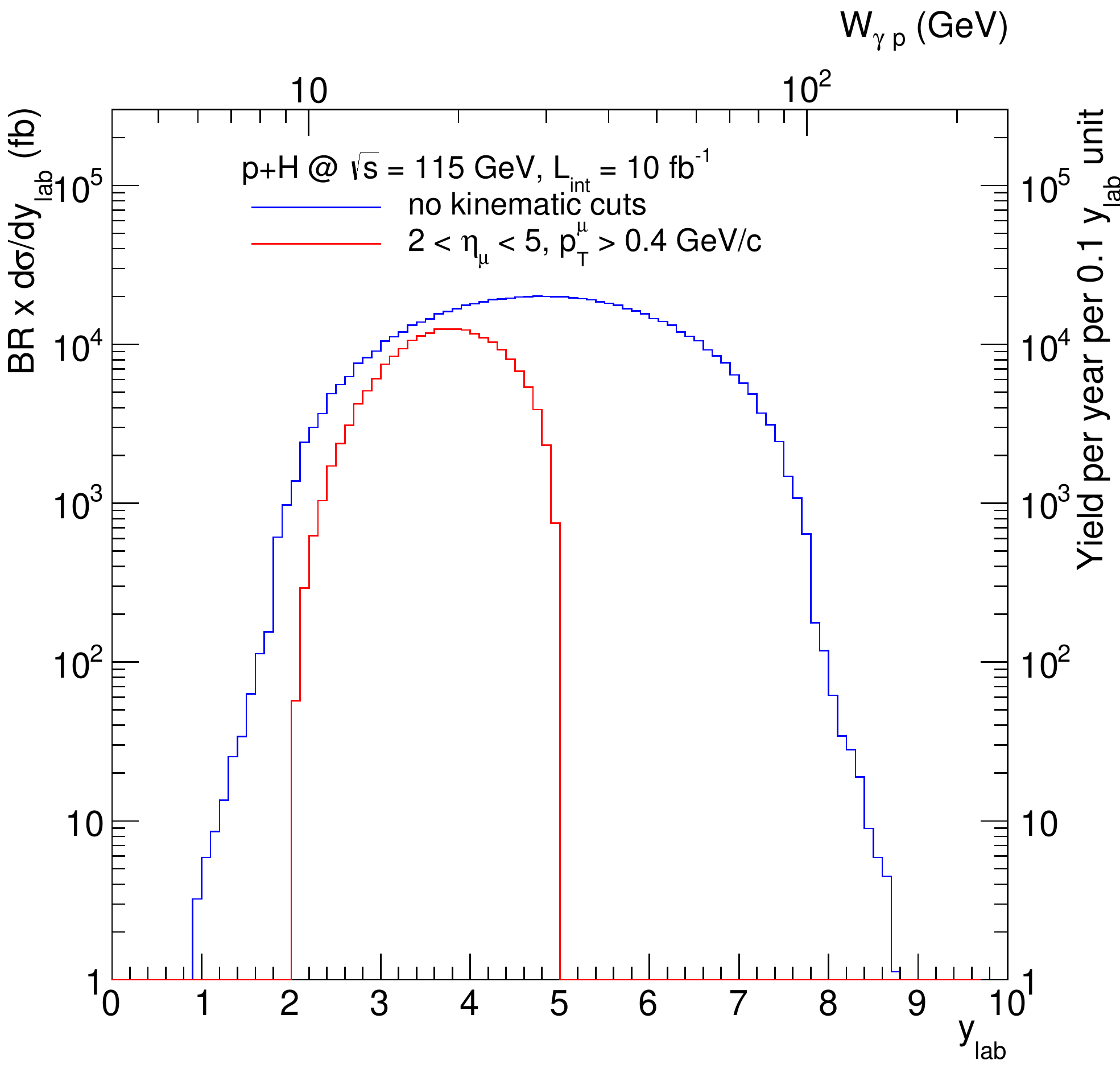}
    \hspace{1 true cm}
    \includegraphics[scale = 0.45]{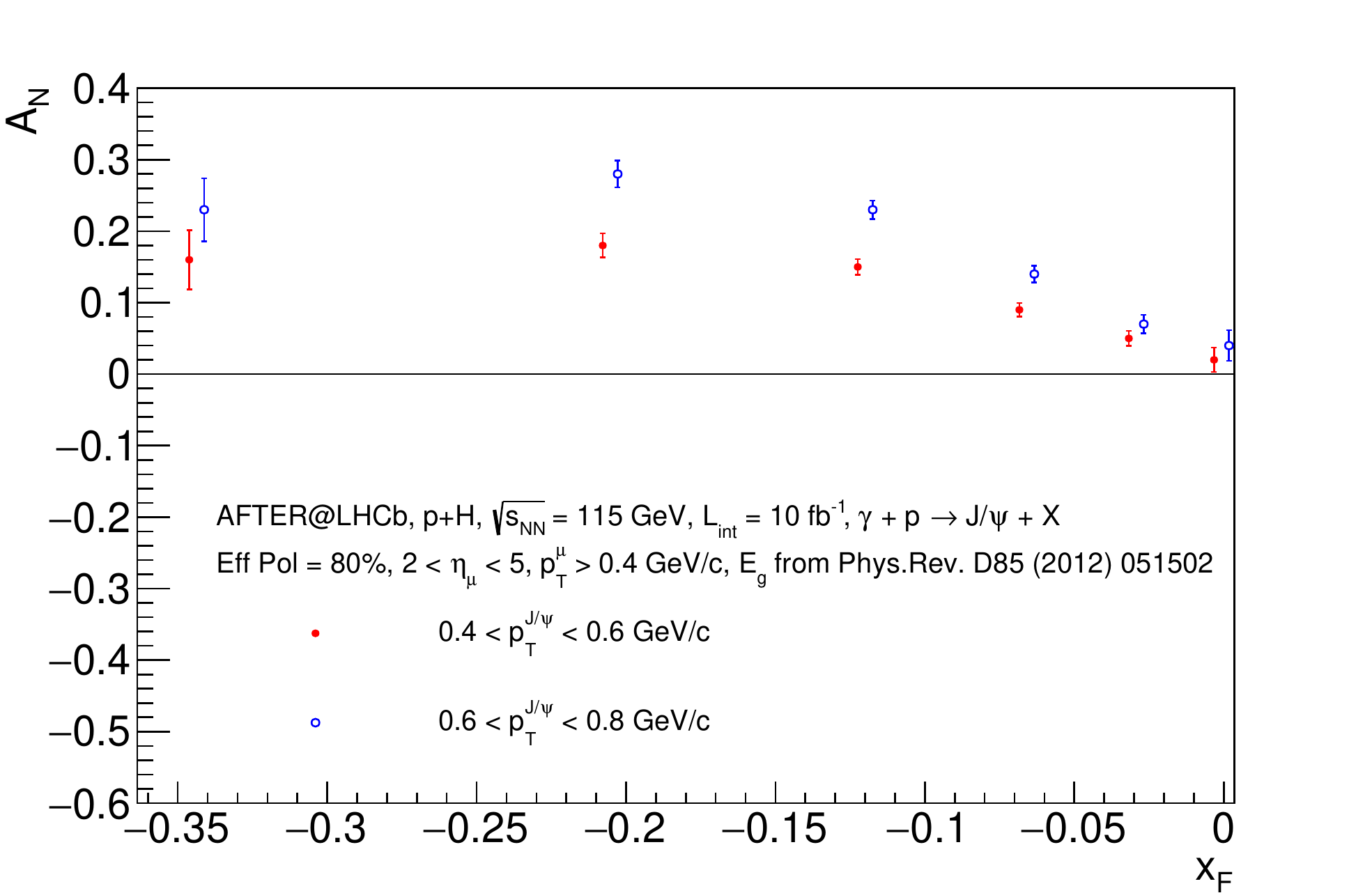}   
    \caption{Rapidity-differential $\jpsi$ photoproduction cross section predicted by the STARLIGHT Monte Carlo generator~\cite{Klein:2016yzr} (Left), and projected $\jpsi$ $A_{N}$ distribution as a function of $x_F$ (Right), for FT $p$H collisions at $\sqrts$~=~115~GeV, assuming the performances of an LHCb-like detector and a polarised target with 80$\%$ effective polarisation. [Figures taken from~\cite{Lansberg:2018fsy}.]}
    \label{fig:FixedTarget}
\end{figure}

\subsection{Accessing Wigner functions through $\Q$-pair production}
\label{sec:dif_wig}

The 1D PDFs, and the 3D TMDs and GPDs, each  describing different aspects of the non-perturbative structure of hadrons, can all be derived from the more general GTMDs~\cite{Meissner:2009ww,Lorce:2013pza}. There are several compelling reasons to study GTMDs. Firstly, they contain more physics content than that encoded in the TMDs and GPDs, and thus allow an exploration of physics that is lost in taking the TMD/GPD kinematic limits. Secondly, GTMDs can be related, via Fourier transformations, to Wigner functions, which are the quantum-mechanical version of classical phase-space distributions encountered in statistical physics. Partonic Wigner functions may allow for a hadron tomography in 5D phase space~\cite{Belitsky:2003nz,Lorce:2011dv}. Thirdly, certain GTMDs can unravel unique correlations between the parton orbital motion and the spin of hadrons~\cite{Lorce:2011kd,Hatta:2011ku,Lorce:2014mxa}. Fourthly, there is a particular GTMD that is related to the Sivers TMD (see also Section~\ref{sec:polarised}). By establishing the equivalence between GTMDs and QCD-odderons at small $x$, the authors in~\cite{Boussarie:2019vmk} have shown that it is possible to access the gluon Sivers TMD through exclusive $\pi^{0}$ production in  \textit{unpolarised} $ep$ scattering. This finding goes against the traditional belief that the Sivers function can only be accessed via a transversely polarised target.

\begin{figure}[h!]
\includegraphics[width =  0.46\textwidth]{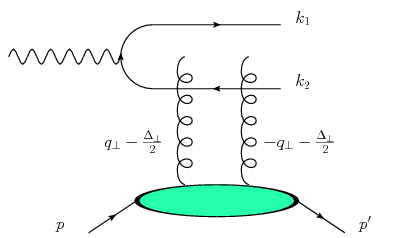}
\hspace{1cm}
\includegraphics[width =  0.38\textwidth]{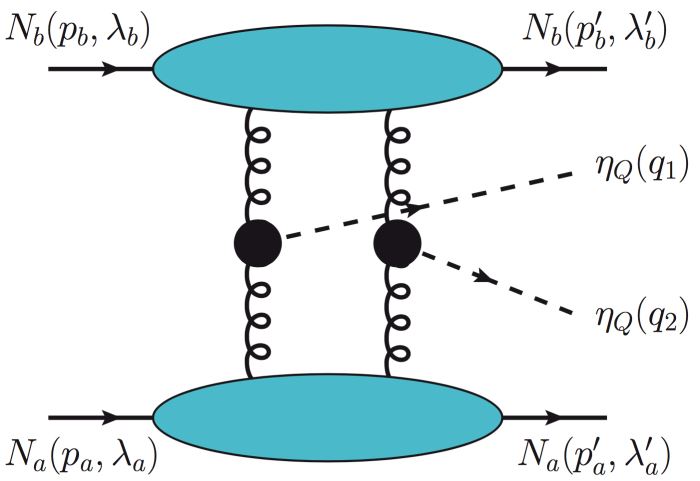}
\caption{Left: Leading-order Feynman diagram for exclusive dijet production in lepton-nucleon/nucleus scattering. Right: Leading-order Feynman graph for the exclusive double quarkonium production in nucleon-nucleon collisions. The perturbative subprocess $gg \rightarrow \eta_{Q}$ is computed in the CSM. The $\eta_{Q}$ meson has a transverse momentum that is determined by the (intrinsic) transverse momentum of the gluons through which it is produced.}
\label{fig:fd_wig}
\end{figure}

For a long time, it was questionable whether GTMDs could be measured at all. The authors of~\cite{Hatta:2016dxp} were the first to suggest the measurement of gluon GTMDs through exclusive dijet production in lepton-nucleon/nucleus collisions at small $x$ (\cf{fig:fd_wig}, Left). Given that the GTMDs depend on the transverse vectors $\vec{q}_{T}$ (partonic transverse momentum) and $\vec{\Delta}_{T}$ (the Fourier transform of the impact-parameter $\vec{b}_{T}$), it is possible to parameterise the angular correlation between these two vectors by a symmetric and an antisymmetric part. The latter, known as the elliptic distribution, has a characteristic $\cos(2\phi)$ angular modulation, where $\phi$ is the angle between the relative transverse momenta of the dijets and the recoiling nucleon/nucleus. This angular modulation is similar to the observed elliptic flow phenomenon in relativistic heavy-ion collisions~\cite{Zhou:2016rnt,Hagiwara:2017ofm,Iancu:2017fzn}.
This pioneering work gave impetus to the field of GTMDs and subsequently many other interesting ideas were put forward~\cite{Hatta:2016aoc,Ji:2016jgn,Hagiwara:2017fye,Bhattacharya:2017bvs}, among which is the ability to access gluon GTMDs in exclusive photoproduction in \pA\ collisions.

In~\cite{Bhattacharya:2018lgm} it was shown that the production of two pseudoscalar quarkonia (\eg\ $\eta_c\eta_c$, $\eta_c\eta_b$ or $\eta_b\eta_b$) in exclusive hadronic collisions, where both hadrons remain intact, is a direct probe of GTMDs for gluons at moderate $x$ (\cf{fig:fd_wig}, Right).  
In that work, an observable sensitive to gluon (canonical) orbital angular momentum was identified. A similar idea came out in~\cite{Boussarie:2018zwg}, where the authors proposed to access gluon GTMDs at small $x$ via the production of pairs of $C=+1$ quarkonia ($\eta_c$ or $\chi_{cJ}$) in exclusive \pp\ collisions, where one of the protons breaks up after the collision. Although the latter has a larger count rate, there may be contamination from NRQCD CO contributions. An alternative suggestion would be to consider either a combination of a $\jpsi$ and a $C$-even quarkonium~\cite{FengYuan}, or possibly the associated production of two $\jpsi$ in \pp collisions at the LHC, as opposed to two $C$-even quarkonium states. The production mechanisms for these processes with one or two $\jpsi$, in the GTMD-type of kinematics, is expected to be more complicated. Nonetheless, given the experimental ease with which $\jpsi$ states can be detected, theoretical efforts in this direction are surely warranted.

From the experimental side, exclusive pairs of charmonia have been studied in \pp\ collisions by the LHCb experiment~\cite{Aaij:2014rms}, but the statistical precision is insufficient to extract the cross section as a function of the \pT\ of the quarkonia. The CMS collaboration recently performed~\cite{CMS:2020ekd} a preliminary measurement of exclusive dijets in PbPb collisions: the $\cos(2\phi)$ modulation between the sum of the transverse momenta of the jets and their difference was extracted, which for the first time provides the information relevant for the determination of the GTMDs. This first experimental access to GTMDs can be extended in the future to similar measurements but for pairs of quarkonia.
As discussed for PDFs and GPDs, dedicated measurements during the HL-LHC should thus provide access to the Wigner distributions of the proton.

\newcommand{\sT}{{\scriptscriptstyle T}}
\newcommand{\C}{{\cal C}}
\newcommand{\fone}{f_1^g}
\newcommand{\hone}{h_1^{\perp g}}
\newcommand{\fonet}{\tilde{f}_1^g}
\newcommand{\honet}{\tilde{h}_1^{\perp g}}
\newcommand{\kTsqav}{\langle k_\sT^2 \rangle}
\newcommand{\Qav}{\langle M_{\Q\Q} \rangle}
\newcommand{\Qavy}{\langle M_{\psi\psi} \rangle}
\newcommand{\qTy}{P_{\psi\psi_T}}
\newcommand{\qTu}{P_{\Upsilon\Upsilon_T}}
\newcommand{\qTq}{P_{\Q\Q_T}}
\newcommand{\CS}{{\rm CS}}
\newcommand{\cnf}[1]{\langle \cos({#1}\phi_\CS) \rangle}
\newcommand{\koneT}{\bm{k}_{1\sT}}
\newcommand{\ktwoT}{\bm{k}_{2\sT}}
\newcommand{\seff}{\s_{\text{eff}}}
\newcommand{\Cff}{{\C[\fone\fone]}}
\newcommand{\Chh}{{\C[w_2\hone\hone]}}
\newcommand{\Cfh}{{\C[w_3\fone\hone]}}
\newcommand{\Chf}{{\C[w_3'\hone\fone]}}
\newcommand{\Cwhh}{{\C[w_4\hone\hone]}}
\newcommand{\Snp}{{S_{{\rm NP}}}}
\newcommand{\bmax}{{b_{T_{\max}}}}
\newcommand{\blim}{{b_{T_{\lim}}}}
\newcommand{\bT}{\bm{b}_\sT}
\newcommand{\dif}{{\mathrm{d}}}
\newcommand{\kn}{k_{\sT}}
\newcommand{\konen}{k_{1\sT}}
\newcommand{\ktwon}{k_{2\sT}}
\newcommand{\qn}{q_\sT}
\newcommand{\bn}{b_\sT}

\section{Transverse-Momentum-Dependent effects in inclusive reactions\protect\footnote{Section editors: Miguel G. Echevarria, Vato Kartvelishvili.
}}
\label{sec:spin}

The multi-dimensional structure in momentum space of a nucleon has recently attracted much interest, as a possible source of many observable effects in hadronic interactions and, more generally, as a way of improving our understanding of QCD. This structure can be parameterised in terms of several objects, which encode different correlations between the momentum and spin of a parton and its parent nucleon. In simple terms, these are three-dimensional generalisations of the usual one-dimensional, collinear PDFs or FFs, but with a dependence on the parton transverse momentum, \kT. The way to introduce and define such generalisations is a subject of intense investigations, and debates, within the community~\cite{Angeles-Martinez:2015sea}. 
What is at stake here is a crucial step in our understanding of the nucleon 3D structure (in momentum space), and hence in our understanding of colour confinement in QCD. 
In what follows, we will discuss four approaches that address transverse-momentum-dependent and/or spin effects and are all relevant for quarkonium studies at the HL-LHC:
\begin{itemize}
\item The TMD factorisation framework, applicable both in unpolarised and polarised collisions, in which the TMDs have a concrete definition in terms of gauge-invariant operators and properties such as QCD evolution; 
\item The High-Energy (HE) or \kT factorisation framework, designed to account for HE effects (large $\sqrt{s}$). 
Besides the transverse momentum of the initial partons, \kT, this formalism also considers their virtualities, which naturally becomes relevant in this limit;
\item The collinear twist-3 (CT3) factorisation framework, which is an extension of collinear factorisation to treat polarised parton/nucleon collisions, and which matches TMD factorisation in the large-\kT limit;
\item The Generalised Parton Model (GPM), a phenomenological model meant to extend collinear factorisation with functions accounting for the Sivers effect both in the quark and gluon sectors.
\end{itemize}

It should be clear to the reader that these approaches are not meant to be considered on an equal footing: 
the GPM computations are restricted to polarised collisions, but more importantly they are essentially descriptive. 
Yet, they can be very useful to check various hypotheses about the underlying phenomena generating the spin asymmetries. 
CT3 predictions go further with a deeper connection to the QCD properties but are based on collinear considerations where the transverse-momentum effect are integrated over in higher-twist correlators. HE factorisation, only applied to unpolarised collisions so far, is first designed to treat new effects at large \sqrts.  As such, care should be taken when using its predictions when \sqrts\ is not very large, in particular for systems or conditions where TMD factorisation is a priori not applicable.
Indeed, the latter, while being probably the most inclusive in terms of phenomena generated by the \kT\ of the partons, is also the most restrictive in terms of applicability owing to its ambition to be the most rigorous.

The purpose of this section is to outline the recent progress regarding quarkonium production in processes where the transverse-momentum-dependent gluon effects enter, and how the HL-LHC can contribute to this emerging research domain.

The TMD factorisation framework is briefly introduced in Section~\ref{sec:TMD_factorisation}, followed by a discussion in Section~\ref{sec:TMDchallenges} on several specificities and open issues related to the treatment of quarkonium production, while HE factorisation is treated in Sections~\ref{sec:HEfactorisation} and~\ref{sec:HEchallenges}.
Section~\ref{sec:unpolarised} focuses on  various-quarkonium production processes in unpolarised \pp\ collisions within the TMD factorisation framework, with a special focus on the unpolarised and the linearly-polarised gluon TMDs, $f_1^g$ and $h_1^{\perp g}$. 
In Section~\ref{sec:beyond_TMD}, we address the complex issue of factorisation-breaking effects or, more generally, effects beyond TMD and HE factorisations, and discuss some easily measurable processes where they can be studied.
Finally, in Section~\ref{sec:polarised}, collisions with polarised nucleons are considered; these become measurable at the HL-LHC with a polarised target in the FT mode, allowing one to measure STSAs in quarkonium production to probe \eg the gluon Sivers effect accounted for by the TMD and CT3 factorisations and the GPM.

\subsection{TMD factorisation in the gluon sector}
\label{sec:TMD_factorisation}
In the last few years, the field of TMDs has taken a large leap forward. 
Both the theoretical framework~\cite{GarciaEchevarria:2011rb,Echevarria:2012pw,Echevarria:2012js,Echevarria:2014rua,Collins:2011zzd,Echevarria:2015uaa,Scimemi:2018xaf} and the phenomenological analyses (see \eg~\cite{DAlesio:2014mrz,Echevarria:2014xaa,Bacchetta:2015ora,Bacchetta:2017gcc,Anselmino:2016uie,Scimemi:2017etj,Bertone:2019nxa,Echevarria:2020hpy,Bury:2020vhj}) have developed, including new, higher-order perturbative calculations (see \eg~\cite{Gutierrez-Reyes:2019rug,Gutierrez-Reyes:2018iod,Vladimirov:2017ksc,Echevarria:2016scs,Echevarria:2015byo,Echevarria:2015usa,Bacchetta:2018dcq}).
This progress, however, has been made mainly in the quark sector, with the gluon sector lagging behind due to the difficulty in cleanly probing gluons in high-energy processes.

Gluon TMDs at the leading twist, first analysed and classified in~\cite{Mulders:2000sh}, are shown in \ct{tab:gluon_TMDs}, in terms of both the polarisation of the gluon itself and of its parent hadron. 
The distribution of unpolarised gluons inside an unpolarised hadron, $f_1^g$, and of circularly polarised gluons inside a longitudinally polarised hadron, $g_1^g$, correspond (\ie\ are matched at large \kT through an operator product expansion) to the well-known collinear unpolarised and helicity gluon PDFs respectively. 
The distribution of linearly-polarised gluons in an unpolarised parton, $h_1^{\perp g}$, is particularly interesting, since it gives rise to spin effects even in collisions of unpolarised hadrons, like at the LHC.
The Sivers function, $f_{1T}^{\perp g}$, which encodes the distribution of unpolarised gluons in a transversely-polarised nucleon, has a very important role in the description of STSAs.
There is a classification analogous to \ct{tab:gluon_TMDs} for quark TMDs, and also for both quark and gluon TMD FFs, which are as relevant as TMD distributions for processes which are sensitive to the role of transverse dynamics of partons in the fragmentation process.

{
\renewcommand{\arraystretch}{1.7}

 \begin{table}[hbt!]
\centering
 \hspace{1cm} gluon polarisation \\ \vspace{0.1cm}
 \rotatebox{90}{\hspace{-1.5cm} nucleon polarisation} \hspace{0.1cm}
 \begin{tabular}[c]{|m{0.5cm}|c|c|c|}
 \hline 
 & $U$ & circular & linear \\
 \hline 
 $U$ & $f_{1}^{g}$ & & \textcolor{blue}{$h_{1}^{\perp g}$} \\
 \hline	
 $L$ & & $g_{1}^{g}$ & \textcolor{red}{$h_{1L}^{\perp g}$} \\
 \hline	
 $T$ & \textcolor{red}{$f_{1T}^{\perp g}$} & \textcolor{blue}{$g_{1T}^{g}$} & \textcolor{red}{$h_{1}^{g}$}, \textcolor{red}{$h_{1T}^{\perp g}$} \\
 \hline
  \end{tabular}
 \caption{Gluon TMD PDFs at twist 2. 
 $U$, $L$, $T$ describe unpolarised, longitudinally polarised and transversely-polarised nucleons. 
 $U$, `circular', `linear' stand for unpolarised, circularly polarised and linearly-polarised gluons. 
 Functions in blue ($h_{1}^{\perp g}$, $g_{1T}^{g}$) are $T$-even. 
 Functions in black ($f_{1}^{g}$, $g_{1}^{g}$) are $T$-even and survive integration over the parton \kT. 
 Functions in red ($h_{1L}^{\perp g}$, $f_{1T}^{\perp g}$,$h_{1}^{g}$, $h_{1T}^{\perp g}$)   are $T$-odd.}
 \label{tab:gluon_TMDs}
 \end{table}

}

As is the case for quark TMDs, gluon TMDs contain information on the initial- and/or final-state QCD interactions of the incoming hadron. {Different types of gluon TMDs exist, distinguished by the precise structure of the gauge links in their operator definition, which depends on the hard process under consideration: the two most common are the so-called Weizs\"acker-Williams (WW) and dipole (DP) types~\cite{Dominguez:2011wm,Buffing:2013kca,Boer:2016bfj}}. 
The WW type involves either initial- or final-state interactions, while the DP type involves both, so different processes probe different types of gluon TMDs.
Incidentally, the WW type is the gluon TMD  which, in the proper choice of gauge, can be written as the gluon number operator acting on the hadron Fock state, implying the physical interpretation of the TMD as a number density.

{Exploratory} analyses on gluon TMD distributions~\cite{Lu:2016vqu,Mulders:2000sh,Pereira-Resina-Rodrigues:2001eda} were done in the so-called \emph{spectator-model} approach.
Originally developed for studies in the quark-TMD sector~\cite{Bacchetta:2008af,Bacchetta:2010si,Gamberg:2005ip,Gamberg:2007wm,Jakob:1997wg,Meissner:2007rx}, this relies on the assumption that the struck nucleon emits a gluon, after which the remnants are treated as a single spectator particle, taken on-shell.
The power of the spectator-model framework lies in the possibility to concurrently generate all TMD densities at twist-2~(\ct{tab:gluon_TMDs}).
In this context, a novel parameterisation for $T$-even distributions has been recently proposed in~\cite{Bacchetta:2020vty}. 
At variance with previous studies, the spectator mass is allowed to take a continuous range of values weighted by a flexible spectral function, which allows one to effectively reproduce both the small- and the moderate-$x$ behaviour of the TMDs. 
Furthermore, it embodies the effect of $q\bar{q}$ contributions, which are generally neglected by standard spectator models.
These results on the 3D tomography of (un)polarised gluons inside (un)polarised nucleons are part of the effort to gain a deeper understanding of observables sensitive to gluon TMD dynamics.

So far, quarkonium-production observables are one of the most promising tools at our disposal to access gluon TMDs, since they are directly sensitive to gluons.
These processes are quite challenging from the theoretical point of view, because they involve several momentum scales and require dealing with different aspects of QCD, from the formation of heavy-quark bound states to soft-gluon resummation. 
For this reason, they represent a wonderful testing ground of 
our knowledge of QCD.
Indeed, the interest has grown lately, with a number of LO analyses assuming TMD factorisation (see \eg\ \cite{Godbole:2012bx,Boer:2012bt,Godbole:2013bca,Godbole:2014tha,Zhang:2014vmh,Zhang:2015yba,Mukherjee:2015smo,Mukherjee:2016qxa,Mukherjee:2016cjw,Boer:2016bfj,Lansberg:2017tlc,Godbole:2017syo,DAlesio:2017rzj,Rajesh:2018qks,Bacchetta:2018ivt,Lansberg:2017dzg,Kishore:2018ugo,Scarpa:2019fol,Boer:2020bbd,DAlesio:2019gnu})
and others that perform NLO calculations (see \eg\ \cite{Ma:2012hh,Ma:2014oha,Ma:2015vpt}).

Experimental information on gluon TMDs is however very limited. The first attempt~\cite{Lansberg:2017dzg} to fit unpolarised gluon TMD PDF, $f^g_1$, was only made in 2017 and was performed using $J/\psi$ pairs. So far nothing is known on $h^{\perp g}_1$.
The possible extension of this first fit with forthcoming LHC data, as well as other quarkonium channels of interest, will also be discussed.  

\subsection{TMD factorisation in $\Q$ production: challenges and opportunities}
\label{sec:TMDchallenges}

As discussed in Section~\ref{sec:pp}, besides NRQCD, other approaches have been proposed to describe quarkonium production, like the CSM~\cite{Chang:1979nn,Berger:1980ni,Baier:1981uk} or the CEM~\cite{Halzen:1977rs,Halzen:1977im,Fritzsch:1977ay}, and their variations
and extensions~\cite{Ma:2016exq,Haberzettl:2007kj,Lansberg:2005pc,Khoze:2004eu}.
All these frameworks have been routinely used along with collinear factorisation.
Whereas a factorisation proof exists for NRQCD and collinear factorisation, it does not exist at present for the other approaches.
Their combination with TMD factorisation is then potentially even more delicate.
This is why, in this section, we only consider the combination of the NRQCD and TMD factorisations and 
some adjustments are needed to properly deal with the low-\pT region.

If one wishes to predict \pT spectra, NRQCD factorisation is only applicable  when the quarkonium is produced with a relatively large $\pT \gtrsim 2 m_Q$. Intuitively, in this kinematic regime, emissions of soft and ultra-soft gluons from the heavy-quark pair cannot alter the large \pT\ of the quarkonium. 
Ignoring these soft emissions, the quarkonium \pT is then determined by the short-distance reactions. 
At the same time, the infrared (IR) divergences that remain from the hard scattering are absorbed into the non-perturbative LDMEs and collinear PDFs. 
However, when quarkonia are produced with small \pT, large double logarithms arise and need to be resummed.
Indeed, the observed (low) $\pT$ distribution of $\ups$ production at the Tevatron and LHC was found to be consistent with the prediction from a TMD factorisation Ansatz with resummation of the large double logarithms~\cite{Berger:2004cc,Sun:2012vc,Qiu:2017xbx} (even if a simple NLO treatment seems to provide fairly good results as well~\cite{Artoisenet:2008fc}).

In any case, the key point here is that, at low \pT, the soft-gluon factorisation assumption must be abandoned.
In~\cite{Beneke:1997qw, Fleming:2003gt, Fleming:2006cd}, the quarkonium production in $\ell p$ collisions and $e^+e^-$ annihilation was studied in the endpoint region, which is sensitive to soft radiations exactly where NRQCD factorisation breaks down.
It was found that promoting the LDMEs into quarkonium shape functions is necessary to accurately account for soft radiation from the heavy-quark pair.
Similarly, for the TMD spectrum of quarkonia, TMD-shape functions are needed to be able to rigorously derive the relevant factorisation theorems at low \pT.

The degrees of freedom for studying such low-\pT processes were introduced in the context of soft-collinear effective theory in~\cite{Echevarria:2019ynx, Fleming:2019pzj}, where it was shown that the cross section for quarkonium production at low \pT involves a new kind of non-perturbative object besides the TMDs, which can be seen as the 3D extension of the well-known NRQCD LDMEs.
These TMD-shape functions, like the LDMEs, scale with the relative velocity, $v$, of the heavy quark-antiquark pair in the quarkonium rest frame.
Therefore, the factorisation turns out to be a simultaneous expansion in the relative quark-pair velocity $v$ and $\pT/(2m_Q)$.

Currently, a few of open questions remain with regard to the factorisation:
\begin{itemize}
\item 
The double expansion in $\pT/(2m_Q)$ and the heavy-quark pair relative velocity, $v$, allows for a priori sub-leading contributions in one expansion parameter, which might be enhanced in the other.
Thus the reorganisation of terms in the cross section becomes non-trivial, and a potential contribution of higher-twist TMDs and TMD-shape functions cannot be discarded.

\item 
This approach involves a summation over the various colour and angular-momentum configurations that contribute to the formation of the bound state. This might spoil the factorisation in \pp\ collisions when CO states are produced.
This is due to the so-called Glauber or Coulomb exchanges, which are a subset of soft gluons that can entangle initial and final states and thus prevent the factorisation.
At the moment, such a factorisation has only been established for $\eta_Q$ production~\cite{Echevarria:2019ynx}, where the CS state dominates the production process, following the NRQCD velocity-scaling rules.
It might be extended to other processes dominated by the CS channel, like di-quarkonium or associated production.
This represents an opportunity to study effects in QCD that connect long and short distance physics.
\item 
In addition to these issues specific to quarkonium production, one should keep in mind that, in hadronic collisions, the final state must not explicitly involve {coloured} objects for TMD factorisation to apply~\cite{Collins:2007nk,Collins:2007jp,Rogers:2010dm,Rogers:2013zha,Gaunt:2014ska,Schwartz:2018obd}. 
Thus, it is not supposed to hold for $\psi$ or $\Upsilon$ hadroproduction in the CSM
where the quarkonium is necessarily produced along with a hard gluon. 
On the one hand, if this hard gluon is not observed, the connection between the quarkonium \pT and the initial-parton \kT is lost. 
On the other, if, instead, one proposes to measure the associate production of $\psi$ or $\Upsilon$ with a jet (or hadron), the observed final state is coloured, and the colour flow arising from the reaction becomes so entangled that it prevents one from deriving a factorised form for the cross section 
(see however~\cite{Boer:2014lka}).
In other words, and as already mentioned in the first point, Glauber exchanges play a role and spoil the factorisation.

\end{itemize}

\subsection{The HE factorisation framework}
\label{sec:HEfactorisation}

The aim of High-Energy (HE) factorisation --also called $k_T$ factorisation or the Parton Reggeisation Approach (PRA)-- is to go beyond collinear factorisation by resumming corrections to the hard-scattering coefficient which are enhanced by powers of $\log(1/z_{\pm})$ when $z_{\pm}$ gets small. 
As $z_{\pm}=q_{\pm}/k_{\pm}$, with $q_{\pm}$ the light-cone components of the momentum of the studied final state and $k_{\pm}$ those of the initial parton, such corrections indeed get large at high \sqrts. 
The general HE factorisation formula for the inclusive cross section $d\sigma$ of a hard process in $pp$ collisions~\cite{Gribov:1984tu,Collins:1991ty,Catani:1994sq} can be outlined as a double convolution (in the momentum fraction $x_i$ and in the transverse momentum $k_{Ti}$ of both incoming partons) of a partonic cross section $d\hat{\sigma}_{ij}$ with two unintegrated PDFs (UPDFs) or gluon densities (UGDs).

When one refers to uPDFs, one considers that they are obtained from the convolution of an evolution factor ${\cal C}_{ij}$, which performs the resummation and satisfies some version of the %
BFKL equation~\cite{Fadin:1975cb,Kuraev:1976ge,Kuraev:1977fs,Balitsky:1978ic}, and a collinear PDF $f_{j/p}$.
The complete cancellation of the $\mu_F$ dependence will happen only if the collinear PDFs with small-$x$-resummed DGLAP evolution are used, see \eg~\cite{Altarelli:1999vw,Ball:2017otu}. 
This however does not prevent one from using the usual PDFs if the observable under consideration shows more sensitivity to the transverse momenta $k_{T1,2}$ than to the $x$ dependence of the UPDF.

On the contrary, as far as the concept of UGD is concerned, no collinear input is implied. These are written as a convolution of the BFKL gluon Green's function and a non-perturbative proton \emph{impact factor} which is meant to be determined by data. They have been the subject of intense studies since the early days both in exclusive and inclusive channels. Originally employed in the study of DIS structure functions~\cite{Hentschinski:2012kr,Hentschinski:2013id}, the UGD has  been studied through exclusive diffractive vector-meson leptoproduction~\cite{Anikin:2009bf,Anikin:2011sa,Besse:2013muy,Bolognino:2018rhb,Bolognino:2018mlw,Bolognino:2019bko,Bolognino:2019pba,Celiberto:2019slj} measured at HERA~\cite{Aaron:2009xp,Adloff:2002tb}, single--bottom-quark production~\cite{Chachamis:2015ona} at the LHC, inclusive forward Drell--Yan di-lepton production~\cite{Motyka:2014lya,Brzeminski:2016lwh,Motyka:2016lta,Celiberto:2018muu} measured {by}  LHCb~\cite{LHCb:2012fja}{, and exclusive $\psi$ and $\ups$ photoproduction~\cite{Bautista:2016xnp,Garcia:2019tne,Hentschinski:2020yfm}}.
Recent analyses on the diffractive electroproduction of $\rho$ mesons~\cite{Bolognino:2018rhb,Celiberto:2019slj} have corroborated the underlying assumption~\cite{Ivanov:1998gk} that the small-size dipole-scattering mechanism is at work, thus validating the use of the UGD formalism, which holds when the observable \pT is large.

In contrast to TMD factorisation, HE factorisation has the advantage not to be limited to the low-$\qT$ region (compared to the relevant hard scale of the process).
Indeed, large values of $\pT$ also contribute to the region $z_{\pm}\ll 1$ if additional radiation is highly separated in rapidity from the observed system. 
Radiative corrections of this kind become important with increasing $\sqrts$, since more phase space for such emissions opens up. 
This difference in their range of applicability is often a source of confusion and debate, especially because sometimes the acronym TMD is also used outside the scope of TMD factorisation. 
In our discussion, when such objects are discussed we will always use their names like $f^g_1$ or $h^{\perp g}_1$.

However, HE factorisation has its own theoretical shortcomings compared to TMD factorisation. 
In general, not all corrections beyond Next-to-Leading Logarithmic approximation\footnote{The N${}^{k}$LL-approximation in the context of HEF is defined as the resummation of terms $\sim \alphas^{n+k}\log^n(1/z_{\pm})$.} can be taken into account by the standard HEF formulation.
This can be traced back to the fact that, even at the leading power in the HE limit, $z_{\pm}\ll 1$, QCD amplitudes only admit a factorisation in terms of matrix elements of multiple light-like Wilson lines with a complicated colour structure or, equivalently, in terms of multi-Reggeon exchanges in the $\hat{t}$-channel (see \eg~\cite{Caron-Huot:2013fea} for a review). 
In order to take all such contributions arising from multi-Reggeon exchanges into account, results from the CGC formalism can be incorporated~\cite{Altinoluk:2019fui} in a factorised formula inspired by TMD factorisation. 
However, in the phenomenology at the leading twist, it is usually assumed that the largest N${}^{k\geq 1}$LL-corrections can still be represented by an effective UPDF that takes into account both DGLAP and BFKL effects. 
Numerous recipes to obtain such UPDF can be found in the literature, such as Kimber--Martin--Ryskin--Watt (KMRW) UPDF~\cite{Kimber:2001sc,Watt:2003mx,Watt:2003vf}, Collins--Ellis--Bluemlein UPDF~\cite{Collins:1991ty,Blumlein:1995eu}, Parton-Branching Method~\cite{Martinez:2018jxt} and many more.

The coefficient function $d\hat{\sigma}$ at LO in $\alphas$ and at leading-power in $z_{\pm}$ can be understood as a partonic cross section involving off-shell (Reggeised) initial-state partons with virtualities $k_{1,2}^2=-{\bf k}_{T1,2}^2$. 
For simple processes, such as $g^\star(k_1)+g^\star(k_2) \to Q\bar Q$, it can be computed by usual QCD Feynman Rules with the following replacement for the polarisation vectors of initial-state gluons: $\varepsilon^\mu(k_{1,2})\to k^\mu_{T1,2}/|{\bf k}_{T1,2}|$. 
However, there is no analogous simple rule for off-shell quarks in the initial state. 
In addition, for more general QCD processes, such coefficient function will not be gauge-invariant. 
The coefficient function for any subprocess can be computed to any order in $\alphas$ using the effective field theory (EFT) for multi-Regge processes in QCD~\cite{Lipatov95, LipatovVyazovsky, AntonovFRs} and its gauge invariance is guaranteed by construction within the EFT. 
The formalism of~\cite{vanHameren:2012if,vanHameren:2013csa,vanHameren:2016kkz} is equivalent to the EFT at tree level. 
Hereafter we will refer to all these approaches, like \kT factorisation and Parton Reggeisation Approach (PRA)~\cite{Karpishkov:2019vyt}, as HE factorisation.

\subsection{High-Energy factorisation in $\Q$ production: challenges and opportunities}
\label{sec:HEchallenges}

The HE factorisation coefficient functions for inclusive heavy-quarkonium production in NRQCD at LO were first computed in~\cite{Hagler:2000dd,Hagler:2000eu,Kniehl:2006sk,Kniehl:2006vm} and the relevance of the gluon off-shellness in $\chi_{c1}$ production to {lifting} the Landau-Yang suppression was first highlighted in~\cite{Hagler:2000dd}.    
LDMEs from recent fits on hadroproduction data~\cite{Saleev:2012hi, Nefedov:2013qya,Baranov:2019lhm,Karpishkov:2020wwe} are comparable to those obtained at NLO in collinear factorisation, especially for the LDME of the ${}^3S_1^{(8)}$ state, while the LDMEs of ${}^3P_J^{[8]}$ and ${}^1S_0^{[8]}$ states turn out to have the same order of magnitude as in collinear factorisation, but often with an opposite sign.
This is because LO HE factorisation calculations do not take into account NLO corrections due to final-state radiation effects.

Recently, HE factorisation has been used together with the formalism of CS Light-Front Wave Functions (LFWFs)~\cite{Babiarz:2020jkh,Babiarz:2019mag} and the Improved CEM (ICEM)~\cite{Cheung:2018upe, Cheung:2018tvq, Maciula:2018bex} to describe {the} bound-state formation. 
The CS LFWF calculation shows an interesting discrepancy with the strict non-relativistic approximation (see \eg Figs.~10 and~11 of~\cite{Babiarz:2020jkh}), which points towards potentially large relativistic corrections. 
The ICEM calculation somewhat counter-intuitively predicts mostly unpolarised production of charmonia~\cite{Cheung:2018upe} and bottomonia~\cite{Cheung:2018tvq} at high $\pT$, unlike \eg the NRQCD-factorisation-based predictions of~\cite{Kniehl:2016sap}. 
This disagreement uncovers some interesting aspects of the physics of heavy-quarkonium polarisation in the ICEM and its interplay with HE factorisation that deserve further study.
  
All the calculations mentioned above have been performed at LO. 
So far, no NLO quarkonium studies exist in HE factorisation, which are far more complex than in collinear or TMD factorisations. 
However, such NLO computations would be in some respects equivalent to an NNLO accuracy for collinear factorisation, which are in fact not yet available for heavy-quarkonium production in none of the aforementioned production models.
With such NLO computations at our disposal, it will also become possible to quantitatively characterise the region of applicability of HE factorisation in quarkonium production, where NLO corrections would be under control.
 
As regards advances towards first NLO computations, the reader is guided to~\cite{Nefedov:2019mrg} for progress in the computations of loop corrections, where the recent progress towards the automation of the computation of gauge-invariant HE factorisation amplitudes reported in~\cite{vanHameren:2017hxx} would certainly be beneficial for the completion of the real-emission-correction computations. 
Exploratory NLO calculations have recently been successfully performed~\cite{Nefedov:2020ecb,Hentschinski:2020tbi} and these show that one can overcome the problem of large unphysical NLO corrections found, for instance, in BFKL-based computations.
All these developments make NLO HE factorisation calculations possible in the near future, with the aim of describing more accurately a variety of observables related to single and associated production of heavy quarkonia in different quarkonium-production models.  
Confronting the results of these calculations with HL-LHC data, which will briefly be discussed in Section~\ref{sec:beyond_TMD}, will allow one to learn more about, on the one hand, the quarkonium-production mechanisms and, on the other, the relevance of HE phenomena in these reactions.

\subsection{Unpolarised TMD studies with $\Q$ at the HL-LHC }
\label{sec:unpolarised}

As already mentioned, inside an unpolarised proton one can define two independent gluon TMD densities: the unpolarised $f_{1}^{g}$ and the linearly-polarised $h_{1}^{\perp g}$  distributions~\cite{Mulders:2000sh,Meissner:2007rx,Boer:2016xqr}.
Being time-reversal even ($T$-even), these TMDs can be nonzero even in (sub)processes where neither initial-state nor final-state interactions are present. 
However, like all other TMDs, they are affected by such interactions, which can render them process-dependent and even hamper factorisation.

The distribution of linearly-polarised gluons has attracted much attention in the last few years. 
It corresponds to an interference between $+1$ and $-1$ gluon helicity states, which can be different from zero if the gluon \kT is taken into account. 
If sizeable, this TMD can affect the \pT distributions of scalar and pseudoscalar particles produced in the final state, such as, for instance, $H^0$ bosons or $C$-even charmonium and bottomonium states.
Interestingly, it turns out that at small $x$, the linearly-polarised distribution may reach its maximally allowed size, bounded by the unpolarised-gluon density~\cite{Mulders:2000sh}. 
Moreover, linearly-polarised gluons can also be generated perturbatively from unpolarised quarks and gluons inside the proton~\cite{Nadolsky:2007ba,Catani:2010pd}. 
This determines the large-\kT tail of the distribution~\cite{Sun:2011iw}.

{From the experimental point of view, in contrast to quark TMDs, almost nothing is known about gluon TMDs, due to the lack of processes, like single-inclusive DIS (SIDIS) and Drell--Yan pair production, that directly probe them.}
A Gaussian-shape extraction of the unpolarised gluon TMD has recently been performed, based on the LHCb measurement of the \pT spectra of $\jpsi$ pairs~\cite{Lansberg:2017dzg}, the first of its kind. 

Many proposals have been put forward to access TMDs in \pp\ collisions, mainly by looking at azimuthal asymmetries and \pT distributions for quarkonium  production.

The quarkonium processes for which one can hope TMD factorisation to hold -- with NRQCD properly modified -- are
\begin{itemize}
    \item $p\,p\to \eta_{c,b}+ X$~\cite{Boer:2012bt},
    \item $p\,p\to \jpsi+ \gamma+ X$ and $p\,p\to \Upsilon+ \gamma+ X$~\cite{Dunnen:2014eta},
    \item $p\,p\to \jpsi + \ell\, \bar \ell + X$ and $p\,p\to \ups+ \ell\, \bar \ell + X$~\cite{Lansberg:2017tlc},
    \item $p\,p\to \eta_c + \eta_c+ X$~\cite{Zhang:2014vmh},
    \item $p\,p\to \jpsi+  \jpsi+ X$ and $p\,p\to \ups +\ups+ X$~\cite{Lansberg:2017dzg,Scarpa:2019fol},
\end{itemize}   
at LO in $v$, thus only considering the CS contributions.
The reason to focus on these CS processes is to avoid the presence of final-state interactions which, together with the initial-state interactions present in \pp\ collisions, would lead to the breakdown of TMD factorisation~\cite{Collins:2007nk,Collins:2007jp,Rogers:2010dm,Rogers:2013zha,Gaunt:2014ska,Schwartz:2018obd}. 
The case of $p\,p\to \chi_{0c, b}+ X$ or  $p\,p\to \,\chi_{2 c,b}+ X $~\cite{Boer:2012bt} is particular since CS and CO appear at the same order in $v$, which is a likely source of complication. 
As such, we will come back to it in Section~\ref{sec:beyond_TMD} when discussing considerations beyond the strict TMD factorisation.

Among these quarkonium reactions, we should make a distinction between single and associated production. 
Whereas the former {is} probably simpler to analyse, it does not  allow the scale of the process to be tuned by increasing the invariant mass of the produced system. Consequenlty, there is not much room for TMD factorisation to apply as one is forced to be in the region $\pT \lesssim 2 m_Q$.
In addition, single-quarkonium production only provides an indirect way to probe $h_{1}^{\perp g}$ through \pT\ modulations, as {it} does not offer the possibility of accessing the azimuthal asymmetries generated by the linearly-polarised gluons. 
Finally, during the HL-LHC period, these single-quarkonium cross sections, though much larger than {for} associated production, will be extremely complicated to measure with the ATLAS and CMS detectors in the applicability region of TMD factorisation. In contrast,  the increased luminosity available at the HL-LHC will make the associated-production channels more accessible. 
However, single low-\pT\ quarkonia can probably be studied in the much less hostile environment of FT-LHC by the LHCb and ALICE detectors.  All these aspects will be addressed in the following three subsections.

At this stage, it is important to note that the unpolarised and linearly-polarised gluon distributions to be extracted from the above-mentioned reactions, which correspond to the WW distributions in the small-$x$ limit, are expected~\cite{Boer:2016fqd} to be the same as those entering (open and closed) heavy-quark-pair production in $ep$ collisions. 
This represents an important test of the universality of the gluon TMDs inside unpolarised protons, which can only be performed by comparing data from \pp\ and $ep$ colliders. 
On the other hand, the consideration of processes where the TMD factorisation is not supposed to hold can be very valuable in advancing our understanding of long-distance correlations in QCD, by quantifying the actual role of the expected factorisation-breaking contributions. 
This will also be addressed in Section~\ref{sec:beyond_TMD}.

\subsubsection{Single low-\pT  $C$-even $\Q$ production}
\label{sec:TMDssinglequarkonium}

Single-quarkonium production offers the possibility of constraining both unpolarised and linearly-polarised gluon TMDs~\cite{Boer:2012bt}, even if the hard scale is set by the mass of the bound state and thus the room for TMD factorisation to work is limited. 
Leaving aside the complications of TMD-shape functions pointed out in~\cite{Echevarria:2019ynx, Fleming:2019pzj}, which should be properly taken into account to perform quantitatively consistent phenomenological analyses, the analysis of their (low) \pT spectra up to roughly half their mass can of course give information about the \kT dependence of the unpolarised TMD $f_{1}^{g}$ at the scale 3 GeV (10 GeV) for the charmonium (bottomonium), but also on the distribution of the linearly-polarised gluons, $h_1^{\perp g}$, which modulate the quarkonium \pT spectrum. Estimations of these modulations and of the $x$ range where they can be accessed at the LHC in the collider and FT modes are given in~\ct{t:processes1}. Estimated rates are also given to illustrate that they can be measured, provided that the detectors can cope with the background at low \pT.

\begin{table}[hbt!]
\begin{center}\renewcommand{\arraystretch}{1.2}
\resizebox{\textwidth}{!}{
\begin{tabular}{c|c|c|c|c|c}
$\Q$& \shortstack{expected yield \\ ($\sqrt{s}=115$~GeV)} & \shortstack{expected yield \\ ($\sqrt{s}=14$~TeV)} & \shortstack{$x_2$ range \\ ($\sqrt{s}=115$~GeV)} & \shortstack{$x_2$ range \\ ($\sqrt{s}=14$~TeV)} & Low-\pT modulation \\
\hline
\hline
$\eta_{c}$ & ${\cal O}(10^{5\div 6})$ & ${\cal O}(10^{6\div 7})$ & $0.02\div 0.5$ & $10^{-6}\div 3\!\cdot\!10^{-5}$ & $0\div 80 \%$~\cite{Boer:2012bt,Signori:2016jwo} \\
\cline{1-5}
$\eta_{b}$ & ${\cal O}(10^{1\div 2})$ & ${\cal O}(10^{3\div 4})$ & $0.1\div 1$ & $5\!\cdot\!10^{-6}\div 10^{-4}$ & $0\div 80 \%$~\cite{Boer:2012bt,Signori:2016jwo,Echevarria:2015uaa} \\
\cline{1-5}
$\chi_{c0}(1P)$ & ${\cal O}(10^{3\div 4})$ & ${\cal O}(10^{4\div 5})$ & $0.02\div 0.5$ & $10^{-6}\div 3\!\cdot\!10^{-5}$ & $0\div 80 \%$~\cite{Boer:2012bt}  \\
\cline{1-5}
$\chi_{c2}(1P)$ & ${\cal O}(10^{5\div 6})$ & ${\cal O}(10^{6\div 7})$ & $0.02\div 0.5$ & $10^{-6}\div 3\!\cdot\!10^{-5}$ & $<1\%$~\cite{Boer:2012bt} \\
\cline{1-5}
$\chi_{b0}(nP)$ & ${\cal O}(10^{1\div 2})$ & ${\cal O}(10^{3\div 4})$ & $0.1\div 1$ & $5\!\cdot\!10^{-6}\div 10^{-4}$ & $0\div 80 \%$~\cite{Boer:2012bt} \\
\cline{1-5}
$\chi_{b2}(nP)$ & ${\cal O}(10^{2\div 3})$ & ${\cal O}(10^{4\div 5})$ & $0.1\div 1$ & $5\!\cdot\!10^{-6}\div 10^{-4}$ & $<1\%$~\cite{Boer:2012bt} \\ \hline
\end{tabular}
}
\caption{%
Expected $\pT$ modulations generated by $h_{1}^{\perp g}$ for a selection of quarkonium-production observables, along with the expected yields and $x_2$ ranges derived from $x_2=M e^{-y_\cm}/\sqrt{s}$ for a rapidity coverage $-2.8<y_\cm<0.2$ for FT mode at $\sqrt{s}=115$~GeV and $2<y_\cm<5$ for collider mode at $\sqrt{s}=14$~TeV.}
\label{t:processes1}
\end{center}
\end{table}

However, we should stress that, in principle, TMD factorisation is only supposed to hold for $\eta_Q$ production. 
The measurement of scalar and tensor $\chi_{c,b}$ states is essential to get a complete picture, but their low-\pT spectra are subject to specific factorisation-breaking effects~\cite{Ma:2014oha} (see Section~\ref{sec:beyond_TMD}).

\subsubsection{$\Q+\Q$ production}
\label{sec:psi-psi-TMDs}

Pair-production of quarkonia at the LHC is in large majority from gluon fusion~\cite{Qiao:2009kg,Lansberg:2013qka}, even down to the energy of the FT mode~\cite{Lansberg:2015lva,Hadjidakis:2018ifr}. 
Thus, they enable the study of the gluon content of the proton with low contamination from quark-induced contributions. 
As seen in Section~\ref{sec:pp} and to be discussed again in Section~\ref{sec:dps}, the hadroproduction of quarkonium pairs can be initiated~\cite{Kom:2011bd,Lansberg:2014swa,Lansberg:2019adr} by SPS or by DPS.

Only the SPS component is of interest here to probe TMDs in gluon fusion. 
It is thus important to control the potential contamination from DPS. 
At low rapidity separations, $\Delta y_{\Q\Q}$, and when the invariant mass of the pair, $M_{\Q\Q}$, increases, the relative contribution of DPS gets so low that it becomes a minor source of uncertainty. 
Near the threshold, \ie\ the region so far measured by LHCb~\cite{Aaij:2016bqq}, the DPS contribution should be subtracted, which calls for a good understanding of its kinematic distribution.

In addition, increasing $M_{\Q\Q}$, as in the data samples of ATLAS~\cite{Aaboud:2016fzt} and CMS~\cite{Khachatryan:2014iia}, allows one to probe higher transverse-momentum of the pair, $P_{\Q\Q_\sT}$, while remaining in the region of applicability of TMD factorisation. $\jpsi$-pair and $\ups$-pair production were already studied several times by LHC collaborations with various setups, although these studies were not designed for the extraction of information regarding gluon TMDs {(see Section~\ref{sec:onium_pair_pp})}.
Increasing the samples of $\jpsi$ pairs would allow for the measurement of double- or even triple-differential cross sections, which are much more suitable for the extraction of gluon TMDs without diluting their effects. 
More data on di-$\Upsilon$ production would allow one to probe gluon TMDs at similar masses of the pair, but in a different system with different feed-down, DPS or $v$-correction contamination.

To highlight the importance of measuring azimuthal modulations, it is instructive to note that the differential cross section of the process of $\Q\Q$ production via gluon-gluon fusion has the general form~\cite{Lansberg:2017dzg}: 
\begin{align}&
\frac{d \sigma}{d M_{\Q\Q} d Y_{\Q\Q} d^2 \bm{P}_{\Q\Q_\sT} d \Omega} 
\propto \frac{\sqrt{M_{\Q\Q}^2-4 M_\Q^2}}{s M_{\Q\Q}^2}
\times  \nonumber\\
&
\Bigl\{F_1\, \Cff+F_2\, \Chh+\cos(2 \phi)\left(F_3\, \Cfh\right.\Bigr. 
\Bigl.\left.+F_3'\, \Chf\right)+\cos(4 \phi)\, F_4\, \Cwhh\Bigr\}\,,
\label{TMDxsect}
\end{align}
where the angular variables in $d\Omega=d\cos\theta d\phi$ are defined in the Collins--Soper frame and describe the spatial orientation of the back-to-back
pair in this frame.
For vector-quarkonium-pair production, the hard-scattering coefficient $F_2$ remains negligible over the whole phase space ($F_2/F_1< 0.01$). 
Thus, $\dif\sigma/\dif\qTq$ is not modulated by $h_1^{\perp g}$ and its measurement gives direct access to $\fone$.

\begin{figure}[htpb!]
\centering
{\includegraphics[width=9.cm]{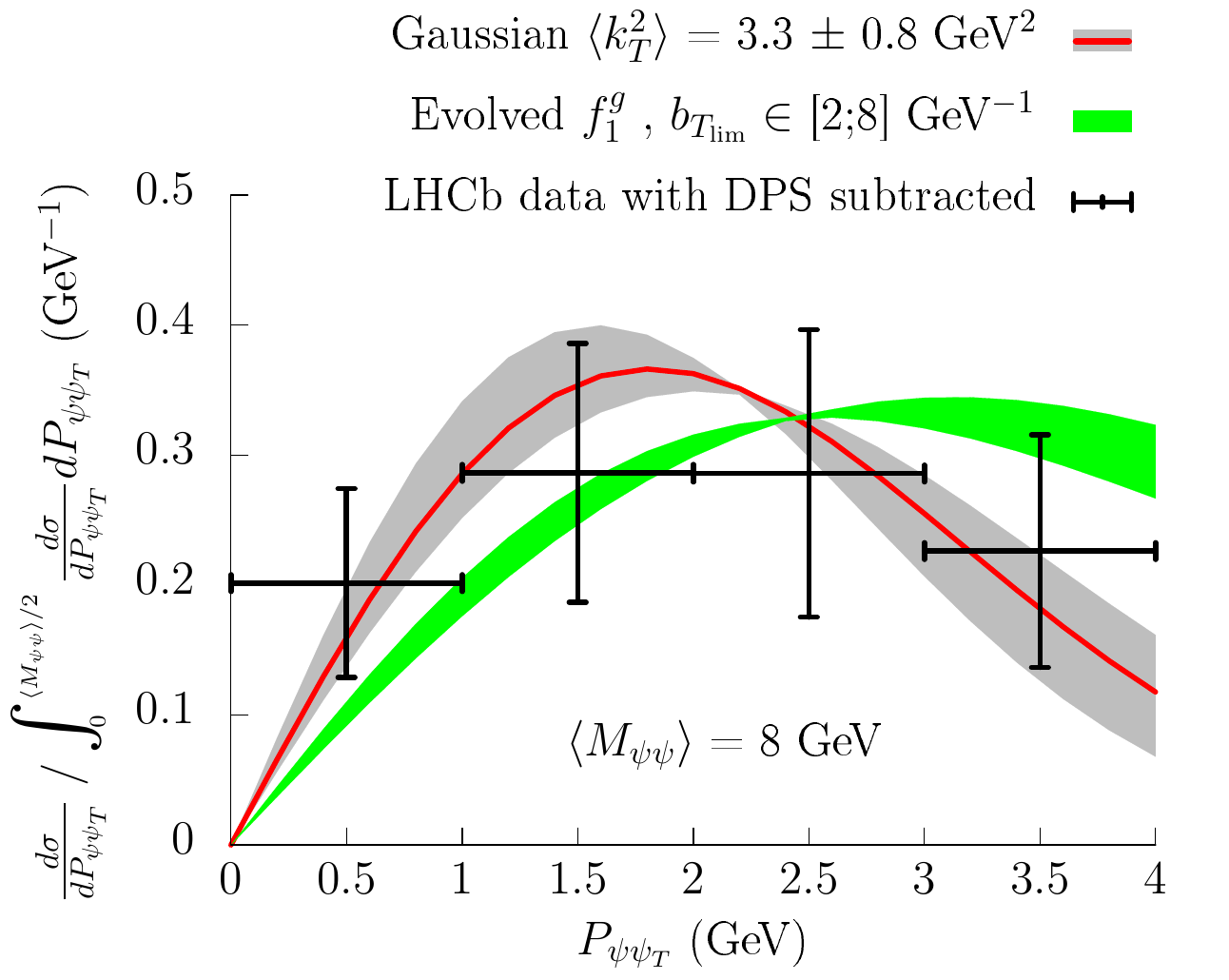}}
\caption{Comparison of the \textit{normalised} $\qTy$-spectrum for $\jpsi$-pair production at $M_{\psi\psi}$ = 8~GeV  computed using two models of the gluon TMDs with that measured by LHCb.
[Figure taken from~\cite{Scarpa:2019fol}]}
\label{fig:qtCf1f1}
\end{figure}

As for the azimuthal asymmetries, they can be conveniently studied by defining
\begin{equation}
\cnf{n}=\frac{\int \dif \phi_{\mathrm{CS}} \cos(n\phi_\CS) \dif\sigma}{\int \dif \phi_\CS \dif\sigma}\,\label{cnf}.
\end{equation}
In fact, $\cnf{2,4}$ represent half the relative magnitude of the corresponding $\phi_\CS$-asymmetries in~\ce{TMDxsect} with respect to the azimuthally-{independent} part, and thus they are directly connected to $h_1^{\perp g}$.

\begin{figure}[htpb!]
\centering
\subfloat[]{\includegraphics[width=0.35\textwidth]{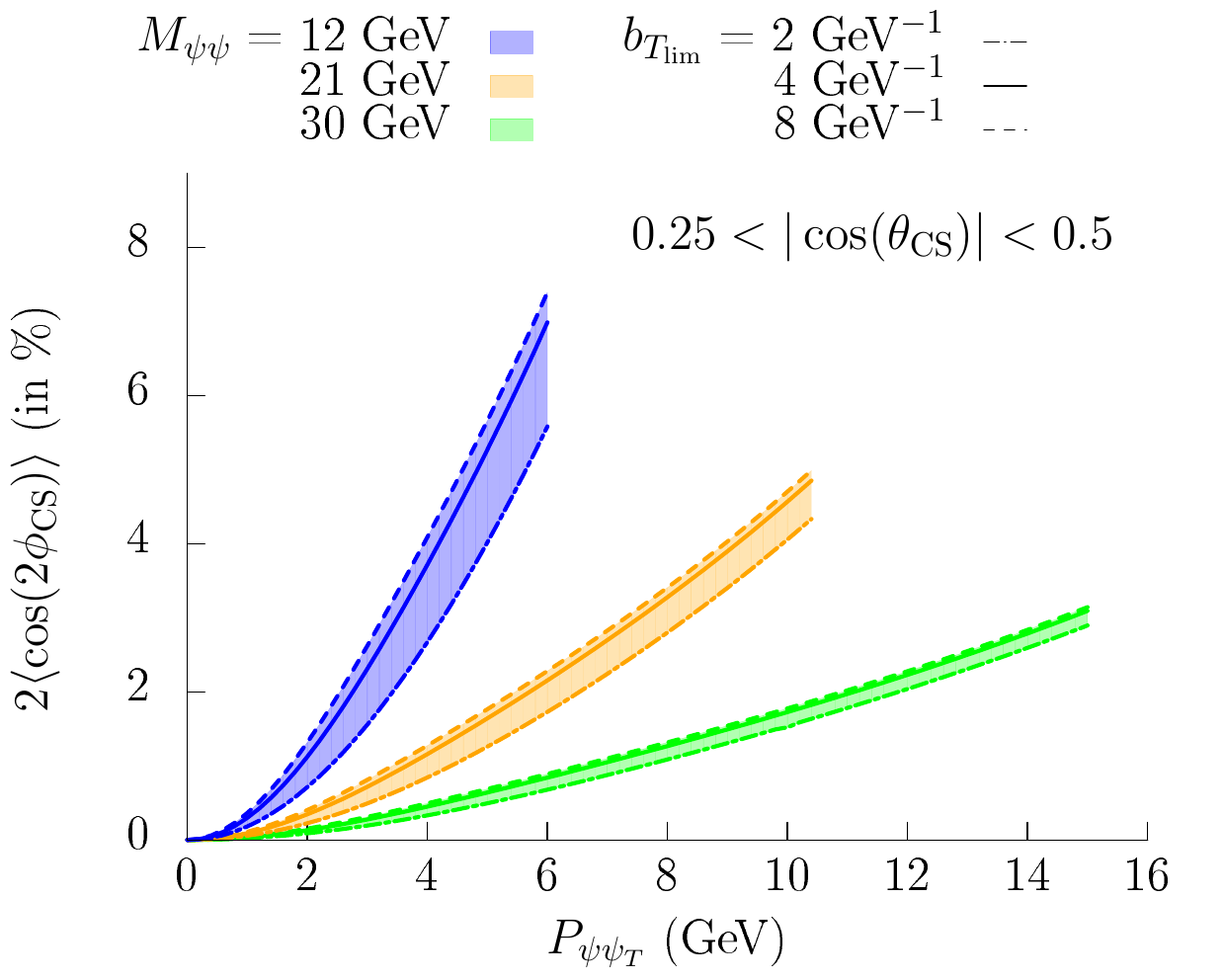}\label{fig:R2_forward_psi}}
\subfloat[]{\includegraphics[width=0.35\textwidth]{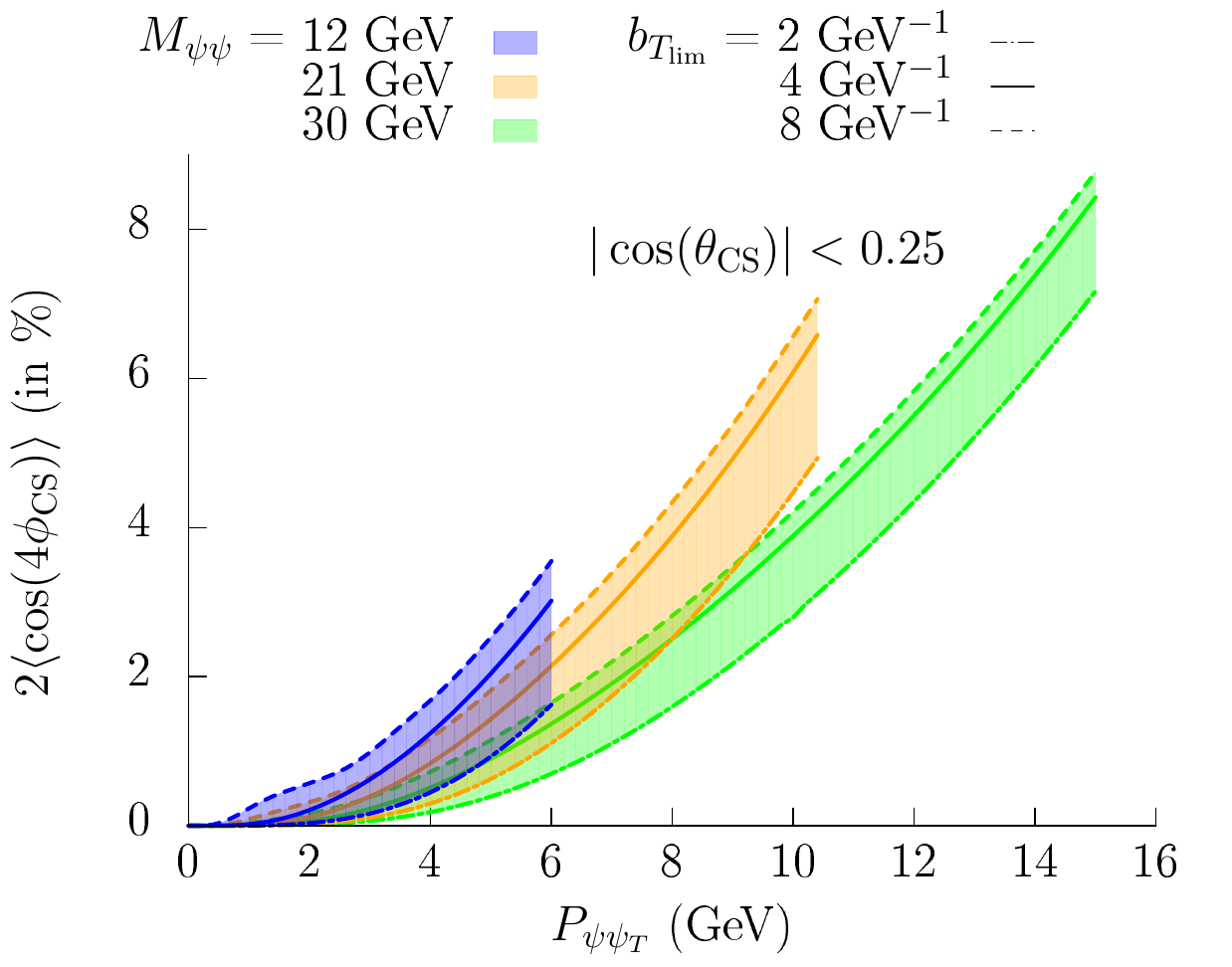}\label{fig:R4_central_psi}}\\
\subfloat[]{\includegraphics[width=0.35\textwidth]{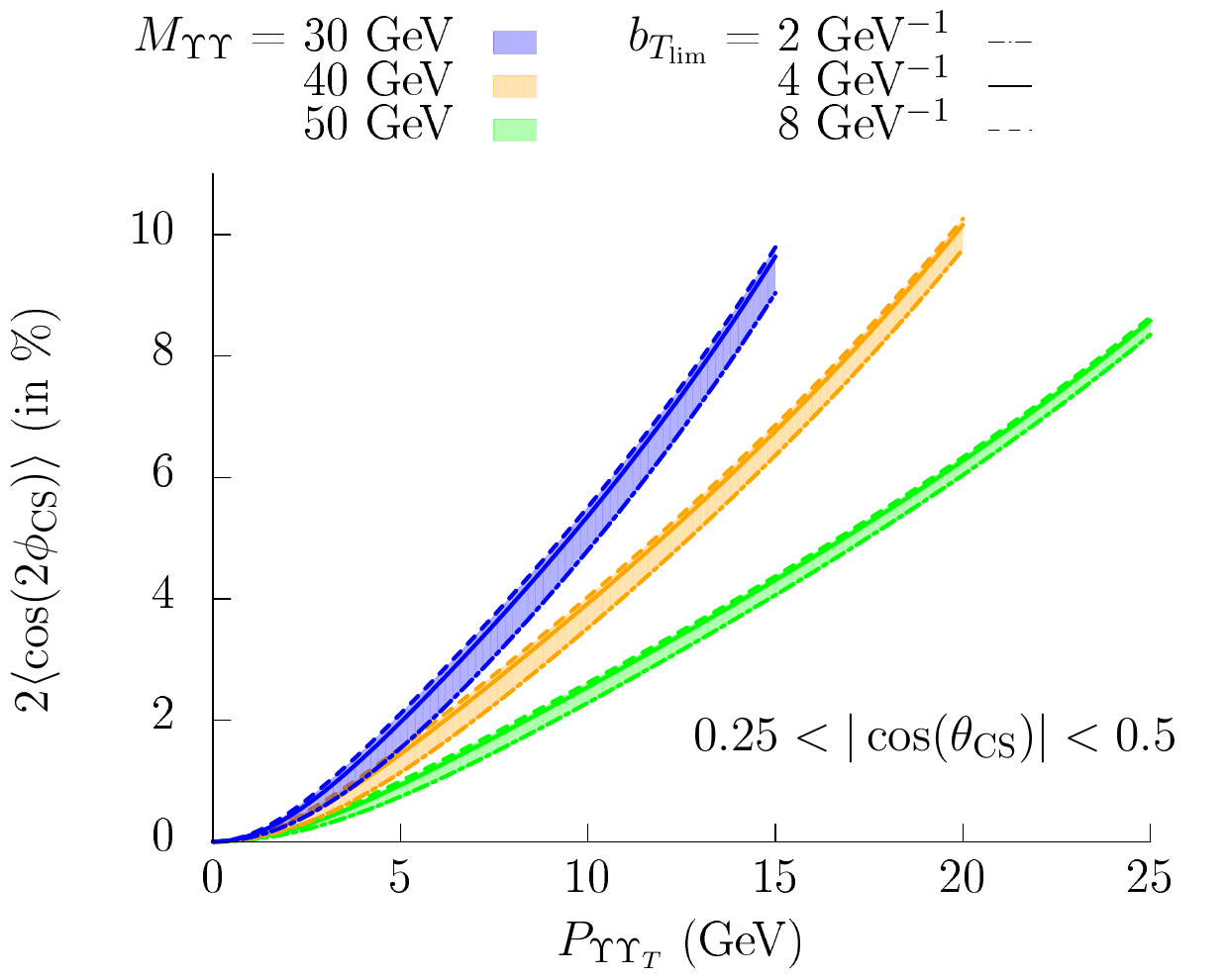}\label{fig:R2_forward_Ups}}
\subfloat[]{\includegraphics[width=0.35\textwidth]{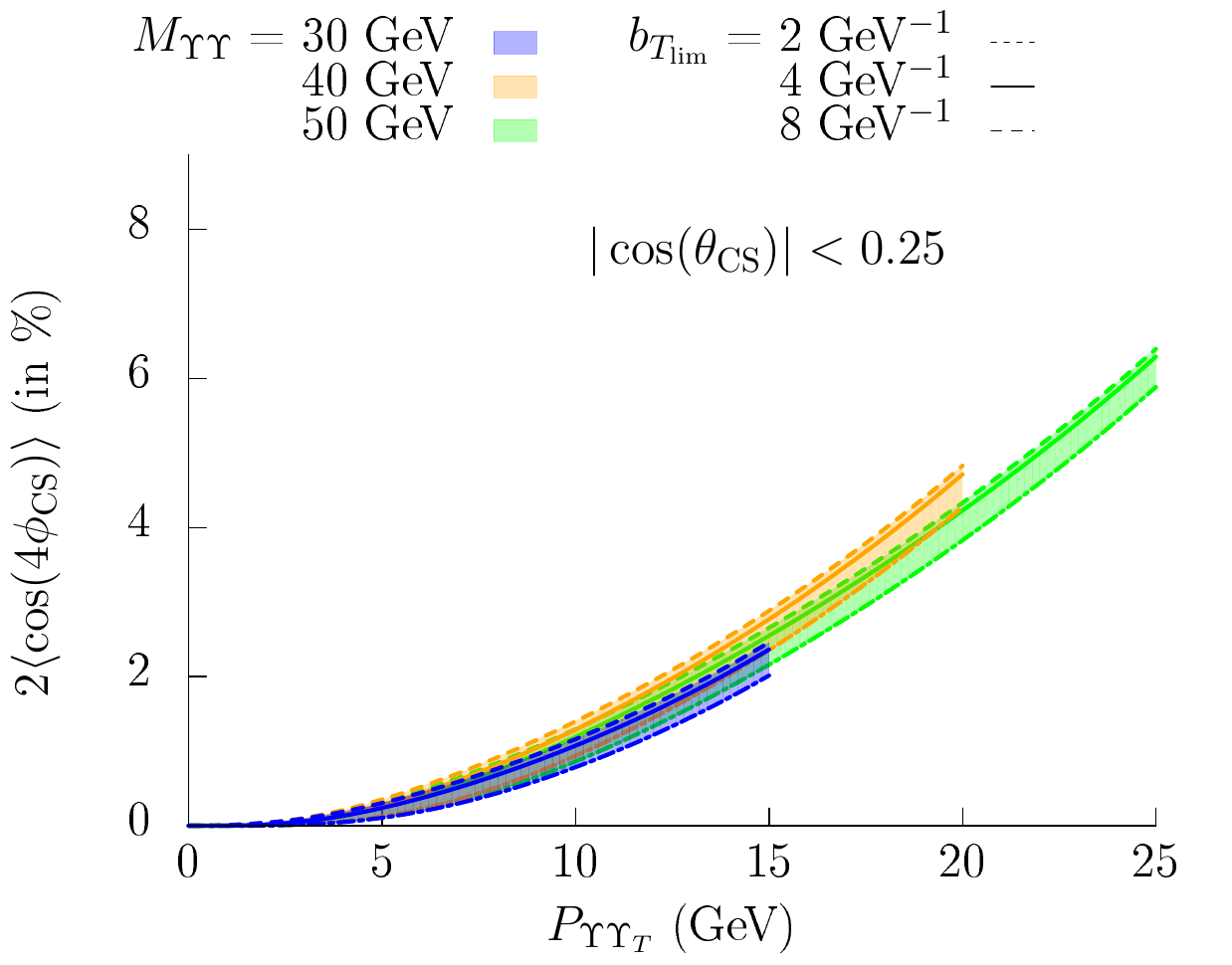}\label{fig:R4_central_Ups}}
\caption{Azimuthal asymmetries for di-$\jpsi$ (a,b) and di-$\Upsilon$ (c,d)  production as functions of $\qTq$: 
(a,c)  $2\cnf{2}$
for $0.25<|\cos(\theta_\CS)|<0.5$,  and (b,d) $2\cnf{4}$ at $|\cos(\theta_\CS)|<0.25$.
The results are presented for $M_{\psi\psi}$ = 12, 21 and 30 GeV and for $M_{\Upsilon\Upsilon}$ = 30, 40 and 50~GeV, for $\blim$ = 2, 4 and 8 GeV$^{-1}$. 
[Figure taken from~\cite{Scarpa:2019fol}]}
\label{fig:asym}
\end{figure}

The normalised $\qTy$ spectra for di-$\jpsi$ production computed using a Gaussian-based TMD model~\cite{Lansberg:2017dzg} or using an evolved TMD~\cite{Scarpa:2019fol} are compared on \cf{fig:qtCf1f1} to the LHCb data~\cite{Aaij:2016bqq}, from which the DPS is subtracted assuming that they are fully uncorrelated. 
The data considered  are for $\qTy < M_{\psi\psi}/2$ with $\langle M_{\psi\psi}\rangle \simeq 8$~GeV. 
The Gaussian-based TMD model fits the data best with a width $\kTsqav$ of the order of 3~GeV$^2$. 
Such a large value is a consequence of TMD evolution increasing the intrinsic momentum of the gluons entering the hard scattering. 
The spectrum using evolved TMDs is plotted for widths $\blim$ of a Gaussian nonperturbative Sudakov factor between 2 and 8 GeV$^{-1}$. 
The lower bound corresponds to the conventional limit with the perturbative region, while the higher bound matches the diameter of the proton.
While the computation with evolution can account for the LHCb spectrum, the lack of a double differential measurement in $\qTy$ and $M_{\psi\psi}$ does not allow the TMD evolution to be constrained.

The relative size of azimuthal asymmetries in $\jpsi$- and $\Upsilon$-pair production are presented in \cf{fig:asym} as a function of $\qTy$, for two ranges of rapidity difference ($|\cos(\theta_\CS)|<0.25$ corresponds to central production, while $0.25<|\cos(\theta_\CS)|<0.5$ corresponds to forward production), different values of $M_{\psi\psi}$ and for $\blim$ in the range [2;8] GeV$^{-1}$. 
Asymmetries reach magnitudes of 8 to 10\% at larger $\qTq$ at central rapidities for both $\jpsi$- and $\Upsilon$-pair production.

Much larger data samples to be collected at the HL-LHC will measure $\qTq$ distributions, allowing for a proper fit of $\fone$ at different scales. 
Additionally, they will allow for a measurement of the azimuthal asymmetries, which could be as large as 10\% and  which would tell if indeed $h_1^{\perp g}$ is non-zero. 
Other studies of quarkonium-pair production are discussed in Section~\ref{sec:beyond_TMD}.

\subsubsection{$\Q+\gamma$ production}
\label{sec:jpsigammaTMD}

Besides vector-quarkonium-pair production, the study of a vector quarkonium produced in association with an isolated photon is another very promising way to access the distribution of both the \kT and the polarisation of the gluon in an unpolarised proton in \pp\ collisions at the LHC~\cite{Dunnen:2014eta}. 
Despite its likely smaller cross section compared to quarkonium-pair production, it is likely to be less prone to factorisation breaking effects (see Section~\ref{sec:beyond_TMD}), while it shows a very similar capability in accessing $h_1^{\perp g}$.

The differential cross section for the production of $\Q+\gamma$ ($\Q=\jpsi,\ups$) via gluon-gluon fusion has the same general form as for di-onia:
\begin{align}\label{separatecrosssect}
&\frac{d\sigma}{dM_{\gamma\Q}dY_{\gamma\Q}d^2{q}_Td\Omega} \propto
\frac{M_{\gamma\Q}^2-M_\Q^2}{s M_\Q^3\,M_{\gamma\Q}^3}
              \left\{           F_1\mathcal{C}[f_1^gf_1^g] +
              \cos(2\phi) \;F_3\mathcal{C}[w_3f_1^{g}h_1^{\perp g}] +
           \cos(4\phi) \;F_4\mathcal{C}[w_4h_1^{\perp g}h_1^{\perp g}]     \right\},
\end{align}
where \qT is the transverse momentum of the quarkonium-photon pair and the angular variables in $d\Omega=d\cos\theta d\phi$ are defined in the Collins-Soper frame~\cite{Dunnen:2014eta}. 
Like for di-onia, the first term in the curly brackets corresponds to the contribution from unpolarised gluons described by $f_1^g$, while the second and third terms contain the linearly-polarised gluon TMD function $h_1^{\perp g}$ and bring in some azimuthal modulations. 

While in~\cite{Dunnen:2014eta} the amplitudes of $2\phi$ and $4\phi$ modulation terms were found to be of comparable size, more realistic simulations suggest that it may be safer to concentrate on the $4\phi$ term, which is less likely to be mimicked by typical acceptance requirements of the general-purpose LHC detectors on muons and photons.
It was found that the $4\phi$ modulation is larger at small values of $\cos^2\theta$, and a cut at $\cos^2\theta=0.1$ allows one to separate the low-$\cos^2\theta$ area where the $4\phi$ modulation is enhanced from the high-$\cos^2\theta$ area where it is suppressed.

In the absence of experimental data, we have found it useful to perform some feasibility studies to extract the $4\phi$ modulation. 
In what follows, the process is simulated for \pp\ collisions at 13~TeV using the \pythia~8 generator, with $h_1^{\perp g}=0$. 
In order to emulate the effects of a possible non-zero $h_1^{\perp g}$, each event is assigned a weight proportional to the expression in the curly bracket in \ce{separatecrosssect}). 
For such pioneering investigations, it is sufficient to mimic evolution effects by assuming a simple Gaussian dependence of the unpolarised gluon distribution $f_1^g$ on the transverse momentum of the gluon ${k}_T$~\cite{Lansberg:2017dzg, Boer:2011kf}, $f_1^g(x,{k}_T^2)={G(x)}/{\pi\langle{k}_T^2\rangle}\exp{\big(-{{k}_T^2}/{\langle{k_T^2}\rangle}}\big)$, where the collinear gluon distribution function is given by $G(x)$, and $\langle{k_T^2}\rangle$ is assumed to be independent of $x$. 
A model-independent positivity bound is used to restrict possible parameterisations for $h_1^{\perp g}$~\cite{Mulders:2000sh}: ${{k}_T^2}|h_1^{\perp g}(x,{k}_T^2)|\leq{2M^2}f_1(x,{k}_T^2)$. 
Following Refs.~\cite{Boer:2011kf, Lansberg:2017dzg}, `Model 1' is defined by  $h_1^{\perp g}(x,{k}_T^2)=\frac{M^2G(x)}{\pi\langle{k_T^2}\rangle^2}\exp{\big(1-{{k}_T^2}/{r\langle{k_T^2}\rangle}\big)},$ while in `Model 2'  $h_1^{\perp g}(x,{k}_T^2)$   is chosen to saturate the positivity bound. 
According to \ce{separatecrosssect}), for an ideal experiment with full acceptance, the dependence on $\phi$ in the absence of gluon polarisation is expected to be flat, while in the case of non-zero gluon polarisation, a $\phi$ modulation appears with the magnitude proportional to the magnitude of $h_1^{\perp g}$.
However, the kinematics of a typical general-purpose LHC detector such as ATLAS or CMS suggests that the minimum $\pT$ of an identified muon is around 4~GeV, which implies $\pT^{\jpsi} > 8$~GeV and hence requires a cut $\pT^\gamma > 8-9$~GeV to produce a $\pT$-balanced final state where \qT is smaller than, say, $M_{\Q\gamma}/2$. 
These cuts cause a significant non-trivial distortion of the observed $\phi$ distribution, which complicates the extraction of the $\phi$-modulated terms. 
It was observed that this distortion is almost independent of $\cos\theta$, and one can use the ratio of the differential cross sections with low and high $\cos^2\theta$ to largely eliminate the kinematic distortion and help to extract the $4\phi$-modulated contribution. 

The comparison between unweighted and weighted distributions of the ratio of differential cross sections with low $\cos^2\theta < 0.1$ and high  $\cos^2\theta > 0.1$ is shown in~\cf{fig:ratiooverlay}. 
The distributions are fitted with a Fourier series truncated after $\cos4\phi$. 
The dashed blue line shows the unweighted result, which assumes $h_1^{\perp g}=0$, while the solid red lines in~\cf{fig:ratiooverlay}a and~\cf{fig:ratiooverlay}b correspond, respectively, to Model 1 and Model 2 defined above.
\begin{figure}[htbp!]
\centering
\subfloat[]{\includegraphics[height = 5.4cm, keepaspectratio]{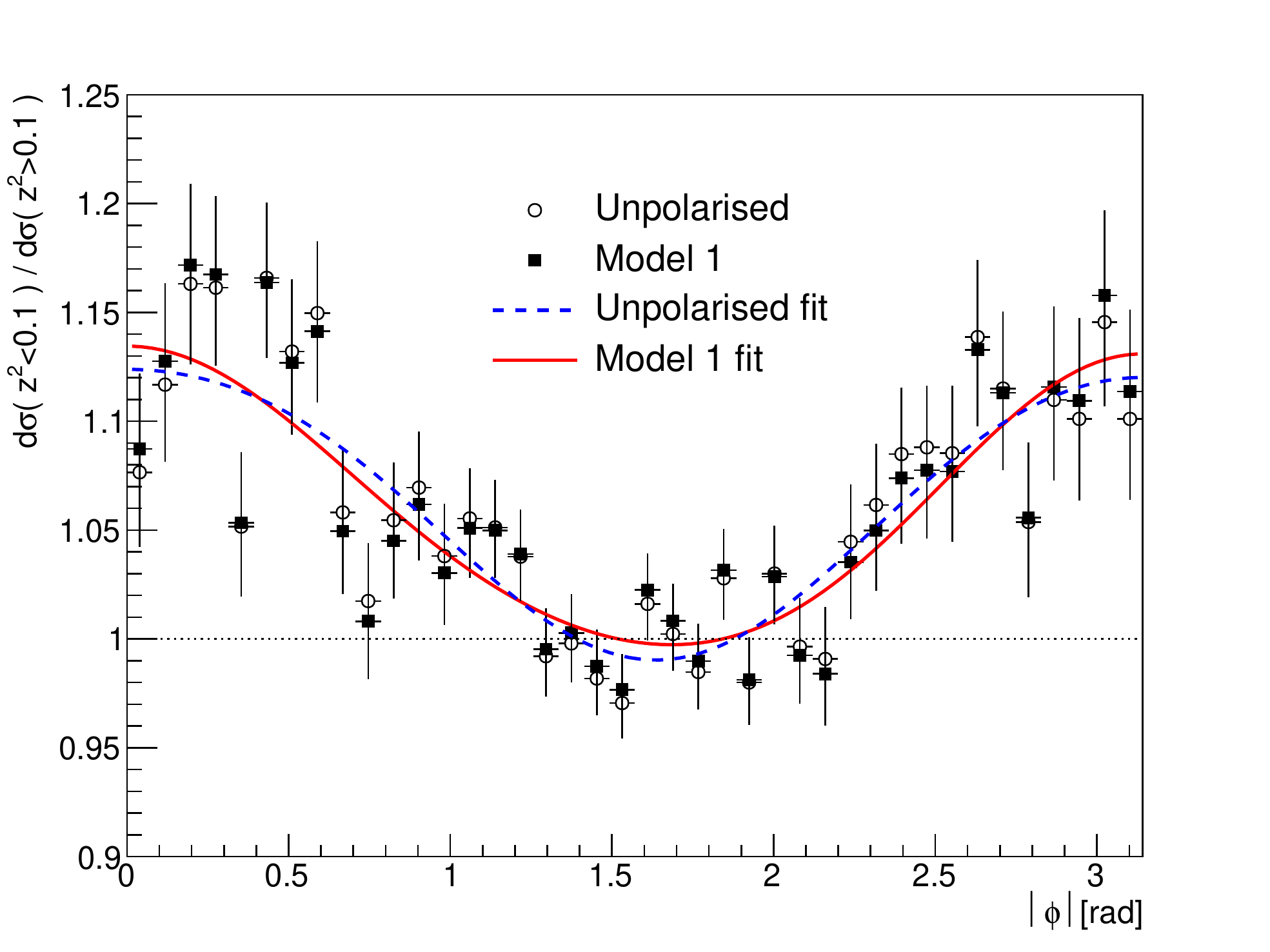}}
\subfloat[]{\includegraphics[height = 5.4cm, keepaspectratio]{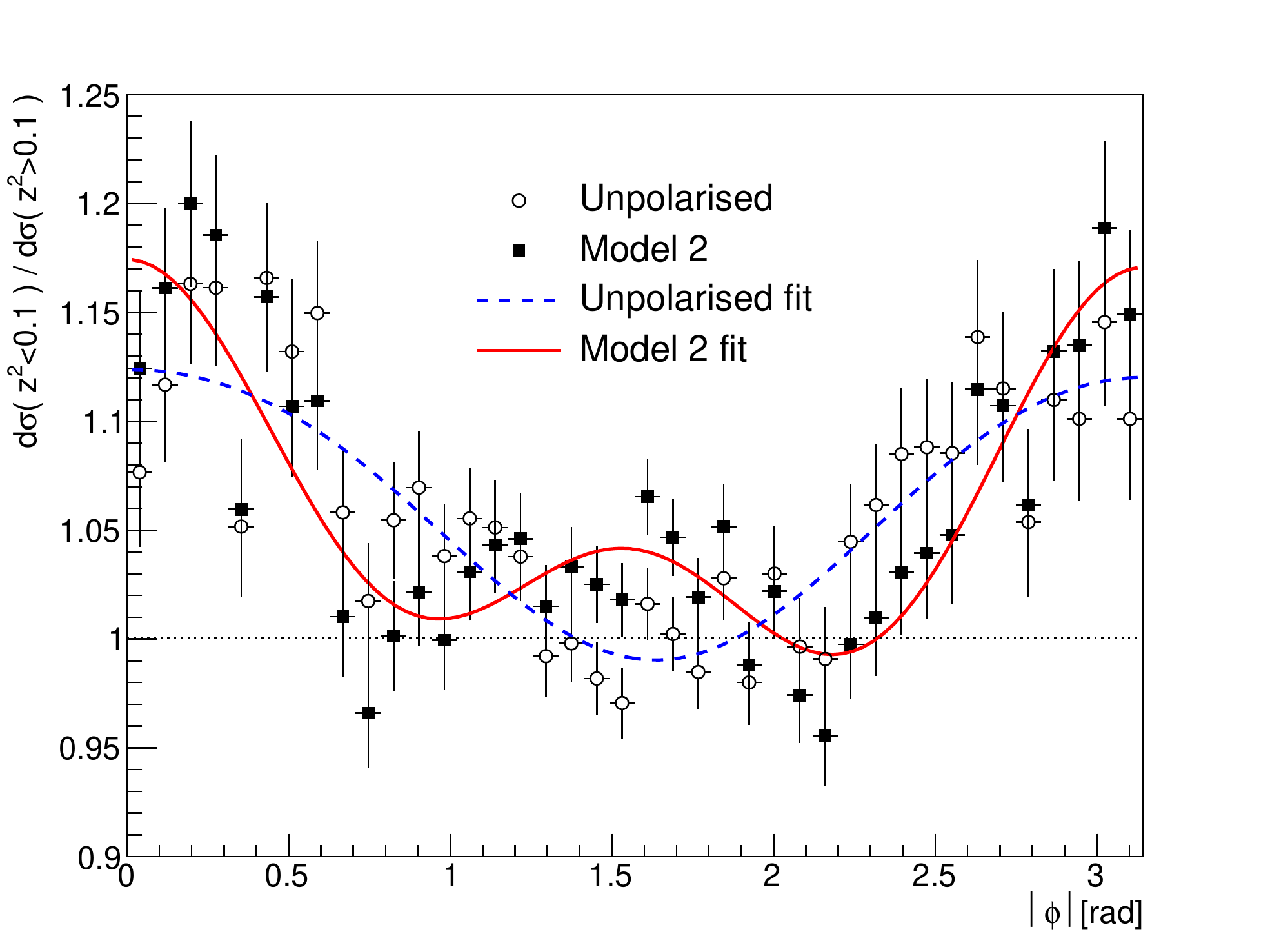}}
\caption{
The ratios of differential cross sections for events with $z^2\equiv \cos^2\theta<0.1$ over the events with $\cos^2\theta>0.1$. 
On both plots, the open points describe the unweighted distribution, corresponding to $h_1^{\perp g}=0$, with the fit shown as the dashed blue line. 
The solid points describe the weighted distributions, with the fits shown as solid red lines, for Model 1 (a) and Model 2 (b) described in the text. 
}
\label{fig:ratiooverlay}
\end{figure}

For the level of statistics of this Monte Carlo sample, which roughly corresponds to an integrated luminosity of 100/fb at 13~TeV (with no account for detection efficiency), the change in the coefficient of the $\cos4\phi$ modulation term relative to the unweighted case, $\Delta P_4$,  is non-significant for Model 1 at $\Delta P_4(M1) = (9\pm6) 10^{-3}$, but should be reliably measured if the gluon TMDs are described by Model 2, with $\Delta P_4(M1,2) = (50\pm6)10^{-3}$. 
An increase in the integrated luminosity by a factor of 100 should allow one to reach the sensitivity needed for Model 1 even for a detection efficiency of $\sim20\%$. 
A similar picture is expected to be obtained for prospects with the CMS detector, whereas dedicated simulations are clearly needed to assess whether one could venture to even lower \pT\ values with the LHCb detector.

\subsection{Beyond and in between TMD and HE factorisations}
\label{sec:beyond_TMD}

Quarkonia are nearly always produced by gluons at the LHC and, as such, their \pT spectra are, more or less directly, sensitive to the gluon distribution in the transverse plane. However, even in unpolarised hadronic collisions, many phenomena come into play when one wishes to study this connection. Depending on the theoretical formalism one employs to approach this relationship between the dynamics of the gluon and that of quarkonia, different effects are emphasised.

As was previously alluded to, TMD factorisation is expected to have a restrictive range of applicability, both in terms of kinematics (\pT should be smaller than the hard scale, the usual invariant mass of the observed system) and processes (no colour flow in the final state in hadronic collisions). On the contrary, HE factorisation is much more inclusive in terms of processes but, being designed to account for HE effects, it may be inaccurate or simply miss some phenomena when the collision energy is finite. When put in the context of quarkonium production, for which the mechanisms at work are not even an object of consensus, it is not surprising that the situation quickly gets intricate. In this section, we will simply attempt to correlate some possible future measurements at the HL-LHC with theoretical objectives. It should be clear that these are not necessarily absolutely rigorous, completely achieved nor objects of consensus.  

Let us first start with ideas of quarkonium  measurements inspired by considerations from TMD factorisation with NRQCD for which specific factorisation-breaking effects can be identified. The first on the list of course is that of single  $J/\psi$ or $\Upsilon$ as a function of \pT, which have been routinely  measured at colliders for thirty years. These have been investigated assuming the validity of TMD factorisation within NRQCD with CO contributions~\cite{Mukherjee:2016cjw} as well as the CEM~\cite{Mukherjee:2015smo}. 

There are two issues here concerning why TMD factorisation should in principle not apply. First, if one {focuses} on the leading-$v$ contributions, thus to the CSM, both  $J/\psi$ and $\Upsilon$ are produced with a gluon. If its momentum is integrated over, the connection between the final-state measured momenta and the initial-state ones is lost.  Second, if one considers the sub-leading $v$ contributions from CO, which at low \pT are enhanced by one power of \alphas, the colour flow is so entangled that one cannot expect to derive a factorised formula for the hadronic cross section. However, it does not prevent the analysis of data along the lines of what a would-be TMD factorised cross section predicts and, then, to attempt to extract information on TMDs. Along these lines, it would be very interesting to compare the low-\pT spectra of the vector and pseudoscalar states, \ie for $\pT < M_\Q/2$. Such data exist for the vector states, but not yet for the pseudoscalar ones. 
If they are found to be different, caution will be needed before attributing this difference either to \pT modulations from $h_1^{\perp g}$ or, simply, to factorisation-breaking effects beyond factorised TMDs expected for these processes.

The same remark can be made for the $\chi_{Q}$ states. Different \pT modulations from $h_1^{\perp g}$ are expected~\cite{Boer:2012bt} between the scalar and the tensor states. They are however also subject to factorisation-breaking effects owing to their CO content~\cite{Ma:2014oha}. It may nevertheless happen that these effects could be related by HQSS between both these $\chi_{Q}$ states. As what regards the pseudovector state, $\chi_{Q1}$, according to NRQCD, its arbitrary\footnote{The trade-off between the CO and CS component in a $P$-wave quarkonium is set by the unphysical NRQCD scale.} CO content would normally allow its production by the fusion of two on-shell gluons. LHCb data however show~\cite{Aaij:2013dja} a $\chi_{c2}/\chi_{c1}$ ratio steadily rising when \pT approaches zero in accordance with the Landau--Yang theorem, but in disagreement with the NRQCD expectations. In other words, the impact of CO is not as expected. In view of this, one should certainly not refrain from incorporating the $\chi_{Q}$ states in a global TMD survey {for} fear of factorisation-breaking effects due to their CO content.  %

The HE factorisation framework provides further motivations for such studies of low-\pT $\chi_{Q}$ states and, particularly, of ratios such as  $\chi_{c2}/\chi_{c1}$. At the LHC, according to HE factorisation, this should also not show the observed Landau-Yang enhancement, this time not because of CO, but because the pseudo-vector $\chi_{c1}$ can be produced by two gluons when at least one is off-shell\footnote{We recall that the gluon virtuality is expected to increase for decreasing $x$ according to HE factorisation.}. Similarly, one may also want to compare the pseudoscalar and tensor \pT spectra. Clearly, what is at stake then is the correlation between the off-shellness of the initial gluons and their fractional momenta, rather than their possible linear polarisation. This illustrates how a single observable can highlight two different phenomena in two different formalisms.  Of course, {if observed, the question on how they are connected could be asked}.

Similarly there remain further aspects of the connection between the virtualities and the \kT\ of the initial gluons that can be studied in associated production of quarkonia. For instance, calculations in HE factorisation for inclusive double charmonium hadroproduction are discussed in Section~\ref{sec:dps} and provide, even at LO, a reasonable account of the $\qTy$ spectra, which is connected to $\fone$ as shown in Section~\ref{sec:psi-psi-TMDs}. In TMD factorisation, a first attempt to connect the size of the azimuthal modulations generated by $h_1^{\perp g}$ to the quarkonium polarisations was made in~\cite{Scarpa:2020sdy}. It would be interesting to see how, in HE factorisation, the quarkonium polarisation evolves with energy and understand how it is correlated to the initial-gluon virtualities. More generally, quarkonium-pair production, which should be studied even more widely at the HL-LHC, likely represents a very versatile laboratory in {which to} analyse possible dualities between HE and TMD factorisations.

\subsection{Single transverse-spin asymmetries at the HL-LHC in FT mode}
\label{sec:polarised}

STSAs, or $A_N$, are defined as\footnote{Note that another direction, such as the transverse momentum of the produced particles, is needed. (see \eg~\cite{Anselmino:2009st} for more details).}
\begin{equation}
A_{N} = \frac{1}{{\cal P}_{\text{eff}}} \frac{\sigma^{\uparrow} - \sigma^{\downarrow}}{\sigma^{\uparrow} + \sigma^{\downarrow}}
\,,
\end{equation}
where $\sigma^{\uparrow\,(\downarrow)}$ is a differential cross section produced with a nucleon polarised upwards (downwards) and ${\cal P}_{\text{eff}}$ is the effective polarisation.
Large STSAs were observed for the first time in 1976, in $\Lambda^0$ production in FT $p$Be scattering at Fermilab ~\cite{Bunce:1976yb}, and have been seen in many other experiments since then. 
When considering only the scattering of quarks or gluons, STSAs are expected to scale with the quark mass and the \cm\ energy as $A_N\sim m_q/\sqrt{s}$, as was shown in the seminal paper~\cite{Kane:1978nd}. 
This prediction is many orders of magnitude smaller than the experimental observation, hence the explanation for large STSAs has to be found beyond the perturbative realm of QCD. 

Two different theoretical mechanisms have been proposed, both relating STSAs to the structure of hadrons in terms of QCD.
The first mechanism, called the collinear twist-3 (CT3) approach~\cite{Efremov:1981sh,Efremov:1983eb,Qiu:1991pp}, is valid in the presence of one hard scale. 
An example would be the single inclusive production of a light meson in \pp\ scattering, where the hard scale is provided by the large \pT of the meson. 
The STSAs is then due to quark-gluon-quark or triple-gluon correlators, which are the sub-leading (in the scale) twist-3 extensions of the usual collinear PDFs (with fragmentation-type twist-3 correlators also being relevant~\cite{Kanazawa:2014dca}).

The second mechanism takes place within TMD factorisation, and is therefore valid in the presence of two ordered hard scales: a small and a large one. In this framework, large STSAs are  caused by the distribution of unpolarised partons inside the transversely-polarised hadron, parameterised by the Sivers TMD PDF $f_{1T}^{\perp}$~\cite{Sivers:1989cc} or by the fragmentation of a transversely-polarised parton into an unpolarised light meson, as parameterised by the Collins TMD FF $H_1^\perp$~\cite{Collins:1992kk}. Note that in the kinematic region where $\pT$ approaches the hard scale $M$, the TMD framework maps smoothly to the collinear regime, see \eg~\cite{Ji:2006ub,Collins:2016hqq,Echevarria:2018qyi}.

Finally, a phenomenological approach is the Generalised Parton Model (GPM)~\cite{DAlesio:2007bjf}, in which the Sivers and Collins mechanisms are applied even in single-scale processes, keeping track of the transverse-momentum exchanges in the partonic scattering.
This approach has proven to be quite successful in phenomenological analyses, although one should be careful when extracting conclusions about the involved TMDs and the underlying physics.
In any case, it can be used to give a fair estimate of STSAs in single-scale processes, where the analysis in the proper twist-3 framework becomes a real challenge, due to the many involved and still unconstrained twist-3 functions.

Below   STSAs in different quarkonium-production processes are discussed, in the context of a future FT experiment at the HL-LHC~\cite{Hadjidakis:2018ifr}, which could perform these measurements by polarising a target.

\subsubsection{Vector $\Q$ production}

In this subsection, STSAs in the $p^\uparrow p \to \Q + X$ process are discussed. %
As mentioned above, such processes are strictly speaking not TMD factorisable and do not therefore directly probe the TMD $f_{1T}^{\perp g}$~\cite{Boer:2015vso}, which encapsulates the Sivers effect believed to generate these STSAs, in the absence of the Collins effect from the fragmentation of the hadron.

However, within the GPM, which is an effective model where spin and intrinsic transverse-momentum effects are taken into account, such STSAs are treated as if they were factorisable in terms of an analogous object which is denoted, in what follows, as the Gluon Sivers function (GSF), in order to account for the gluon Sivers effect. 

A more sophisticated extension of the GPM, the Colour-Gauge-Invariant GPM (CGI-GPM)~\cite{Gamberg:2010tj,DAlesio:2017rzj}, can also be considered, where effects from initial-state and final-state interactions, in the one-gluon approximation, are encapsulated in the GSF modelling. {Within the CGI-GPM, there are two independent GSFs, acting as phenomenological counterparts to the previously mentioned WW and DP $f_{1T}^{\perp g}$ in Section~\ref{sec:TMD_factorisation}, here denoted by $f$- and $d$-type, respectively.} 
In this respect, one can in principle also address the process dependence of the GSF. 

Concerning the quarkonium-production mechanism, both the CSM and NRQCD can be considered since factorisation-breaking effects are put aside. All details about such phenomenological studies can be found in~\cite{DAlesio:2017rzj,DAlesio:2019gnu,DAlesio:2020eqo}. In what follows, some selected results which support future experimental studies at the LHC will be shown.

\begin{figure}[htbp!]
\centering
\includegraphics[trim = 2.5cm .15cm 2.5cm 0cm, clip, height = 6.5cm, keepaspectratio]{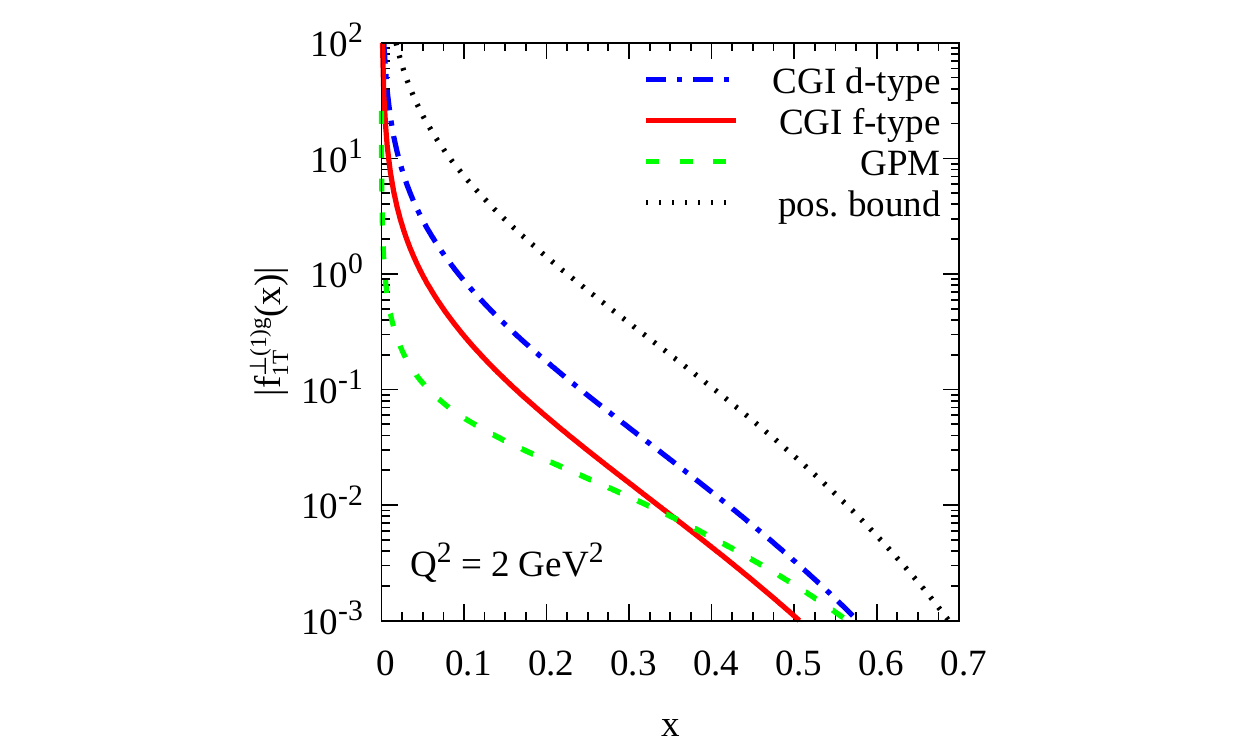}
\caption{ Upper values for the first $k_T$ moments of the GSFs at $Q^2 = 2$ GeV$^2$~\cite{DAlesio:2018rnv}.}
\label{fig:1stm_AN}
\end{figure}

A first attempt to constrain these effective GSFs both within the GPM and the CGI-GPM approaches, from mid-rapidity pion and $D$-meson STSA data from RHIC~\cite{Adare:2013ekj,Aidala:2017pum}, was presented in~\cite{DAlesio:2015fwo,DAlesio:2018rnv}. 
\cf{fig:1stm_AN} shows the extracted upper bounds of the first $k_T$ moment of the GSFs. Note that this quantity is a necessary ingredient for the evolution of the GSF itself. 
As a matter of fact, the obtained GSF allows for a fairly good description of the available $\jpsi$ STSA data (almost compatible with zero)~\cite{Aidala:2018gmp}, even if no definite conclusion can be drawn owing to the possible presence of non-factorisable contributions, the feed-down from states that could depend in a different manner on the GSF, and the still rather large experimental uncertainties in a restricted domain in $x$.
It is thus extremely important to extend this analysis to more quarkonium states and to the
kinematics reachable at the FT-LHC as discussed in~\cite{Kikola:2017hnp,Hadjidakis:2018ifr}.

\begin{figure}[htbp!]
\centering
\includegraphics[trim = 1.cm 0cm 1cm .25cm, clip, height = 5.5cm, keepaspectratio]{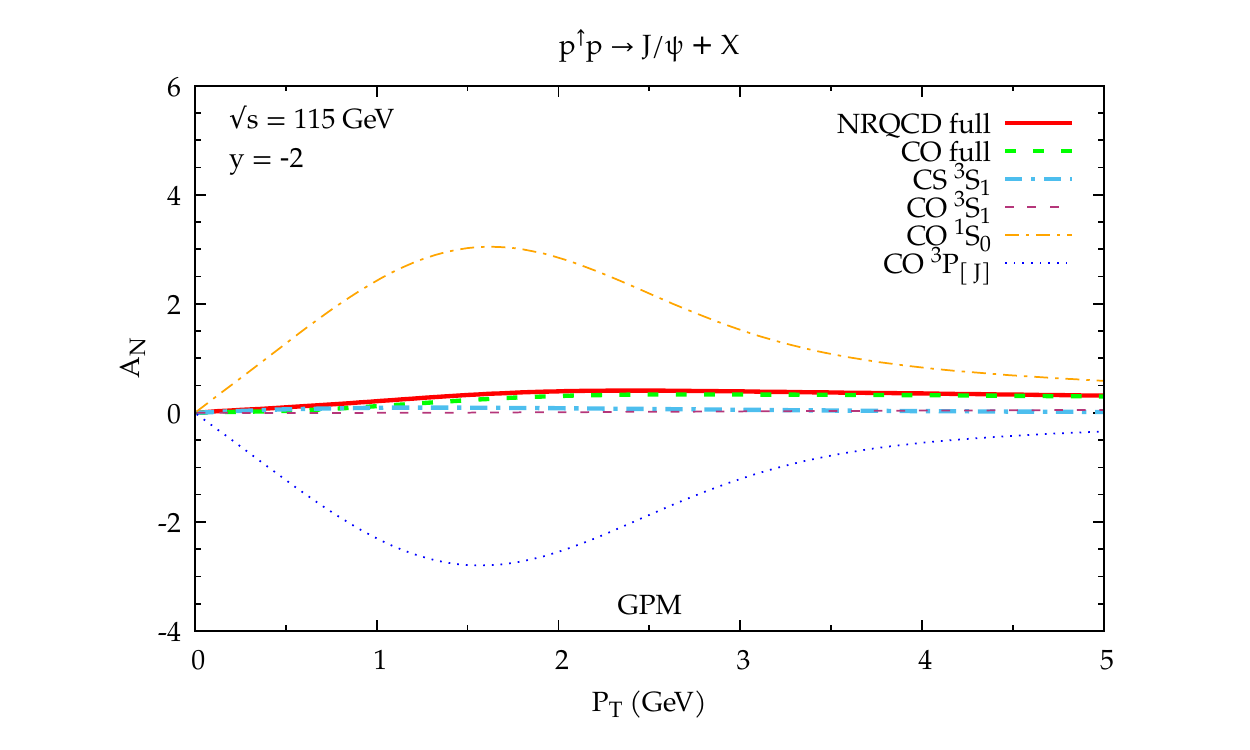}
\includegraphics[trim = 1.cm 0cm 1cm .25cm, clip, height=5.5cm, keepaspectratio]{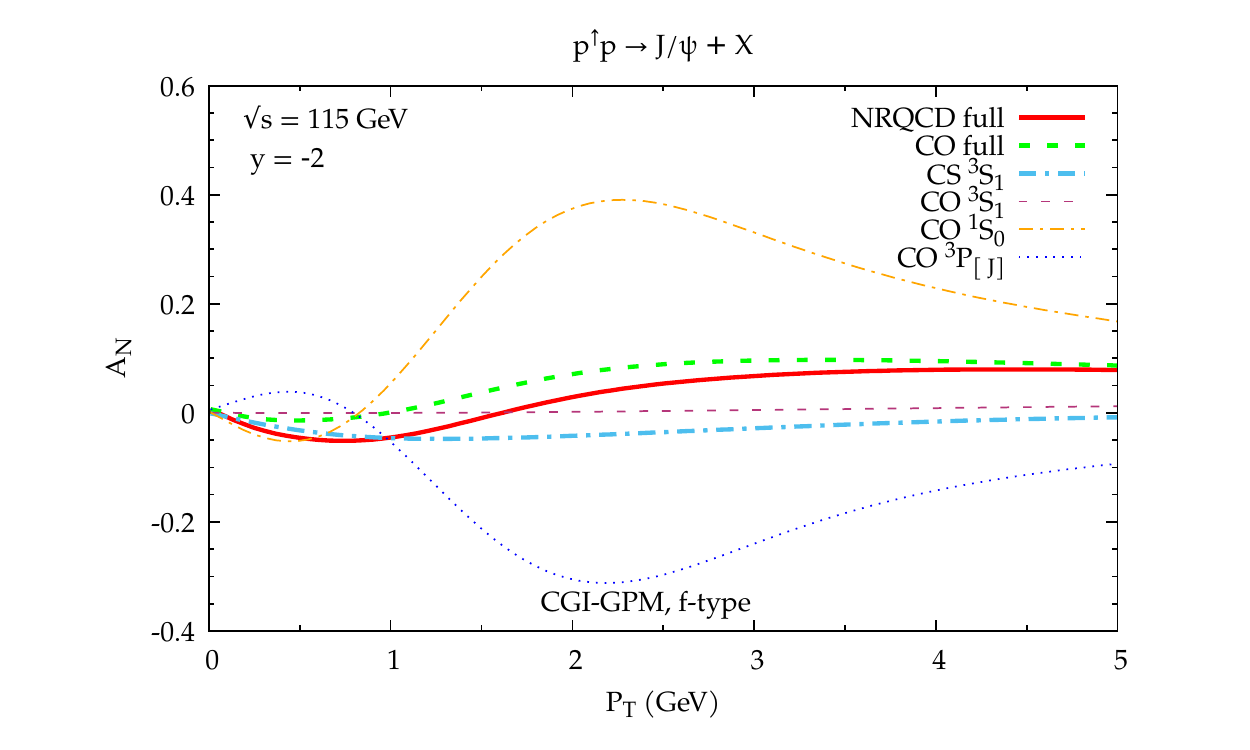}
\caption{Maximised values for $A_N$ vs.~$P_T$ for the process $p p^\uparrow \to \jpsi + X$ at $\sqrt s=115$ GeV and $y_{\rm cm}=-2$ within the GPM (left panel) and the CGI-GPM (right panel) approaches~\cite{DAlesio:2019gnu,DAlesio:2020eqo}. The full result (red solid lines) together with its wave decomposition (see legend) are shown.
}
\label{fig:AN_GPM-CGI_waves}
\end{figure}

\cf{fig:AN_GPM-CGI_waves} shows estimates for the STSA in the FT mode at the LHC, $A_N$, obtained by \emph{maximising} the Sivers effect within the GPM (left panel) and the CGI-GPM $f$-type (right panel) approaches. 
The maximised quark contributions as well as those from the $d$-type GSF (not shown) are compatible with zero. 
For completeness, the contribution from each wave to the full result is shown (see legend). 
Notice that some contributions within the GPM are larger than one. 
Since the denominator of the STSA includes all terms (entering with relative signs), while the numerator considers only a specific wave state. 
The overall result (red solid lines) is, as expected, smaller than one. 
A full comparison between two different mechanisms for quarkonium production (CSM vs.~NRQCD) and two effective TMD schemes (GPM vs.~CGI-GPM) for the maximised $A_N$ is presented in \cf{fig:AN_GPM-CGI}. These results, where no previous information on the GSF has been used, illustrate the potential role of such a dedicated phenomenological study.

\begin{figure}[htbp!]
\centering
\includegraphics[trim = 1.cm 0cm 1cm .25cm, clip, height = 5.5cm, keepaspectratio]{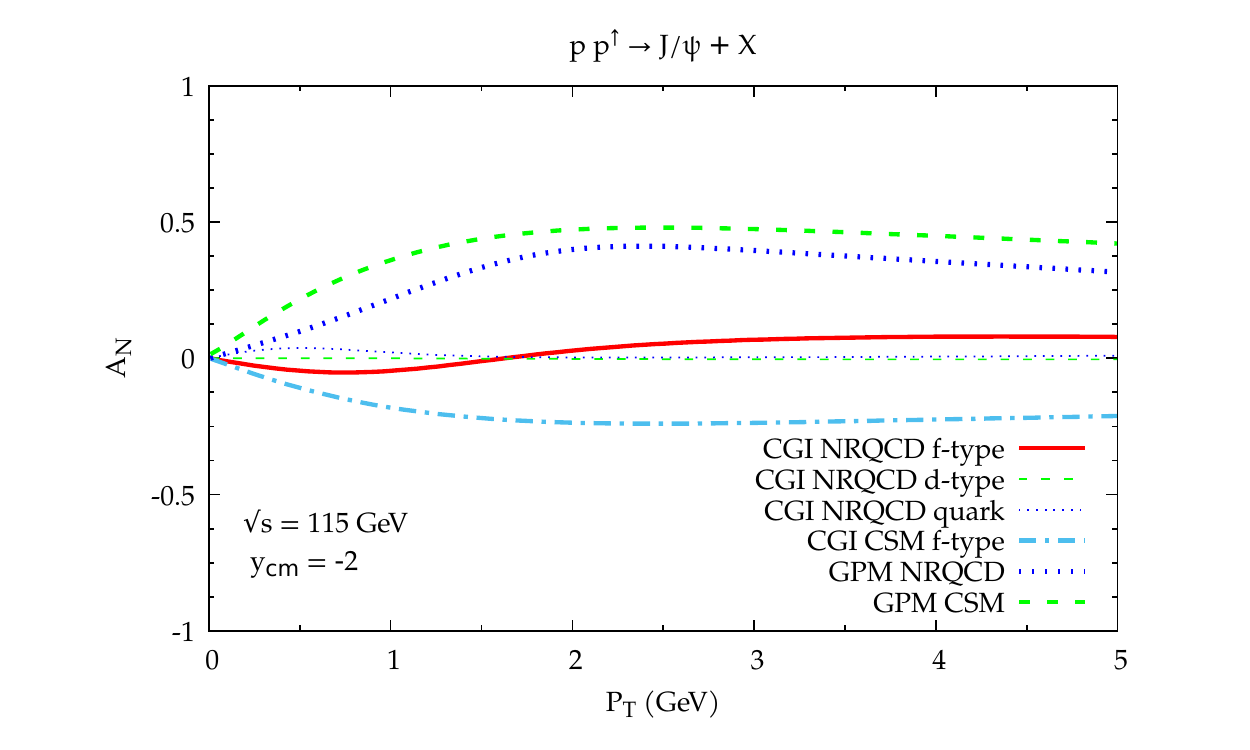}
\includegraphics[trim = 1.cm 0cm 1cm .25cm, clip, height=5.5cm, keepaspectratio]{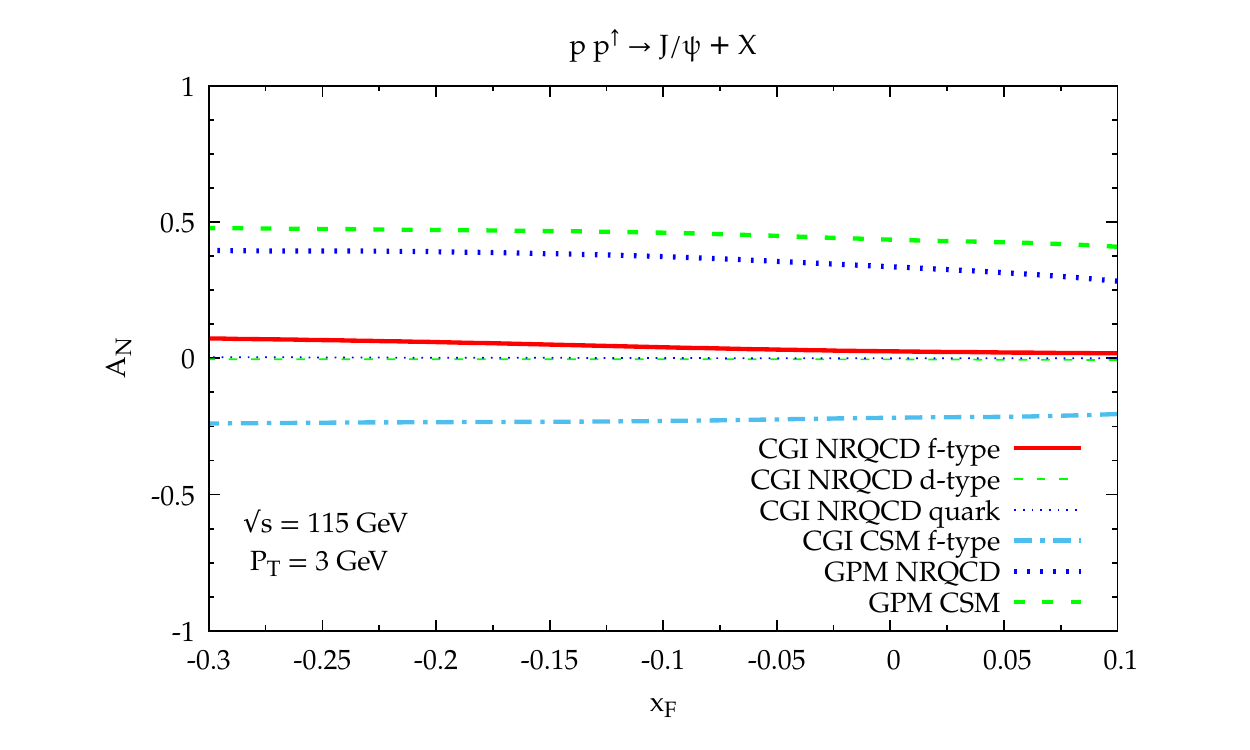}
\caption{Maximised values for $A_N$ for the process $p p^\uparrow \to \jpsi + X$ at $\sqrt s=115$ GeV at fixed $y_{\rm cm}=-2$ vs.~$P_T$ (left panel) and at fixed $P_T=3$~GeV vs.~$x_F$ (right panel), obtained adopting the CGI-GPM and GPM approaches, within the CSM and NRQCD~\cite{DAlesio:2019gnu,DAlesio:2020eqo}.}
\label{fig:AN_GPM-CGI}
\end{figure}

\begin{figure}[htbp!]
\centering
\includegraphics[width=0.5\textwidth]{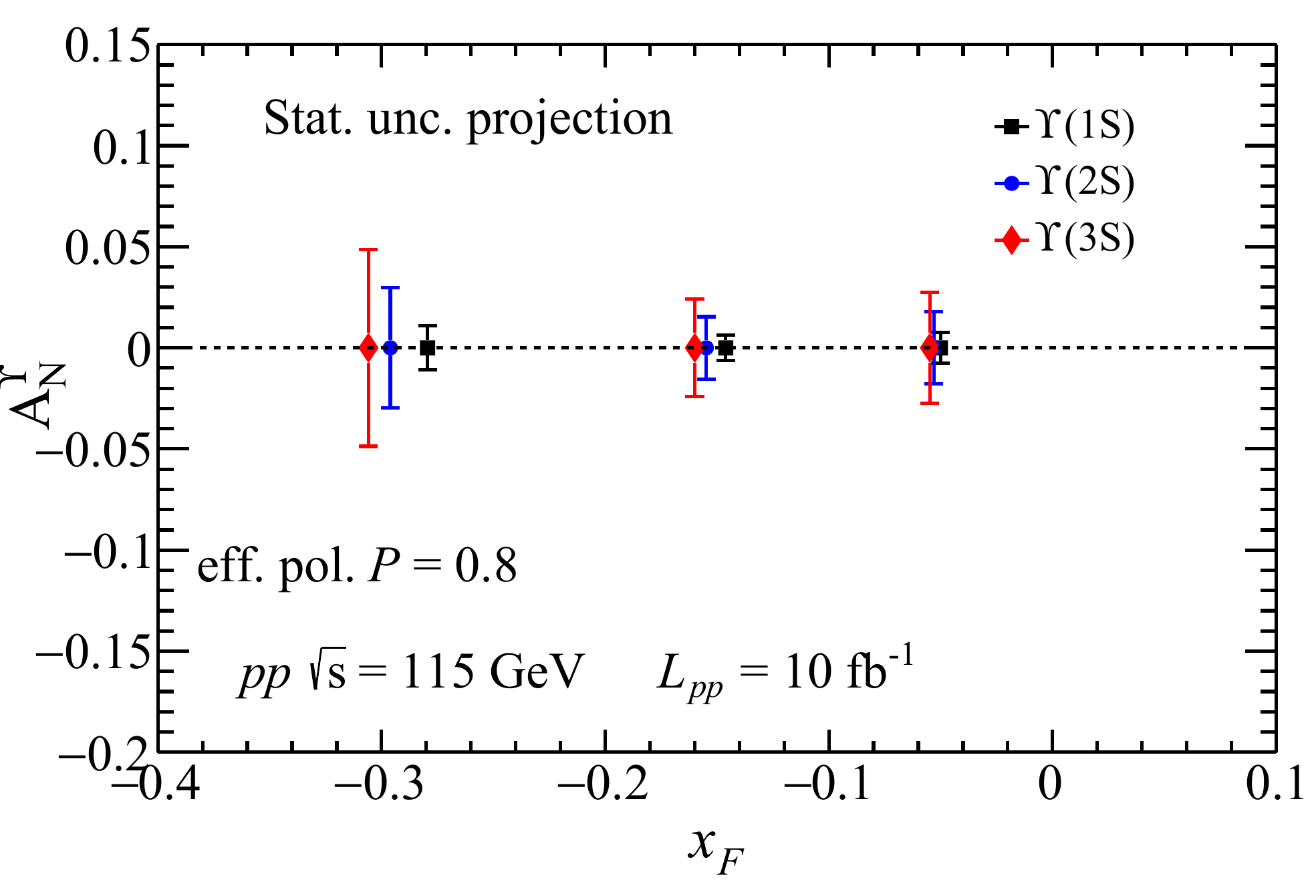}
\caption{Statistical-precision projections for $\Upsilon(nS)$ $A_{N}$ as a function of $\xF$. 
The quarkonium states are assumed to be measured in the di-muon channel with a LHCb-like detector. 
The signal and the background are calculated in fast simulations that take into account the performance of the LHCb detector~\cite{Massacrier:2015qba,Kikola:2017hnp}.
[Figure taken from~\cite{Hadjidakis:2018ifr}]}
\label{fig:An:VectorOnium}
\end{figure}

\begin{figure}[hbt!]
\centering
\includegraphics[width=0.5\textwidth]{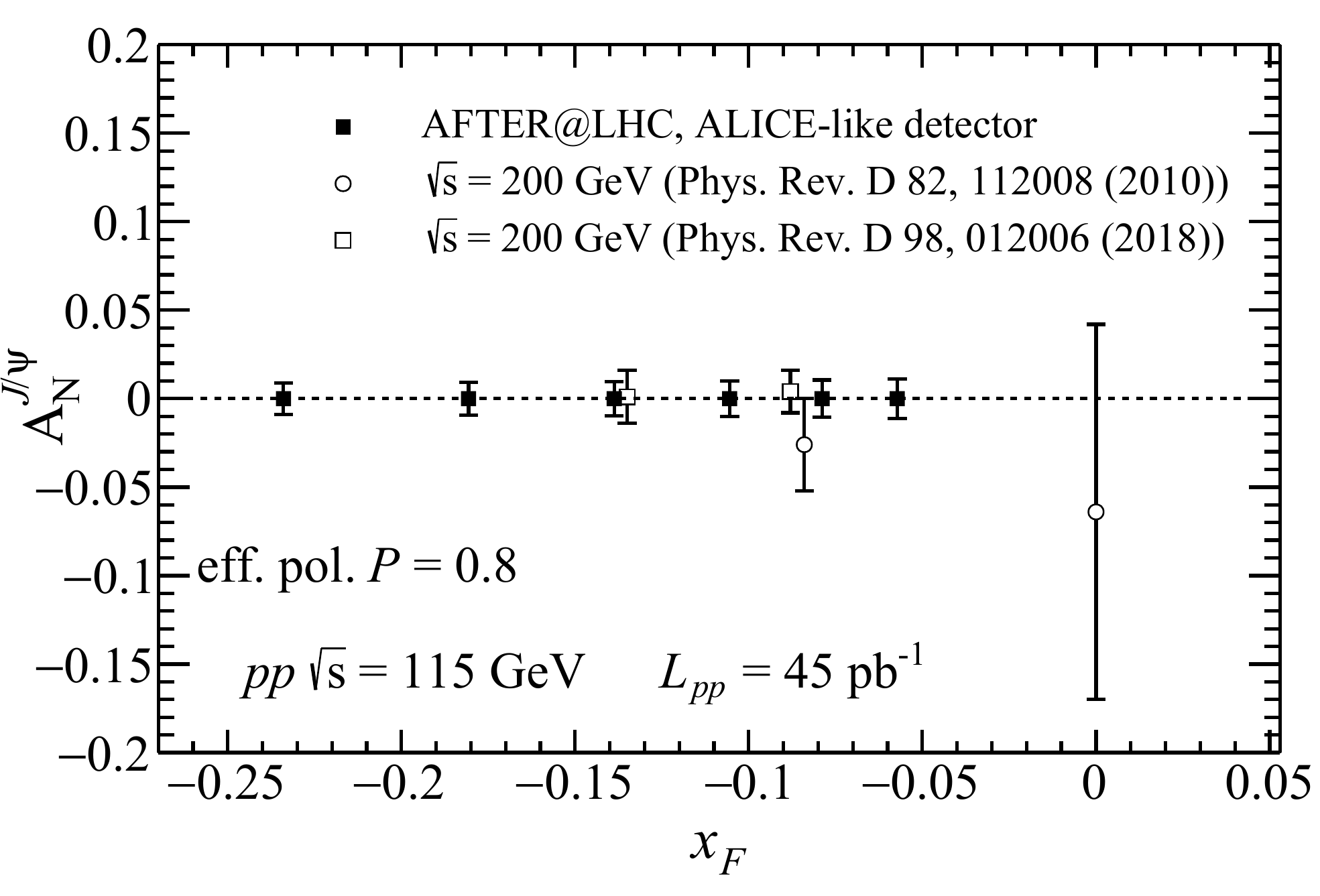}
\caption{Statistical-precision projections for $J/\psi$ $A_{N}$ as a function of $\xF$ compared to the existing measurements~\cite{Adare:2010bd,Aidala:2018gmp}. The \jpsi\ di-muon spectrum is assumed to be measured in the Muon Spectrometer of the ALICE detector, with the target located at the nominal interaction point ($z_{\mathrm{target}}\approx 0)$.
The signal and the background are extrapolated at $\sqrtsnn=115$~GeV from the ALICE measurements in~\cite{Adam:2015rta}.
[Figure taken from~\cite{Hadjidakis:2018ifr}]}
\label{fig:An:VectorOnium:ALICE}
\end{figure}

In~\cite{Kikola:2017hnp,Hadjidakis:2018ifr}, first studies of the projected uncertainties on $A_N$ were performed.
\cf{fig:An:VectorOnium} shows the estimated statistical uncertainties at the FT-LHC for $A_N$ as a function of $\xF$ in $\Upsilon$ production in $pp^{\uparrow}$ collisions at $\sqrt{s} = 115$~GeV for an LHCb-like detector with 10 fb$^{-1}$ of luminosity, while \cf{fig:An:VectorOnium:ALICE} shows the expected statistical precision for $A_N$ in \jpsi production with an ALICE-like detector for $pp^{\uparrow}$ collisions with 45 pb$^{-1}$ of luminosity. 
The expected $\Upsilon$, \jpsi and background yields were extrapolated from the \jpsi-rapidity spectrum and the signal-to-background ratios of~\cite{Adam:2015rta} with the procedure described in~\cite{Kikola:2017hnp}. 
The signal-to-background ratio at 115 GeV is 1.2 and an efficiency of 13\% was assumed~\cite{Abelev:2014qha}.
The projected uncertainties, on the order of a few percent, can certainly help in constraining the GSF and the related twist-3 correlators, investigating different phenomenological approaches and entering a more quantitative phase in the study of gluon TMDs.

\subsubsection{$C$-even $\Q$ states}

The production of $C$-even quarkonium states has recently attracted 
great attention both theoretically and experimentally (see Section~\ref{sec:pp}). %
With a detector similar to LHCb, STSAs for $\chi_c$, $\chi_b$ and $\eta_c$ could be measured {at low \pT} in the FT mode, as suggested by several studies of $\chi_c$ states in the busier collider mode down to a $\pT$ as low as 2 GeV~\cite{LHCb:2012af,LHCb:2012ac}.
The first study of inclusive $\eta_c$ production above $\pT=6$~GeV was performed by LHCb together with non-prompt $\eta_c(2S)$ production~\cite{Aaij:2016kxn}. 
Such prompt studies can clearly be carried out by LHCb~\cite{Lansberg:2017ozx}. Indeed, given the lower combinatorial background at lower energies and the fact that the cross section for pseudoscalar charmonium production is similar to that of the vector ones, the low-$\pT$ region should be in reach. It may also be the case for $\eta_b$ production~\cite{Lansberg:2020ejc}, which offers a slightly wider range of applicability for TMD factorisation in terms of the \pT range.

The measurement of STSAs of $C$-even quarkonium states would give a clean access not only to CT3 tri-gluon correlators~\cite{Schafer:2013wca}, but also to $f_{1T}^{\perp g}$ and the GSF of the GPM, if the low-$\pT$ region can be measured. Such processes would offer an oppotunity for comparisons between these frameworks. Estimations of both $\eta_Q$ and $\chi_Q$ STSAs from the CT3 formalism are however not yet available, nor is any robust information on $f_{1T}^{\perp g}$. \ct{t:processes1} presents some yield estimations and the expected $x$ ranges that can be accessed.

\subsubsection{STSAs in associated $\Q$ production}

Associated-production channels~\cite{Dunnen:2014eta,Boer:2014lka,Lansberg:2015hla,Signori:2016jwo,Signori:2016lvd,Boer:2016bfj,Scarpa:2019fol}, where a quarkonium is produced along with another particle (\eg\ another quarkonium, a photon, a lepton-pair, etc.), represent a very useful tool to access $f_{1T}^{\perp g}$ of TMD factorisation, the  GSF of the (CGI-)GPM and the related tri-gluon correlators for CT3 factorisation. With the possibility to scan over the invariant mass of the observed system, one gets an interesting handle on the scale evolution of the Sivers effect. In addition, these associated-production channels enlarge the range of processes (with gluon-sensitive colourless final states) where TMD factorisation is expected to apply, offering various options to verify the universality of the extracted TMDs. A problem however is that such processes usually have small cross sections at RHIC and FT-LHC energies, and this makes their study very challenging and probably requires high luminosity.

\begin{figure}[hbt!]
\centering
\includegraphics[width=0.49\textwidth]{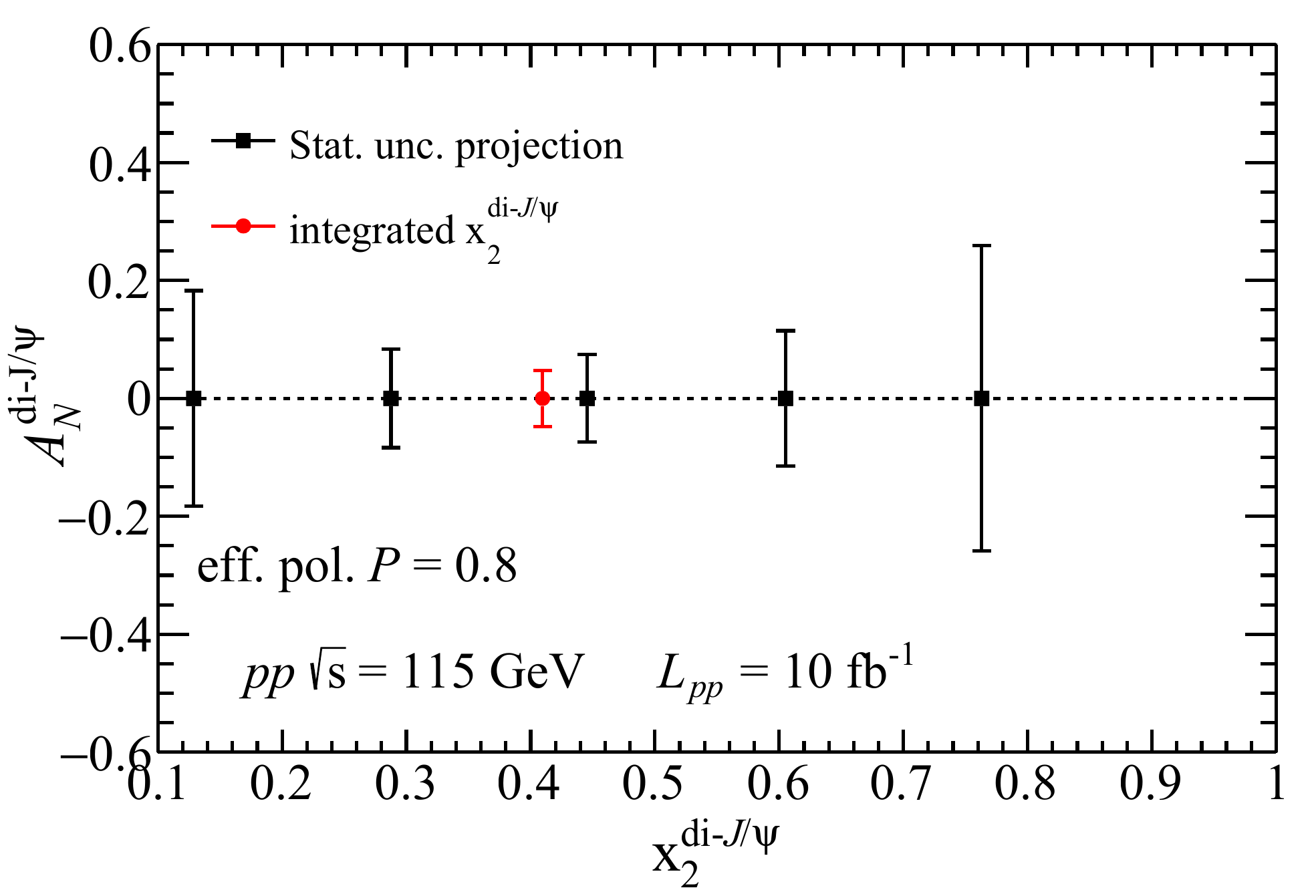}
\hspace{0.1cm}
\includegraphics[width=0.49\textwidth]{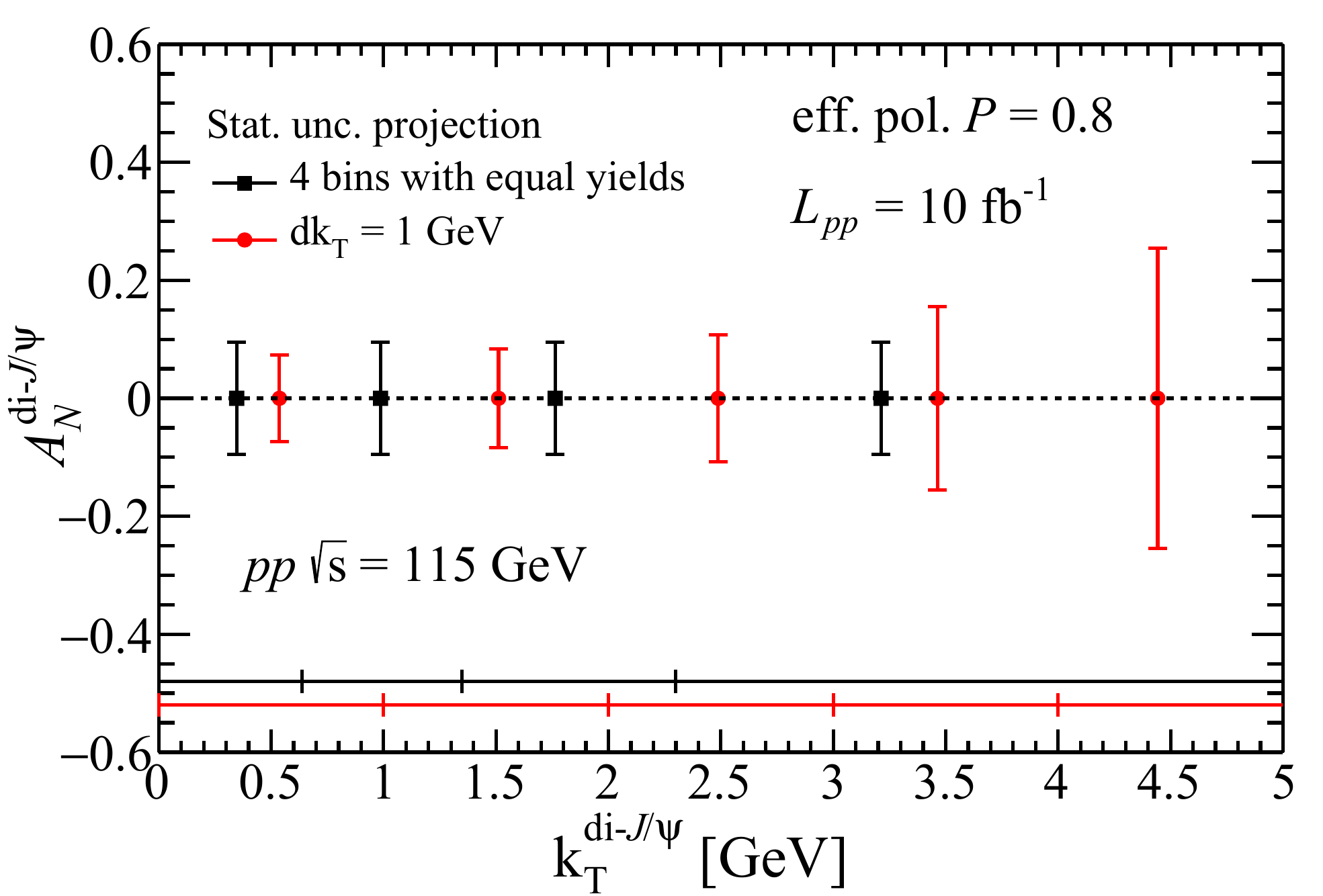}
\caption{Statistical-precision projections for di-$\jpsi$ $A_N$ as a function of (a) $\xF$ and (b) the pair $\kT$ with a LHCb-like detector. 
The horizontal lines in (b) denote the width of the $\kT$ bins used for the calculations.
[Figure taken from~\cite{Hadjidakis:2018ifr}.]}
\label{fig:An:diJpsi}
\end{figure}

Di-$\jpsi$ production is certainly one of the most promising channels since the yields are not too small at the FT-LHC~\cite{Lansberg:2015lva} and the measurement is clearly feasible (unlike, for example, di-$\gamma$ studies). Furthermore, the feed-down contamination is limited to \psip~\cite{Lansberg:2014swa,Lansberg:2015lva}, which probe $f_{1T}^{\perp g}$ in the same way.
\cf{fig:An:diJpsi} shows the expected  statistical precision for $A_N$ obtainable from di-\jpsi\ production at the FT-LHC with the LHCb detector as a function of the transverse momentum of the pair, $\kT$, and the corresponding $x_2$. Two scenarios are considered for the analysis of $A_{N}$ as a function of $\kT$: bins with a fixed width of 1~GeV\ ($d\kT = 1$~GeV, red points) and bins containing equal yields (black points).
Here, the $\kT$ dependence is modelled as a Gaussian distribution with width $\sigma = 2$~GeV. 
The $x_2$-integrated $A_{N}$ will allow for the determination of the STSA with a few percent precision, and the $A_N(\kT)$ will give access to the $\kT$-dependence of the gluon Sivers TMD up to $\kT \approx 4$~GeV, which is not accessible anywhere else.

\section{Proton-nucleus collisions\protect\footnote{Section editors: Michael Winn, Bertrand Duclou\'e. 
}}
\label{sec:pa}
\subsection{Introduction}

Quarkonium production in \pA collisions is studied in several contexts at the LHC. Traditionally, it is used as a baseline for the investigation of quarkonium production in \AaAa collisions, where the production of heavy quark-antiquark bound states with different binding energies contains information about the properties of the final-state deconfined medium (see Section~\ref{sec:aa}). In the absence of any other initial- or final-state effects, any changes to the yield in \pA collisions compared to \pp collisions is attributable to intrinsic modifications of the PDFs in the nucleus with respect to the free proton PDFs. Quarkonium production at the LHC is thereby an interesting probe of the nuclear PDFs (nPDFs) in both the collider and FT operations. Quarkonium production in the collider mode gives access to very low parton momentum fractions, down to below~$x\approx 10^{-5}$, up to still relatively large energy scales (\cf{fig:plane}). This extreme kinematic region has not been constrained by accurate measurements at any heavy-ion facility so far,
and hence deserves further attention. The FT mode allows probing large partonic momentum fractions $x>0.3$, another region where only loose constraints exist at present. 

In addition, \pA collisions provide an ideal arena to explore the dynamics of heavy-quarks in cold-nuclear matter. At collider energies, the non-perturbative hadronisation of the heavy-quark pair is factorisable from its production~\cite{Andronic:2015wma}, %
allowing the study of some universal features of cold-nuclear effects. Initial-state multiple interactions of a colliding parton inside a heavy ion together with final-state multiple scatterings of the produced heavy-quark pair before it exits the nucleus can interfere, which also modifies the yield in \pA collisions compared to \pp collisions (see Section~\ref{sec:pa_cnm_models}).
However, the soft-gluon interaction between the produced heavy-quark pair and the parton spectators of the colliding beams in both \pp and \pA collisions could break the factorisation between the production and the bound-state formation that dominates at lower $\pT$. Such a factorisation breaking can be studied using the CIM (see Sections~\ref{sec:pp-xyz} \& \ref{pAcom}). Systematic theoretical treatments of the multiple scattering between the heavy-quark pair and the traversed nucleus are urgently needed for predictions for the HL-LHC \pA programme~\cite{Albacete:2016veq,Albacete:2017qng}.

Furthermore, the heavy-ion-physics programme at the LHC has shown smooth system-size evolutions of various key QGP signatures. They appear for large final-state particle multiplicities, but they extend also towards lower particle-multiplicity environments in \pPb\ and \pp\ collisions. Quarkonium is interesting due to its role as a signature of the QGP creation, as well as its heavy mass providing an additional dimension in the investigation of these phenomena. 
Strong nuclear modifications of quarkonium production were observed in \pA\ collisions at the LHC~\cite{Andronic:2015wma}, but whether or not they can be ascribed to the modification of the nuclear wave function~\cite{Ferreiro:2013pua,Vogt:2015uba}, or to energy loss, or to other mechanisms including final-state phenomena associated usually to \AaAa collisions, is not yet resolved. Hence, the question is open as to whether quarkonium and heavy-flavour observables are a tool to constrain the hadronic wave function or whether they inform us about final-state parton collectivity; it may also depend on the specific observable. 

In addition to collider data, quarkonium production in the LHC FT mode, pioneered by LHCb~\cite{Aaij:2018ogq}, implemented for higher luminosity in Run~3 for LHCb and  investigated for further upgrades in LHCb and ALICE, opens up new possibilities to study QCD phenomena at large $x$ in nuclei with unprecedented detail and precision~\cite{Brodsky:2012vg,Lansberg:2012kf,Massacrier:2015qba,Hadjidakis:2018ifr,Kusina:2019grp}.  %

All the open questions outlined above need to be addressed in order to deepen our understanding of the QGP properties and of hadron structure at high energies. New quarkonium studies, enabled by higher luminosities in \pA data at the HL-LHC and related theory progress, are reported in the following.

\begin{figure}[h!]
    \centering
    \includegraphics[width=0.6\textwidth]{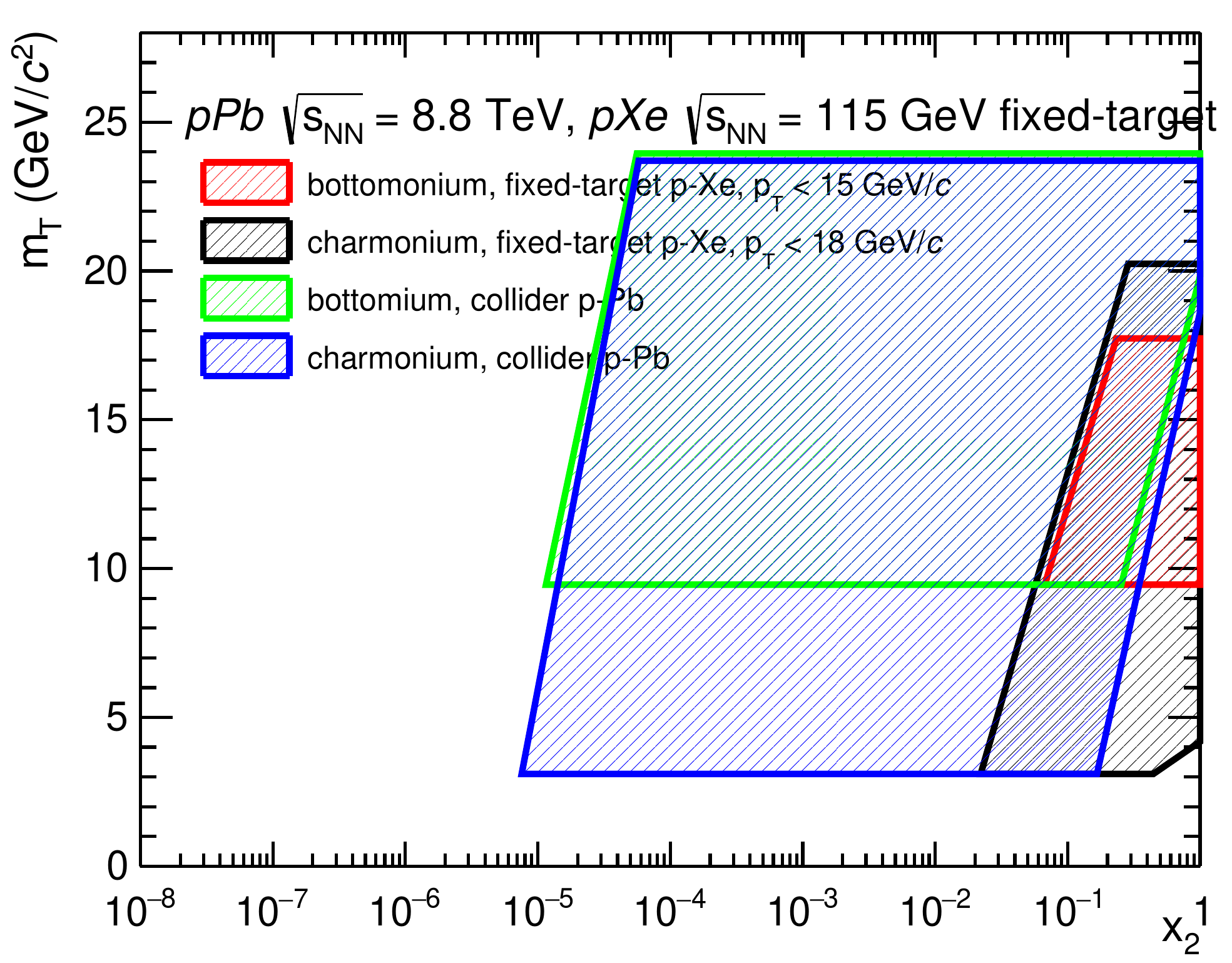}
    \caption{Kinematic plane $M_T$ vs $x_2$ at the LHC in collider and FT mode. The \pT range is limited to kinematic regions where large enough data samples can be collected. For simplicity, small gaps within the experimental acceptance at small momentum in the collider mode are omitted. FT coverages are taken from~\cite{Hadjidakis:2018ifr}.
    }
    \label{fig:plane}
\end{figure}

\subsection{Cold-nuclear-matter effects in $\Q$ production}
\label{sec:pa_cnm}

In view of using \pA collisions as a baseline for \AaAa collisions or as a testing ground for the initial-state partonic structure of the nucleus --\ie\ using nuclear modifications in \pPb\ collisions compared to \pp\ collisions to study the so-called cold-nuclear-matter effects--, one relies on several theoretical models to capture the origin of these effects based on different assumptions.

\subsubsection{Theoretical models: setting the scene}
\label{sec:pa_cnm_models}

\paragraph{Collinear factorisation and nPDFs.} %
The use of nPDFs for calculations involving \pA collisions relies on the assumption that one can factorise \pA\ scattering cross sections into a hard component --the partonic cross section-- and soft components --the PDFs and fragmentation functions (FFs). These soft components are supposed to be universal, meaning in particular that the nPDFs would not depend on the process under study or the collision system (\eg $\ell A$ or \pA).
Even though there is still no proof that factorisation is applicable to collisions involving nuclei, there are few doubts that it applies for processes like Drell--Yan pair production in \pA\ collisions. Yet, this remains the first and probably most important assumption made when performing nPDF fits using \pA\ data, and it indeed deserves further dedicated studies. In addition, it is assumed that the nPDFs encapsulate all the nuclear effects at play. At FT energies, it is very possible that the energy loss can play a visible role in Drell--Yan production whereas, at collider energies, this effect is negligible. %

Existing data usable to fit nPDFs are limited  and the resulting uncertainties are significant, which makes it difficult to perform tests of the factorisation hypothesis. Future LHC \pA\ data for processes probing different scales and incoming parton flavours can be crucial in this respect: since the DGLAP equation governs the evolution from low to high scales and couples quarks with gluons, the inclusion of a given data set in global fits would also have an impact on the description of another. However, disentangling the impact of a possibly violated factorisation assumption is  not trivial at the moment, due to the fact that we have not reached yet an accurate comprehensive description of all cold-nuclear-matter effects. The simultaneous analysis of future LHC data, as well as of those that will be available at the EIC, will certainly help to develop a more complete picture in this respect.

While proton PDF fits have reached a high level of sophistication and precision~\cite{Accardi:2016ndt}, modern nPDF fits~\cite{Eskola:2009uj,deFlorian:2011fp,Kovarik:2015cma,Eskola:2016oht,Khanpour:2016pph, Walt:2019slu,AbdulKhalek:2019mzd} are not yet as precise, for basically three reasons: 
\begin{enumerate}[(i)]
    \item much less data are available from $\ell A$ and \pA\ collisions compared to from $\ell p$ and \pp, and these data cover more restricted ranges in $x$ and the scale;
    \item nPDFs require the further determination of their $A$ dependence, where $A$ is the atomic number of the nucleus, and consequently more data to obtain a similar precision; 
     \item  
since the physics of \pA\ collisions is more complex than that of \pp\ collisions, due to the possible presence of multiple nuclear matter effects, one has to be more cautious when enlarging the data sets used in fits to new reactions.
\end{enumerate}
Whereas the accuracy of proton PDF fits is at NNLO (and further work has started towards N$^3$LO accuracy), a similar level of accuracy for unpolarised nPDF fits is desirable, but so far has only been achieved in fits exploiting nuclear DIS data~\cite{Walt:2019slu, AbdulKhalek:2019mzd}. In the NNLO fit of~\cite{Khanpour:2016pph}, besides DIS data, FT \pA\ Drell--Yan data have been used, covering larger $x$ values, up to $x \approx 1$. 
Unfortunately, data on FT $\ell A$ DIS can only constrain nPDFs down to $x \approx 6 \cdot 10^{-3}$.
In contrast, the minimum $x$ probed in \pA\ collisions at the LHC extends to significantly lower values, depending on the \cm, the kinematic cuts, and the measured final states.
Therefore, the tendency to use collider data obtained in \pA\ collisions to constrain nPDFs has recently increased, mainly thanks to newly available hard-scattering high-precision measurements from the \pPb\ runs at the LHC.

\paragraph{The Colour-Glass Condensate (CGC).} At high collision energy (or small $x$ values), parton densities inside hadrons become so large that non-linear effects such as gluon recombination become important, which can lead to a saturation of the gluon density. The CGC effective field theory
~\cite{McLerran:1993ni,McLerran:1993ka,McLerran:1994vd,Iancu:2002tr,Iancu:2003xm,Gelis:2010nm,Kovchegov:2012mbw}
provides a convenient framework to describe processes in this regime. In this formalism, a \pA collision can be seen as the scattering of a dilute projectile (the proton) on a dense target (the nucleus). Because the gluon density in a nucleus scales roughly like $A^{1/3}$, such non-linear effects are enhanced in \pA compared to \pp collisions. The Nuclear Modification Factor (NMF) $\RpA$ is thus a useful observable to study saturation effects. In particular, the CGC provides a natural explanation for the decreasing behaviour of $\RpA$ as a function of rapidity observed in forward $\jpsi$ production at the LHC, which probes very small $x$ values: since at a given $x$ the gluon density in a nucleus is larger than in a proton and thus closer to saturation, it will not increase as quickly with decreasing $x$ (or increasing rapidity). %
While this general trend was present in the first predictions for this process at the LHC in the CGC formalism~\cite{Fujii:2013gxa} and was subsequently confirmed experimentally~\cite{Abelev:2013yxa}, it turned out that the measured $\RpA$ was much larger than the predicted one. The origin of this discrepancy can be attributed to approximations related to the initial conditions for the gluon density. Indeed, while in the CGC formalism the evolution of the gluon density as a function of $x$ is fully (perturbatively) determined by the Balitsky--Kovchegov equation~\cite{Balitsky:1995ub,Kovchegov:1999yj}, the initial conditions of this evolution, expressed at moderate $x_0 \approx 0.01$, involves non-perturbative dynamics that cannot be computed. The initial conditions for a proton target can be extracted by a fit to HERA DIS data, but the lack of similar high-precision low-$x$ data for nuclei makes some modelling mandatory in the case of proton-nucleus collisions. In~\cite{Fujii:2013gxa}, this was done by taking $Q_{s 0, A}^2=c Q_{s 0, p}^2$, where $Q_{s 0, p}$ and $Q_{s 0, A}$ are the initial saturation scales of the proton and nucleus, respectively, and $c \sim A^{1/3}$. In practice, the value of $c$ was varied between 4 and 6. However a study looking at the $A$ dependence of $F_2$ measured by the NMC collaboration~\cite{Arneodo:1996rv} found that that data is best described with $c$ values between 1.5 and 3~\cite{Dusling:2009ni}. Another way to extrapolate the initial condition of a proton to a nucleus is to use, as in~\cite{Lappi:2013zma}, the optical Glauber model, which assumes that at $x=x_0$ the high-energy probe scatters independently off the nucleons, which are distributed according to the standard Woods-Saxon distribution~\cite{dEnterria:2020dwq}. These two methods were shown to lead to results in good agreement with experimental data~\cite{Ducloue:2015gfa,Ma:2015sia,Fujii:2015lld}.

\paragraph{Coherent energy loss.} Another possible nuclear effect is medium-induced gluon radiation via multiple scatterings of an incoming probe in the target nucleus. In~\cite{Arleo:2010rb}, it was shown that, for long formation times, the interference between initial- and final-state emissions can lead to an energy loss which is proportional to the incoming particle energy. This is in contrast with the Landau--Migdal--Pomeranchuk (LPM) effect~\cite{Wang:1994fx}, at play for short formation times, that shows an energy dependence that is at most logarithmic and is expected to be small at LHC energies. This fully coherent energy loss (the coherent action of all scattering centres in the medium at large formation times) requires both initial and final states to be coloured (in this model, quarkonium production is assumed to proceed via the splitting of an incoming gluon into a quark-antiquark pair in a colour octet state). Under these assumptions the energy-loss spectrum is found to depend  only on one free parameter, the transport coefficient of the nuclear medium $\hat{q}$. A fit to E866 data~\cite{Leitch:1999ea} leads to a value of $\hat{q}=0.075$~GeV$^2$/fm~\cite{Arleo:2012rs} which can then be used to make predictions for other energies. The main output of the calculation is the energy loss probability distribution that, when convoluted with the \pp\ cross section (evaluated at a shifted energy corresponding to the energy loss), leads to the \pA\ cross section. To reduce the number of assumptions and parameters (especially in the case of quarkonia for which the production mechanism is not well understood), in~\cite{Arleo:2012rs} the cross section in \pp collisions is not calculated but instead obtained by fitting experimental data using a simple functional form. However, it is not clear how such an energy loss depends on the production mechanism or on the quantum numbers of the produced quarkonia. Under these assumptions, the resulting prediction for the NMF of $\jpsi$ production was later found to be in good agreement with the measurements at the LHC, both at forward and backward rapidities.

\paragraph{Global view.}
As shown in \cf{fig:jpsiRpA}, several calculations, based on different theoretical models, are able to describe the experimental measurements of the $\jpsi$ NMF as a function of rapidity in \pPb\ collisions at the LHC. The question of which physical mechanisms are responsible for these observed nuclear modifications is still under debate, noting that these effects are not all mutually exclusive. 
In particular, the suppression of quarkonium production at forward rapidities can be caused by shadowing, or by coherent energy losses, or by heavy-quark absorption in the nucleus, or by a combination of more than one of these effects. To ascertain which effects are responsible for the observations, simultaneous measurements of different heavy-flavour probes in \pA\ and \AaAa\ collisions covering the same small $x$ values, in collisions at the same $\sqrtsnn$, can help to {disentangle} them.

\begin{figure}[h!]
	\centering
	\includegraphics[width=0.6\textwidth]{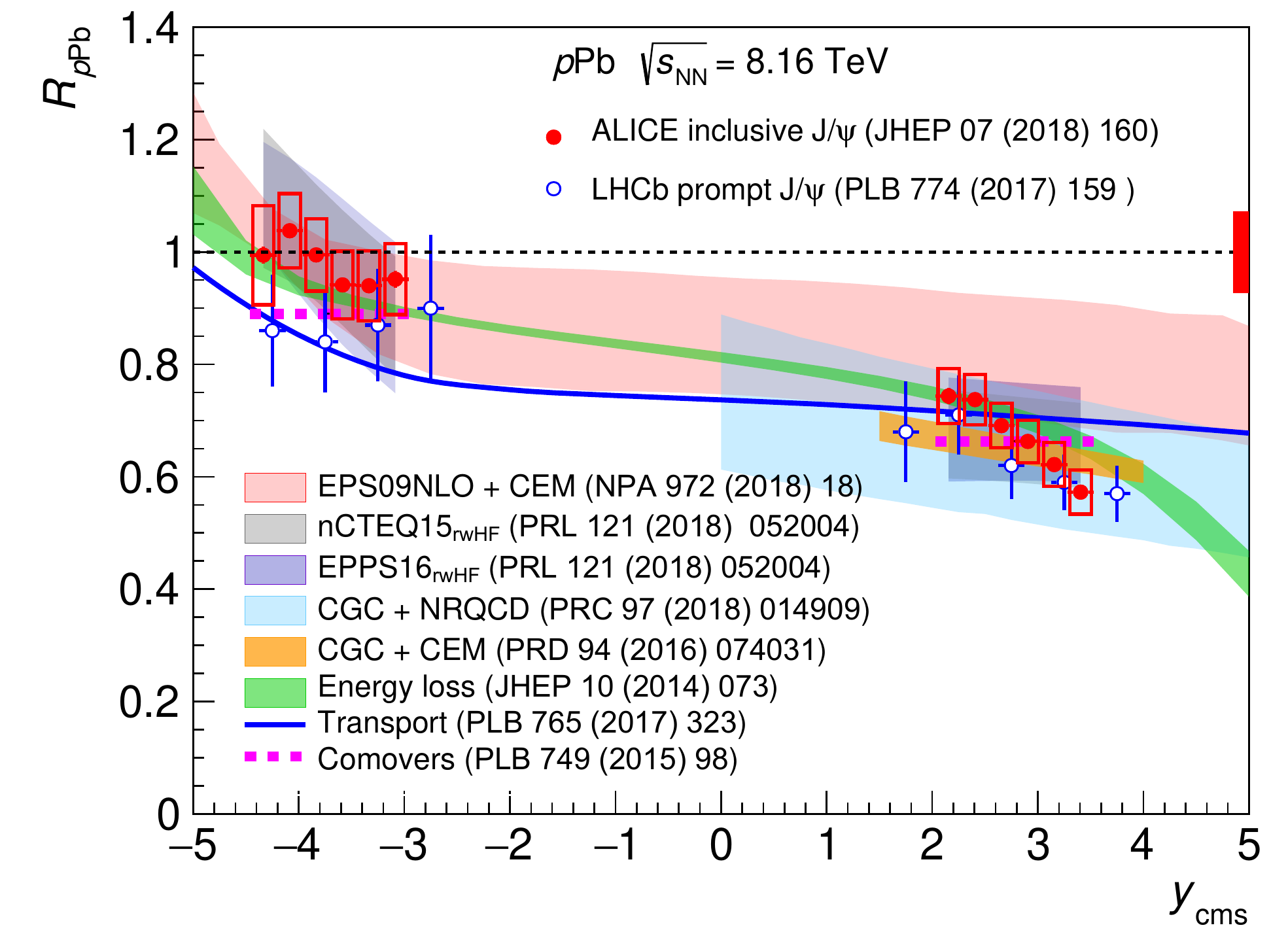}
	\caption{Comparison of the ALICE~\cite{Acharya:2018kxc} and LHCb~\cite{Aaij:2017cqq} measurements of the NMF of $\jpsi$ production in \pPb\ collisions at $\sqrtsnn=8.16$ TeV with several model calculations~\cite{Ferreiro:2014bia,Kusina:2017gkz,Albacete:2017qng,Ma:2017rsu,Ducloue:2016pqr,Arleo:2014oha,Chen:2016dke}. Note that the curves labelled nCTEQ15 and EPPS16 are obtained after reweighting the corresponding nuclear PDF sets using LHC heavy-flavour data. [Figure taken from~\cite{Acharya:2018kxc}.] 
	}
	\label{fig:jpsiRpA}
\end{figure}

It is also worth noting some essential issues in the model calculations. Firstly, the fundamental mechanisms of the heavy-quark pair production in \pA\ collisions could be quite different from that in \pp\ collisions. In particular, when $\pT\lesssim m_Q$, and when one specifically addresses $\pT$-dependent quantities as opposed to integrated ones, the perturbative QCD collinear factorisation approach for quarkonium production is not necessarily the most reliable theoretical approach (see \eg~\cite{Brambilla:2010cs,Lansberg:2019adr} for a review).
As discussed in Secs.\,\ref{sec:quarkonium-pt-pp} and \ref{sec:spin}, the transverse momentum dependent (TMD) factorisation framework should take over the collinear factorisation for heavy-quark pair production when $\pT \ll m_Q$, which makes the nPDFs effect unclear. 

For evaluating nuclear effects in the TMD factorisation approach, one must clarify how to include nuclear size or $A^{1/3}$-enhanced power corrections~\cite{Accardi:2004be,Qiu:2001hj,Qiu:2003vd} into the leading-twist TMD factorisation approach~\cite{Kang:2008us} although, in general, the power corrections in hadronic collisions cannot be factorised beyond the sub-leading power~\cite{Qiu:1990xy,Qiu:1990cu}. Besides, one must understand how to incorporate the nuclear dependence into the non-perturbative TMD distributions~\cite{Kang:2012am}. Interestingly, it has been clarified that the leading-twist TMD factorisation framework can be recovered by getting rid of higher body scattering corrections in the CGC framework~\cite{Altinoluk:2019fui,Altinoluk:2019wyu,Fujii:2020bkl}. Therefore, some precautions are required to compare nPDFs with parton saturation effects. Higher-twist effects can be studied by considering ``clean'' processes, such as Drell--Yan production in \pA\ collisions, and semi-inclusive nuclear DIS at the EIC. See Section~\ref{ex:disc} for experimental prospects.
It has been argued that if the quarkonium is produced at a very forward rapidity, the hadronisation of the pair takes place outside of the colliding heavy ion (see \eg~\cite{Arleo:1999af} and references therein).
Multiple scattering of the produced pair in the nuclear medium could enhance its invariant mass so much (beyond the $D\bar{D}$ or $B\bar{B}$ mass threshold) to prevent the pair from binding and leading to threshold-sensitive suppression~\cite{Qiu:1998rz}. Some model calculations have been carried out along this line in the CGC framework~\cite{Ma:2017rsu}. This threshold-sensitive suppression could be caused by multiple scattering of the pair in the colliding ion or by exchanging soft gluons with co-moving spectator partons of the beam, also known as the comover mechanism. One could investigate this threshold suppression effect on top of the energy loss effect. So, a careful examination  is required  in order to  distinguish the modification of the hadronisation mechanism in nuclear medium from the modification of the initial PDFs, or higher twist effects, and so forth. See Section~\ref{sec:hadronisation-pA} for a discussion about hadronisation.

\subsubsection{Improved constraints on nPDFs from LHC data}

Given the caveats related to possible confounding factors involved in \pA\ collisions already discussed, the usage of quarkonium to improve our knowledge on nPDFs depends on whether or not the nPDF effect is the dominant one. Even though such a dominance may not be straightforward to establish, one should note that the quarkonium data sets are large, relatively precise and thus constraining under this working hypothesis that can be falsified if tensions with other data sets appear.

The first \pPb\ LHC data included into global nPDF fits~\cite{Eskola:2016oht} are from inclusive $W$ and $Z$ boson production measurements, constraining valence and sea quarks down to $x \approx 10^{-3}$ {at high scale}, and di-jet data, with direct sensitivity to the gluon distribution  (whereas inclusive DIS is sensitive to gluons only indirectly, through scale violations) down to $x \approx 5 \cdot 10^{-3}$. 
Including data on charmonium and bottomonium production, as well as open-charm production in the LHC kinematics, can further extend the range in $x$ and the scale. In particular, in~\cite{Kusina:2017gkz}, using LHC heavy-flavour data at \cm energies up to 8~TeV to reweight two nPDF fits (EPPS16 and nCTEQ15) led to evidence for shadowing effects at $x \approx 10^{-5}$ and for antishadowing at $x \approx 10^{-1}$. It was also shown that the inclusion of these data has a strong impact on the gluon nPDF uncertainties, with a reduction of at least a factor of two with respect to the nominal values.

This first very promising analysis encourages further developments. In particular, the matrix-elements that have been used in the study correspond to a purely phenomenological parametrisation, instead of a calculation from a Lagrangian. It would be interesting to check to which extent the results of the study are confirmed by adopting a more sound theoretical description of the considered processes, once it becomes available. This development is not straightforward because of the present limitations of the NRQCD framework, which has not been able to explain all charmonium observables simultaneously. %
In addition,  it was stressed in~\cite{Kusina:2017gkz,Kusina:2018pbp,Kusina:2020dki} that the nPDF constraints set by the LHC \pPb\ heavy-flavour data crucially depend on the value of the factorisation scale, $\mu_F$, at which the reweighting is performed. This should always be kept in mind when these constraints are discussed, especially for the charm(onium) cases, where the constraints are stringent, while the ambiguity from the scale choice is significant.

Moreover, since the analysis of~\cite{Kusina:2017gkz} is based on a reweighting technique, the compatibility with other data (\eg\ Drell--Yan data and DIS data) included in the original nPDF fits has only been assessed a posteriori, by verifying that the goodness-of-fit ($\chi^2$) is not worsened by the inclusion of the additional heavy-flavour data. It would be desirable to go beyond this approach, by including all the data at the same time, \ie\ from the very beginning, in the nPDF fit. This would allow an homogeneous treatment of all parameters in the theory predictions, ensure they are fully consistent with each other, and keep track of correlations between different sets of data. In this first reweighting study, the  data from $D$, $J/\psi$, $B$ and $\ups$ were considered separately. In the case of $D$ and $J/\psi$, it was however noted that the constraints for strong shadowing were very similar, which is a first sign of the universality expected if the nPDF factorisation holds~\cite{Kusina:2017gkz} (See also~\cite{Eskola:2019bgf}). 

It remains an open question whether the antishadowing effect obtained in~\cite{Kusina:2017gkz} is a consequence of the data included, or simply depends on having imposed the momentum sum rule. In order to explore this further, it is important to consider data at various $x$ values, \eg\ in different rapidity ranges, including the antishadowing peak region, as well as overlaps between different data sets, covering slightly different $x$ ranges.

It was also noted~\cite{Kusina:2017gkz} that LHC bottomonium data (as well as $B$ data) are affected by much smaller scale uncertainties but that the current very large experimental uncertainties do not yet allow the gluon nPDF to be constrained.  It is expected that with forthcoming data, in particular those on bottomonium production, the nPDF constraints will further improve~\cite{Citron:2018lsq}. The statistical uncertainties for ALICE and LHCb from Run-3 and -4 data will shrink by about a factor of 5 compared to the 2016 data at $\sqrtsnn=8.16$~TeV, systematic uncertainties will also improve  due to the better precision with which control channels can be determined.
ATLAS and CMS will reduce the statistical uncertainties by about a factor of 2 or 3. These numbers do not include further improvements that may be achieved due to better resolution, acceptances, or efficiencies.

The high-luminosity phase of the LHC should also allow for precise measurements of the feed-down contributions from $\chi_c$ and $\psip$ states to $\jpsi$, and from $\chi_b$, $\upsp$ and $\upspp$ to $\ups$, that are fundamental aspects for reliable predictions of the absolute cross sections for $\jpsi$ and $\ups$ production and to control the impact of final-state effects.%

\subsubsection{\pA collisions at the FT-LHC: high-precision input for global nPDF fits}

In addition to collider-mode data, FT data at the LHC, using various nuclear targets (He, Ar, Ne, Xe, \dots) at relatively low $\sqrtsnn$, can provide crucial constraints for global nPDF fits, especially in the large-$x$ region. These data can be regarded as complementary to the \pA collider measurements because they allow for the study of NMFs  down to small values of $A$. Using relatively light targets is very important not only to understand nuclear-matter effects for increasing system sizes, but also for high-energy-cosmic-ray astrophysical applications~\cite{Zenaiev:2019ktw, Bhattacharya:2016jce}, bearing in mind that the extrapolation from the region of large $A = 208$ (for Pb) to the region of $A \lesssim 20$ is delicate at large $x$. Along the sames lines, a high-luminosity collider run with oxygen will be particularly useful~\cite{Citron:2018lsq}.

In the LHC FT kinematics, one is sensitive to the EMC effect (see~\eg \cite{Arneodo:1996rv}), which is seen to be linked to short-range nucleon correlations~\cite{Hen:2016kwk}. These are strongly nucleus-dependent, and thus one cannot rely on a simple $A$-dependent parameterisation like $\sigma_\pA=\sigma_{pp} \times A^\alpha$ is indeed not suitable in the EMC region.
We also stress the fundamental importance of having a H target to derive a precise enough \pp baseline reference at \sqrtsnn=115~GeV in order to extract \RpA, which can then be compared to theoretical models. 

The high-intensity LHC beams offer unique opportunities for HL \pA data taking in the FT mode as detailed in~\cite{Hadjidakis:2018ifr}. In the kinematic regime accessible with ALICE\footnote{for which a FT upgrade is under consideration.} and LHCb\footnote{for which a FT programme exists with an ongoing upgrade.}, the $x_2$ region to be probed by charmonium and bottomonium production spans from 0.02 up to 1. Quarkonium production in this energy range may be subject to several effects besides those of the nPDFs. One effect is the energy loss (see \eg~\cite{Arleo:2018zjw}); another is the dissociation of quarkonium due to secondary interactions within the nucleus that could explain the suppression patterns at the SPS and partially at RHIC~\cite{Vogt:1999cu,Satz:2005hx,Ferreiro:2008wc,Ferreiro:2011xy}. The separate quantification of these nuclear modifications will require a systematic study of open heavy flavour, quarkonium, and Drell-Yan production over the kinematically available phase space accessible via measurements with large integrated luminosity and varying nuclear targets. Once these ambiguities are lifted based on a careful analysis of the various dependencies, quarkonium production offers unique access to parton densities in the EMC-effect region of the nuclei, where strong nuclear effects were observed for quark degrees of freedom, but where the gluonic modifications remain largely \textit{terra incognita}. 

In the collider mode, this region can also be probed with top quark production~\cite{dEnterria:2015mgr}, which is severely statistically limited, and more promisingingly, with di-jet measurements as already started with CMS data~\cite{Chatrchyan:2014hqa,Sirunyan:2018qel,Citron:2018lsq}.
Compared to these collider-mode data, FT quarkonium data probe very different hard scales and energies but similar $x$ values. These measurements are thus complementary since, as explained in Section~\ref{sec:pa_cnm_models}, their description in collinear factorisation is connected via DGLAP evolution. Performing both kinds of measurements would thus lead to more stringent constraints on nPDFs and potentially allow one to test the factorisation hypothesis.

\begin{figure}[htpb!]
    \centering
    \includegraphics[width=0.48\textwidth]{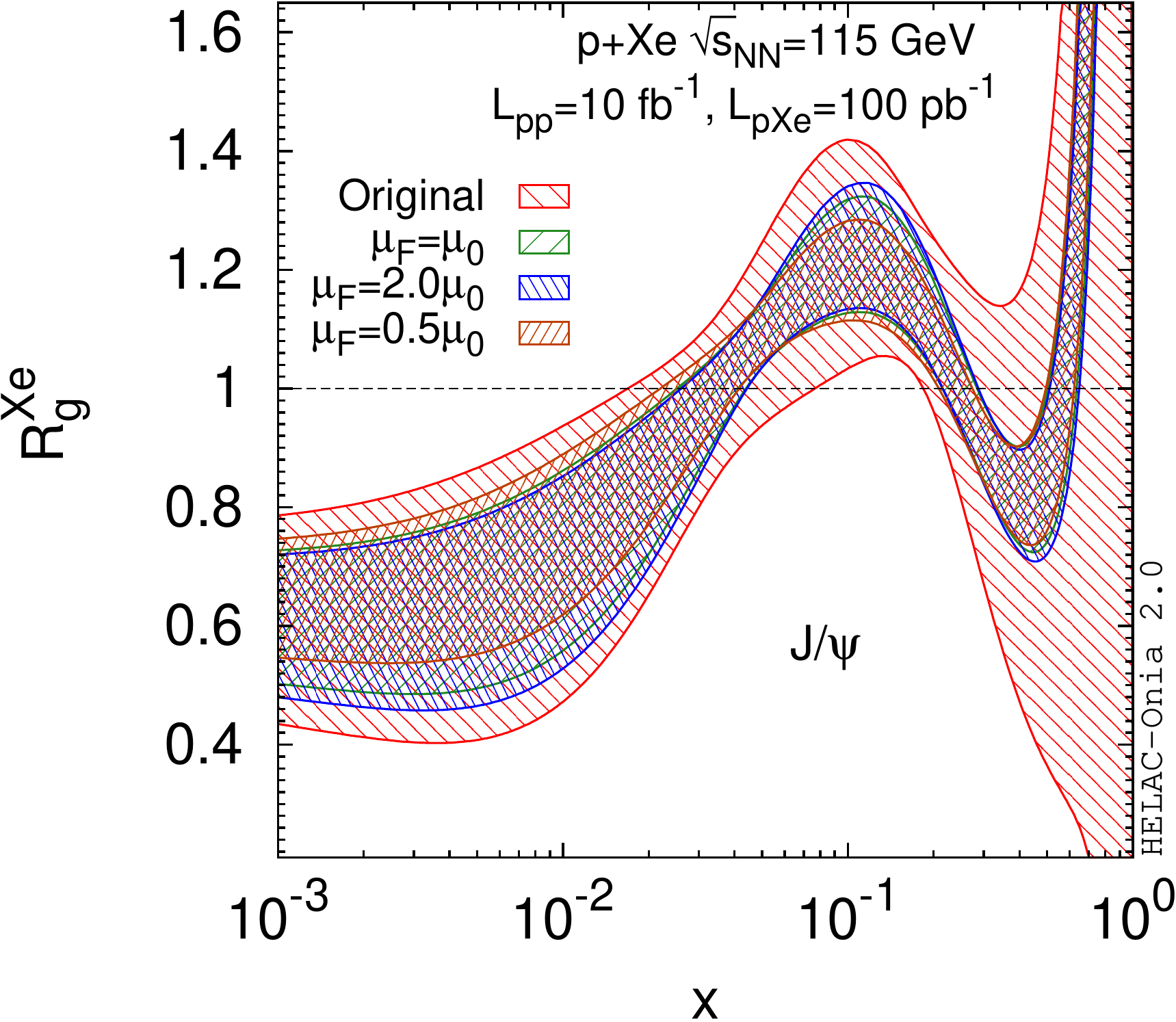}\hspace{0.2cm}
\raisebox{6pt}{\includegraphics[width=0.48\textwidth]{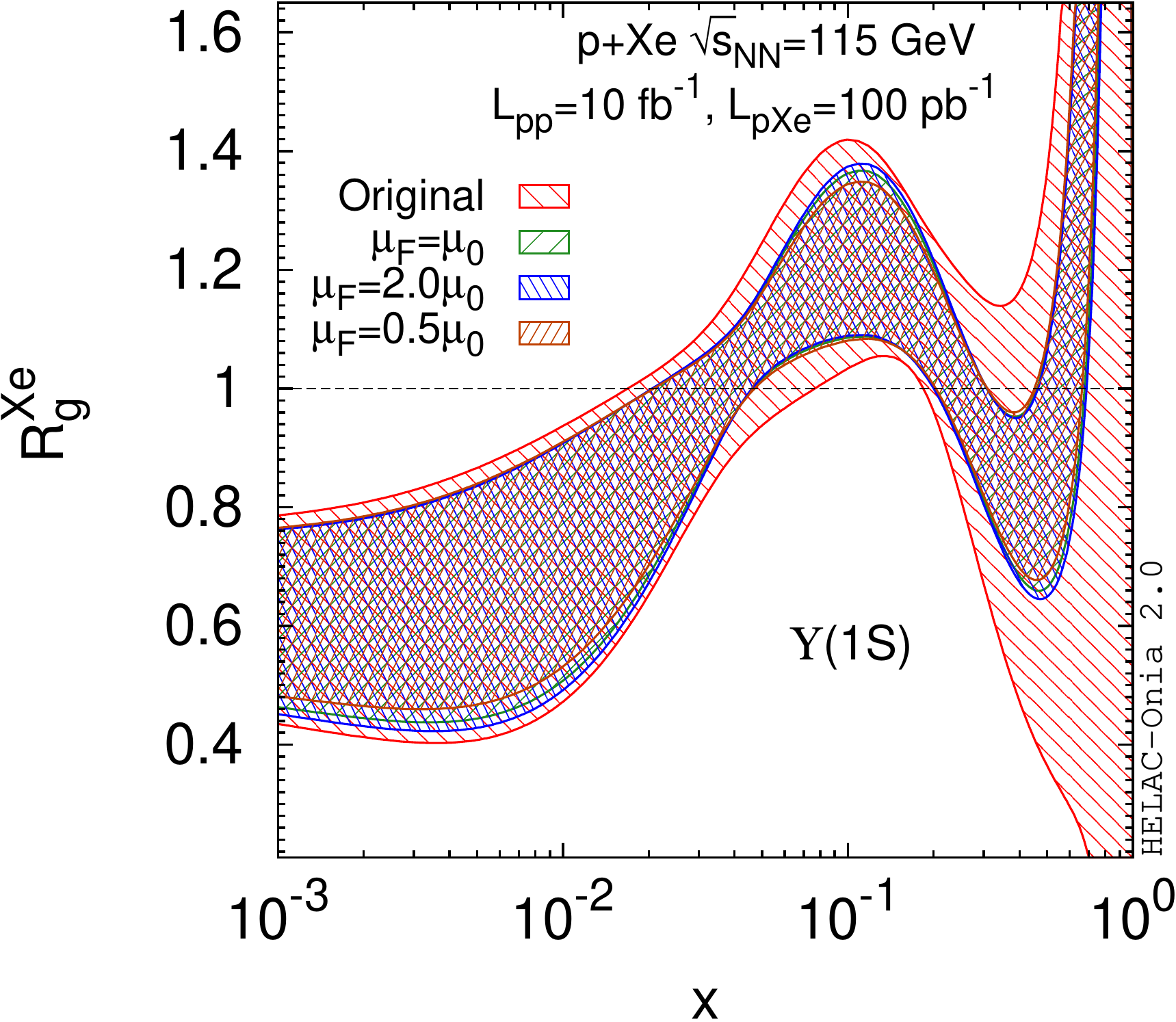}}
    \caption{Ratio of the xenon nuclear over the proton PDFs, showing the impact of a nPDF reweighting according to generated pseudodata for prompt $\jpsi$~production~(Left) and $\ups$~(Right) for integrated luminosities of 10~fb$^{-1}$ (for \pp collisions) and 100~pb$^{-1}$ (for $p$Xe collisions) in a LHCb-like setup. A strong reduction in the nPDF uncertainty is visible in the EMC region $0.3<x<0.7$. [Figures taken from~\cite{Hadjidakis:2018ifr}].}
    \label{fig:pAFT}
\end{figure}

Several possible experimental setups will allow for high-luminosity \pA data-taking in ALICE and LHCb~\cite{Hadjidakis:2018ifr}. Gaseous targets as pioneered with the upgraded SMOG setup~\cite{LHCBSMOG}, starting operation in 2021, will enable the collection of high-luminosity \pp\ data. %
Fig.~\ref{fig:pAFT} illustrates the constraining power of \jpsi and $\ups$ production for the nPDFs in the EMC region based on the yearly  luminosities achievable with LHCb and a gaseous FT setup, assuming that nPDF effects are the dominant mechanism behind nuclear modifications or that other nuclear effects are subtracted, which {\it de facto} means that various probes are measured altogether to disentangle these different effects.

\subsubsection{Developments in theory and phenomenology of QCD at low $x$ and its connection to $\Q$ production}

Most phenomenological studies in the CGC formalism performed recently have relied on the LO approximation, with a subset of NLO corrections related to the running of the $\alphas$ coupling. In particular, this includes  calculations of forward $\jpsi$ production in \pA collisions that have been compared to LHC data~\cite{Fujii:2013gxa,Ducloue:2015gfa,Ma:2015sia}. Given that $\alphas$ is not very small, especially at the low scales probed in $\jpsi$ production, theoretical uncertainties on cross sections are large which is one of the reasons why many studies focus on ratios such as the NMF. The full NLO corrections are expected to be sizeable and must be taken into account for realistic applications and reliable comparisons with experimental data, not only on ratios but also on cross sections. Calculations in this formalism rely on factorisation into a process-dependent hard part, or impact factor, which can be computed perturbatively, and unintegrated gluon distributions, whose $x$ evolution is governed by the Balitsky-Kovchegov (BK) equation, and are assumed to be process-independent (although not all processes probe the same distributions~\cite{Dominguez:2011wm}). Intense efforts are ongoing to extend the accuracy of the CGC framework to NLO, which requires the NLO expressions of both the BK equation and process-dependent impact factors.

The NLO BK equation was first presented in~\cite{Balitsky:2008zza}, and the first numerical solution to this equation showed that it is unstable because of very large and negative NLO corrections enhanced by collinear logarithms~\cite{Lappi:2015fma}. This instability was not unexpected since it already appeared with the NLO BFKL equation (the linearised version of the BK equation) and was cured by resumming these logarithms to all orders, restoring the convergence of the perturbative expansion. Similar resummations were proposed in the non-linear BK regime~\cite{Beuf:2014uia,Iancu:2015vea}, apparently leading to a stable evolution~\cite{Iancu:2015joa,Albacete:2015xza,Lappi:2016fmu}. However, it was later found that while the evolution itself is stable, it is affected by a very large resummation scheme dependence that spoils the predictability of the calculations~\cite{Ducloue:2019ezk}. This was traced back to the fact that the studies in~\cite{Beuf:2014uia,Iancu:2015vea} considered the evolution as a function of the projectile rapidity and not the target rapidity, which is the natural variable in this context and simply corresponds to $\ln 1/x_\text{Bj}$ in the case of DIS. Considering the evolution as a function of target rapidity, and performing the resummation in this variable leads to a stable evolution, a small resummation-scheme dependence, and a good agreement with HERA DIS data at small $x$~\cite{Ducloue:2019jmy,Beuf:2020dxl}.

The other important source of higher-order corrections is the process-dependent impact factor. The first process for which the NLO impact factor, necessary for saturation studies, was derived is single inclusive light-hadron production~\cite{Chirilli:2011km,Chirilli:2012jd}. The numerical implementation of these expressions~\cite{Stasto:2013cha} showed that the NLO cross section turns negative when the produced-particle \pT is on the order of the saturation scale, which is precisely the regime where the CGC formalism should apply. {A new formulation of the factorisation at NLO~\cite{Iancu:2016vyg}, which relaxes some approximations made in~\cite{Stasto:2013cha} and which leads to a factorisation that is non-local in rapidity}, was shown to lead to physical results~\cite{Ducloue:2017mpb}. It was also found that a very similar problem appears with the recently-derived NLO impact factor for inclusive DIS~\cite{Beuf:2017bpd} and that it can be solved in the same way~\cite{Ducloue:2017ftk}.

Despite recent progress in the two directions (evolution and impact factors), there is at the moment no phenomenological study taking into account both sources of NLO corrections in \pA\ collisions, which would be necessary for more reliable comparisons with data. In addition, only a few impact factors have been computed up to NLO, and these only consider massless quarks as additional complications arise due to finite mass effects. Therefore it is difficult to expect comparisons of NLO calculations with LHC quarkonium data in the near future. However there are better prospects for relatively simpler processes such as forward light hadron~\cite{Chirilli:2011km,Chirilli:2012jd,Stasto:2013cha,Stasto:2014sea,Iancu:2016vyg,Ducloue:2017mpb} or isolated photon~\cite{Benic:2016uku,Benic:2018hvb} production, which can also probe very small $x$ values in the target.

The experimental study of a wider variety of processes at the LHC is also important in view of the current limitations related to the modelling of the initial conditions for a nucleus (see the discussion in Sec.~\ref{sec:pa_cnm_models}). Future \pA\ data on isolated photon and Drell--Yan production
would provide direct constraints on the nuclear wave function, and allow one to test whether these various processes can be described with a single set of parameters. Theoretical calculations for these processes are also under better control, as they are not affected by the uncertainties related to the hadronisation mechanism of quark-antiquark pairs into quarkonia.

Future data on light hadrons production would also be valuable. While absolute cross sections for this process are affected by rather large fragmentation functions uncertainties~\cite{dEnterria:2013sgr}, the NMF is a more robust observable and the comparison with other processes could help to discriminate between different nuclear suppression mechanisms. For instance, in the CGC approach, the suppression at forward rapidities is similar for light hadrons and Drell-Yan production~\cite{Ducloue:2017kkq,Ducloue:2017zfd}. On the other hand, in the coherent energy loss model, the first process shows a sizeable suppression~\cite{Arleo:2020eia,Arleo:2020hat} while the NMF for Drell-Yan is unity at forward rapidity since one considers the production of a colourless object~\cite{Arleo:2015qiv}. The comparison between quarkonium and Drell-Yan suppression was advocated in~\cite{Arleo:2015qiv} as a possible way to discriminate between initial- and final-state effects (see the discussion in the following section). However light hadrons production may be a more reliable example than quarkonium of a process sensitive to coherent energy loss, since it is generally better understood.

\subsubsection{Discrimination of different nuclear effects with new measurements}
\label{ex:disc}
In order to resolve the ambiguity between the energy loss and the nuclear modification of the nuclear wave function, electromagnetic measurements, like Drell--Yan pair ~\cite{LHCb:2018qbx} or photon~\cite{ALICE-PUBLIC-2019-005} production, in particular at forward rapidities, involving hard scales similar to quarkonium production can be crucial.
The authors of~\cite{Arleo:2015qiv} proposed to study the ratio of the NMFs of $\jpsi$ over Drell-Yan production as a function of rapidity. They showed that this ratio has a very different behaviour in calculations employing nPDFs, where it remains close to or above unity, and in the coherent energy loss model, where it decreases quickly (see~\cf{fig:DY} (Right)). However, it should be noted that a full cancellation of the scale uncertainties is assumed in these different processes, which may not be justified and needs to be investigated with further studies. A calculation of this ratio in the CGC formalism shows that it is relatively flat and close to unity~\cite{Ducloue:2017zfd}. Therefore this observable could help to discriminate between  models based on the energy loss and on nuclear modifications of parton densities, with or without saturation. The NMF of Drell-Yan production itself is also of great interest: since this process is not subject to coherent energy loss, this ratio is unity in the energy-loss model at forward rapidities, where isospin effects are negligible~\cite{Arleo:2015qiv}. On the contrary, recent nPDF fits such as nCTEQ15~\cite{Kovarik:2015cma} and EPPS16~\cite{Eskola:2016oht} show a rather strong shadowing at small $x$ values, which would lead to a NMF significantly below unity. This can be seen for example in \cf{fig:DY} (Left), where the central value for the LHCb projection was obtained using the EPPS16 NLO set. It has to be emphasised that here the nPDF parameterisations for the gluon distribution assume the same functional shape as for the sea quarks, which is currently difficult to test with the limited data available. Future, more accurate data could invalidate this assumption and show the need for new nPDF fits with more flexible parameterisations. In view of this, one is entitled to question the discriminative power of such ratios of NMF  advocated in~\cite{Arleo:2015qiv}. Yet, a measurement as precise as possible of the Drell-Yan process in this kinematic regime is welcome as an important and clean probe of the nPDFs in a range where gluons dominate the partonic content. 

\begin{figure}[htpb!]
    \centering
  \raisebox{6pt}{  \includegraphics[width=0.52\textwidth]{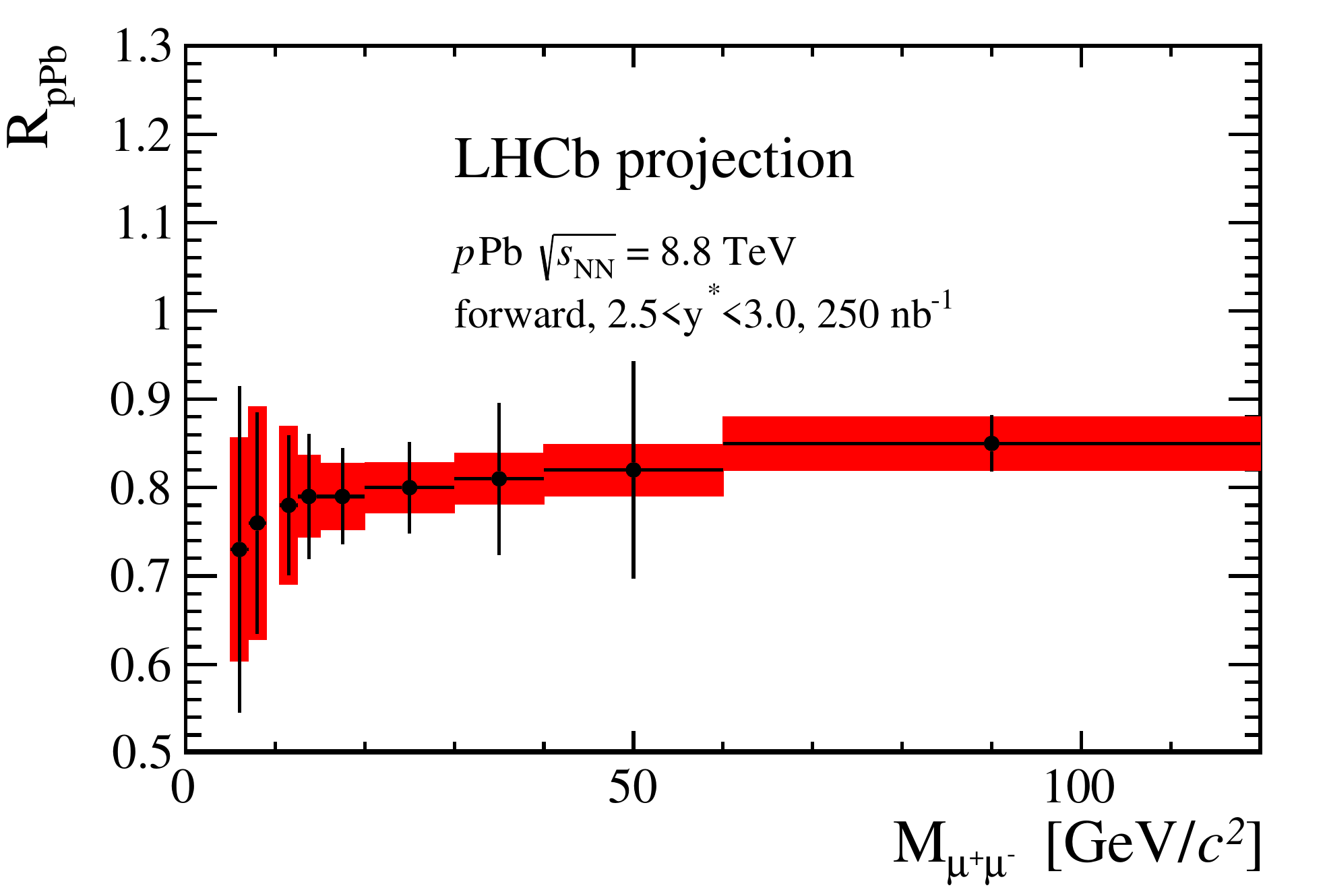} }
    \includegraphics[width=0.45\textwidth]{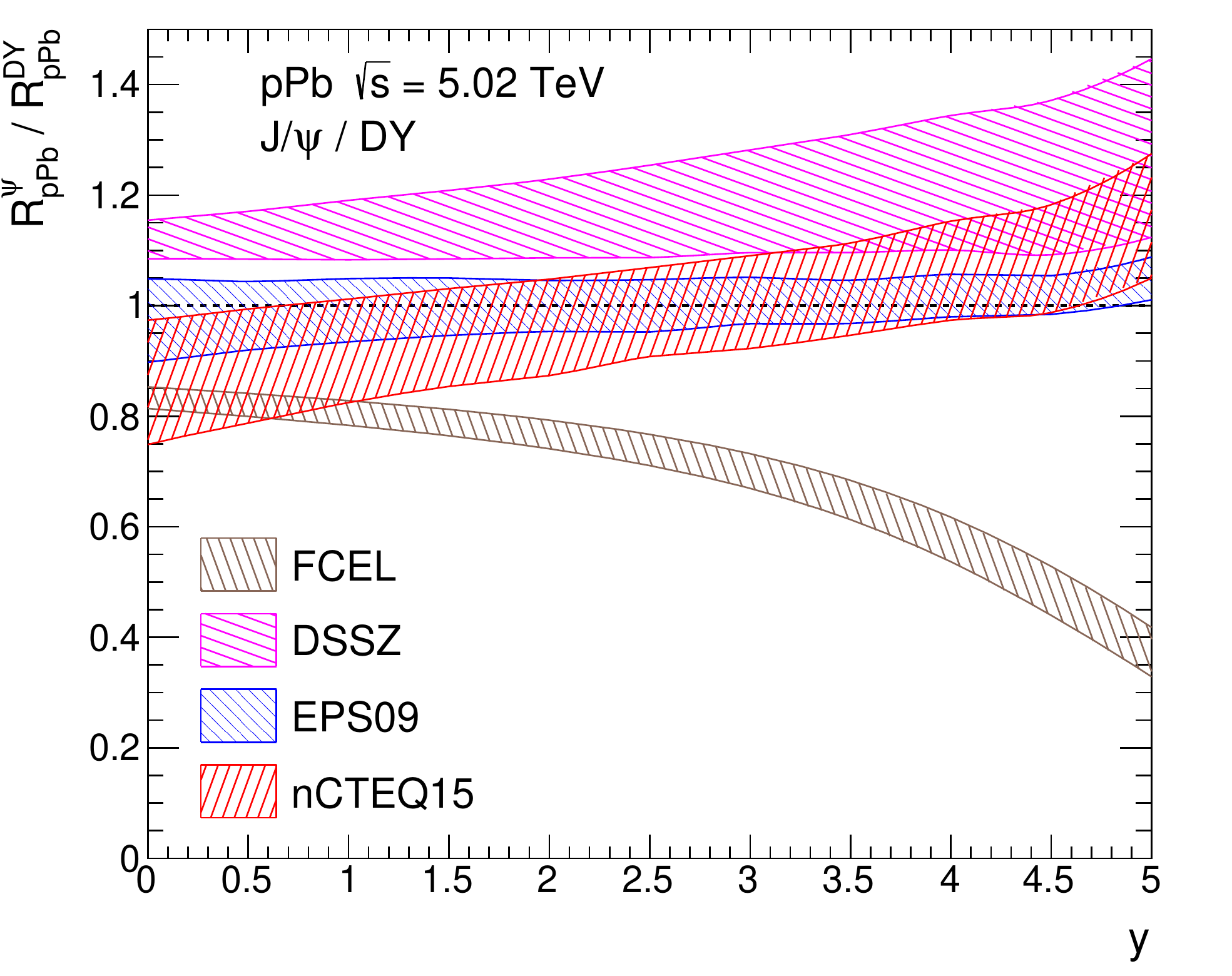}
    \caption{Left: Projection for the measurement of \RpPb for Drell-Yan pair production  with the anticipated integrated luminosity for Run 3 and Run 4 at LHCb, see~\cite{LHCb:2018qbx} for details. [Figure taken from~\cite{LHCb:2018qbx}]. Right: Double ratio of the NMFs for Drell-Yan and quarkonium production in \pPb\ collisions for various nPDFs (see the text for the assumptions used). [Figure adapted from~\cite{Arleo:2015qiv}]. %
    }
    \label{fig:DY}
\end{figure}

Since these measurements access an uncharted domain of the nuclear wave function at low $x$, clearly beyond the reach of the future EIC, they should be pursued with high priority for the quest of non-linear parton evolution, as pointed out in the CERN Yellow Report~\cite{Citron:2018lsq}. Such measurements would also provide new data to be included in global fits of nPDFs, which are at present poorly constrained at low $x$ and scales due to a lack of existing data in this region. This would help to reduce the corresponding theoretical uncertainties and thus lead to a better discrimination with other models.

The studies performed for Drell-Yan measurements illustrate the need for large data samples, as can be seen from the relatively large error bars in \cf{fig:DY} (Left), in all bins except that containing the $Z$ boson. 
The systematic uncertainties, dominated by the knowledge of the background, in particular at low invariant masses, could also be further reduced by improved precision. The CMS collaboration recently made a step in this direction, extending the Drell-Yan measurement down to $M_{\mu\mu}\approx 15$~GeV requiring a $\pT$ greater than 15~GeV for the leading muon and 10~GeV for the sub-leading muon in the central rapidity acceptance of CMS ($|\eta_\mu|<2.4$)~\cite{CMS-PAS-HIN-18-003}. The current precision of the measurement cannot discriminate between different scenarios in this central rapidity region and at still rather high lepton-pair momenta.
In order to make full use of the \pA\ measurements at the HL-LHC, it will also be necessary to improve the theory in order to tame the impact of the factorisation-scale uncertainty that can become the largest of all the theoretical uncertainties, in particular in the regime of charm production~\cite{Kusina:2017gkz,Kusina:2020dki}. 

Our discussion of the aforementioned models has, until now, only concerned vector quarkonium states. 
In parallel, specific efforts are needed both on the theory and experimental sides to compare these models to measurements of \eg\ $\chi_c$ and $\eta_c$. This would allow one to further test the underlying theoretical assumptions. LHCb pioneered the measurement of the hidden-over-open-charm ratio~\cite{Aaij:2017gcy}, but with uncertainties that were too large to see differences with the value obtained in \pp\ collisions. 
More precise measurements are therefore required and could provide information on the magnitude of effects that act differently on quarkonium and open-charm production\footnote{However, the interpretation of such ratios should always be made while considering the possible non-cancellation of theoretical uncertainties such as those from the unphysical renormalisation and factorisation scales.}. These measurements are also of prime interest in the context of heavy-ion-like suppression patterns observed for the excited states as well as considerations related to their azimuthal anisotropies. 

In addition, a comprehensive set of precise quarkonium measurements in \pA collisions including polarisation data may provide better constraints than when considering only \pp\ data in data-theory comparisons and allow for additional consistency checks. For example, in the CGC+NRQCD approach, each intermediate state is suppressed in a different way~\cite{Ma:2015sia}. Including \RpA data in global fits could be a way to try to set more stringent constraints on the LDMEs and to further test NRQCD, although this, of course, would be at the expense of the predictivity of the CGC+NRQCD approach in \pA collisions.

\subsection{Flow-like phenomena in $\Q$ production}

\subsubsection{Theoretical prospects}

Traditionally, \pA\ collisions were considered to produce final states without any QGP, and were therefore used as a tool to study cold-nuclear-matter effects as well as a baseline reference for QGP effects in \AaAa\ collisions. However, the discovery of collective phenomena in high-multiplicity \pp~\cite{Khachatryan:2010gv} and \pPb~\cite{CMS:2012qk,Abelev:2012ola,Aad:2012gla} collisions at the LHC has led to a change of paradigm. The presence of long-range azimuthal correlations observed for particles produced in such events, quantified by the second harmonic coefficient $v_2$ with respect to the reaction plane, were found to be similar to those found in \AaAa\ collisions, where they are conventionally interpreted as a signature of an anisotropic hydrodynamic flow built up in the QGP. Various mechanisms have been proposed to explain the experimental observations. These include the formation of a hydrodynamically expanding mini-QGP and a initial-state gluon saturation within the CGC formalism.

In this context, the case of quarkonia is of particular interest. Precise measurements in \PbPb collisions at the LHC showed a significant $\jpsi$ flow~\cite{ALICE:2013xna,Khachatryan:2016ypw,Acharya:2017tgv,Aaboud:2018ttm,Acharya:2020jil}. Even though the experimental data are not entirely consistent with the predictions of a parton-transport model~\cite{Du:2018wsj}, the observed flow is believed to originate from the recombination of thermalised charm quarks within the QGP volume or at the phase boundary at low $\pT$, and from the path-length-dependent colour screening and energy loss at high $\pT$. Within the QGP scenario, the $\jpsi$ flow is expected to be practically negligible in \pPb\ collisions. At low $\pT$, the incomplete thermalisation of the charm quarks during the short-lived QGP phase and the small number of initially produced $c\bar{c}$ pairs would result in negligible recombination effects. The small system size is not expected to produce significant path-length dependent effects either. The ALICE and CMS collaborations performed measurements of the inclusive and prompt $\jpsi$ flow in \pPb\ collisions~\cite{Acharya:2017tfn,Sirunyan:2018kiz}. The $\jpsi$ mesons are reconstructed %
via their di-muon decay channel. The flow measurements are performed using associated mid-rapidity charged-particle yields per $\jpsi$ trigger. The contribution from recoil jets is suppressed by a subtraction of the yields in low-multiplicity collisions from those in high-multiplicity collisions. The second harmonic coefficient of the azimuthal distribution of the produced $\jpsi$ is then extracted assuming a factorisation of the flow coefficients of the $\jpsi$ and of the associated charged particles. 
The ALICE  and CMS data clearly indicate significant $\jpsi$ $v_2$ values  approaching the values measured in central \PbPb\ collisions, while the transport model predict smaller values (\cf{Quarkonia_v2_pPb}, Left).
Alternative calculations developed within the CGC approach (see Section~\ref{sec:v2_IS}) result in $\jpsi$ $v_2$ values consistent with the experimental data. Nevertheless, it is worth noting that in this scenario the heavy-quark momentum space anisotropy is not correlated with the spatial anisotropy of the initial state of the collision, while the measurements of $\jpsi$ $v_2$ are done with respect to the charged-particle bulk, and the various LHC and RHIC measurements in small and large collisions indicate that the flow coefficients of the bulk are driven by the initial-state collision geometry~\cite{Nagle:2018nvi}.

\subsubsection{Experimental prospects}

The future LHC data from Runs 3 and 4 will allow for a significant improvement in the precision of the $\jpsi$ flow measurement in \pPb\ collisions. \cf{Quarkonia_v2_pPb} (Right) shows the projection of the expected precision of the ALICE measurement using an integrated luminosity of 500 nb$^{-1}$. The Muon Forward Tracker (MFT) detector~\cite{CERN-LHCC-2013-014,CERN-LHCC-2015-001}, which is part of the 2020--21 ALICE upgrade, will allow the separatation of the prompt $\jpsi$ contribution from the {feed-down} of $B$-hadron decays. A comparable precision is likely to be reachable also in the dielectron channel at mid-rapidity, because of the upgraded central-barrel readout system that will allow one to record basically the same integrated luminosity as the MFT system. The CMS experiment will also be able to improve significantly the precision of the measurement due to a more than six-fold increase of the integrated luminosity compared to that in Run 2. The wide combined rapidity coverage of these measurements will shed light on the origin of the significant $\jpsi$ $v_2$. At the same time, more differential theory predictions  are needed.

The parton-transport model predicts higher $\psip$ $v_2$ values compared to those of $\jpsi$ (\cf{Quarkonia_v2_pPb}, Left), while within the CGC-based model, $\psip$ and $\jpsi$ $v_2$ values are expected to be practically the same. Taking into account the charmonium-signal significance, the expected
statistical uncertainty of $\psip$ $v_2$ is at least one order of magnitude larger than for $\jpsi$, and thus significantly larger than the $\psip$ $v_2$ enhancement predicted by the transport model. Nevertheless, given the sizeable suppression of $\psip$ with respect to $\jpsi$ in the Pb-going direction, the measurement of $\psip$ $v_2$ at forward and backward rapidities is quite interesting in itself.

\begin{figure}[htpb!]
\centering
\includegraphics[width = 0.54\textwidth]{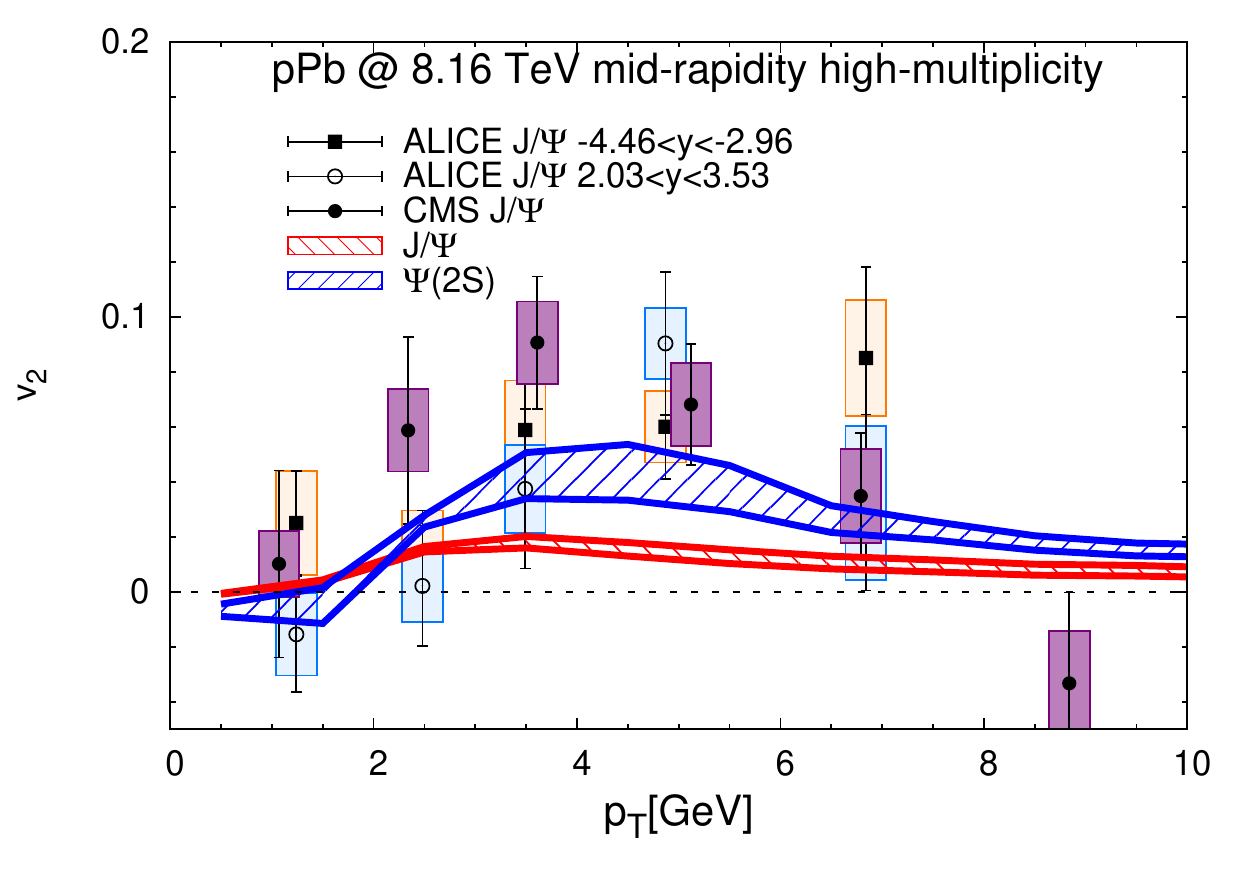}
\raisebox{5pt}{ \includegraphics[width = 0.42\textwidth]{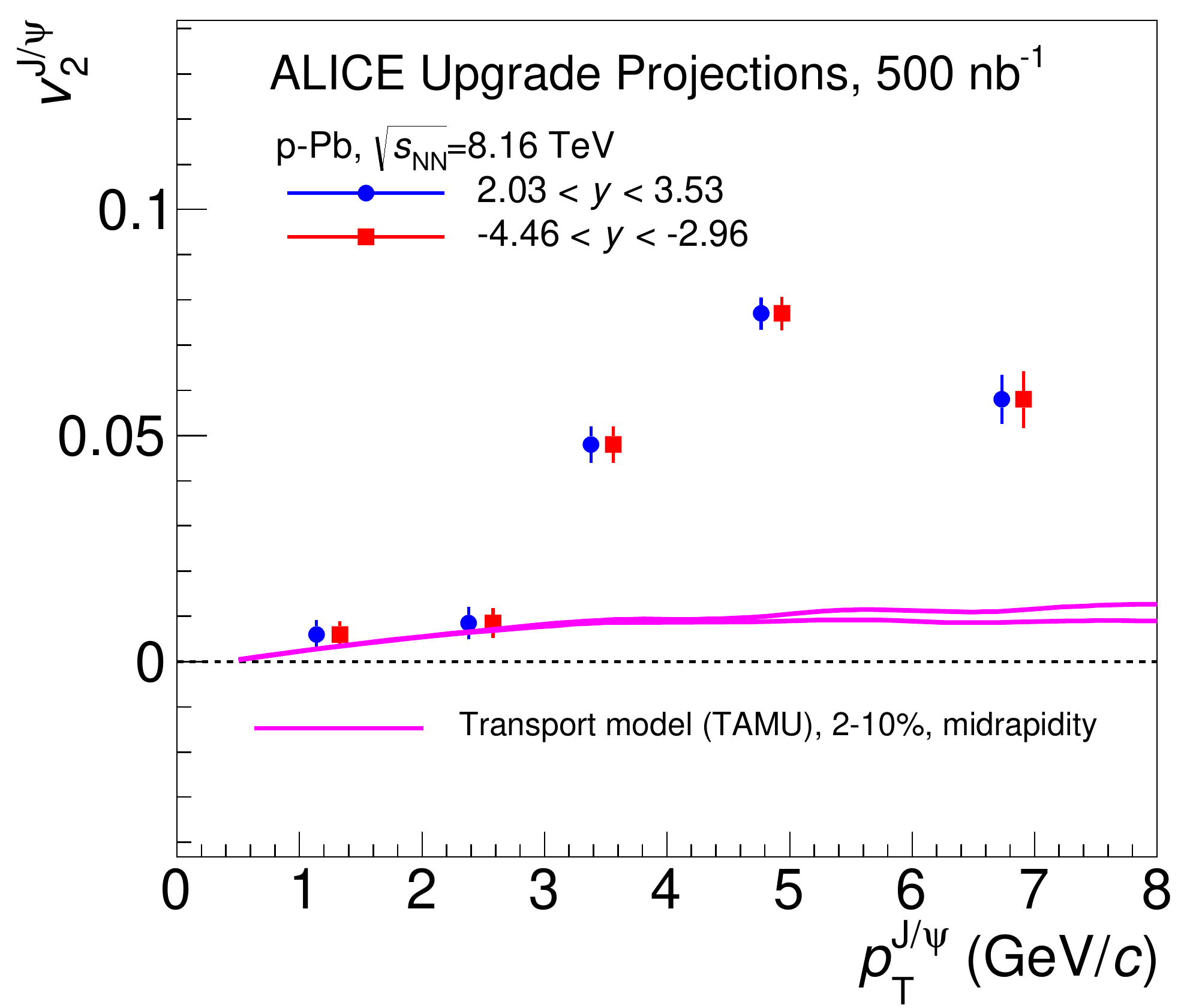} }
    \caption{Left: Comparison between parton-transport-model calculations of {the} $\jpsi$ and $\psip$ azimuthal flow versus $\pT$ (curves) and published ALICE~\cite{Acharya:2017tfn} and CMS~\cite{Sirunyan:2018kiz} data on $\jpsi$ flow (data points) [Figure taken from~\cite{Du:2018wsj}]. Right: Projection of the expected precision of $\jpsi$ $v_2$ to be measured by ALICE in \pPb\ collisions  with an integrated luminosity  of 500~nb$^{-1}$)~\cite{Citron:2018lsq}. [Figure taken from~\cite{ALICE-PUBLIC-2019-001}].}
  \label{Quarkonia_v2_pPb} 
\end{figure}

The measurement of $\ups$ flow will be of particular interest, but a realistic extrapolation of the expected precision with the upcoming luminosities awaits the first measurements by CMS or ATLAS which are better positioned than ALICE for these measurements thanks to larger acceptances and partially better resolutions.

\subsubsection{Azimuthal anisotropies from initial-state effects}
\label{sec:v2_IS}

As can be seen from \cf{Quarkonia_v2_pPb}, current calculations based on parton-transport models seem to not be able to generate enough flow to match the data on $\jpsi$ $v_2$ at the LHC due to the large particle mass and the short-lived system. On the other hand, a recent calculation based on the CGC framework has shown that momentum space anisotropies for heavy mesons can directly be generated  by the \pPb\ collision process itself, without the need for a QGP-droplet phase~\cite{Zhang:2019dth,Zhang:2020ayy}, leading to a good agreement with LHC data (\cf{fig:v2_jpsi_upsilon}). This requires that the gluon density in the nucleus {be} large enough for non-linear QCD dynamics to be relevant, which is the case in high-multiplicity collisions at the LHC. By contrast with the heavy-ion-like flow paradigm, these anisotropies are not connected to the initial spatial anisotropies of the system. 

In this model, one considers that both a quark (which serves as a reference to evaluate the $v_2$) and a gluon (which then splits into a $Q\bar{Q}$ pair) from the projectile proton participate in the interaction with the nucleus. The quark and gluon are assumed to be initially independent, but they can become correlated due to colour interference arising when they interact with the dense gluonic-background fields in the nucleus.
In this initial-state-effect-only picture, the mass dependence of these intrinsic QCD anisotropies is not the same for quarkonia and open production, due to completely different production mechanisms.
In the first case, the $Q$ and $\bar{Q}$ must be produced close to each other in order to form a quarkonium state, and therefore the correlation of the $Q\bar{Q}$ pair with the reference quark is driven by the kinematics of the gluon before the $g \to Q\bar{Q}$ splitting. On the other hand, in the case of open production, one of the heavy quarks is integrated over and the distance between the $Q$ and $\bar{Q}$ can be arbitrarily large, which allows for some mass dependence. 
Along the same lines, it was predicted  that the azimuthal anisotropies for open bottom would be much smaller than for open charm, which was confirmed by experimental data~\cite{Sirunyan:2020obi}, but that the azimuthal anisotropies for $\ups$ would be essentially identical to those for $\jpsi$ because both are directly driven by the gluon kinematics. A future measurement of the $\ups$ $v_2$ at the LHC to either confirm or invalidate this prediction would therefore provide an important test for this model and could give more insight into the origin of azimuthal anisotropies in quarkonia produced in \pA collisions. %
 
\begin{figure}[htpb!]
	\centering
	\includegraphics[width=0.5\textwidth]{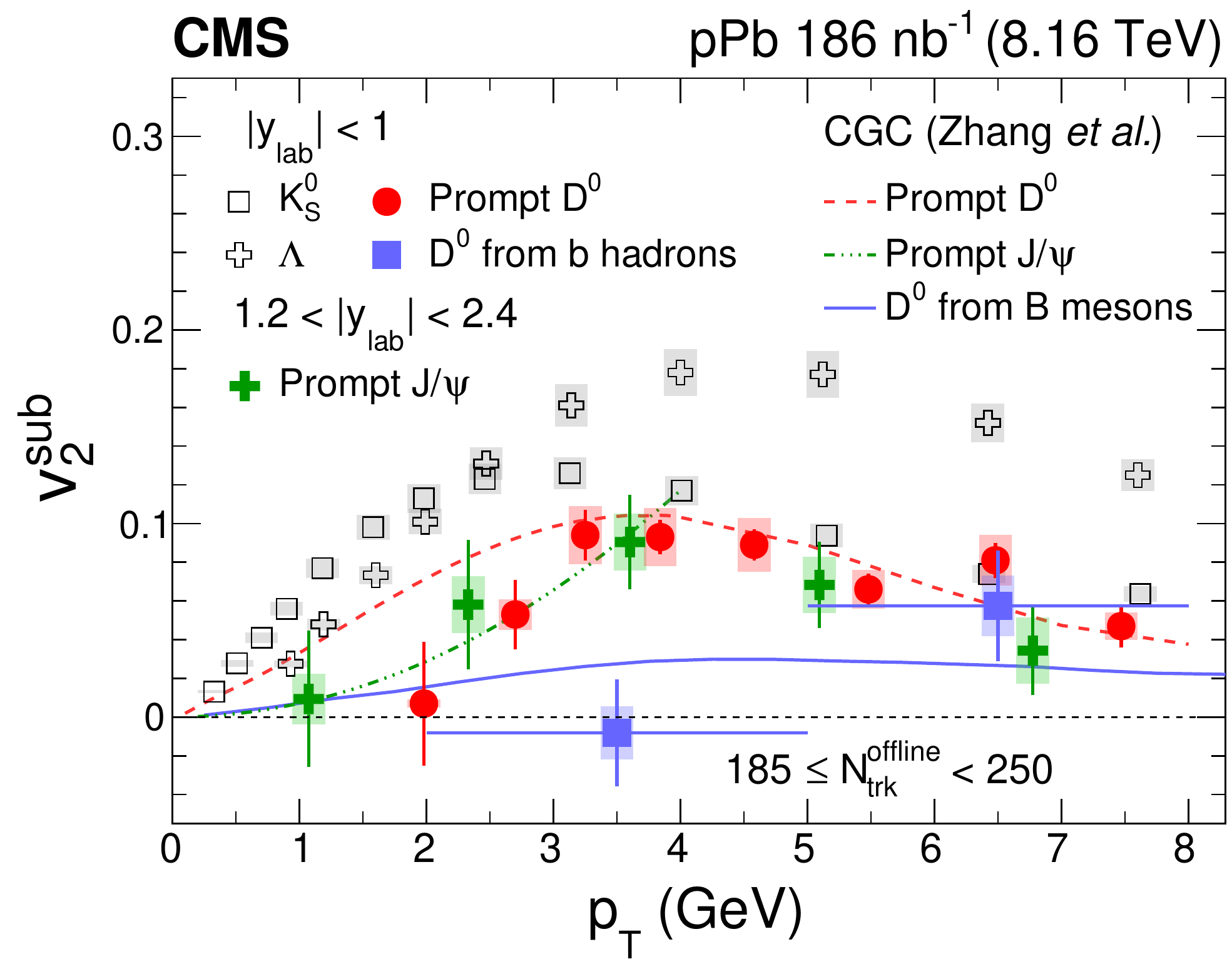}
	\caption{Comparison of the CGC calculation of $\jpsi$ and open-heavy-flavour $v_2$ coefficients versus $\pT$~\cite{Zhang:2019dth} with CMS data~\cite{Sirunyan:2020obi}. [Figure taken from~\cite{Sirunyan:2020obi}].}
	\label{fig:v2_jpsi_upsilon}
\end{figure}

\subsection{$\Q$-hadronisation studies}
\label{sec:hadronisation-pA}

\subsubsection{Theoretical status}

\label{pAcom}
When studying effects on quarkonium production, an extended analysis must include both initial- and final-state effects. However, it is probably more phenomenologically appropriate  to distinguish between effects that impact the whole charmonium (or bottomonium) family with the same magnitude from those which are expected to impact differently the ground and the excited states. 
Among those effects equally affecting all the states of a given family, one naturally finds the initial-state effects, since they act on a pre-quarkonium state, which is not yet fixed when the effects are at work. On the contrary, final-state effects can affect each quarkonium state differently.

Former measurements of the production rates of excited and ground charmonia in \pA collisions at lower energies by the E866/NuSea~\cite{Leitch:1999ea} and NA50~\cite{Alessandro:2006jt} collaborations revealed a stronger suppression of the excited state near $x_F \approx 0$. At such low energies, this dissimilarity has been straightforwardly interpreted as the effect of a stronger breakup of the \psip meson in interactions with the primordial nucleons contained in the colliding nucleus, the so-called nuclear absorption.

However, at higher energies, the quarkonium formation time is expected to be larger than the nucleus radius.
This is due to the large boost between the nucleus (and its nucleons) and the pair. At rest, a $c\bar{c}$ or $b\bar{b}$ pair takes 0.3–0.4~fm to hadronise, but this time has to be considered in the rest frame of the nucleus. The formation time is thus dilated by a large Lorentz-boost factor, which at LHC energies\footnote{except for extremely large Pb-going rapidities.} results in times orders-of-magnitude\red{-}longer than the Pb nucleus radius. In other words, the spectator nucleons of the nucleus cannot discriminate ground and excited quarkonium states, since they cross each others too early.

An alternative explanation has been proposed to be at play in \pA collisions, based on the interaction of the nascent quarkonia with some particles that are produced in the collision and which happen to travel along with the heavy-quark pair. This implies that such {\it comovers}, those particles with a similar rapidity as the quarkonium state, can continue to interact well outside of the nuclear volume up to the moment where the quarkonium is fully formed. Since the excited states are larger, the comover dissociation affects them more strongly than the ground states, which explains the observed difference between them even at high energies.
This effect is naturally more important when the densities of particles are larger. As such, it increases with the centrality and, for asymmetric collisions such as \pA, it will be stronger in the nucleus-going direction. Along the same lines, the same effect should be at work for any colliding system. In particular, the excited states should also dissociate more in high-multiplicity \pp\ collisions (see Section~\ref{sec:pp-xyz}).

In the CIM~\cite{Ferreiro:2014bia}, initial-state and final-state effects are treated separately, respectively via nPDFs and the aforementioned interaction with comoving particles. The behaviour of the excited-over-ground-state ratio is then naturally driven by the comover suppression since the nPDF effects cancel in the ratio.
The rate equation that governs the density of quarkonia is just a Boltzmann equation that depends on the density of particles and their break-up cross section. It has been proposed in~\cite{Ferreiro:2018wbd} that the break-up can be evaluated using an empirical  formula that accounts for the geometrical size and the binding energy of the different states. It can be applied to all the states and yields a natural explanation for the experimental data on excited-over-ground states (\cf{fig:UpsipPb8TeVLHCb}). 

\begin{figure}[htpb!]
    \centering
    \includegraphics[width=0.4\textwidth]{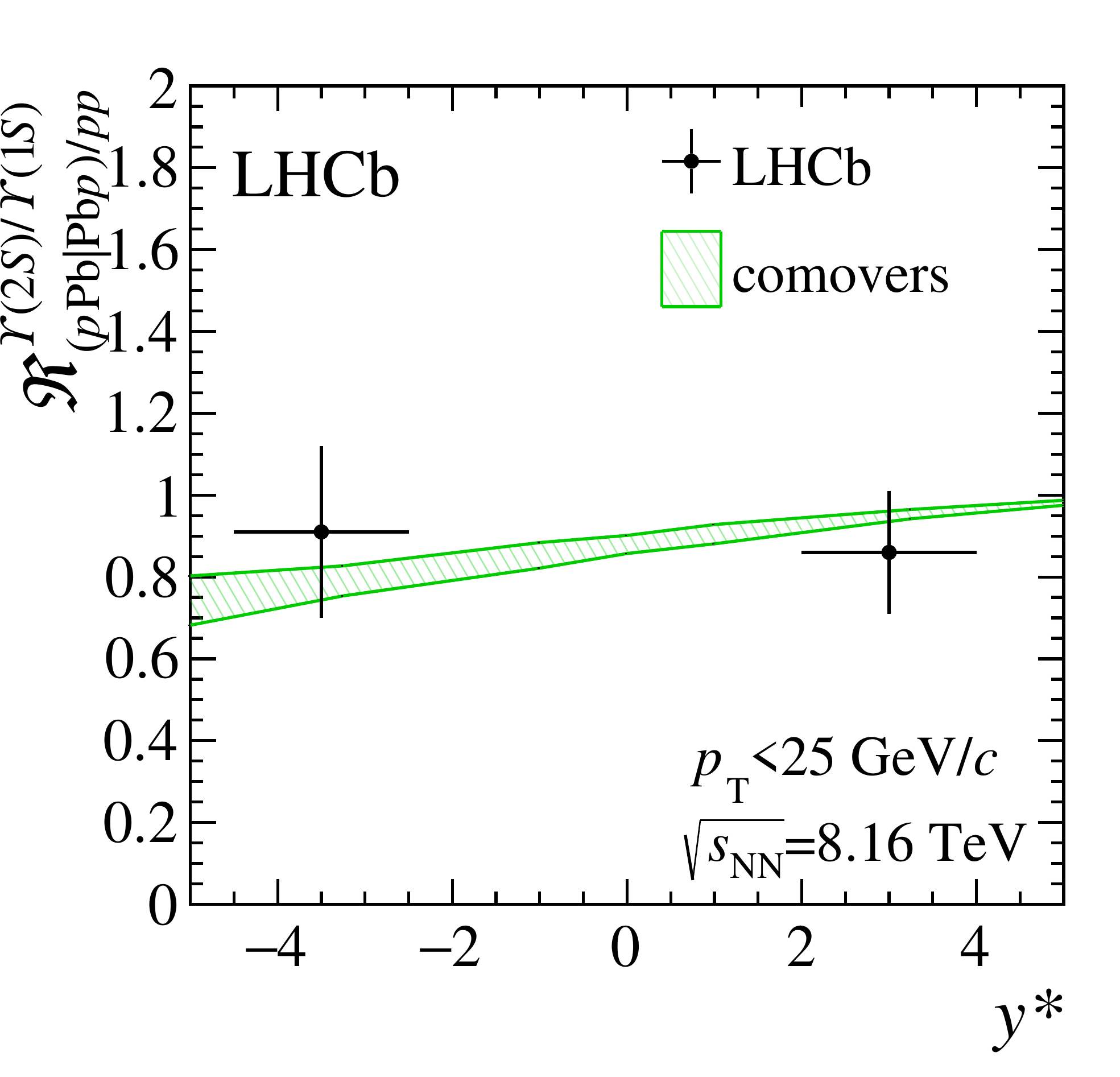}
	\hspace{1.2cm}
	\includegraphics[width=0.4\textwidth]{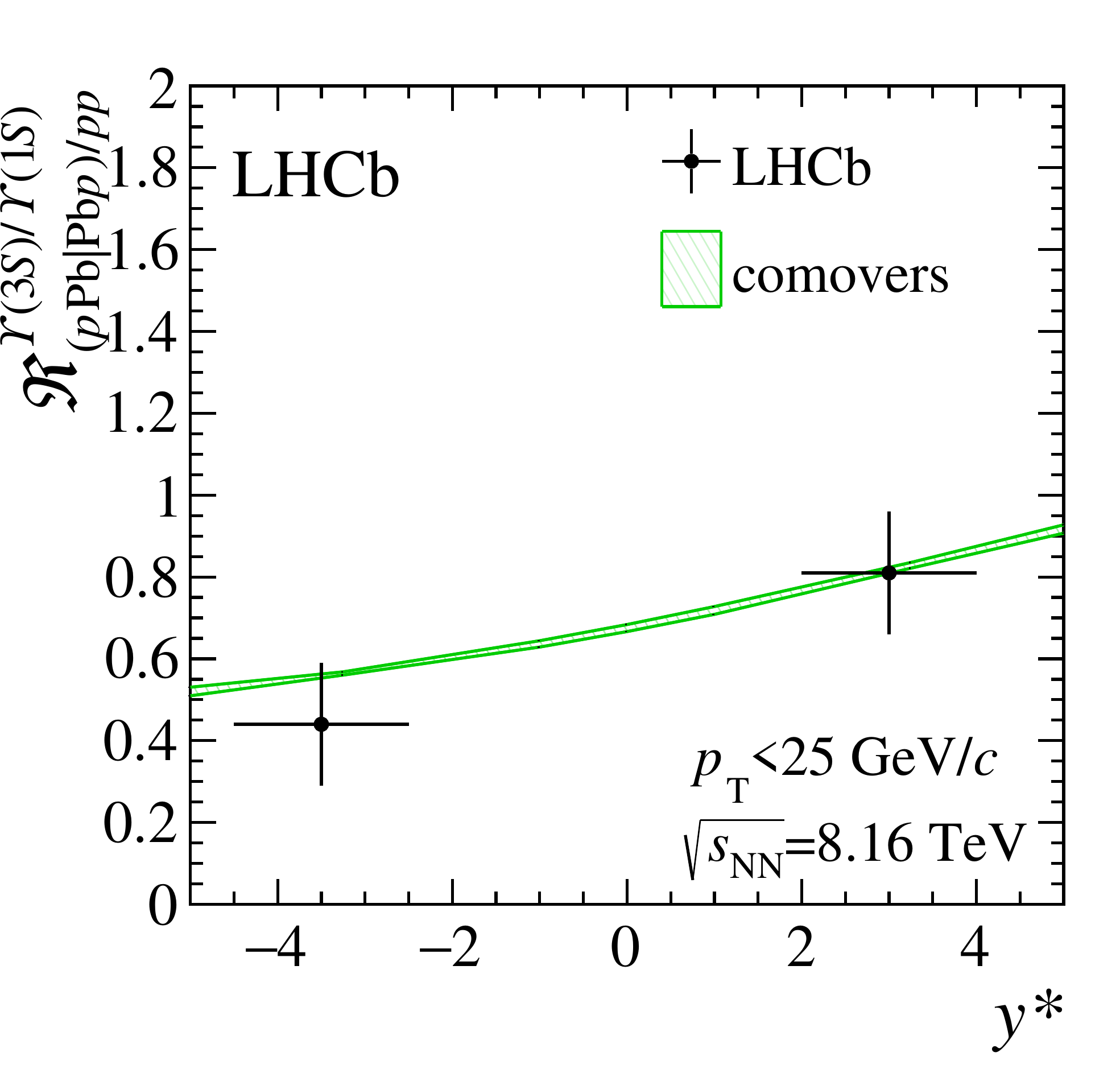}
    \caption{Double ratios for $\upsp$ (Left) and $\upspp$ (Right) as a function of laboratory rapidity in \pPb\ collisions at $\sqrts = 8.16$~TeV. The bands correspond to the theoretical prediction for the CIM~\cite{Ferreiro:2018wbd} compared to LHCb data~\cite{Aaij:2018scz}. [Figures taken from~\cite{Aaij:2018scz}].}
    \label{fig:UpsipPb8TeVLHCb}
\end{figure}
In~\cite{Ma:2017rsu}, the breaking of factorisation was examined for $\jpsi$ and $\psip$ production at forward rapidities in the CGC framework by modelling the threshold-sensitive suppression due to the comover interactions. The factorisation effectively holds for $\jpsi$ production in minimum bias \pA\ collisions, but it may break in high-multiplicity events, even though the $\jpsi$ is a strongly-bound system. The multiple rescattering effects in the CGC framework can describe high-multiplicity events due to the sizeable semi-hard saturation scale~\cite{Ma:2018bax,Ma:2018wdl}, although the underlying physics behind such phenomena is still not very clear (see Section~\ref{sec:oniaparticlemultiplicity}). In high-multiplicity \pA\ collisions, the nuclear-enhanced-comover-rescattering effect could lead to the modification of the $\jpsi$ production rate due to a  threshold-sensitive suppression~\cite{Qiu:1998rz}. For larger quarkonium states, like $\psip$ and $\chic$, both the semi-hard and soft comover rescatterings before the bound-state formation should be essentially of the same size in high-multiplicity \pp\ collisions. Further investigations are thus welcome to examine the duality between the CIM and these qualitative expectations from the CGC framework.

With the increased luminosity available in the upcoming HL-LHC, the measurement
of production rates for multiple quarkonium states, with
different physical sizes and binding energies, offers an excellent tool for probing
the time scale of the evolution of heavy quark-antiquark
pairs into bound colourless quarkonium states.
The study of the yields of excited-over-ground quarkonia as a function of the charged-particle multiplicity, both in \pp and \pA collisions, will help {to} clarify the hadronisation mechanism at play. In particular, the measurement of different quarkonium states with good precision should lead to better model constraints that test its applicability. Estimates of the precision with which the yields can be determined in the different experiments are given in the following section.

In addition, much progress in heavy-meson spectroscopy has taken place in the last years at the LHC. Recently, the LHCb collaboration presented results on the relative production rates of promptly produced $X$(3872) over $\psip$ as a function of particle multiplicity, given by the total number of charged particle tracks reconstructed in the VELO detector for the forward pseudorapidity region, $2 < \eta < 5$ in \pp collisions at 8~TeV~\cite{Aaij:2020hpf}. 
This ratio is found to decrease with increasing multiplicity. Hadronisation mechanisms, as described above, can provide insights on the nature of these exotic states~\cite{Esposito:2020ywk}.

\subsubsection{Experimental prospects}
\label{pAhadro}

The NMFs of excited vector-meson states in \pA collisions show significantly stronger suppression than for the ground states, as measured by the ALICE~\cite{Acharya:2019lqc, Acharya:2020giw, Abelev:2013yxa, Abelev:2014zpa, Abelev:2014oea, Adam:2015iga, Adam:2015jsa, Adam:2016ohd,Acharya:2018yud}, ATLAS~\cite{Aaboud:2017cif}, CMS~\cite{Sirunyan:2017mzd,Sirunyan:2018pse}, and LHCb~\cite{Aaij:2018scz,Aaij:2013zxa,Aaij:2014mza,Aaij:2016eyl} collaborations at the LHC. In order to solidify the experimental findings, it will be crucial to analyse larger statistical samples. %

The ratio between two different quarkonia is a useful quantity to directly compare  the relative suppression of both states between various experiments as a function of different kinematic variables since many systematic uncertainties, both in experiments and theory predictions, cancel in the ratio. We focus here on the expectations for $\psip/\jpsi$ and $\ups(nS)/\ups(1S)$ ratios in \pPb\ collisions in ALICE. These kinds of ratios have already been measured at the LHC~\cite{Acharya:2020wwy, Acharya:2019lqc, Aaij:2018scz}, but the statistical uncertainties are $20\%$ or larger.  
All projections use a total integrated luminosity of 500~nb$^{-1}$ of \pPb\ collisions, split in two sub-samples of equal size for the two possible beam orientations. 

The left panel of \cf{Psi2S_by_JPsi_UpsinS_by_UPsi1S} shows the projection for the $\psip/\jpsi$ ratio as a function of $y_{\cm}$ with the \pPb\ data expected to be collected at $\sqrts = 8.8$~TeV at the HL-LHC, compared to the current Run-2 results at $\sqrts = 8.16$~TeV. This $\psip/\jpsi$ ratio is currently found to have a significance (in units of standard deviations) of 2.9$\sigma$ and 0.9$\sigma$ below the same ratio measured in \pp\ collisions at $\sqrt{s}$ = 8.16~TeV at backward and forward rapidity, respectively. The statistical uncertainty is reduced by a factor of $\sim$5 using the same luminosity increase assumed for other ALICE projections shown in~\cite{Citron:2018lsq}. Similar expectations for the $\upsp/\ups(1S)$ and $\upspp/\ups(1S)$ ratios as a function of $y_{\cm}$ are shown in \cf{Psi2S_by_JPsi_UpsinS_by_UPsi1S} (Right). The $\upsp/\ups(1S)$  ratio is found, at backward and forward rapidity, to be respectively 1$\sigma$ and 0.8$\sigma$ below the same ratio measured in \pp\ collisions at $\sqrt{s}$ = 8~TeV~\cite{LHCb:2015upsilon}. Correspondingly, the $\upspp/\ups(1S)$ ratio is  respectively 0.4$\sigma$ and 1.6$\sigma$ below the \pp\ ratio. The statistical uncertainty is reduced by a factor of $\sim$5.

\begin{figure}[h!]
  \centering
  \includegraphics[width=0.47\textwidth]{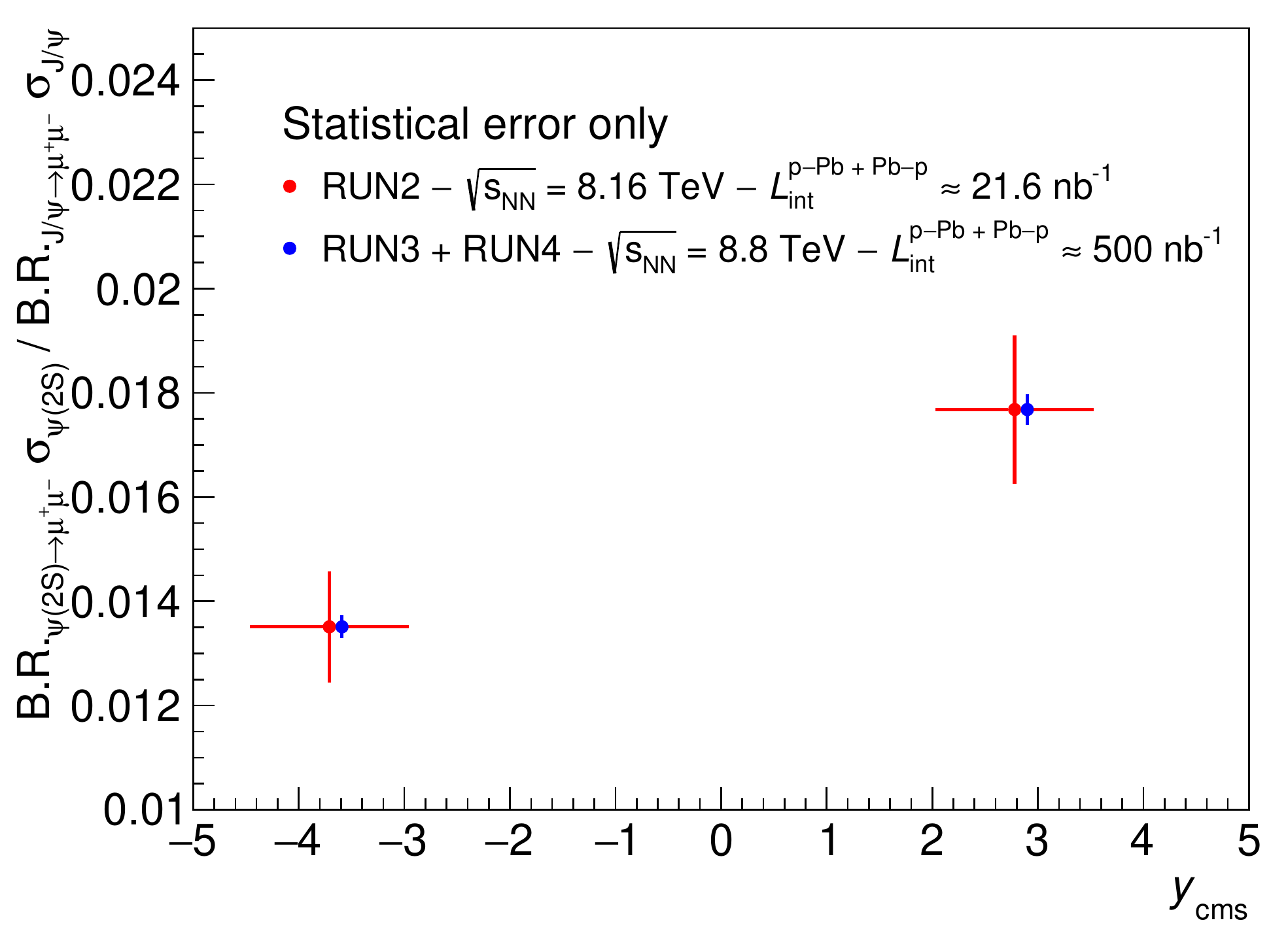}
  \hspace{0.5cm}
  \includegraphics[width=0.47\textwidth]{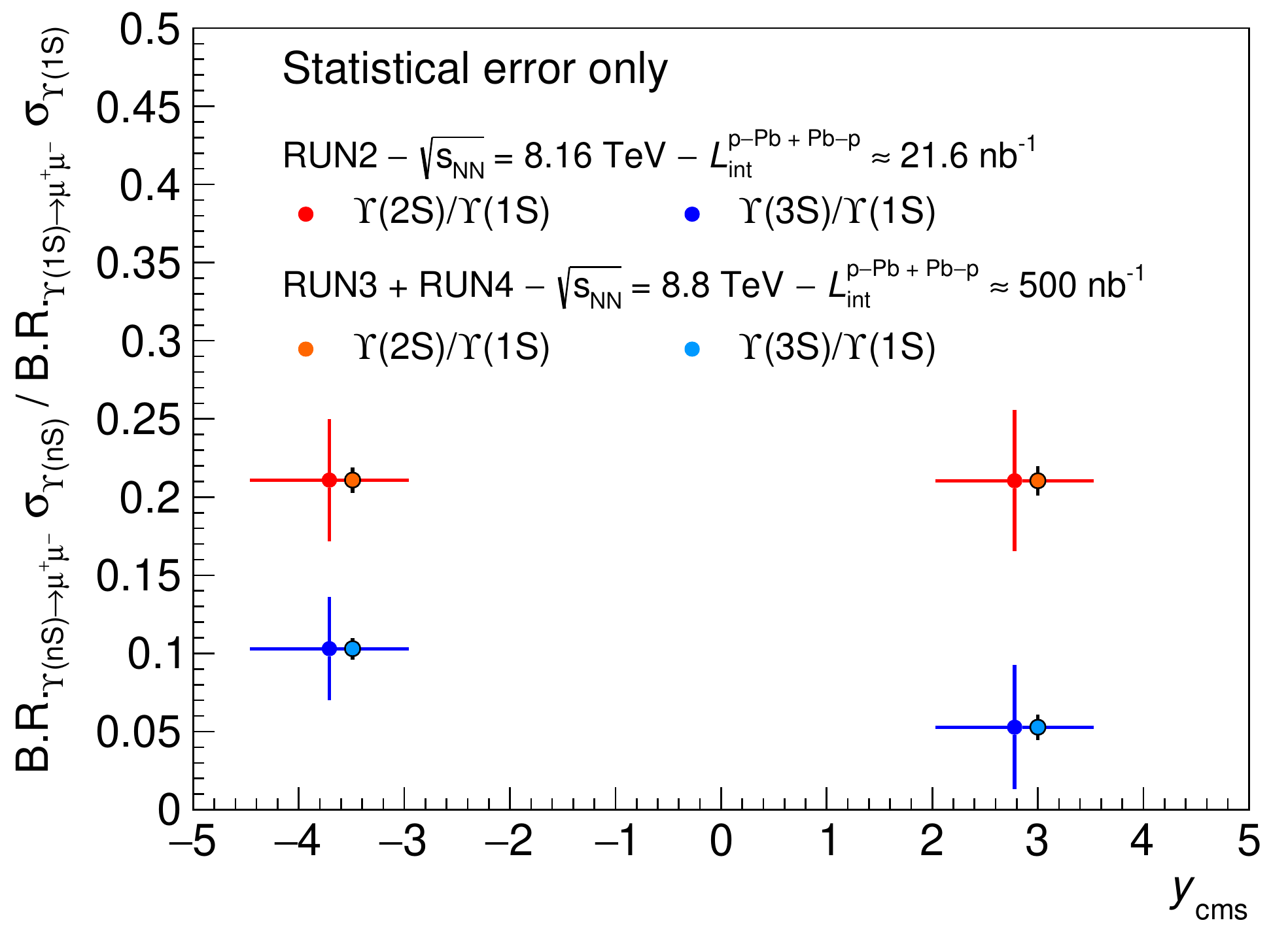}
  \caption{Projections for the 
  $\psip/\jpsi$ (Left) and  $\ups(nS)/\ups(1S)$   (Right) ratios as a function of the $\cm$ rapidity 
$y_\cm$ in \pPb\ collisions at $\sqrts = 8.8$~TeV compared to the current Run-2 results at $\sqrts = 8.16$~TeV. [Figures taken from~\cite{ALICE-PUBLIC-2020-008}].
} 
  \label{Psi2S_by_JPsi_UpsinS_by_UPsi1S} 
\end{figure}

The systematic uncertainties on the signal extraction, which are most relevant for the ratio measurements, are also expected to decrease significantly with the improved knowledge of the background distributions, but 
precise numbers are difficult to estimate at the moment. The reduction of uncertainties will allow one to clearly establish the suppression of the excited states and to perform precise model comparisons to advance our understanding of the physical origin of the observed nuclear modifications. Similar improvements of statistical precision can be expected for LHCb thanks to a similar luminosity ratio between the already available data and the planned luminosity for the 2030s in \pPb\ collisions. For ATLAS and CMS, a reduction of the statistical uncertainties by about a factor 2 can also be expected according to~\cite{Citron:2018lsq}. 
The estimates provided here are purely statistical: further improvements can be expected from better instrumentation in terms of acceptances, efficiencies, and resolution thanks to the foreseen upgrades that will be fully beneficial for detector-occupancy conditions in \pA  collisions, which are well below those of the \pp data with orders-of-magnitude-larger pileup.

\begin{figure}[h!]
    \centering
    \includegraphics[width=0.47\textwidth]{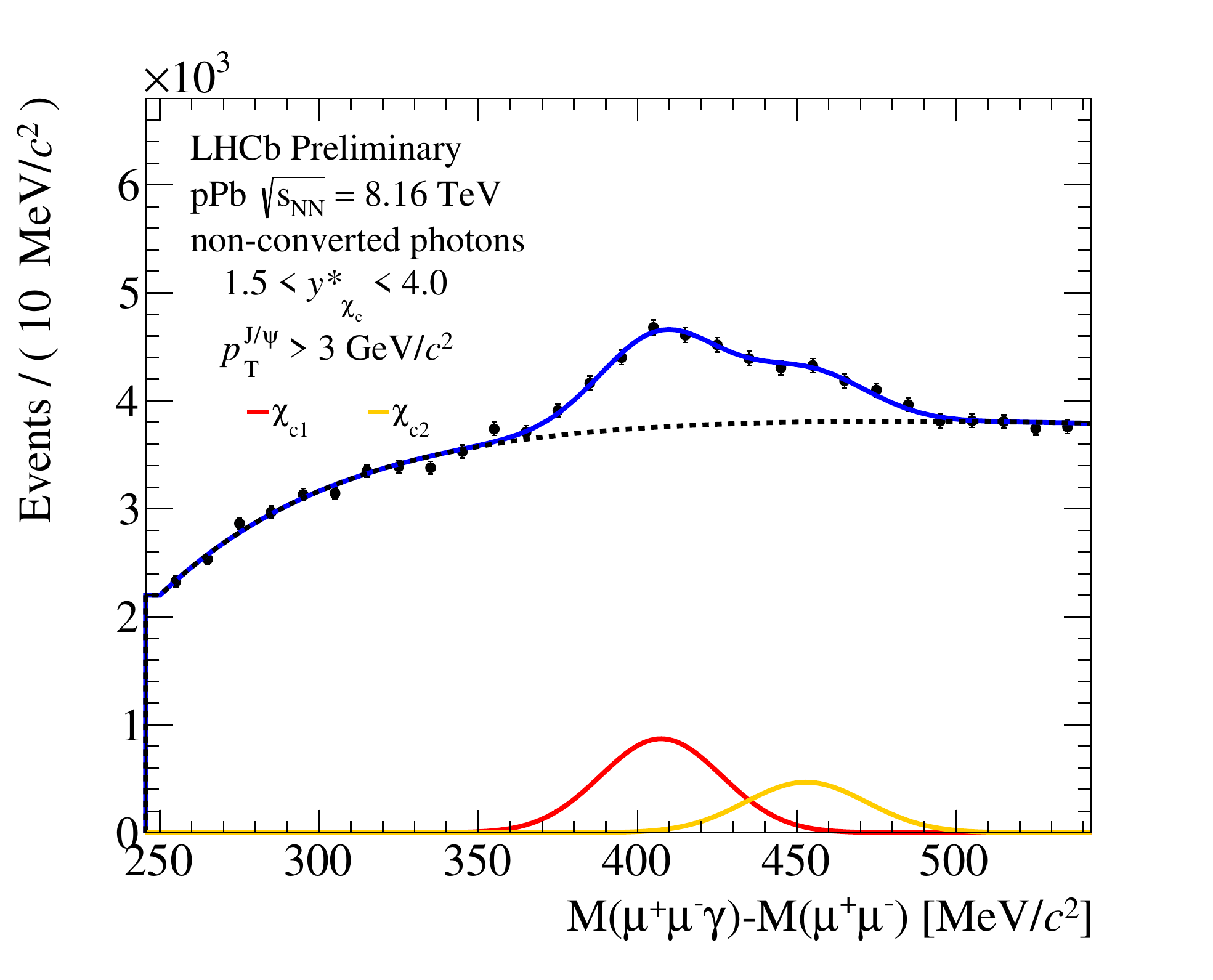}
    \includegraphics[width=0.47\textwidth]{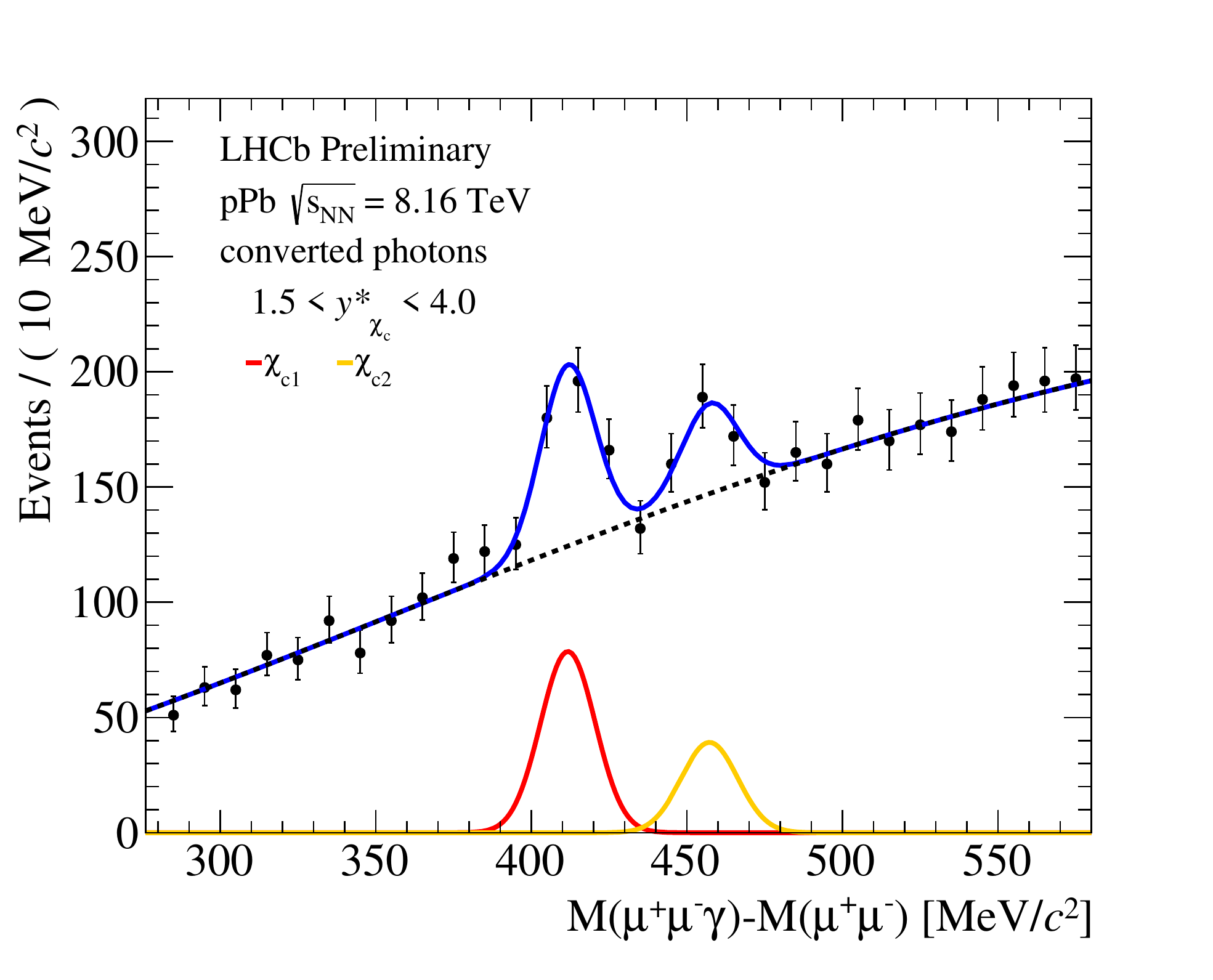}
    \caption{Mass difference $\Delta M = M(\mu^+\mu^-\gamma)- M(\mu^+\mu^-)$ measured at forward rapidity by LHCb in the 2016 \pPb\ data (integrated luminosity $13.6\pm0.3$~$\mu$b$^{-1}$)~\cite{LHCB-FIGURE-2019-020}. Left: Reconstruction via calorimetric photons for $\pT^{\jpsi}>3$~GeV. Right: Reconstruction via converted photons, integrated over $\pT$. The distributions are integrated over $1.5<y^\star<4$ where $y^\star$ is the laboratory rapidity. These plots demonstrate the feasibility of first measurements at the LHC in \pA collisions. [Figures taken from~\cite{LHCB-FIGURE-2019-020}]. %
    }
    \label{fig:chicpA}
\end{figure}

In addition to the vector states, the $P$-wave states of charmonium, $\chi_c$, could provide complementary information to that shown in~\cite{Abelevetal:2014cna} based on the models outlined in \cite{Andronic:2003zv, Zhao:2011cv}, in particular due to their strong suppression predicted in \AaAa\ collisions. In \pp\ collisions, LHCb carried out measurements of $\chi_{c1}/\chi_{c2}$~production ratios~\cite{LHCb:2012ac} and of the $\chi_c$ to $\jpsi$~cross section ratio~\cite{Aaij:2013dja} based on an integrated luminosity of 36~pb$^{-1}$ at 7~TeV. Similar measurements were performed by CMS~\cite{Chatrchyan:2012ub} and ATLAS~\cite{ATLAS:2014ala} at $\sqrt{s}=7$~TeV with an integrated luminosity of $\sim$4.5~pb$^{-1}$. The LHCb measurements appear complementary for the study of nuclear modifications in \pPb\ collisions, given the lower \pT region explored compared to the CMS ($7<\pT<$~25~GeV) and ATLAS ($10<\pT<$~30~GeV) ones. All these measurements are based on the decay channel $\chi_c \to \jpsi \gamma $ reconstructing the $\jpsi$ in the di-muon channel and the photon in the calorimeter. Measurements of cross-section ratios of $\chi_c$~production have been performed %
based on the reconstruction of the converted photon in the tracker~\cite{Aaij:2013dja}. 
The planned increased \pPb\ integrated luminosity at the HL LHC should allow measurements in these decay channels with similar precision as in \pp~collisions. The data already recorded %
demonstrate the reconstructibility of the decay channels of interest~\cite{LHCB-FIGURE-2019-020} as shown in \cf{fig:chicpA}.  Recently, LHCb~\cite{Aaij:2017vck} and the BESIII~\cite{Ablikim:2019jqp} collaboration observed the decay of $\chi_c \to \jpsi \mu \mu$. With the upcoming very high-luminosity data takings beyond Run 3, this decay channel could be used as well for cross section measurements despite its very small branching fraction of the order of $10^{-4}$ for $\chi_{c1}$ and $\chi_{c2}$~\cite{Ablikim:2019jqp}.  %

%

%

%
%

%
%
%
%
%

%
%
%
%
%
%
%
%
%
%
%
%

%

%
%
%
%
%
%
%

%
%
%

%
%
%

%

%

%

%

%

%

%

%

%
%
%

%

%
%

%
%
%

%
%
%
%
%
%
%
%
%
%

%

%
%
 
%

%

%

%

%

%


\section{Nucleus-nucleus collisions\protect\footnote{Section editors: \'{E}milien Chapon, Pol-Bernard Gossiaux.
}}
\label{sec:aa}
\newcommand{\lb}{\Big{\lbrack}}
\newcommand{\rb}{\Big{\rbrack}}
\newcommand{\lp}{\Big{(}}
\newcommand{\rp}{\Big{)}}
\newcommand{\lbc}{\Big{\lbrace}}
\newcommand{\rbc}{\Big{\rbrace}}
\newcommand{\nn}{\nonumber}
\newcommand{\Bvert}{\Big{\vert}}
\newcommand{\Rangle}{\Big{\rangle}}
\newcommand{\Langle}{\Big{\langle}}
\newcommand{\ve}{\lambda}
\newcommand{\raa}{\ensuremath{R_{\textrm{AA}}}\xspace}
\newcommand{\vtwo}{\ensuremath{v_{2}}\xspace}
\newcommand{\vthree}{\ensuremath{v_{3}}\xspace}
\newcommand{\Npart}{\ensuremath{N_{\rm part}}\xspace}

\subsection{Introduction and context}
\label{AAintro}

A hot and dense state of matter, the quark gluon plasma (QGP), is  created in \PbPb collisions at the LHC. The quarkonia provide natural probes to study its properties, since the heavy quarks are created early in the collision and, as bound states, they are sensitive to a large variety of initial- to final-state effects. The in-medium modifications to the fundamental force between two static colour charges can be investigated via changes in the quarkonium spectroscopy. The heavier, looser bound states should typically be the most suppressed in heavy-ion collisions. In the cooling of the fireball, the quarkonia can also be ``(re)generated'' through the recombination of individual heavy quarks and antiquarks in the medium. Such an effect is more pronounced for charmonia than bottomonia due to the larger number of \ccbar pairs present in the QGP.

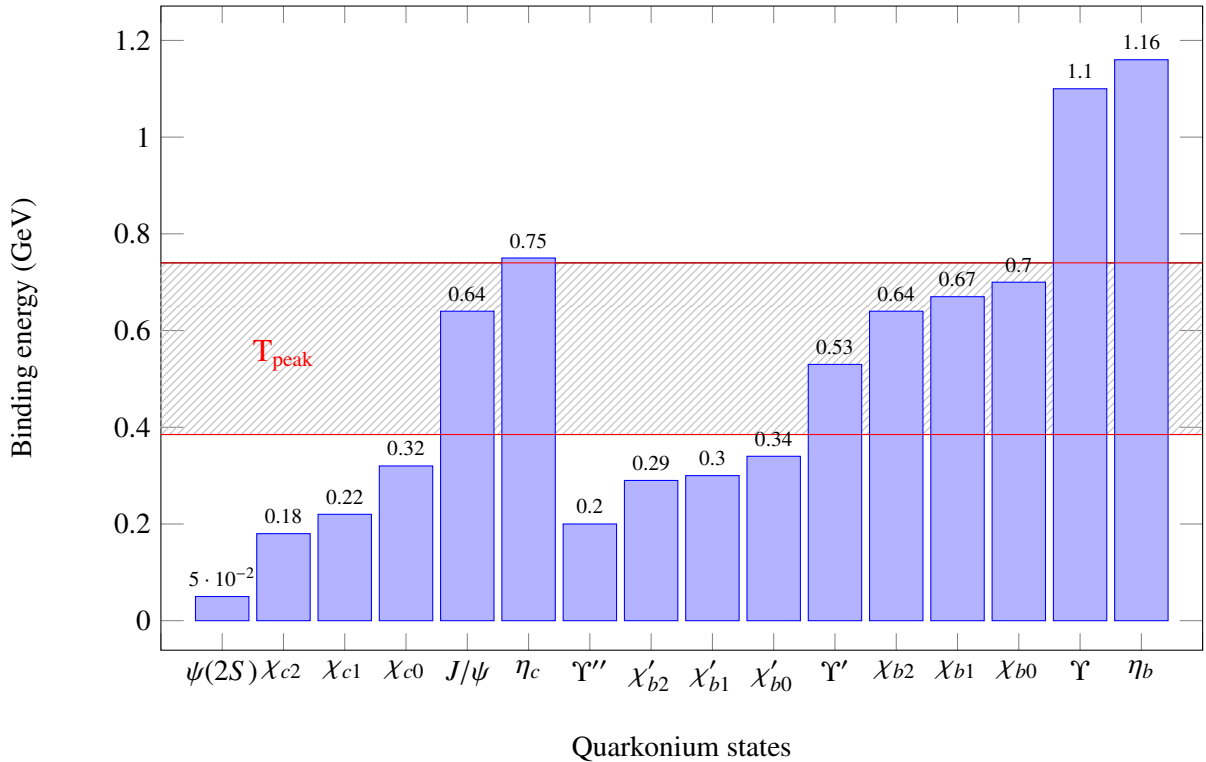
\begin{figure}[h!]
\centering
\begin{tikzpicture}
  \begin{axis}[
  xscale=2.0,
  yscale=1.5,
	xtick={0,1,2,...,17},
    xticklabels={ ,\psip, $\chi_{c2}$, $\chi_{c1}$, $\chi_{c0}$, \jpsi, $\eta_c$,
    			$\Upsilon^{\prime\prime}$, $\chi_{b2}^{\prime}$, $\chi_{b1}^{\prime}$, 
        		$\chi_{b0}^{\prime}$, $\Upsilon^{\prime}$, $\chi_{b2}$, $\chi_{b1}$, $\chi_{b0}$, $\Upsilon$,
        		$\eta_b$},
    xmin=0, xmax=17,
    xlabel=Quarkonium states,
    ylabel=Binding energy (GeV)]
	\addplot [pattern color=gray!50, pattern=north east lines] coordinates {(0, 0.385) (17,0.385) (17,0.740) (0, 0.740)};
	\addplot [draw=blue,fill=blue!30!white,
    	ybar,xtick=data,font=\scriptsize,
        nodes near coords,
    	nodes near coords align={vertical}] coordinates {  
	(1,0.05)
    (2,0.18)
    (3,0.22)
    (4,0.32)
    (5,0.64)
    (6,0.75)
    (7,0.20)
    (8,0.29)
    (9,0.30)
    (10,0.34)
    (11,0.53)
    (12,0.64)
    (13,0.67)
    (14,0.70)
    (15,1.10)
    (16,1.16)
};
	\addplot [draw=red] coordinates {(0, 0.385) (17,0.385)};
    \addplot [draw=red] coordinates {(0, 0.740) (17, 0.740)};
   	\node [font=\large,color=red] at (axis cs:2,0.55) {T$_{\rm peak}$};
  \end{axis}
  \end{tikzpicture}
\caption{\label{fig:quarkonia_states} 
Binding energies of several quarkonium states in the vacuum~\cite{Satz:2005hx} along with an estimation for the QGP peak temperature, $T_{\rm peak}$, in \PbPb\ collisions at \sqrtsnn = 2.76~TeV~\cite{Adam:2015lda} (shaded area).}
\end{figure}

The quarkonium production in heavy-ion collisions has long been proposed to be directly related to the temperature of the produced QGP~\cite{Matsui:1986dk}. \cf{fig:quarkonia_states} illustrates the power of quarkonium spectroscopy in measuring the medium temperature, compared to conventional methods using the slope of photon yields at intermediate $\pT\approx 1$--4~GeV~\cite{dEnterria:2005jvq,Adare:2009qk}. The peak temperature from the thermal-photon spectrum strongly relies on the unknown QGP formation time assumed in the model estimations. The reported direct-photon spectra may also include a large contribution from meson bremsstrahlung~\cite{Linnyk:2015tha}. Some caveats should be kept in mind when using quarkonia as a QGP thermometer: 
\begin{description}
\item (i) for a given family, the measurement of the relative quarkonium yields is supposed to be insensitive to initial-state effects, but a validation is desirable via the measurements of the quarkonium modification versus rapidity, in order to scan over the fractional momentum $x$, in a controlled particle-multiplicity and collision-energy environment;
\item (ii) it is still a question which temperature the quarkonium states probes: the peak temperature, the initial temperature or some average temperature;
\item (iii) it is also under debates how important the quarkonium breaking during the hadronic phase is; 
\item (iv) for the charmonia, the recombination competes with the suppression in heavy-ion collisions and these effects should be disentangled.
\end{description}
It is also useful to take the opposite approach: fix the temperature to a value favoured by other measurements, and test the quarkonium dynamics and interactions given this assumed thermodynamic properties of the bulk.

The main features of the nuclear modification factor (\raa), the ratio of production in \AaAa to \pp\ accounting for the amount of nuclear overlap, for \jpsi mesons at the LHC are already well measured after ten years of operation. A smaller suppression relative to the lower RHIC collision energies has been found~\cite{Abelev:2012rv} 
especially at low \pT (below approximately 5 GeV)~\cite{Abelev:2013ila,Adam:2016rdg} and has been attributed to  regeneration. A greater supppresion is observed at higher \pT~\cite{Aaboud:2018quy,Sirunyan:2017isk}, which is consistent with stronger dissociation effects from the bigger, longer-lived QGP at the LHC. The behaviour of \raa at high \pT at the LHC (up to 50\,GeV or beyond), on the other hand, is still uncertain. There are hints  of an increasing trend in \raa from ATLAS~\cite{Aaboud:2018quy} and CMS~\cite{Sirunyan:2017isk} with the Run-2 data. Data from the HL-LHC will allow one to improve the precision and push to even higher \pT~\cite{CMS-PAS-FTR-18-024}, which will confirm (or disprove) this trend which could be related to parton energy loss~\cite{Arleo:2017ntr} or to jet quenching~\cite{CMS-PAS-HIN-19-007}. Further studies are needed, however, as the energy-loss phenomenology yields tensions with the hierarchy of nuclear modifications of ground and excited states~\cite{Makris:2019ttx}. A projection of the high \pT reach from CMS is shown in \cf{fig:Jpsi_pTvsLumi}. %

\begin{figure}[h!]
    \centering
    \includegraphics[width=0.48\textwidth]{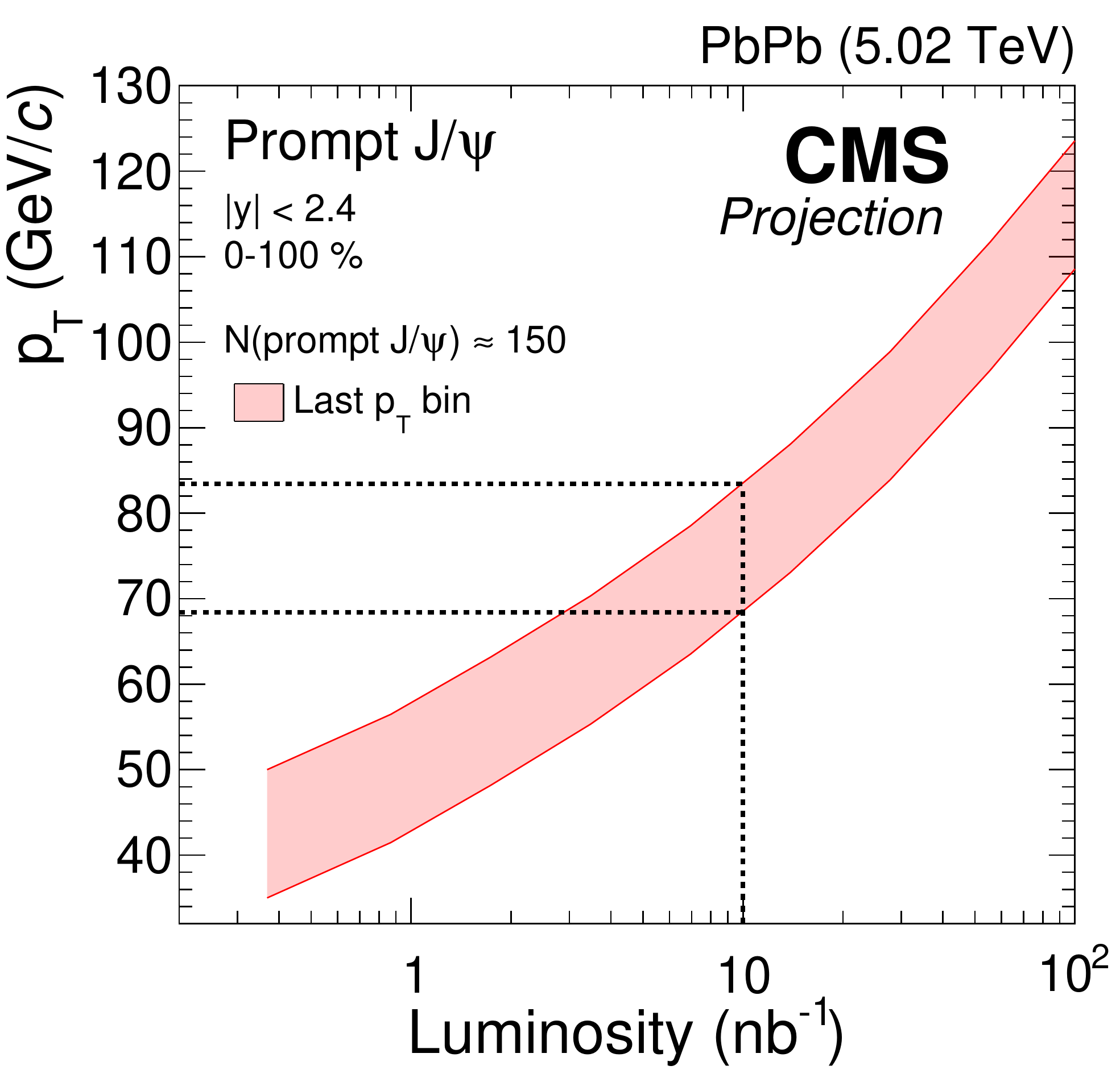}
    \hspace{0.3cm}
    \includegraphics[width=0.48\textwidth]{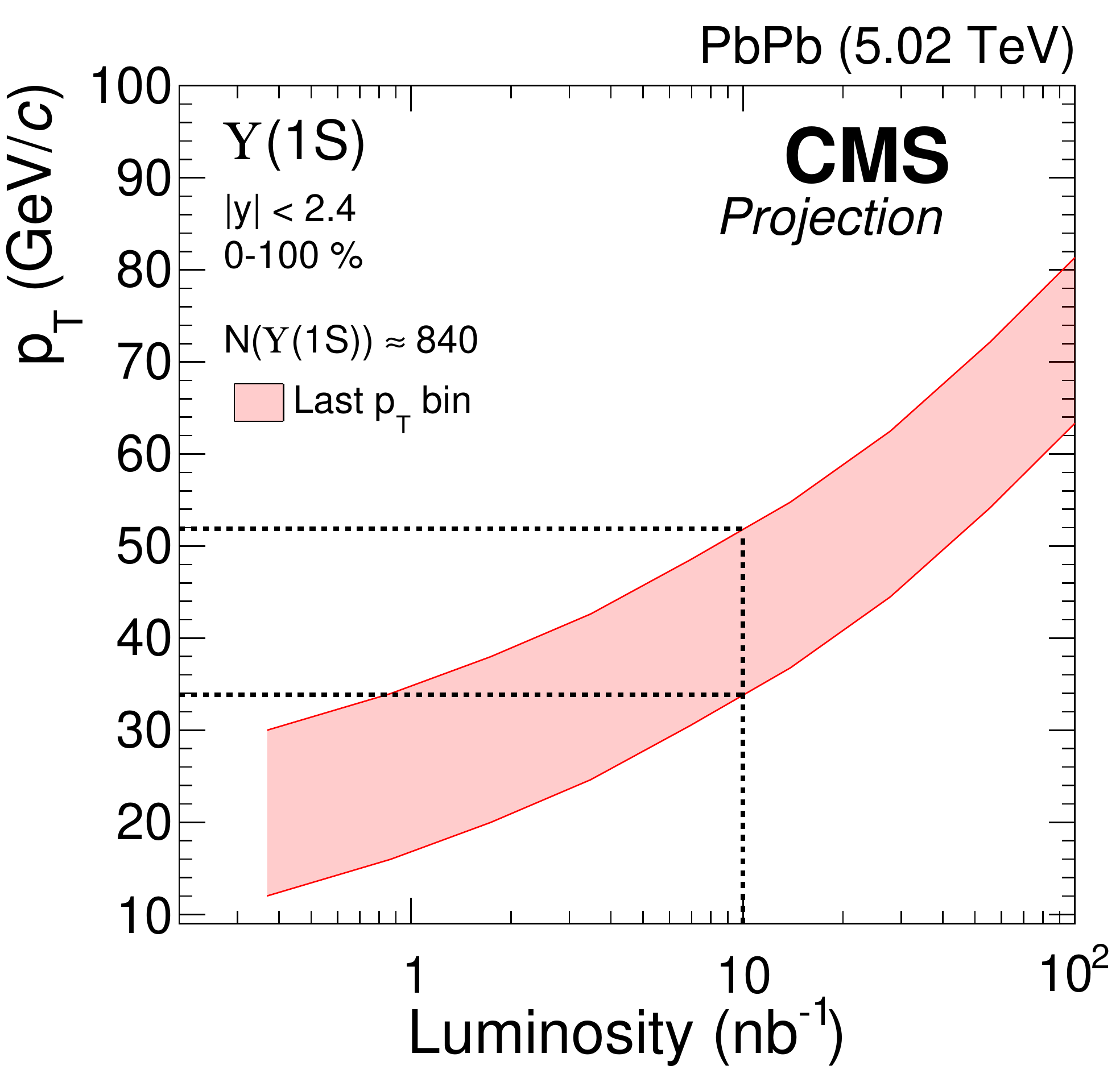}
    \caption{Prompt \jpsi (Left) and \ups (Right) [$\pT^{low}$,$\pT^{up}$] boundaries for the highest \pT bin as a function of the luminosity in the CMS experiment~\cite{CMS-PAS-FTR-18-024}. The boundaries are chosen in such a way that the number of quarkonia in the bin for the corresponding luminosity equals the number of mesons found in the last \pT bin of the analysis with a luminosity of 368~\invmub, as used for existing measurements~\cite{Sirunyan:2017isk,Sirunyan:2018nsz}, keeping the width of this last \pT bin fixed. The projection for the expected luminosity of 10~\invnb, roughly matching the expectation for the HL-LHC (see \ct{tab:yrlumis}), is highlighted with dashed lines. [Figure  from~\cite{CMS-PAS-FTR-18-024}]}
    \label{fig:Jpsi_pTvsLumi}
\end{figure}

In particular, information about excited-quarkonium production in heavy-ion collisions is very limited at the moment~\cite{Sirunyan:2016znt,Sirunyan:2018nsz,Aaboud:2018quy,Acharya:2018mni,Adam:2015sia}, and further data on the excited states will be crucial. Given the large feed-down contributions, corresponding to about half of the \ups yield at large \pT for instance~\cite{Lansberg:2019adr,Andronic:2015wma,Aaij:2014caa},  the yield of excited states is difficult to determine and this directly reflects on the precision of the model predictions for \jpsi or \ups. In addition, the charmonium ground state, \etac, remains unmeasured.

Anisotropic pressure gradients in the QGP, which are a consequence of the non-spherical (elliptic, to first order) shape of the overlap region between the colliding nuclei, induce anisotropies in the azimuthal distribution of the final particle momenta, including the so-called elliptic flow. This flow is characterised by the second order \vtwo of the Fourier expansion of this distribution. The \vtwo of \jpsi mesons has been measured to be non-zero~\cite{Aaboud:2018ttm,Acharya:2020jil}, which is qualitatively well reproduced by the transport models at low \pT where the \jpsi \vtwo relates to the thermalisation of the charm quarks and their interaction with the hydrodynamic expansion of the medium. However, at higher \pT, the agreement is not good. As shown in \cf{fig:projvtwo}, a precise measurement of \jpsi \vtwo up to higher \pT, complementary to $\raa(\pT)$, will be instrumental in understanding the charmonium-production mechanisms in heavy-ion collisions. It will allow their behaviour at high \pT to be related to the energy loss via a path-length dependence effect, by distinguishing between the long and short axes of the ellipsoid-like medium. Higher orders of the anisotropic flow of quarkonia are becoming accessible, such as \vthree~\cite{Acharya:2020jil}, and more precise measurements of these will provide further information about quarkonium production and transport at low \pT.

\begin{figure}[htbp!]
    \centering
    \includegraphics[width=0.6\textwidth]{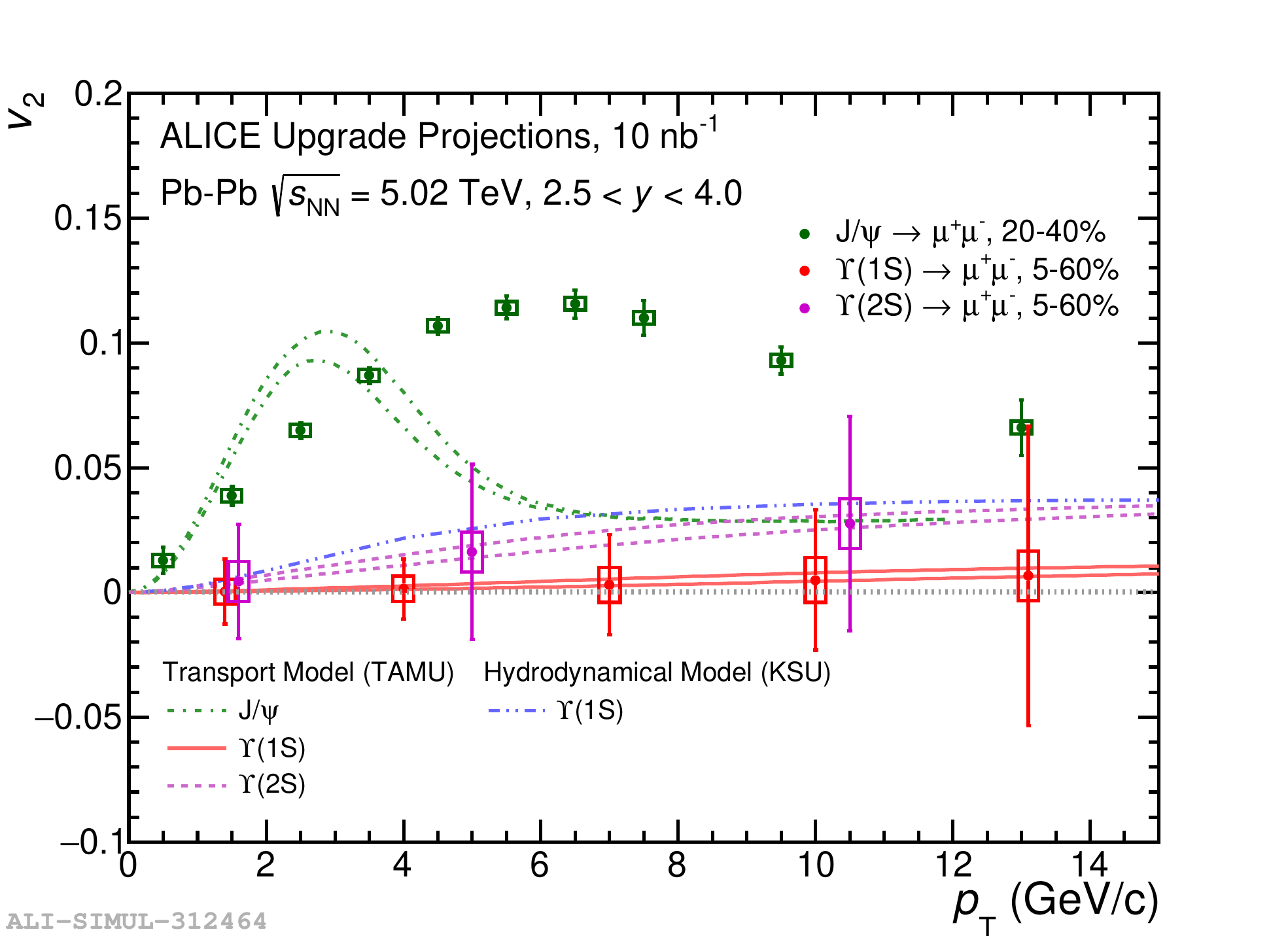}
    \caption{Projections for the $v_2$ coefficient as a function of $\pT$ for the \jpsi, \ups and \upsp mesons in \PbPb\ collisions at $\sqrtsnn~=~5.02$~TeV in the ALICE experiment, assuming the predictions from the transport model of~\cite{Du:2017qkv} and compared to an alternative model~\cite{Bhaduri:2018iwr}. [Figure from~\cite{Citron:2018lsq}]}
    \label{fig:projvtwo}
\end{figure}

Measurements of other charmonium states than \jpsi are currently limited to the \psip meson, though these have poor precision. The production of  \psip mesons is found to be much more suppressed than \jpsi, even at relatively high \pT (up to 30\,GeV)~\cite{Sirunyan:2016znt,Aaboud:2018quy}. Data from the HL-LHC will help {to} better understand  \psip production in \PbPb collisions, though a \vtwo measurement may remain challenging. In particular, a precise measurement of the \psip/\jpsi ratio (\cf{fig:psi2spsi_aa})  will test the validity of the statistical hadronisation model~\cite{Andronic:2017pug} compared to dynamical models~\cite{Du:2015wha}.
The measurement of the $P$-wave states, such as \chic, would help complete the picture, but it is experimentally challenging due to the difficulty of reconstructing very-low-\pT photons %
in a heavy-ion environment. A possible option would be to look at $\chi_c \to \jpsi \mu\mu$ decays.
Similarly, the \etac states only have large branching fractions  to hadrons, 
usually with a rather large number of final-state particles, and thus will be very challenging to measure in heavy-ion collisions.

\begin{figure}[h!]
    \centering
    \includegraphics[width=0.5\textwidth]{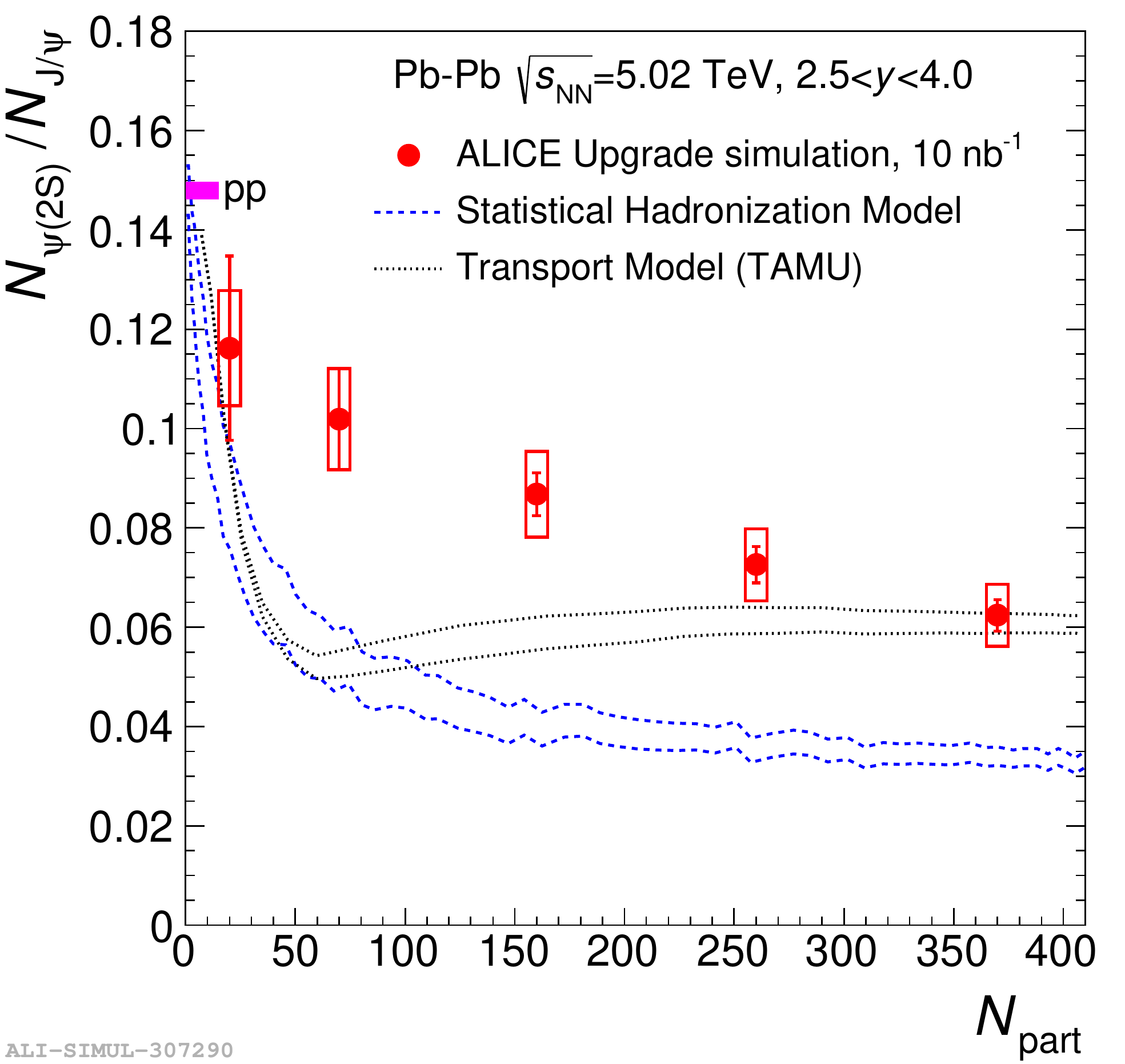}
    \caption{Ratio of the $\psip / \jpsi$ production yields vs. \Npart in \PbPb\ collisions measured by ALICE over $2.5<y<4$~\cite{Abelevetal:2014cna,CERN-LHCC-2013-014}. Model predictions in the transport approach~\cite{Du:2015wha} and from the statistical hadronisation~\cite{Andronic:2017pug} are included. The values of the ratio used for the projections are quasi-arbitrary. [Figure from~\cite{Citron:2018lsq}]}
    \label{fig:psi2spsi_aa}
\end{figure}

Due to the much smaller number of \bbbar compared to \ccbar pairs produced in \PbPb collisions, regeneration is thought to play a much smaller role in the dynamical models for bottomonia than for charmonia, though it may still be significant for the strongly suppressed excited states~\cite{Du:2017qkv,Krouppa:2017jlg}. Indeed, \raa for the \ups states does not feature a significant \pT or rapidity dependence~\cite{Sirunyan:2018nsz}. In addition, there is a strong centrality dependence (with smaller \raa in central collisions), as well as a sequential ordering in the suppression of the different states, with the \upspp being so suppressed that it has not yet been measured significantly in heavy-ion collisions. Data from the HL-LHC will provide higher precision measurements (\cf{fig:upsilon_aa}), up to high \pT, enabling better model discrimination: smaller uncertainties may reveal structures in \raa, which appears flat as a function of \pT with current experimental precision. The \vtwo of \ups mesons in \PbPb collisions has recently been measured for the first time~\cite{Sirunyan:2020qec,Acharya:2019hlv}, though with limited precision. Such studies will be continued at the HL-LHC. Finally, similar comments to their charmonium counterparts above, can be made for  \chib (and \etab) mesons.

\begin{figure}[h!]
    \centering
    \includegraphics[width=0.5\textwidth]{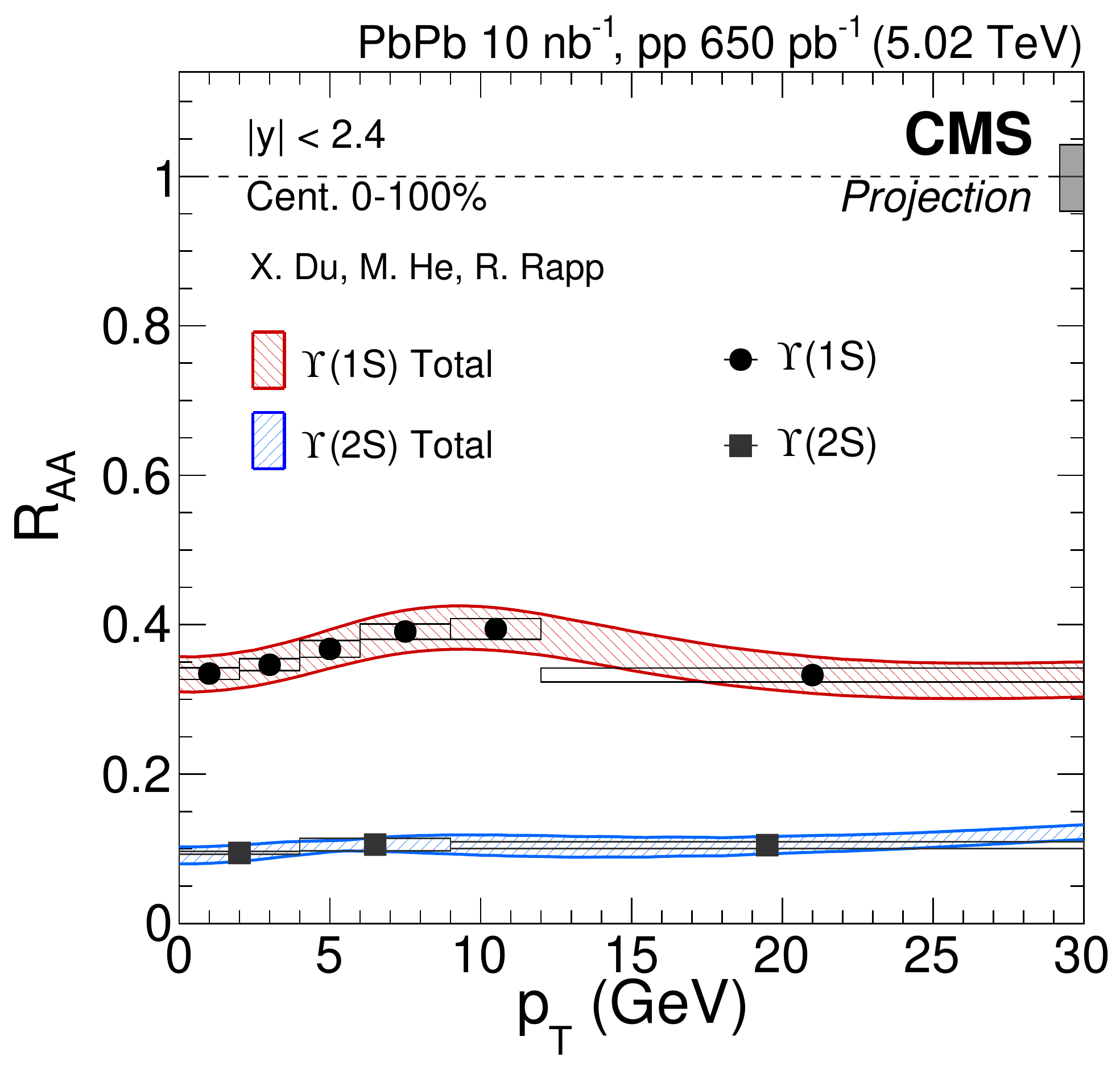}
    \caption{Projected \raa as a function of \pT for \ups and \upsp yields expected at the  CMS~\cite{CMS-PAS-FTR-18-024} experiment, with $10~\invnb$ of \PbPb data and $650~\invpb$ of reference \pp data. [Figure from~\cite{CMS-PAS-FTR-18-024}]
    }
    \label{fig:upsilon_aa}
\end{figure}

The possible complications due to the different production mechanisms in \AaAa collisions (the suppression of the direct production, and the recombination from correlated pairs or the regeneration in the plasma from uncorrelated pairs) could be circumvented by comparing data at several collision energies, \ie running at lower energies than the nominal one. This is one of the advantages of taking data in the FT mode~\cite{Hadjidakis:2018ifr}, as mentioned in Section~\ref{sec:energy_dependence}.

A running scenario for the HL LHC, starting with Run~3 (2021) and the major upgrades of the ALICE and LHCb detectors for heavy-ions, has been proposed in the corresponding CERN Yellow Report (Working Group 5)~\cite{Citron:2018lsq} and is given in \ct{tab:yrlumis}. The listed integrated luminosities have not yet been endorsed by the experiments, the LHC, or CERN. They are reported indicatively in order to give the reader the order of magnitude of what can be expected. The interested reader may refer to~\cite{Bruce:2722753} for a recent study of the expected future performances of the LHC for heavy-ion beams. There is also an interest in running with ion beams in Run~5 and beyond, possibly with lighter ions (such as argon or krypton), to reach higher luminosities and reduce combinatorial background, giving access to rarer states such as $\chib (1P)$. The ALICE~\cite{Adamova:2019vkf} and LHCb~\cite{Bediaga:2018lhg} experiments have started planning upgrades supporting this scenario.

\begin{table}[h!]
    \begin{center}\renewcommand{\arraystretch}{1.5}
\begin{tabular}{llll}\hline
Year & Systems, $\sqrtsnn$ & Duration & $\Lint$ \\
\hline \hline
2021 &  \PbPb 5.5~TeV & 3 weeks & $2.3~\invnb$ \\
     &  \pp\ 5.5~TeV & 1 week & $3~\invpb$ (ALICE), $300~\invpb$ (ATLAS, CMS), $25~\invpb$ (LHCb)  \\
\hline
2022 &  \PbPb 5.5~TeV & 5 weeks & $3.9~\invnb$ \\
     &  OO, $p$O & 1 week & $500~{\rm \mu b}^{-1}$ and $200~{\rm \mu b}^{-1}$ \\
\hline
2023 &  \pPb\ 8.8~TeV & 3 weeks & $0.6~\invpb$ (ATLAS, CMS), $0.3~\invpb$ (ALICE, LHCb) \\
     &  \pp\ 8.8~TeV & few days & $1.5~\invpb$ (ALICE), $100~\invpb$ (ATLAS, CMS, LHCb)  \\
\hline
2027 &  \PbPb 5.5~TeV & 5 weeks & $3.8~\invnb$ \\
     &  \pp\ 5.5~TeV & 1 week & $3~\invpb$ (ALICE), $300~\invpb$ (ATLAS, CMS), $25~\invpb$ (LHCb)  \\
\hline
2028 &  \pPb\ 8.8~TeV & 3 weeks & $0.6~\invpb$ (ATLAS, CMS), $0.3~\invpb$ (ALICE, LHCb) \\
     &  \pp\ 8.8~TeV & few days & $1.5~\invpb$ (ALICE), $100~\invpb$ (ATLAS, CMS, LHCb)  \\
\hline
2029 &  \PbPb 5.5~TeV & 4 weeks & $3~\invnb$ \\
\hline \hline
Run-5 & Intermediate \AaAa & 11 weeks & \eg\ \,Ar--Ar 3--9~$\invpb$ (optimal species to be defined) \\
     &  \pp\ reference & 1 week & \\
\hline
\end{tabular}
\end{center}
    \caption{Indicative running scenarios for different heavy-ion runs at the HL-LHC, with the expected integrated luminosity, as proposed in the CERN Yellow Report~\cite{Citron:2018lsq} (but subject to review by the experiments, LHC, and CERN). The years in the table do not account for modifications of the schedule after the publication of the report, which include a delay of the start of Run~4 (2027 $\to$ 2028) and a delay of the start of Run~3 because of the COVID-19 pandemic (2021 $\to$ 2022).}
    \label{tab:yrlumis}
\end{table}

Further discussion of the physics case and prospects for quarkonium measurements in heavy-ion collisions at the HL-LHC is given in Section~7 of the aforementioned CERN Yellow Report~\cite{Citron:2018lsq}. A few selected topics are discussed below.

\subsection{Recent theory developments}
\label{subsection_rec_th_dev}
Although some concrete predictions for quarkonium yields can be made assuming they are produced according to the laws of statistical physics at the pseudo-transition temperature, most approaches on the market attempt to implement the suppression and (re)generation of quarkonia, advocated in the introduction, through dedicated dynamical transport models. While early approaches were formulated in terms of kinetic equations making use of dissociation rates and cross sections obtained from pQCD calculations or effective models, more recent developments have profited from the concept of imaginary potentials that, in approaches like potential NRQCD (pNRQCD), makes the bridge between those coefficients used in the transport models and lattice QCD (lQCD) calculations. These new developments enable more solid links between the experimental observables to be measured with better precision at HL-LHC runs and the basic fundamental properties of the in-medium $Q\bar{Q}$ interactions.

In parallel, the formalism of open quantum systems (OQS) is nowadays considered, by an increasing part of the theoretical community, as the emerging paradigm that should either supersede semi-classical approaches or, at least, provide the methods to generate corrections to these approaches. This prospect is particularly appealing for HL-LHC runs as well, as the experimental precision needs to be matched by an increased control of the theoretical models. In Section~\ref{AAtransportOQS}, we provide a description of progresses achieved with the OQS formalism as well as its links to Boltzmann transport. 

Among the various theoretical challenges, the correct quantum treatment of the regeneration of low-\pT charmonia, due to numerous $c\bar{c}$ pairs, is a key question for the most central \PbPb collisions at LHC energies and beyond. In Section~ \ref{AAtransportdensityop}, we discuss a recent approach stemming from a direct reduction of the Von Neumann equation, which provides an alternative to semi-classical algorithms deduced for instance in~\cite{Blaizot:2017ypk,Yao:2018nmy} and yields preliminary predictions for \raa and \vtwo of \jpsi.

The production of quarkonia at high \pT is another key issue in the global landscape, as the few present \vtwo predictions from transport models fail to reproduce the experimental data at intermediate \pT, leaving the door open for other mechanisms like that involving energy loss of the $Q\bar{Q}$ pair. However, the modelling of such a situation requires knowledge of the in-medium interaction of a $Q\bar{Q}$ pair at finite velocity, which is still not fully known. In Section~ \ref{AAtransportEFT}, 
we present a summary of recent progresses made in one of the state-of-the-art approaches. 

\subsubsection{Semi-classical transport and open quantum system}
\label{AAtransportOQS}

Modern phenomenological studies of quarkonium production in heavy-ion collisions require consistency in considering static screening, dissociation and recombination in the treatment of hot-medium effects. Semi-classical transport equations such as the Boltzmann equation and the rate equation, which is obtained from the Boltzmann equation by integrating the quarkonium distribution over phase space, have been applied widely and shown to be phenomenologically successful~\cite{ Grandchamp:2003uw,Grandchamp:2005yw,Yan:2006ve,Liu:2009nb,Song:2011xi,Song:2011nu,Sharma:2012dy,Nendzig:2014qka,Krouppa:2015yoa,Chen:2017duy,Zhao:2017yan,Du:2017qkv,Aronson:2017ymv,Ferreiro:2018wbd,Yao:2018zrg,Hong:2019ade,Chen:2019qzx,Yao:2020xzw}.

\begin{figure}[h!]
    \centering
    \includegraphics[width=0.6\textwidth]{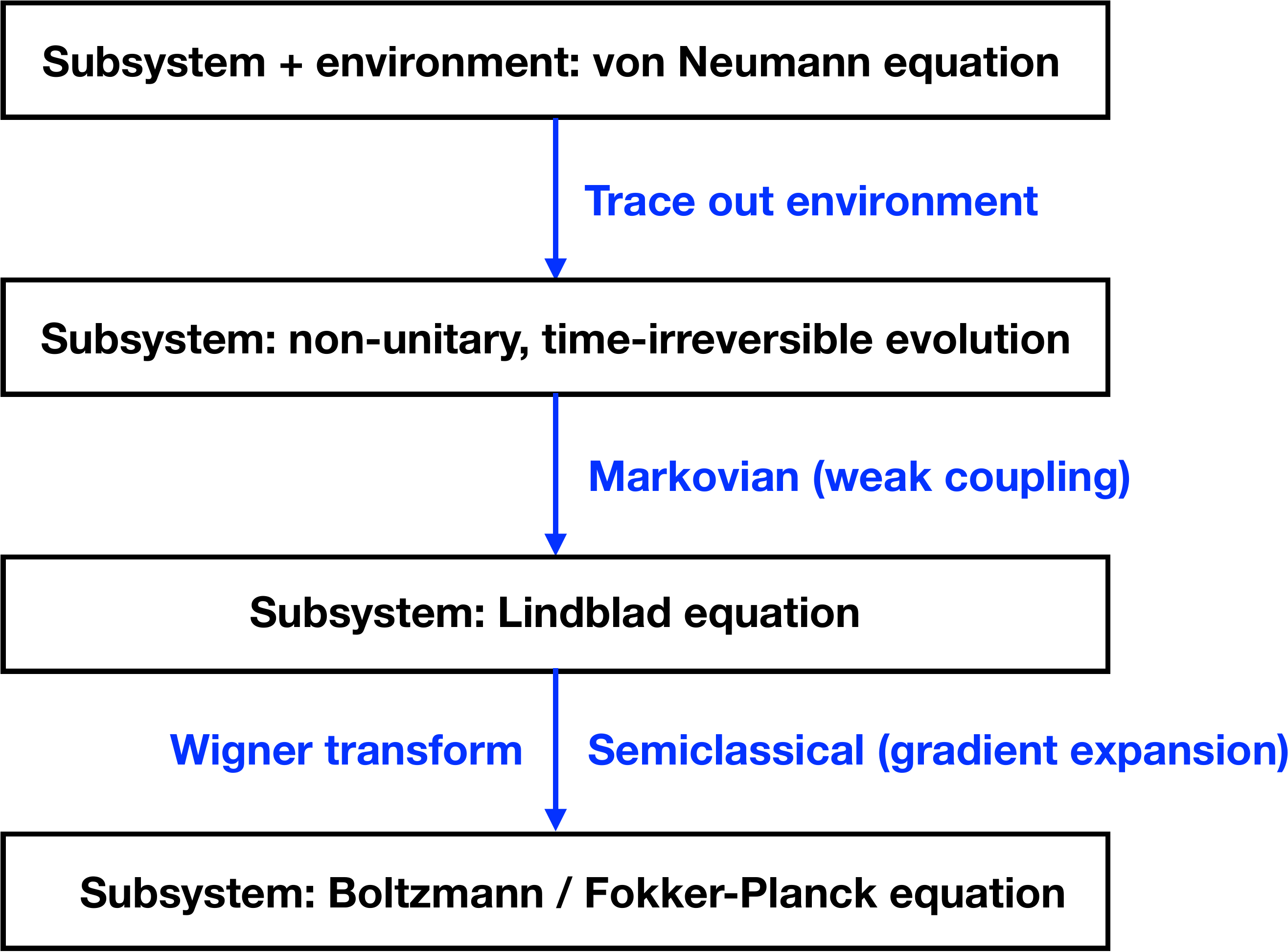}
    \caption{Various approximations made and evolution equations obtained in the derivation of the semi-classical transport equation for quarkonium from the von Newmann equation of closed quantum systems.}
    \label{fig:lindblad_transport}
\end{figure}

Such a phenomenological success of the semi-classical Boltzmann equation has been explained by deriving the transport equation under systematic expansions that are closely related to a hierarchy of scales $M\gg Mv \gg Mv^2 \gtrsim T$~\cite{Yao:2018nmy,Yao:2020eqy}, where $M$ is the heavy-quark mass, $v$ the typical relative velocity between the heavy quark-antiquark pair in a quarkonium state and $T$ the temperature of the plasma. Under this separation of scales, pNRQCD~\cite{Brambilla:1999xf,Brambilla:2004jw,Fleming:2005pd} can be used to simplify the calculations. The starting point of the derivation is the OQS formalism that has recently been used to study quarkonium transport~\cite{Young:2010jq,Borghini:2011ms,Akamatsu:2011se,Akamatsu:2014qsa,Blaizot:2015hya,Katz:2015qja,Brambilla:2016wgg,Brambilla:2017zei,Kajimoto:2017rel,DeBoni:2017ocl,Blaizot:2017ypk,Blaizot:2018oev,Akamatsu:2018xim,Miura:2019ssi,Sharma:2019xum,Brambilla:2019tpt}. In this formalism, the quarkonium is an open subsystem interacting with the thermal QGP. Integrating out the degrees of freedom of the thermal bath results in a non-unitary and time-irreversible evolution equation, which further leads to a Lindblad equation in the Markovian approximation. The Lindblad equation can be shown to become the semi-classical Boltzmann equation in the semi-classical limits after a Wigner transform is applied to the subsystem density matrix (a Gaussian smearing is required to maintain positivity). The Markovian approximation can be justified when the subsystem is weakly interacting with the thermal bath, which is true in our power counting since the quarkonium size is small $rT\sim\frac{T}{Mv} \lesssim v \ll 1$. A schematic diagram of the various approximations and resulting equations is shown in Fig.~\ref{fig:lindblad_transport}. The derivation clearly demonstrates the validity condition of the semi-classical transport equations. Essentially, what should be preserved is the non-relativistic nature of the heavy quarks from the QGP viewpoint. As the \pT\ of the quarkonium increases, the energy of the medium excitation in the rest frame of the quarkonium is boosted, which could ruin the nonrelativistic expansion. Furthermore, the derivation provides a way of systematically including the quantum corrections to the semi-classical transport equation. Alternatively, one can directly solve the Lindblad equation for the quarkonium phenomenology, which in principle is a {quantum} system. Improvements towards a full quantum phenomenological treatment are still needed.

\begin{figure}[t]
    \centering
    \includegraphics[width=0.5\textwidth]{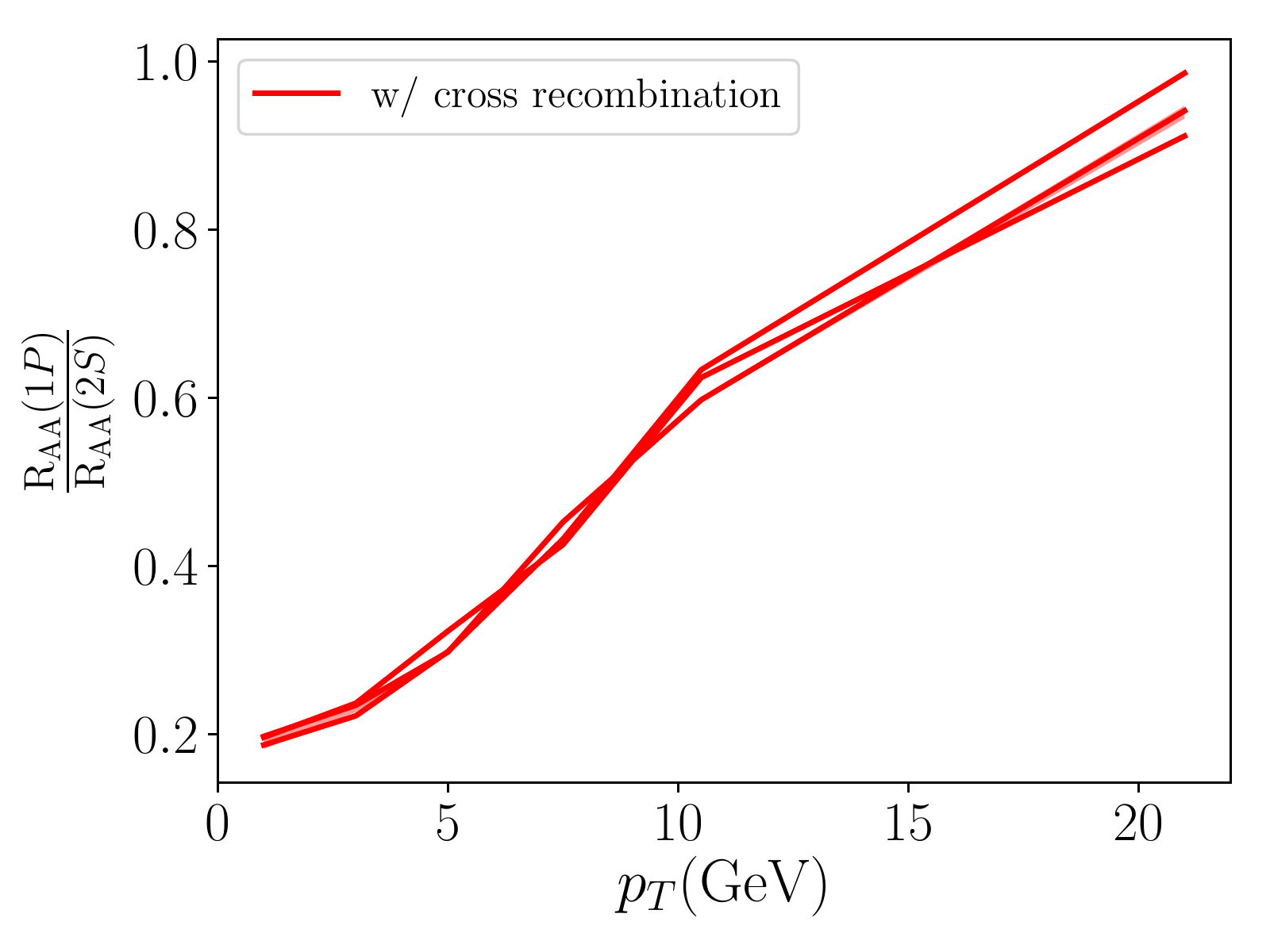}
    \caption{Ratios of \raa$(\chib (1P))$ and \raa(\upsp) as a function of $\pT$ predicted in~\cite{Yao:2020xzw}. Different lines correspond to different choices of parameters. Nuclear PDF effects largely cancel in the ratio.}
    \label{fig:2Svs1P}
\end{figure}

To explore the limits of the semi-classical transport approaches and find solid experimental evidence of quantum effects in the quarkonium transport, experimental data with high precision are needed. For example, precise measurements of the azimuthal angular anisotropy and of \raa of excited quarkonium states such as $\chib (1P)$ and \upspp will be very helpful in distinguishing different semi-classical transport calculations.  A precise measurement of \raa of $\chib (1P)$ is of particular interest~\cite{Yao:2020xzw} since, if the dissociation is the only hot medium effect, then one expects \raa {of} \upsp and of $\chib (1P)$ to be similar (with the value for $\chib (1P)$ slightly higher)~\cite{Krouppa:2015yoa} since the binding energies and sizes of the two states are comparable. However, it is known from recent studies using the OQS formalism that the dissociation is a result of the wave-function decoherence. Due to the decoherence of the original state, say \upsp, a non-vanishing overlap can be developed with other states that exist in the medium (\ie\ the local temperature is below their melting temperature), say $\chib (1P)$. This gives a probability to form another quarkonium state from a dissociating state. This recombination process  (known as ``correlated recombination"~\cite{Yao:2020xzw}) involves a heavy quark-antiquark pair from the same hard vertex (a dissociating quarkonium) and is different from the traditional concept of recombination, which comes from heavy quarks and antiquarks initially produced from different hard vertices.  The existence of such a correlated recombination is 
mandatory for theory consistency and is well-motivated from OQS studies. With correlated recombination, an initial \upsp state may dissociate first and then recombine as a $\chib (1P)$ state and vice versa. The probabilities of both these processes are similar since \upsp and $\chib (1P)$ have similar binding energies and sizes. However, primordially many more $\chib (1P)$ mesons are produced, 
which leads to less suppressed yields of the \upsp than the $\chib (1P)$ state. Consequently, it is of great interest to measure the ratio between the \raa suppression factors of $\chib (1P)$) and \upsp. Calculations that include correlated recombination (which requires some information of the two-particle distribution function of the heavy quark-antiquark pairs) such as~\cite{Yao:2020xzw}, predict the ratio to be about $1/3$ in central collisions while calculations such as~\cite{Krouppa:2015yoa}, which  do not include correlated recombination, give a ratio larger than unity. The contrast is dramatic (for example, compare Fig.~1 of~\cite{Krouppa:2015yoa} and Fig.~7 of~\cite{Yao:2020xzw}). If it is possible to measure this ratio in future experiments, its power for discriminating between models will be high. This ratio calculated in~\cite{Yao:2020xzw} also depends on \pT, and approaches unity as \pT increases, as shown in Fig.~\ref{fig:2Svs1P}. 

The correlated recombination discussed above is not an intrinsic quantum effect and can be accounted for in semi-classical transport equation calculations. With high precision data, it may also be possible to find a common inconsistency between all semi-classical approaches and experimental data, which then hints at the importance of quantum effects in quarkonium in-medium dynamics. In view of the discussion in Section~\ref{sec:pa_cnm} on nPDF effects, one expects these to largely cancel in such a ratio and, in any case, to be negligible compared to the variation shown in \cf{fig:2Svs1P}.

\subsubsection{A density-operator model for $\Q$ production in \AaAa collisions}
\label{AAtransportdensityop}

In this section, a newly developed model is introduced that aims to offer a better understanding of heavy-quarkonium formation in the presence of many $Q\bar{Q}$ pairs (that is, in the LHC conditions) while offering a different approach to the existing ones. In this model, the formation of heavy quarkonia is conceived as a coalescence in phase space, based on composite particle cross sections, following Remler's formalism~\cite{GYULASSY:83,REMLER:81} directly deduced from the Von Neumann's equation.

By computing an effective production rate (including both dissociation and recombination processes), the model is able to keep track of the inclusive formation probability of the quarkonium system with time, taking into account the heavy-quark kinematics and their interaction with medium particles. For a given $\{Q,\bar{Q}\}$ pair $\{1,2\}$, the contribution to the rate is 
\begin{equation}
\Gamma(t)=\sum_{i=1,2}\sum_{j\geqslant 3}\delta(t-t_{ij}(\nu))[W_{Q\overline{Q}}(t+\epsilon)-W_{Q\overline{Q}}(t-\epsilon)],
\end{equation} 
where the sum $j$ reflects the sum over all particles from the bulk, while the $\delta$ factor only acts when one of the members of a given pair ($i=1$ or $i=2$) undergoes a collision with a particle ($j\geqslant 3$). %
The rate expression relies on the Wigner distribution of the quarkonium vacuum states, through the gain $W_{Q\overline{Q}}(t+\epsilon)$ and loss $W_{Q\overline{Q}}(t-\epsilon)$ terms. While the expression only shows the rate contribution from one pair, it can easily be extended to all pairs in the medium by summing over all combinations. In this way, both recombination and dissociation are taken into account inside both the gain and loss terms, which represent the overall contribution of the recombination and dissociation at any given time. The time evolution of the probability $P$ for a quarkonium state formation thereby follows:
\begin{equation}
P(t)=P(t_{0})+\int \Gamma(t)dt\,.
\end{equation}       
In this approach, the heavy quarks do not need to be considered at thermal (or chemical) equilibrium at any stage of the collision. This feature makes it possible to apply it, not only to large systems like \AaAa collisions, but to small systems as well. It also offers concrete perspectives to deal with several particles ($Q\overline{Q}$ pairs) in real-time dynamics. Finally, the model is also quite sensitive to key ingredients of the quarkonium production such as: the primordial, or initial, $Q\overline{Q}$ pair production (with cold nuclear matter effects); the $Q\overline{Q}$ interaction; and the local medium temperature field which is modelled according to the EPOS event generator~\cite{Werner:2013tya} in the present implementation.
In \cf{fig:Remler}, preliminary predictions are provided for \raa and \vtwo as a function of the \jpsi $\pT$ obtained within this operator model, both with and without screened binding interactions.

\begin{figure}[h!]
\centering
\includegraphics[width=0.47\textwidth]{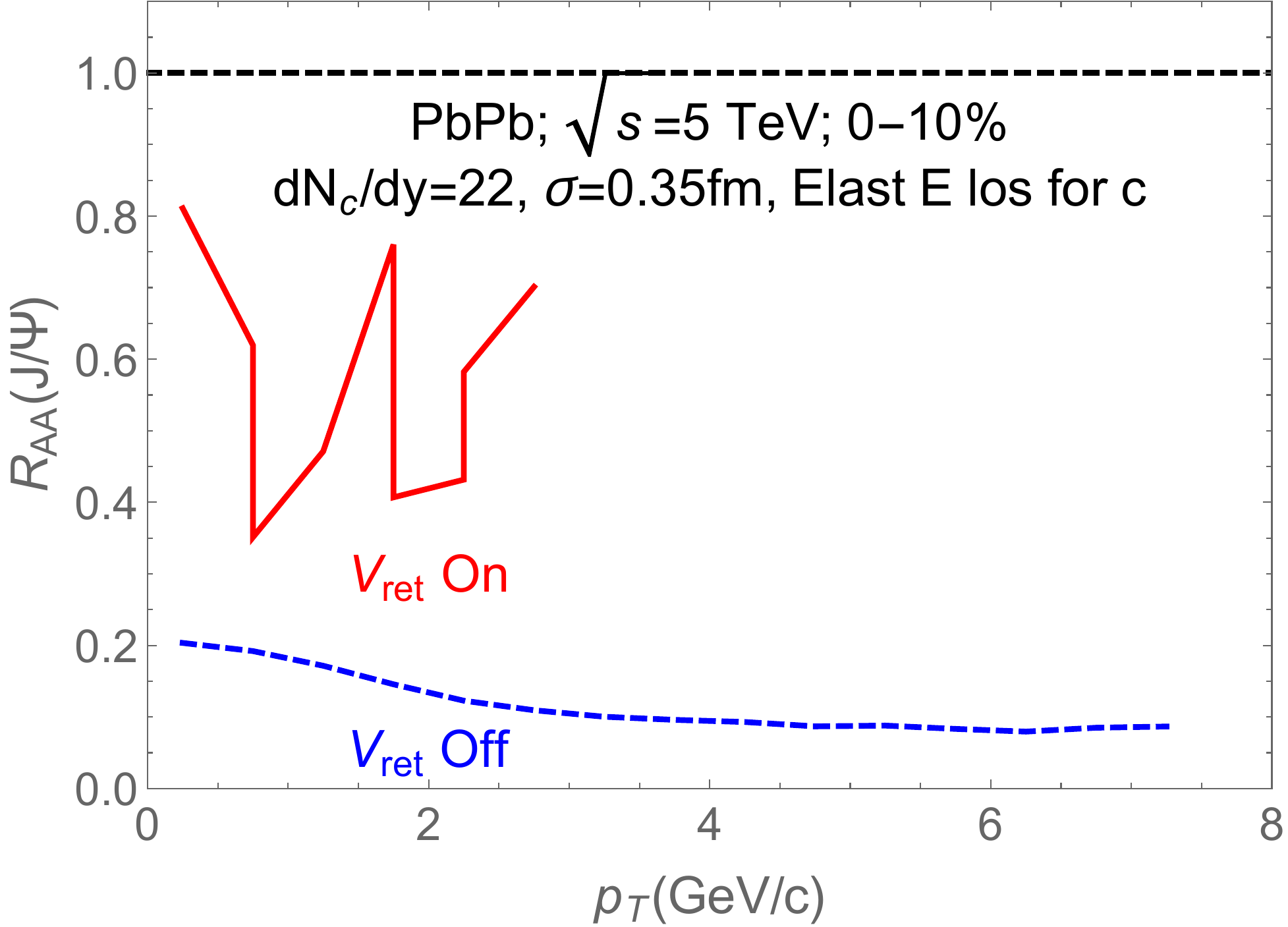}
\hspace{0.2cm}
\includegraphics[width= 0.47\textwidth]{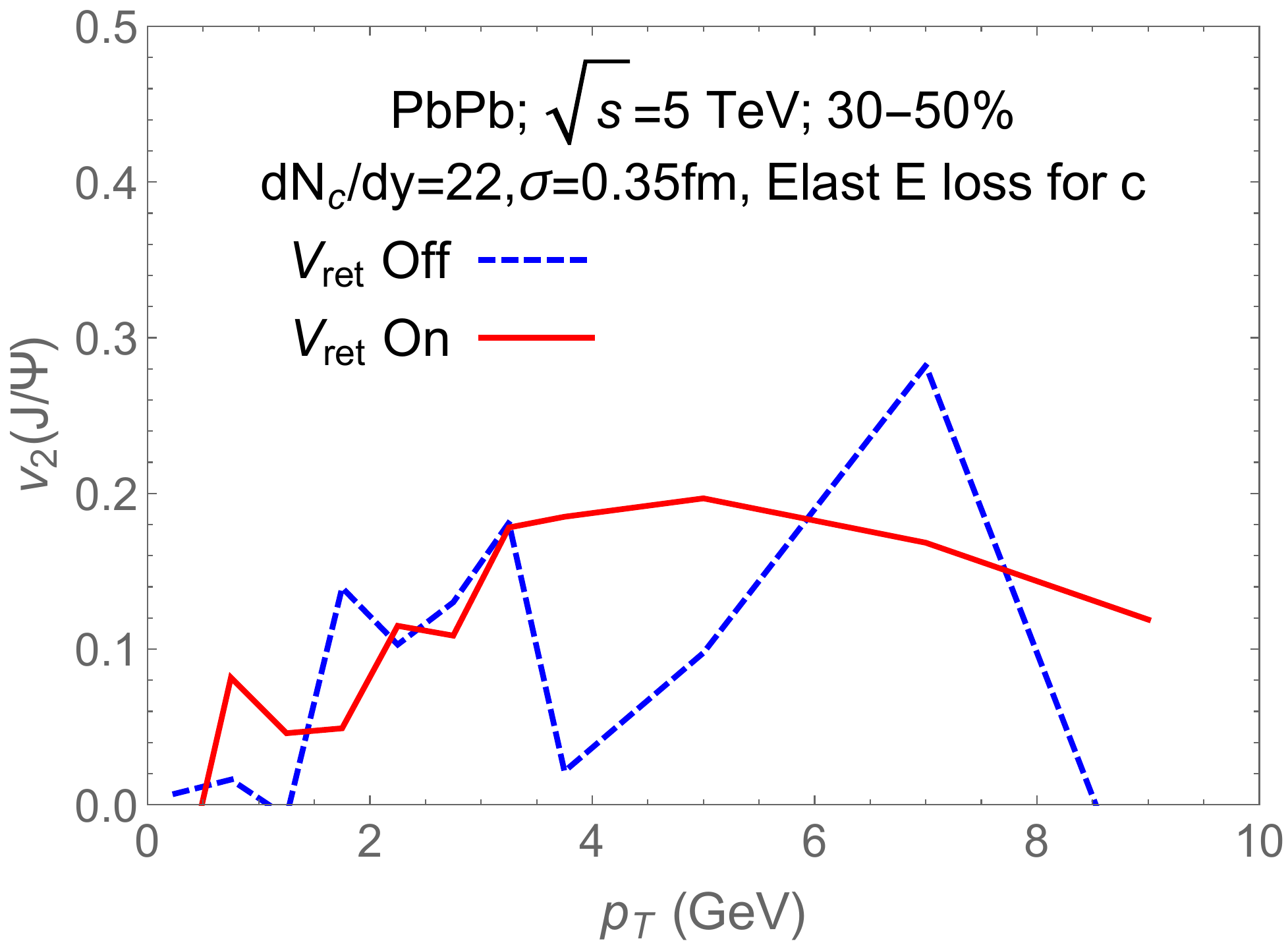} 
\caption{Prediction for \raa (Left) and \vtwo (Right) {of} the \jpsi with the density-operator model for \PbPb\ collisions at \sqrtsnn = 5~TeV. The solid lines corresponds to \ccbar interactions modelled following screened binding potential, while the dashed curves correspond to $c$ and $\bar{c}$ solely interacting with the light quarks and the gluons through elastic scatterings. [The observed oscillations result from numerical fluctuations.]
} 
  \label{fig:Remler}
\end{figure}
In the future, both lQCD calculations and HL-LHC data on \raa and \vtwo of \jpsi will be brought together to constrain the interactions among $Q$, $\bar{Q}$, and medium partons, and then explore in detail the consequences of the model for excited states and higher harmonics like $v_3$, which start to be accessible experimentally~\cite{Acharya:2018pjd}.

\subsubsection{An advanced EFT for $\Q$ in matter}
\label{AAtransportEFT}

In recent years, significant progress in understanding subatomic particle propagation in matter has been reached using modern effective field theories (EFTs). First developed for light partons~\cite{Idilbi:2008vm,Ovanesyan:2011xy,Fickinger:2013xwa}, the Soft-Collinear Effective Theory with Glauber gluons (SCET$_{\rm G}$)  was applied to describe the suppression of inclusive hadrons and jets as well as the modification of jet substructure~\cite{Chien:2015hda,Chien:2015vja,Kang:2017frl}. This approach was subsequently extended to open heavy-flavour~\cite{Kang:2016ofv,Sievert:2019cwq} to understand the production of $D$ mesons, $B$ mesons, and heavy-flavour-tagged jets~\cite{Li:2017wwc,Li:2018xuv}. A logical next step is to start with the theory of quarkonium production,  %
NRQCD and its modern 
formulations~\cite{Bodwin:1994jh,Brambilla:1999xf,Luke:1999kz}, and to introduce interactions with the background nuclear medium~\cite{Makris:2019ttx}.      

The hierarchy of ground and excited quarkonium suppression emphasises the need for such an EFT approach. In order for the traditional energy-loss phenomenology~\cite{Arleo:2017ntr} to contribute significantly to the modification of quarkonium cross sections in QCD matter, the quarkonium formation must happen outside of the medium and be expressed as the fragmentation of partons into the various \jpsi and \ups states. 
This is possible in the recently developed Leading-Power (LP) factorisation approach to QCD~\cite{Bodwin:2014gia} (see also Section~\ref{sec:pp}) although we stress that it is only in the high-\pT range that this LP factorisation is thought to work well.

\begin{figure}[h!]
\centering
\includegraphics[width=0.47\textwidth]{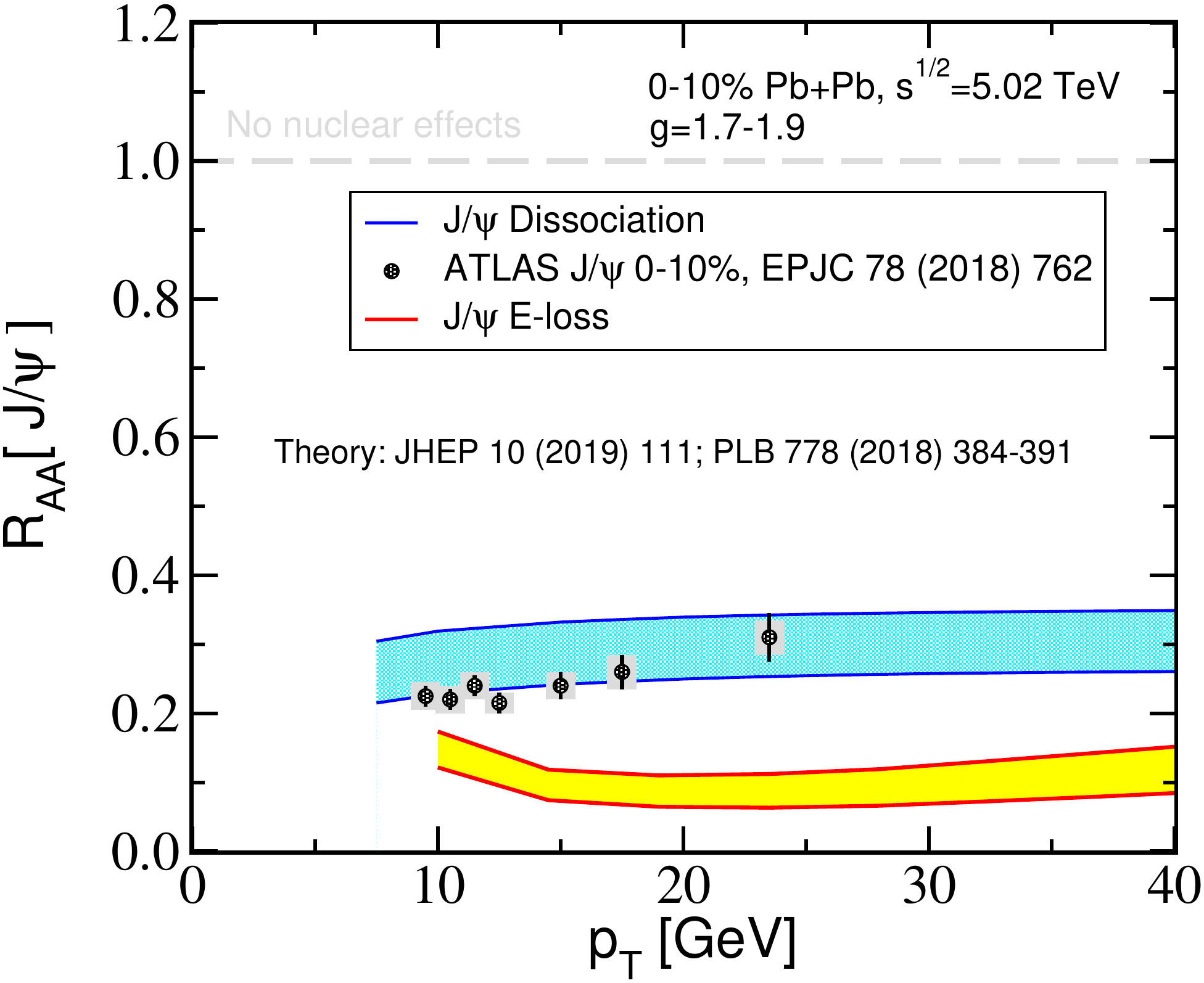}
\hspace{0.2cm}
\includegraphics[width=0.47\textwidth]{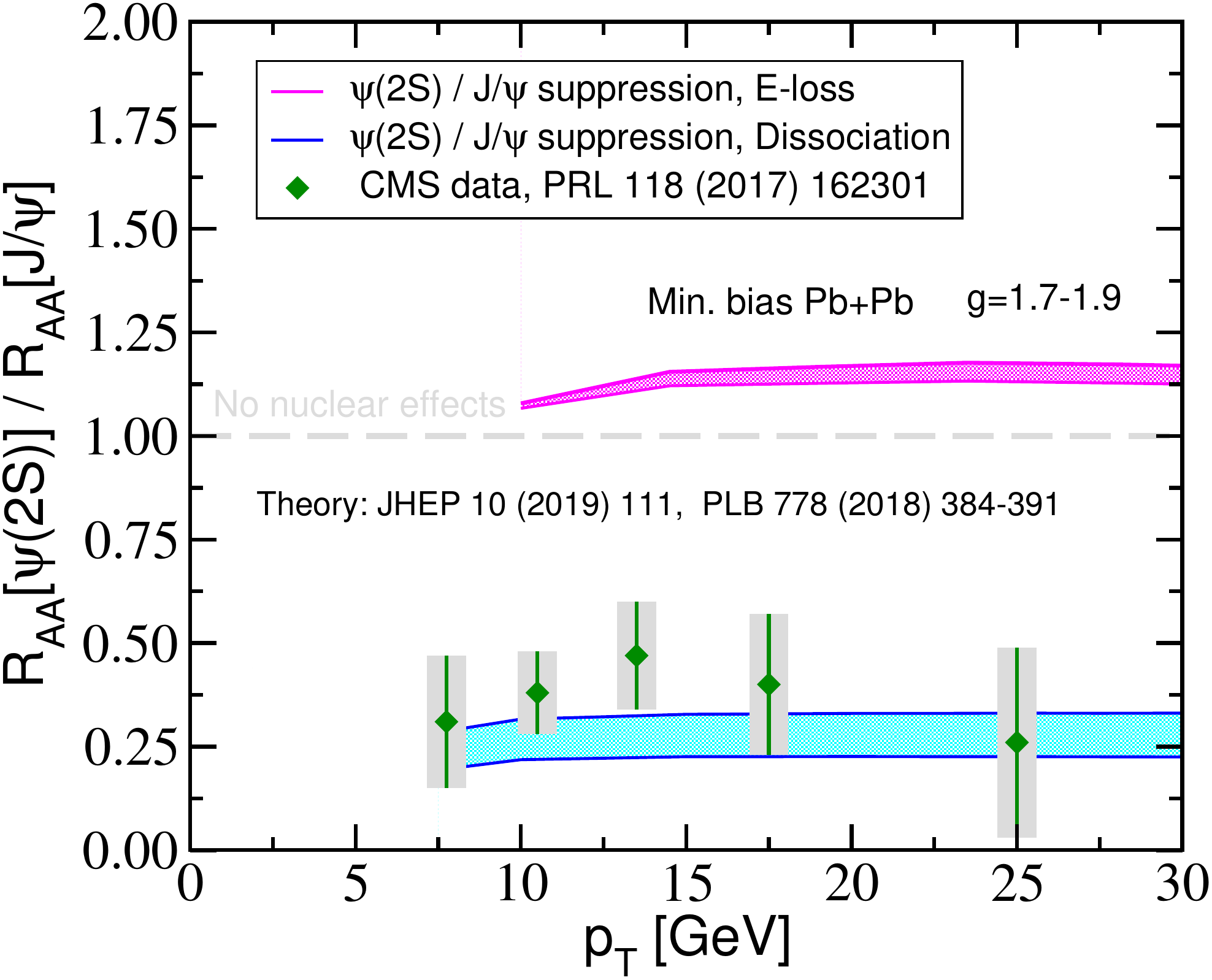} 
\caption{Left: Suppression of the \jpsi production in central PbPb collisions at ATLAS compared to the energy loss (yellow) and EFT (blue) quarkonium dissociation calculations. Right: The double ratio of \psip to \jpsi suppression as a measure of the relative significance of QCD matter effects on  ground and excited states (CMS) compared to the same  energy loss (purple) and EFT (blue) theoretical models. } 
  \label{fig:2Sto1Sptcent}
\end{figure}

As an example, we calculate the baseline \jpsi and \psip cross sections from  LDMEs extracted using LP factorisation in \PbPb collisions at the LHC (Fig.~\ref{fig:2Sto1Sptcent}). The energy-loss evaluation is carried out in the soft-gluon-emission limit of the full in-medium splitting kernels~\cite{Sievert:2018imd,Sievert:2019cwq}, and is well constrained by light-hadron quenching~\cite{Chien:2015hda,Kang:2014xsa}.  The energy loss approach overpredicts the suppression of \jpsi measured by ATLAS~\cite{Aaboud:2018quy} in the range $\pT > 10$~GeV where the computation starts to be applicable. The discrepancy is a factor of 2 to 3 in both minimum bias and central collisions (yellow band in \cf{fig:2Sto1Sptcent} Left).  The most important discrepancy, however, is in the relative medium-induced suppression of \psip to \jpsi as shown in the right panel of \cf{fig:2Sto1Sptcent} (purple band). The energy-loss model predicts smaller suppression for the \psip state compared to \jpsi  and $\raa[\psip] / \raa[\jpsi] \approx 1.1$.  The CMS experimental results~\cite{Sirunyan:2016znt} show that the suppression of the weakly bound \psip is 2 to 3 times larger than that of \jpsi. 

Such a tension between the data and the energy-loss calculations shows  that a formulation of a general microscopic theory of the quarkonium interactions in matter~\cite{Makris:2019ttx,Makris:2019kap,Vitev:2019yig} is necessary.
When an energetic particle propagates in a hot or cold nuclear medium, the interaction with its quasi-particles  is  typically mediated  by off-shell-gluon exchanges -- Glauber or Coulomb gluons.  Their typical momenta depend on the source of in-medium interactions -- collinear, static, or soft.
We construct the Lagrangian of NRQCD$_{\rm G}$ by adding to the velocity-renormalisation-group NRQCD (vNRQCD) Lagrangian the  terms that include the interactions with medium sources through virtual Glauber/Coulomb gluons exchanges. It takes the form:
\begin{equation}
  \mathcal{L}_{\text{NRQCD}_{\rm G}} = \mathcal{L}_{\text{vNRQCD}} + \mathcal{L}_{Q-G/C} (\psi,A_{G/C}^{\mu,a}) + \mathcal{L}_{\bar{Q}-G/C} (\chi,A_{G/C}^{\mu,a})\;,
\end{equation}
where, in the background field method, the effective  $A_{G/C}^{\mu,a}$ incorporate the information about the sources. Here, $\psi$ and $\chi$ are the heavy quark and antiquark fields respectively.  
The leading and subleading correction to the NRQCD$_{\rm G}$ Lagrangian in the heavy-quark sector from virtual (Glauber/Coulomb) gluon insertions, \ie\ $\mathcal{L}_{Q-G/C}$,
are derived using three different methods yielding the same results. We find that at LO the modification in  the  leading Lagrangian, $\mathcal{L}_{Q-G/C}^{(0)}$, is independent of the nature of the quasiparticles of the QCD medium: 
\begin{equation}
  \label{eq:L0-NR}
  \mathcal{L}_{Q-G/C}^{(0)} (\psi,A_{G/C}^{\mu,a})  = \sum_{{\bf p},{\bf q}_T}\psi^{\dag}_{{\bf p}+{\bf q}_T} \lp - g A^{0}_{G/C} \rp \psi_{{\bf p}}\;\; \textrm{(collinear/static/soft)}\; .
\end{equation}
As the quarks (and antiquarks) couple to the time like component of the Glauber/Coulomb field, it is easy to implement the interactions in a background-field approach. 
In contrast, at NLO, the distinction is manifest in the subleading Lagrangian, $\mathcal{L}_{Q-G/C}^{(1)}$~\cite{Makris:2019ttx}.

The dissociation rates for the various quarkonium states in the QGP can be obtained from NRQCD$_{\textrm{G}}$ and incorporated into the rate equations first developed to describe the propagation of open heavy-flavour states in matter~\cite{Adil:2006ra,Sharma:2009hn}. Under the approximation where the transition between states is neglected, the quarkonium transport takes the form derived in~\cite{Sharma:2012dy,Aronson:2017ymv}. The EFT predictions are also shown in \cf{fig:2Sto1Sptcent} (blue bands), and give a much better description of the data than the energy-loss approach. Furthermore, in this limit, the surviving quarkonia are expected to retain the polarisation acquired from their initial production. A measurement that the HL-LHC might explore is that of quarkonium polarisation in nucleus-nucleus collisions (see Section~\ref{Pol-AA}).     

On the theory side, it will be important to extend such predictions to higher \pT, on the order 100~GeV. The increased data sample at the HL-LHC should in principle allow one to check whether the LP factorisation limit and the energy-loss dominance are reached. Furthermore, as \jpsi and \ups data becomes more precise, 
in particular with smaller uncertainties on the relative suppression of excited to ground states, it will be possible look for effects of medium-induced transition between quarkonium states~\cite{Makris:2019ttx}.

\subsection{Opportunities at HL-LHC}
\subsubsection{Studying the collision-energy dependency of $\Q$ production}
\label{sec:energy_dependence}

The LHC high-luminosity program will have the opportunity to explore different collision energies. One of the most interesting possibilities is to explore the current maximum \PbPb energy of $\sqrtsnn=5.02$~TeV and low-energy collisions in the FT mode, similar to the RHIC \cm\ energy range.
The current energy achieved in the FT mode by the LHCb experiment, using the nominal 2.5~TeV Pb beam energies, is $\sqrtsnn=69$~GeV. Quarkonium total cross-sections decrease by approximately a factor 15 in \pp collisions between $\sqrts=5$~TeV and $\sqrts=69$~GeV. Large integrated luminosities in this mode are desirable to compensate for such a difference in yields. The expected integrated luminosities, \Lint, for the FT mode in LHCb, SMOG2, is 20~\invnb in one year of \PbAr\ collisions at $\sqrtsnn=72$~GeV~\cite{Bursche:2649878}, nearly 100 times larger than what was recorded by the same experiment in \PbPb collisions at 5~TeV. Similar estimations have been obtained by the AFTER@LHC study group for different techniques (gas target, solid target with bent-crystal beam splitting or with a dedicated beam line)~\cite{Lansberg:2012kf,Trzeciak:2017csa,Hadjidakis:2018ifr} using both the LHCb and ALICE detectors.

Whereas the freeze-out temperatures are nearly constant in the aforementioned collision energy range, the peak temperature is expected to change by roughly a factor of three when comparing the estimated peak temperatures obtained from direct-photon-yield slopes measured at RHIC~\cite{Adare:2009qk} and at the LHC~\cite{Adam:2015lda}. The same factor is obtained for charged-particle multiplicities as reported in~\cite{Adam:2015ptt}. It is noteworthy in these publications that the peak temperature also depends on the collision centrality. However, the peak temperature shows a more modest variation of around 50\% from peripheral to central events. It would be relevant to measure the modification of the quarkonium spectrum at the same particle multiplicity but distinct collision energies. Such a measurement, made preferably with the same detector, would have a stringent constraint on models that consider the quarkonium breaking by co-moving hadrons~\cite{Hadjidakis:2018ifr}.

The contribution from charmonium (re)generation strongly depends on the collision energy. More than 100 $c\bar{c}$ pairs are produced in a single central \PbPb collision at $\sqrtsnn = 5$~TeV according to the charm cross section published in~\cite{ALICE_charm}. This large number of uncorrelated \ccbar pairs is an abundant source for the charmonium (re)generation process,  which could explain the enhancement of low-\pT \jpsi yields~\cite{Abelev:2013ila}, and to a lesser extent the elliptic flow~\cite{Acharya:2017tgv, Aaboud:2018ttm, Khachatryan:2016ypw} observed in \PbPb collisions at LHC.

Taking the FONLL computation~\cite{Cacciari:1998it} of the charm cross section at $\sqrtsnn = 69$~GeV, 
less than one charm pair per collision is expected on average, leaving no room for charmonium (re)generation. This feature alone makes high-luminosity, low-energy collisions at the FT-LHC a significant opportunity to use (re)generation-free charmonium states to probe the QGP temperature.

Bottomonia are expected not to be affected by (re)generation (from uncorrelated \bbbar pairs) given that, even at the maximum-energy \PbPb collisions at LHC, less than one \bbbar pair is produced on average per collision. Looking at  quarkonium suppression as a function of rapidity and system size (another asset of the versatile FT mode) would permit searches for the onset of QGP effects, and hence put constrains on the in-medium modification of the quarkonium potential. 

\begin{figure}[htbp!]
\centering
\includegraphics[scale=0.38]{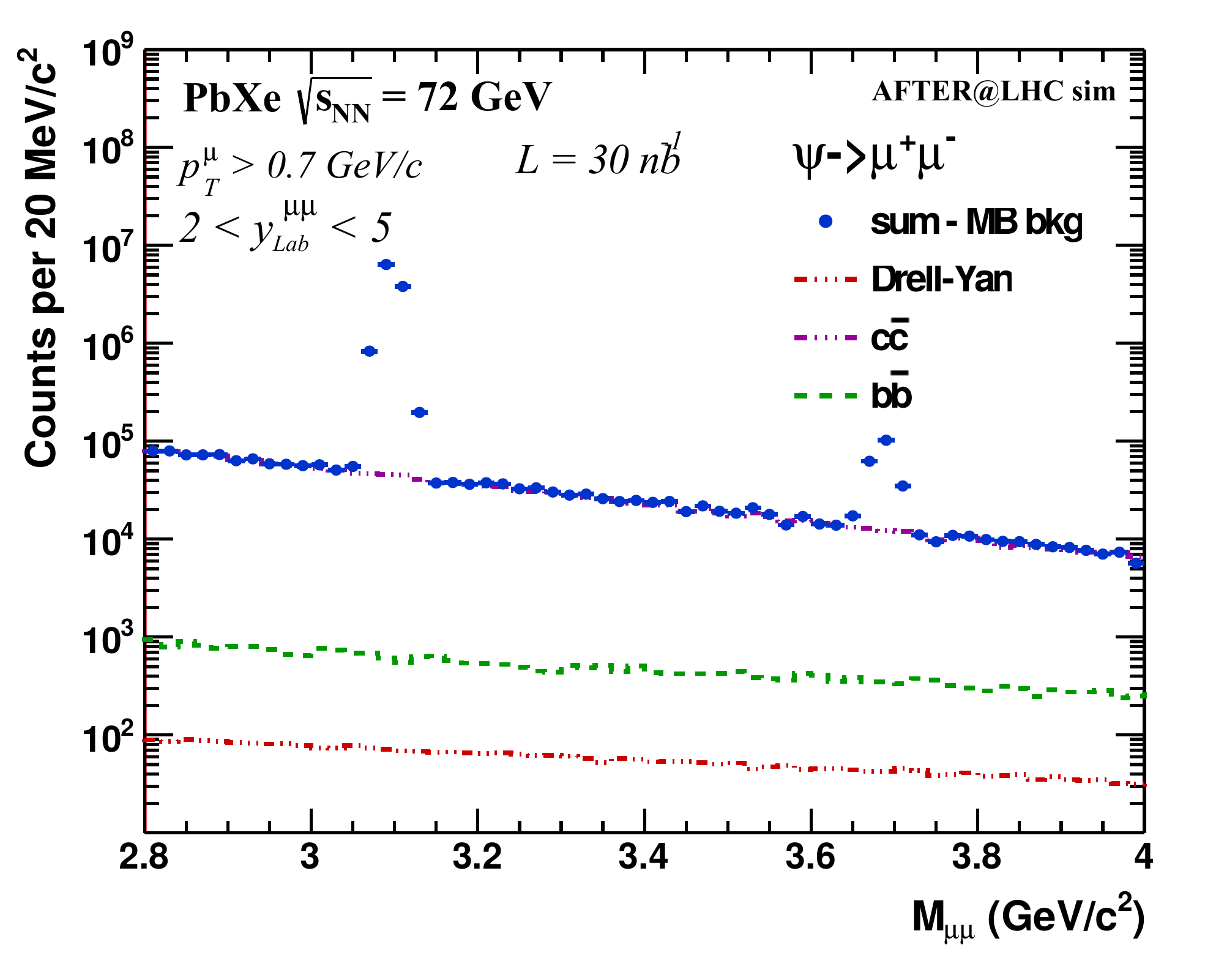}
\hspace{0.3cm}
\includegraphics[scale=0.38]{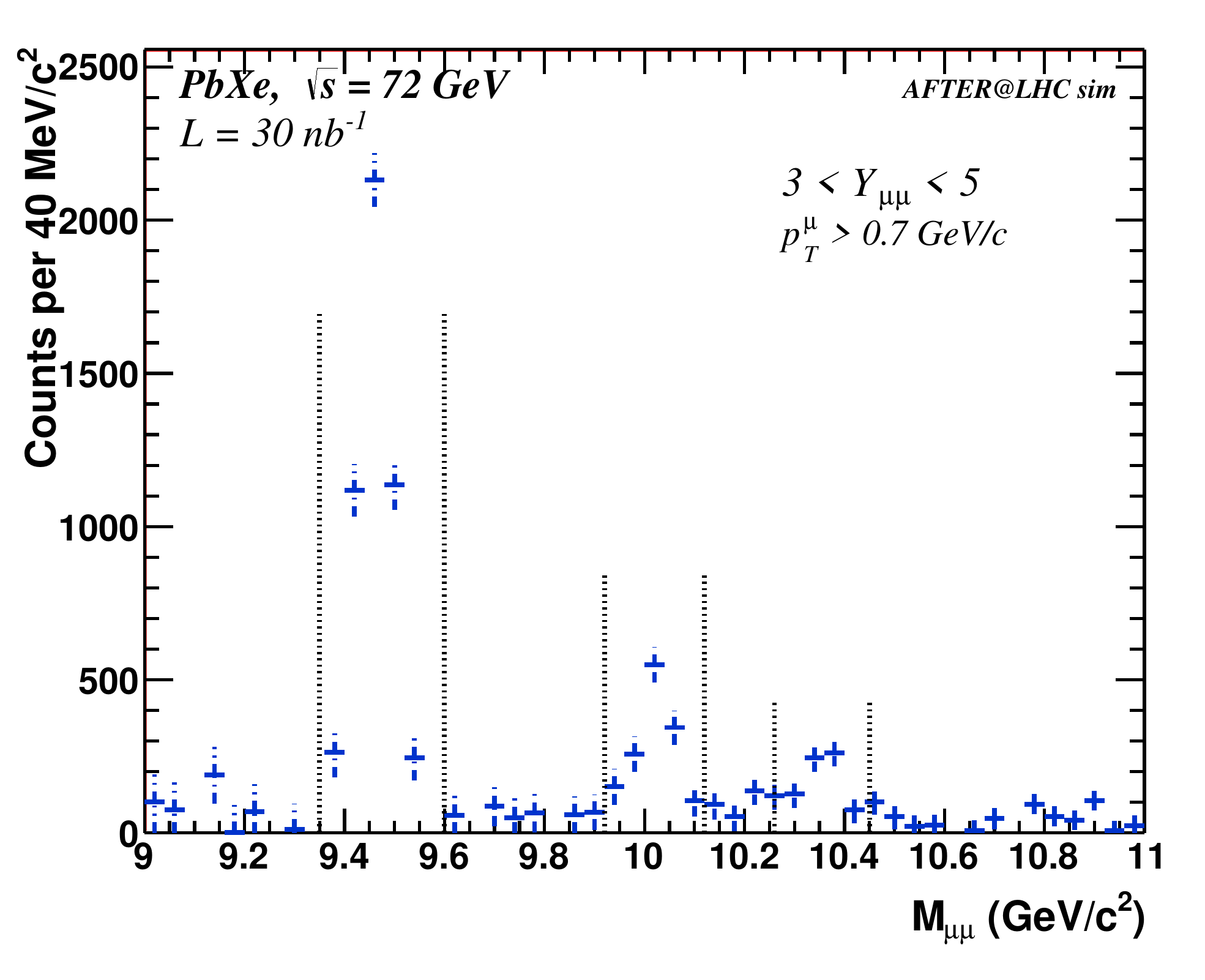}   
\caption{Di-muon invariant mass distribution in the \jpsi, \psip (Left) and $\Upsilon(nS)$ (Right) regions, expected in \PbXe\ FT collisions at $\sqrtsnn = 72$~GeV, for an LHCb-like detector ($2 < y_{\rm lab} < 5$). The combinatorial background is subtracted using like-sign pairs. No nuclear modification are assumed. The integrated luminosity of $\Lint = 30 \invnb$ corresponds to one LHC-year of data-taking for ions (\ie typically one month of data taking), with an LHCb-like detector equipped with a gaseous storage-cell of 1-m length. The maximum luminosity achieved has been limited such that no more than 15$\%$ of the Pb beam is removed by the interaction with the target. [Plots taken from~\cite{Hadjidakis:2018ifr}].}
    \label{fig:FixedTarget-Quarkonium}
\end{figure}

 The current status of the implementation of a FT operation mode within the LHCb detector, and the ongoing technical developments performed towards the achievement of an extended FT physics programme during HL-LHC, both in LHCb and potentially with the ALICE detector, have been discussed in Section~\ref{sec:dif_FT}. Conventional detectors with a coverage of the forward rapidities in the laboratory frame, such as LHCb or the ALICE muon arm, allow scanning of the mid to backward \cm\ rapidity region.  With the HL-LHC luminosities in the FT mode, the yearly charmonium yields in PbXe collisions are expected to be very large, of the order of $\sim$10$^{7}$ \jpsi mesons in LHCb (Fig.~\ref{fig:FixedTarget-Quarkonium}, left) and of the order of a few 10$^{6}$ \jpsi mesons in the ALICE muon spectrometer. 
 
 The understanding of the (dynamical) charmonium suppression (given the negligible (re)generation expected at low \cm\ energy) in the QGP would highly benefit from systematic and precise studies of all excited states. This includes direct \chic measurement, which becomes less challenging at low energy and backward rapidity, thanks to a low-background environment. Projections with an LHCb-like detector of the \psip nuclear modification factor indicate, for instance, that a statistical precision of a few percent could be reached at mid rapidity in the \cm~\cite{Hadjidakis:2018ifr}. In addition to the \psip and \chic measurements, novel observables, such as quarkonium-pair (\jpsi + \jpsi) or quarkonium--heavy-quark correlations ($\jpsi + D$), which require a larger luminosity and acceptances than that achieved in the past at the SPS and RHIC, could be explored for the first time in this energy regime. As can be seen in \cf{fig:FixedTarget-Quarkonium} (Left), the low level of background in the \jpsi region supports the feasibility of the aforementioned studies, especially in the most backward region where backgrounds are smallest. 
 
Figure~\ref{fig:FixedTarget-Quarkonium} (Right) shows projections for the $\Upsilon(nS)$ invariant-mass region, in the di-muon-decay channel, after the combinatorial-background subtraction, in \PbXe\ collisions, with an LHCb-like detector. The yearly \ups, \upsp and \upspp yields are about {$4 \times 10^{3}$, $10^{3}$ and $5 \times 10^{2}$} mesons, respectively. Given the excellent resolution of LHCb, the three states are well separated. The expected statistical precision  on the measurements of \RAA\ {of} each of the three states will be about 7\%, 20\% and 30\% for the \ups, \upsp and \upspp, respectively. Yield projections in the bottomonium sector also exists for the ALICE muon arm. Typically a few hundred \ups mesons will be  collected in \PbXe\ collisions in one year of LHC data taking. The study of the excited states, even for several data-taking years, will remain rather limited with ALICE in the FT mode. Such studies of $\Upsilon(nS)$ suppression, especially with the LHCb detector, will therefore bring crucial new inputs to our understanding of the nature of the hot medium created in this energy regime, complementary to the studies already performed at the LHC (see CMS results~\cite{Sirunyan:2017lzi, Khachatryan:2016xxp, Chatrchyan:2012lxa}). This will allow tests of the different approaches discussed in 
sections \ref{AAintro} and \ref{subsection_rec_th_dev} 
and comparisons with effective models, such as the CIM, that deals with quarkonium suppression and accounts for Landau damping~\cite{Ferreiro:2018wbd}.

\subsubsection{Prospects for $X(3872)$ studies}
\label{sec:aa_x3872}
About 20 years after its discovery~\cite{Choi:2003ue}, the question of whether $X(3872)$ is a molecule, a compact tetraquark, or a hybrid state is still a subject of intense debate. It is thus worth wondering whether heavy-ion experiments can help us understand its nature, in addition to the other exotic $XYZ$ states. Such investigations with heavy ions are needed in parallel to the recent experimental~\cite{Aaij:2020hpf} and theoretical~\cite{Esposito:2020ywk} work related to high-multiplicity \pp\ collisions mentioned in Section~\ref{pAcom} (see also Section~\ref{sec:pp-xyz} for a general discussion of the production in \pp\ collisions). To advance  our understanding using heavy ions, two inputs are  needed:
 precise measurements of the $X(3872)$ yields in heavy-ion collisions and solid theoretical calculations that lead to different results for different underlying structures.  Neither is an easy task. 

On the theory side, many phenomena may affect the production of the $X(3872)$. For example, for low-\pT production, the dissociation and recombination of the $X(3872)$ similar to those of charmonium can happen in the hot QGP (for a compact tetraquark state) or in the hadronic gas (for a molecule). These processes in the hadronic gas are also connected with similar processes in \pp and \pA collisions, though the background hadronic gas densities are different. These reaction rates are poorly understood from first principles. Furthermore, the recombination is sensitive to the total number of charm quarks produced in one event, which has not been precisely determined in heavy-ion collisions. 

At larger \pT, energy loss may also affect the $X(3872)$ yields. Moreover, in order to convert calculations to phenomenology, one needs precise knowledge of the branching ratio of the decay channel of the $X(3872)$  used in the measurement (for example, $J/\psi\pi\pi$), which is also not well known but may be improved with future measurements at $B$ factories. Though the task is difficult, one still hopes that one can do some analyses with precise data. So far, only the ratio between the production yields of the $X(3872)$ and the \psip has been measured by the CMS collaboration in the $\pT =10$--50~GeV range~\cite{CMS-PAS-HIN-19-005}. Since the suppression mechanism  of \psip  is not well understood, it is preferable if the direct yield (rather than the ratio)  of $X(3872)$ can be measured 
as a function of \pT in the soft regions. This would indicate how significant recombination is to the production of the $X(3872)$, since recombination is sensitive to the particle wave function.
This idea is motivated by the important contribution from recombination in charmonium production at low \pT. On the experimental side, the size of the $X(3872)$ data samples needs to be increased in order to carry out more differential measurements.
At the same time, phenomenological calculations assuming different structures of the $X(3872)$ have to be carried out. 
Using the $X(3872)$ production yields in heavy-ion collisions to understand its structure may not be fully
successful by itself, but provides complementary information to measurements in other collision systems.

\subsubsection{${\psi}$ polarisation in ${\PbPb}$ collisions}
\label{Pol-AA}

The question of whether the \jpsi meson is polarised in \AaAa collisions has been addressed by only a few authors~\cite{Gupta:1998ut,Ioffe:2003rd,Faccioli:2012kp}, who have advocated that a modification of the \jpsi polarisation in \AaAa (as compared to \pp) could be due to either the disappearance  of feed-down from higher states   due to the suppression of these states in QGP, or the modification of the \ccbar $\to$ \jpsi conversion mechanism, which would be altered in those collisions, through a modification of the LDMEs at {freeze-out}. To quantify the \chic suppression in \AaAa collisions, in~\cite{Faccioli:2012kp}, it is considered that the {feed-down} from \chic results in the ``blurring'' of the direct \jpsi production in \pp, which would then be recovered in \AaAa. 

However, the first measurement of \jpsi polarisation in \AaAa collisions by ALICE ~\cite{Acharya:2020xko}, although still affected by sizeable uncertainties, does not show a significant modification of the \jpsi polarisation parameters, $\lambda_\theta$, $\lambda_\phi$ and $\lambda_{\theta \phi}$. At present, the first question to be answered is indeed whether the polarisation differs in the \pp\ and \AaAa samples, rather than the actual size of the polarisation in  \AaAa collisions.
 In this context, it would be helpful to look at \RAA as a function of the cosine of the di-lepton polar angle, $\cos \theta$, {since the polarisation directly affects the $\cos \theta$ distribution}: this measurement  may have  greater experimental  precision  than a direct measurement of $\lambda_\theta$.

Details of the quarkonium formation can also be addressed in \AaAa collisions. From the viewpoint of the theoretical modelling, the SCET$_{\rm G}$ approach described in Section~\ref{AAtransportEFT} is an ideal candidate to investigate the polarisation at large \pT, a regime where each directly produced charmonium is expected to have the same polarisation as in \pp collisions,  due to helicity quasi-conservation in the energy loss process, but where the energy loss and suppression affect their relative yields and their subsequent contribution to the lower-lying states.
At smaller \pT, the interactions with the QGP, neglected in most of the previous studies, and the large fraction of the \jpsi yield due to recombination are expected to partly wash away the polarisation of the \ccbar state. On the contrary, the strong magnetic field created in the early QGP stage could enforce a spin alignment of the \jpsi perpendicular to the event plane~\cite{Tuchin:2013ie}. Experimental investigations should therefore be supported by quantitative theoretical predictions that include all these ingredients, and are based on state-of-the-art understanding of \jpsi polarisation in \pp collisions.             

An alternate strategy could consist in measuring the polarisation of prompt \psip in \AaAa collisions, which do not receive any {feed-down}, in kinematic regions where the yield is found to be  polarised in \pp collisions. One may reasonably anticipate a gradual reduction of the polarisation in \AaAa collisions for an increasing centrality 
due to interactions with the QGP constituents. The issue is that, for now,  the \psip has only been found to be (longitudinally) polarised for $\pT > 10$~GeV at forward rapidities~\cite{Aaij:2014qea}. 
In addition, this represents a genuine experimental challenge: the $\psip / \jpsi$ ratio in \PbPb in the di-muon channel is about 1--2\% (3--5\% in \pp)~\cite{Sirunyan:2016znt}. The published ALICE results in \PbPb~\cite{Acharya:2020xko} use $750~\invmub$ of data, while $10~\invnb$ are expected after Runs 3--4. This means that naively a \psip polarisation in \PbPb at the end of Run~4 will still be less precise than the Run-2 \jpsi measurement (not accounting for the much lower signal-over-background for \psip compared to \jpsi, nor for improvements due to detector upgrades). The situation may be different with lighter ions (which may be available for Run~5 or beyond), providing more integrated luminosity, and for which less suppression is expected with respect to \pp. Finally, if such a measurement was to be done by the ATLAS or CMS experiments, the gain in acceptance in rapidity may be compensated by a larger \pT threshold.

\section{Double and triple parton scatterings\protect\footnote{Section editors: David d'Enterria, Tomas Kasemets. %
}}
\label{sec:dps}
\newcommand\cO{{\cal O}}
\newcommand\cN{{\cal N}}
\newcommand\bR{{\cal B}}

\newcommand{\Sig}{\mathcal{S}}
\newcommand{\LumiInt}{\mathcal{L}_{\mbox{\rm \tiny{int}}}}
\newcommand{\Lunits}{cm$^{-2}$s$^{-1}$}

\newcommand{\dtwor}{{d^2r}}
\newcommand{\dtwob}{{d^2b}}

\newcommand{\sigmasps}{\sigma_{{\rm SPS}}}
\newcommand{\sigmadps}{\sigma_{{\rm DPS}}}
\newcommand{\sigmatps}{\sigma_{{\rm TPS}}}

\newcommand{\sigSPS}{\sigma_{{\rm SPS}}}
\newcommand{\sigDPS}{\sigma_{{\rm DPS}}}
\newcommand{\sigTPS}{\sigma_{{\rm TPS}}}

\newcommand{\sigmaeff}{\sigma_{\rm eff}}
\newcommand{\sigmaeffpp}{\sigma_{{\rm eff},pp}}
\newcommand{\sigmaeffpA}{\sigma_{{\rm eff},pA}}
\newcommand{\sigmaeffAA}{\sigma_{{\rm eff},AA}}

\newcommand{\sigmaeffdps}{\sigma_{\rm eff,{DPS}}}
\newcommand{\sigmaefftps}{\sigma_{\rm eff,{TPS}}}
\newcommand{\sigmaeffnps}{\sigma_{\rm eff,{NPS}}}
\newcommand{\sigmaeffdpspA}{\sigma_{{\rm eff,{DPS}},pA}}
\newcommand{\sigmaefftpspA}{\sigma_{{\rm eff,{TPS}},pA}}
\newcommand{\sigmaeffdpsAA}{\sigma_{{\rm eff,{DPS}},AA}
\newcommand{\sigmaefftpsAA}{\sigma_{{\rm eff,{TPS}},A}}}

\newcommand{\sigmaDPSone}{\sigma_{{\rm {DPS,1}}}}
\newcommand{\sigmaDPStwo}{\sigma_{{\rm {DPS,2}}}}
\newcommand{\sigmaDPSthree}{\sigma_{{\rm {DPS,3}}}}

\newcommand{\sigmaTPSone}{\sigma_{{\rm {TPS,1}}}}
\newcommand{\sigmaTPStwo}{\sigma_{{\rm {TPS,2}}}}
\newcommand{\sigmaTPSthree}{\sigma_{{\rm {TPS,3}}}}
\newcommand{\sigmaTPSnine}{\sigma_{{\rm {TPS,9}}}}

\newcommand{\NDPS}{\rm N_{{\rm {{DPS}}}}}

\newcommand{\htr}[1]{{\color{red} #1}}
\newcommand{\htb}[1]{{\color{blue} #1}}
\newcommand{\htg}[1]{{\color{green} #1}}
\newcommand{\prn}[2]{{}^{#1} #2}       %
\newcommand{\prb}[2]{{}^{#1}\! #2}    %
\newcommand{\prl}[2]{{}^{#1}\! #2}     %
\newcommand{\tvec}[1]{\boldsymbol{#1}}

\subsection{Introduction} 

The extended nature of hadrons and their large parton densities when probed at the HL-LHC collision energies, make it very likely to produce simultaneously two or more quarkonium states alone or together with other heavy particles via separate multi-parton interactions in \pp~\cite{Lansberg:2014swa}, \pA~\cite{Strikman:2001gz,Cattaruzza:2004qb,dEnterria:2014lwk,dEnterria:2016yhy}, and \AaAa~\cite{dEnterria:2013mrp,dEnterria:2014lwk} collisions. Double, triple, and in general {$n$-tuple} parton scatterings (DPS, TPS, and NPS, respectively) depend on the degree of transverse overlap of the matter densities of the colliding hadrons, and give access to the phase space distributions of partons inside the proton or nucleus. The study of NPS provides thereby valuable information on the hadronic wave functions describing the correlations among partons in space, momentum, flavour, colour, spin, etc., and their corresponding evolution as a function of collision energy.
In addition, understanding double and triple parton scatterings is of relevance in the study of backgrounds for the associated production of quarkonia plus other hard particles (Section~\ref{sec:oniumassociate}), for rare Standard Model (SM) decays, and/or for searches for new physics in final states with multiple heavy particles.

The pQCD-factorised expression to compute the cross section of a given double parton scattering process in hadron collisions reads
\begin{align}
& \sigma_{\text{DPS}}
  = \left(\frac{\mathpzc{m}}{2}\right)
  \sum_{a_1a_2b_1b_2}\sum_{R}
     \left[\prn{R}{\hat{\sigma}}_{a_1 b_1}\,
     \prn{R}{\hat{\sigma}}_{a_2 b_2}\right]\,
    \otimes \int d^2\tvec{y}\;
     \prb{R}{F}_{b_1 b_2}(x_i,\tvec{y}) \,
     \prb{R}{F}_{a_1 a_2}(\bar{x}_i,\tvec{y}) \,,
\label{eq:dps_xsec}
\end{align}
where $\mathpzc{m}$ is a combinatorial factor to avoid multiple counting of the same process, $\otimes$ denotes a convolution over longitudinal momentum fractions, $\hat{\sigma}_{ab}$ is the partonic cross section for the interaction between partons $a_i$ and $b_i$, while $F_{a_1a_2}$ is the double parton distribution function (dPDF) of two partons inside a proton~\cite{Paver:1982yp}, separated by a distance $\tvec{y}$ and each carrying a longitudinal momentum fraction $x_i$. The sum over $a_i$ and $b_i$ runs over parton flavours and spin, while $R$ runs over the allowed colour representations~\cite{Buffing:2017mqm,Diehl:2017kgu}. 

Section~\ref{sec:DPS_TH} describes the current status of the theoretical DPS calculations based on \ce{eq:dps_xsec}. Often, however, rather than the full calculations, a useful simplistic approximation is employed to estimate the cross sections for the DPS production of two hard particles $H_1$ and $H_2$ from the product of their corresponding single-parton-scattering ($\sigmasps$) values
normalised by an effective cross section $\sigmaeff$ to warrant the proper units of the final result, namely
\begin{equation}
\sigma^{hh' \to H_1+H_2}_{\rm DPS} = \left(\frac{\mathpzc{m}}{2}\right)\, \frac{\sigma^{hh' \to H_1}_{\rm SPS} \cdot
\sigma^{hh' \to H_2}_{\rm SPS}}{\sigmaeff}\,. 
\label{eq:pocketDPS}
\end{equation}
This so-called ``pocket formula'' encapsulates the intuitive result that, in the absence of any partonic correlations, the probability to have two parton-parton scatterings producing two heavy or high-$\pT$ particles (\eg\ quarkonium states) in a given inelastic hadron-hadron collision should be proportional to the product of probabilities to independently produce each one of them. In the extreme case where one assumes that (i) the dPDF factorises as the product of transverse and longitudinal densities, (ii) the longitudinal components themselves reduce to the product of independent single PDFs, (iii) the transverse profile is the same for all partons, and (iv) no other parton correlations are present, the effective cross section is~\cite{Calucci:1997ii,dEnterria:2017yhd}
\begin{equation}
\sigmaeff\equiv\sigmaeffdps=\left[ \int d^2b\,T^2(b)\right]^{-1} \,.
\label{eq:sigmaeff_DPS}
\end{equation}
Consequently, $\sigmaeff$ can be written as a function of the \pp overlap $T(b)$ at impact parameter $b$, computable from the transverse parton-density profile of the proton $\rho(b)$ in a Glauber approach~\cite{dEnterria:2020dwq}.
For conventional transverse parton $\rho(b)$ distributions of the proton, such as those typically implemented in the modern \pp Monte Carlo (MC) event generators \pythia~8~\cite{Sjostrand:2017cdm}, and \herwig~\cite{Seymour:2013qka}, one expects values of $\sigmaeff\approx15-$25~mb. These  $\sigmaeff$ values are smaller than the purely geometric ``soft'' \pp\ cross section of $\sigma_\mathrm{inel}\approx 35$~mb, derived from the electromagnetic radius of the proton, because of the inherent ``centrality bias'' that appears when one or more hard enough parton scatterings is required.

The pocket formula (\ce{eq:pocketDPS}) can be used for \pp, \pA, and \AaAa\ collisions, and its generalisation for the hard production of $n$ sets of particles, denoted $H_i$, in $n$ parton scatterings from the corresponding single parton values can be expressed as the $n^{\rm th}$-product of the corresponding SPS cross sections for the production of each single final-state particle, normalised by the ($n^{\rm th}-1$) power of an effective NPS cross section~\cite{dEnterria:2017yhd}:
\begin{equation} 
\sigma^{hh' \to H_1+\ldots+\,H_n}_{\rm {nps}} = \left(\frac{\mathpzc{m}}{n!}\right)\,
\frac{\prod_i\sigma^{hh' \to H_i}_{\rm SPS}}{(\sigmaeffnps)^{n-1}},
 \;\;\;\mbox{with}\;\;\;
\sigmaeffnps=\left\{\int d^2b \,T^n(b)\right\}^{-1/(n-1)}\,,
\label{eq:pocketNPS}
\end{equation} 
where, again, the second equality holds in the strong assumption of absence of any parton correlations and $\sigmaeffnps$ bears a simple geometric interpretation in terms of powers of the inverse of the integral of the hadron-hadron overlap function $T(b)$ over all impact parameters.

The expressions based on \ce{eq:pocketNPS} applied to quarkonium production in DPS and TPS processes below provide baseline (purely ``geometric'') order-of-magnitude estimates of their expected cross sections by combining (i) SPS cross sections $\sigmasps$, which are either experimentally measured or computed within perturbative QCD \eg\ at NLO or NNLO accuracy today, %
plus (ii) a value of $\sigmaeffnps$, theoretically derived in a Glauber geometric approach, or extracted from experimental measurements. The comparisons of experimental data, and/or more complete theoretical predictions with reduced number of approximations, to the simple cross sections expected from the pocket formula, allow one to assess the corresponding size and impact of parton correlations in the proton (or nuclear) wave functions. Since DPS and TPS cross sections depend on the square and cube of the corresponding SPS cross sections, perturbative processes with large enough SPS cross sections are needed in order to have a visible number of events at the HL-LHC: this is the advantage of using multiple production of quarkonia, over more rare heavy particles such as electroweak bosons, in NPS studies. %

\subsection{Theoretical status of Double Parton Scattering}
\label{sec:DPS_TH}

\subsubsection{Factorisation of DPS cross sections}

\begin{figure}[h!]
\centering
\includegraphics[width=\textwidth]{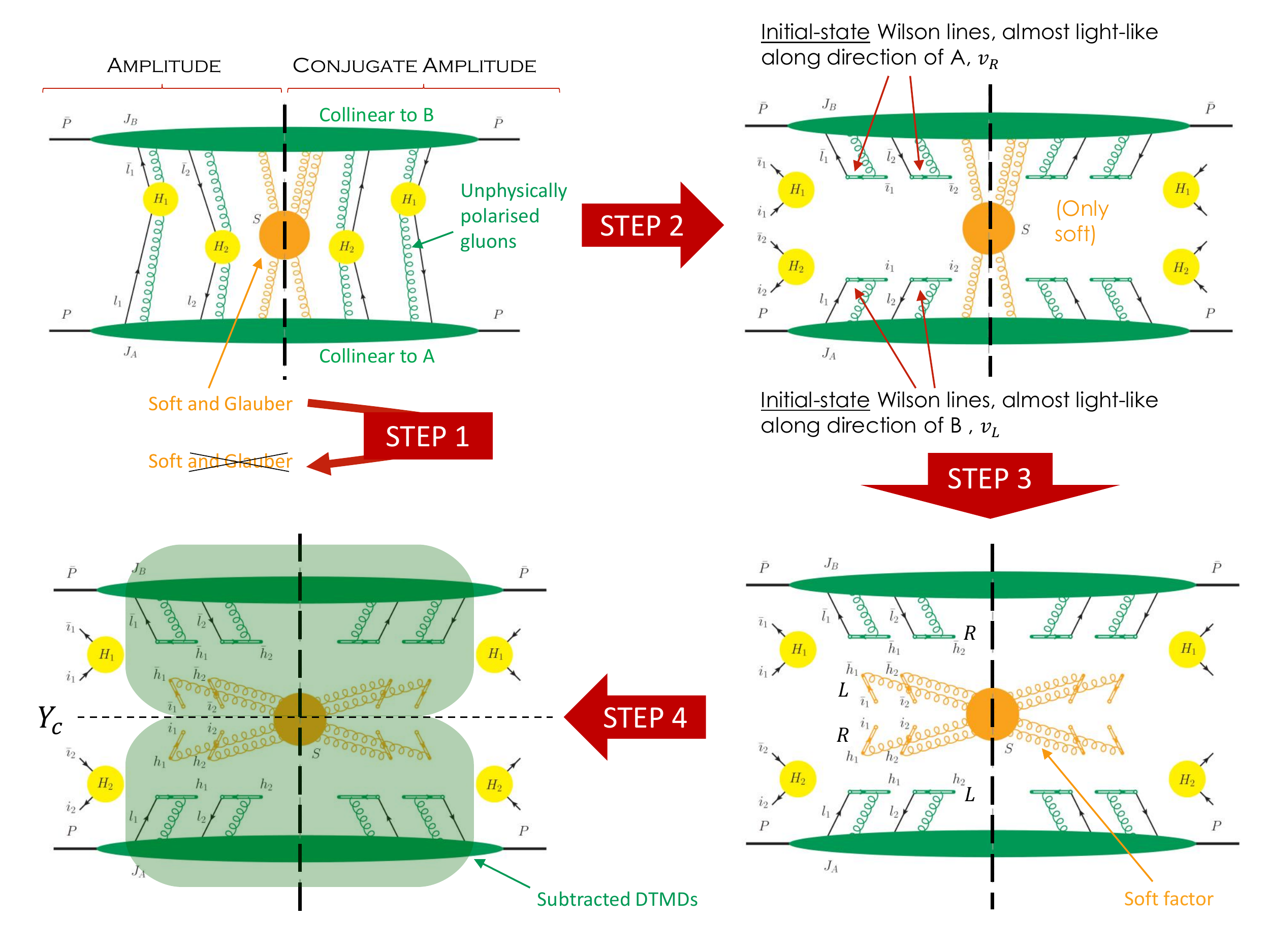}
\caption{Diagrammatic illustration of the steps to achieve a proof of factorisation for the double Drell--Yan cross sections at a given transverse momentum. The green blobs represent the right and left moving proton, yellow blobs the hard interactions and orange blob the soft interactions. The graph is for the cross section, with the vertical line indicating the final state cut. [Figure modified from~\cite{Diehl:2018wfy}].}
\label{fig:DDYfact}
\end{figure}

The theoretical predictions for DPS cross sections rely on the factorisation of the underlying dynamics as a convolution of parton-parton cross sections and dPDFs, as described by \ce{eq:dps_xsec}. A first vital question is whether this factorisation actually holds, up to power-suppressed corrections. A starting point is to look at the production of two colourless systems, \eg of (pairs of) $W$, $Z$ or $H^0$ bosons since, for their SPS production, factorisation was proven for both the total and TMD cross sections in the 1980s~\cite{Bodwin:1984hc, Collins:1985ue, Collins:1988ig, Collins:2011zzd}. In recent years, it has been established that DPS factorisation also holds for the double Drell--Yan (DDY) process both for the case of the total cross section, and for the case where the transverse momenta of the bosons are measured (double TMD, or DTMD, case)~\cite{Diehl:2015bca,Diehl:2016eyd,Diehl:2018wfy,Gaunt:2018eix,Vladimirov:2017ksc, Buffing:2017mqm}. 
The steps that need to be taken to demonstrate factorisation for the latter case are schematically shown in Fig.~\ref{fig:DDYfact}.

Certain contributions to DPS overlap with (loop) contributions to single scattering, and yet others overlap with other more-exotic scattering mechanisms such as higher twist contributions, or DPS-SPS interference. A consistent factorisation framework for DPS should avoid double counting between DPS, single scattering, and other mechanisms. A framework that achieves this, and maintains a description of the DPS part in terms of separate rigorously defined dPDFs for each hadron, was developed in~\cite{Diehl:2017kgu} (other proposals were made earlier \cite{Blok:2011bu, Gaunt:2011xd, Manohar:2012pe, Gaunt:2012dd}, although these did not have this last property). The application of the approach of \cite{Diehl:2017kgu} to the DTMD case is described in~\cite{Buffing:2017mqm,Gaunt:2018eix}.
Whether factorisation holds for other DPS processes, including those involving quarkonium production, is less clear. Whatever factorisation-breaking complications apply for SPS of quarkonium, via the CO channel at a given $\pT$ (Section~\ref{sec:TMDchallenges}), are expected to be carried over to the DPS case. 
\subsubsection{Evolution of dPDFs}

Apart from DPS factorisation, a second key element for the computation of double parton cross sections via \ce{eq:dps_xsec}) is to control the phase space evolution of dPDFs. Double parton distributions $F_{ab} (x_i, y, \mu_i)$, with $i = 1,2$, enter the DPS factorised cross section through the parton luminosities $\mathcal{L}_{a_1 a_2 b_1 b_2} = \int \! \mathrm{d}^2 \tvec{y} \, F_{a_1 a_2} (y) \, F_{b_1 b_2} (y)$, where $\tvec{y}$ is the transverse separation between the two partons. 
A measurement of the dPDFs functional form is at present not yet available, and DPS phenomenological studies rely on model Ans\"{a}tze. However, their scale dependence is well known~\cite{Snigirev:2003cq, Gaunt:2009re, Diehl:2011yj, Blok:2011bu, Manohar:2012jr} and is analogous to the familiar one for PDFs. Double PDFs evolve in energy scale $\mu_i$ according to generalised DGLAP equations
\begin{align} \label{eq:dps:double_dglap}
\frac{d}{d \log \mu^2_1} \, F_{a_1 a_2} (x_i, y, \mu_i) = \left( P^{(1)}_{a_1 c} \underset{1}{\otimes} F_{c a_2} \right) (x_i, y, \mu_i) \, \& \, 
\frac{d}{d \log \mu^2_2} \, F_{a_1 a_2} (x_i, y, \mu_i) = \left( F_{a_1 c} \underset{2}{\otimes} P^{(2)}_{c a_2} \right) (x_i, y, \mu_i)
\,,\end{align}
where $P^{(1,2)}_{ab}$ are the splitting functions, $y=|\tvec{y}|$ and the convolution $\otimes$ is performed in $x_1$ or $x_2$.
This set of equations is valid for the $y$-dependent dPDFs: integration over $y$ introduces an additional inhomogeneous term in \ce{eq:dps:double_dglap}.

\begin{figure}[h!]
\centering
\includegraphics[width=0.42\textwidth,draft=false]{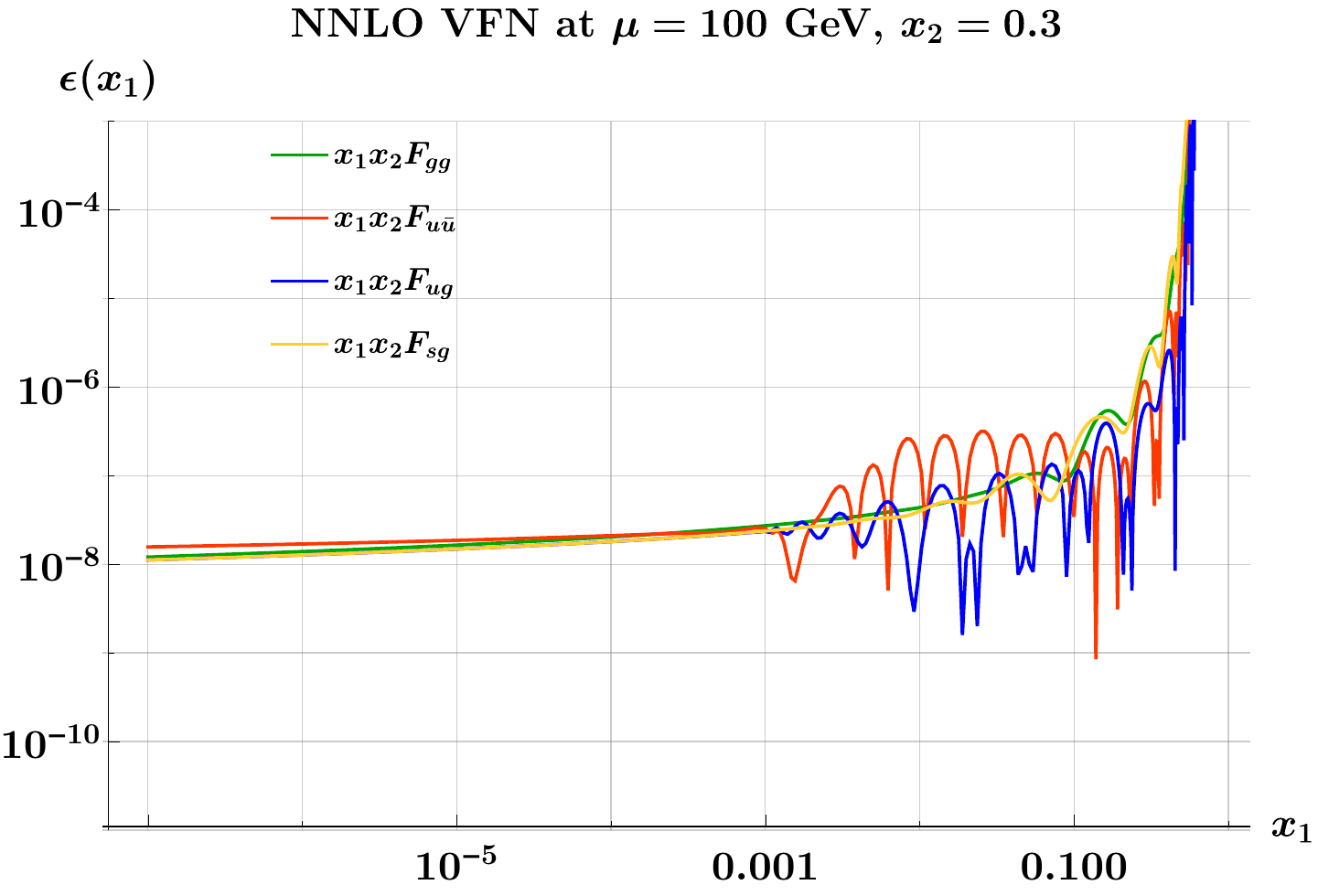}
\includegraphics[width=0.28\textwidth,draft=false]{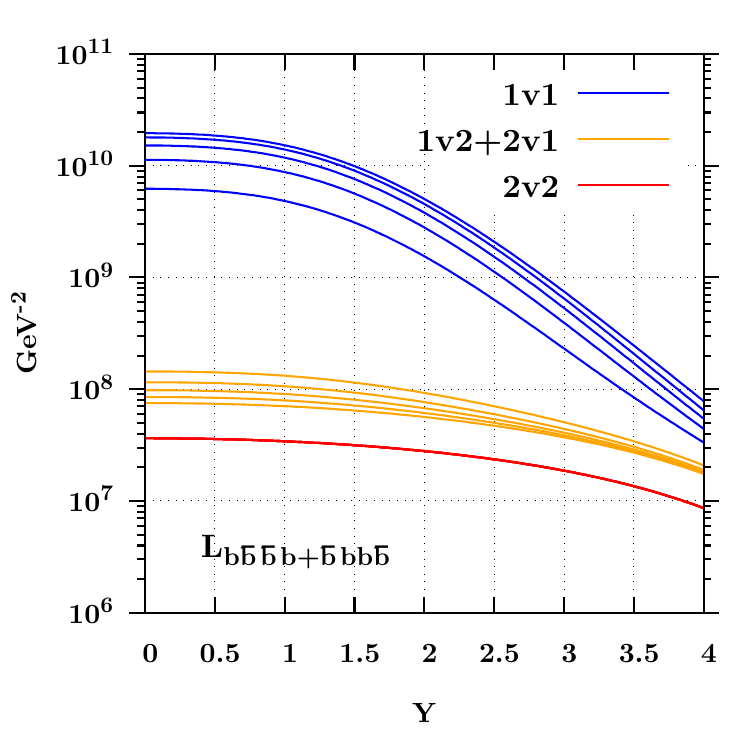}
\includegraphics[width=0.28\textwidth,draft=false]{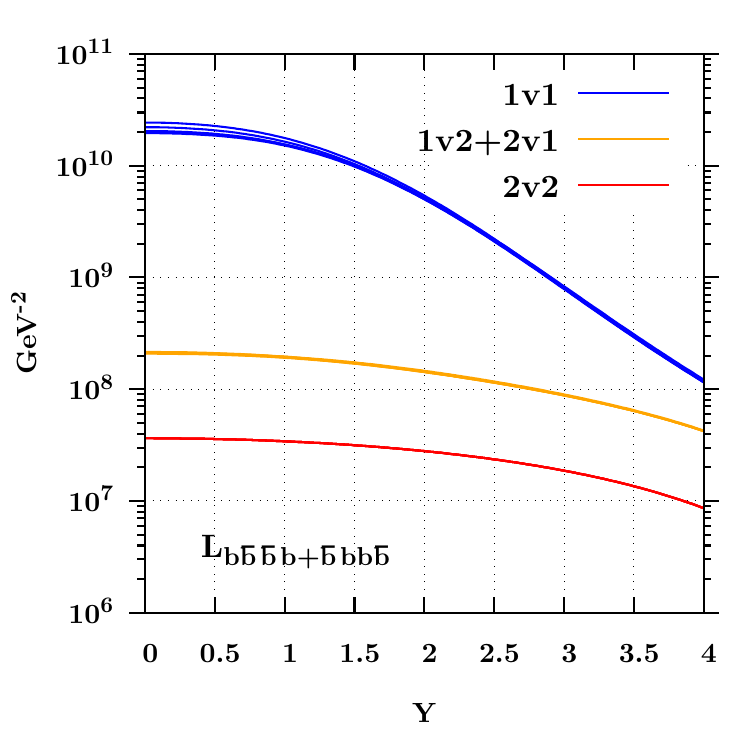}
\caption{Left: Relative accuracy reached by ChiliPDF for a set of dPDFs ($F_{gg}$, $F_{u\bar{u}}$, $F_{ug}$, $F_{sg}$) evolved to $\mu_1 = \mu_2 = 100 \, \text{GeV}$ at NNLO in the variable-flavour-number (VFN) scheme, as a function of $x_1$ at fixed $x_2 = 0.3$.
Centre: LO symmetrised luminosity $\mathcal{L}_{b\bar{b}\bar{b}b}$ for pure splitting (``1v1''), pure product (``2v2''), and mixed (``1v2+2v1'') combinations of dPDF Ans\"{a}tze, at $\mu_1 = \mu_2 = 25 \, \text{GeV}$, and with the two final system rapidities $Y_1 = 0$ and $Y_2 = Y$. The lines correspond to variations of the matching scales $\mu_{c,b} = (1\dots 5)\cdot m_{c,b}$ in the splitting Ansatz. Right: Same as centre plot, but with the inclusion of the gluons splitting into massive $c\bar{c}$ and $b\bar{b}$ pairs. [Left and centre figures are taken from~\cite{Nagar:2019njl}].}
\label{fig:dps:dpds}
\end{figure}

Given their dependence on many parameters, and their $\mathcal{O}(N_f^2)$ multiplicity, dPDFs are complex to handle numerically, both in terms of memory occupation and computational time. The LO double-DGLAP evolution for the $y$-integrated dPDFs was first studied in~\cite{Gaunt:2009re}, and a publicly available dPDF set (GS09 \cite{Gaunt:website}) was provided based on a product Ansatz $F_{ab} = f_a \cdot f_b \cdot \Phi$, where $f_{a,b}$ are regular PDFs, and $\Phi$ is a suppression factor. Recent progress has been made with a new tool called ChiliPDF~\cite{Nagar:2019njl,Diehl:2020uuu}. This tool can solve the double-DGLAP equations up to NNLO, including $\mathcal{O}(\alpha_s^2)$ matching at the flavour transition scales, in a fast and relatively lightweight way, with a working numerical precision below $\mathcal{O}(10^{-4})$ for $x_1 + x_2 < 0.8$ that is safely far beyond theory uncertainties on dPDFs (Fig.~\ref{fig:dps:dpds}, Left). Quark-mass effects can be sizeable for dPDFs. Considering as boundary condition for \ce{eq:dps:double_dglap} the perturbative splitting from PDFs into dPDFs (denoted as ``\textbf{1}'', as opposed to the product Ansatz ``\textbf{2}'' \cite{Diehl:2017kgu}), the inclusion of the heavy-quark masses by matching at the flavour-transition scales
introduces large scale uncertainties on the evolved dPDFs (Fig.~\ref{fig:dps:dpds}, centre). The insertion of the gluon splitting into massive quarks in the ``\textbf{1}'' dPDFs visibly reduces the uncertainties even at LO, as shown in Fig.~\ref{fig:dps:dpds} (Right)~\cite{Diehl:2020vvv}.

\subsubsection{Impact of parton correlations on \texorpdfstring{$\sigmaeff$}{sigma(eff)}}

\begin{figure}[htpb!]
\centering
\includegraphics[scale=0.77]{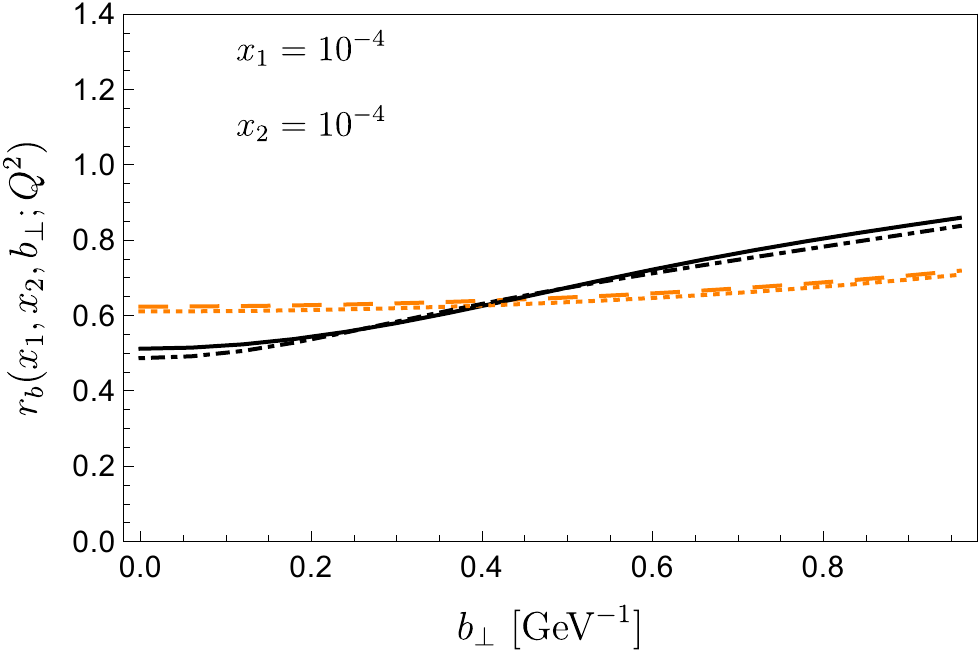} \quad
\raisebox{6pt}{\includegraphics[scale=0.2]{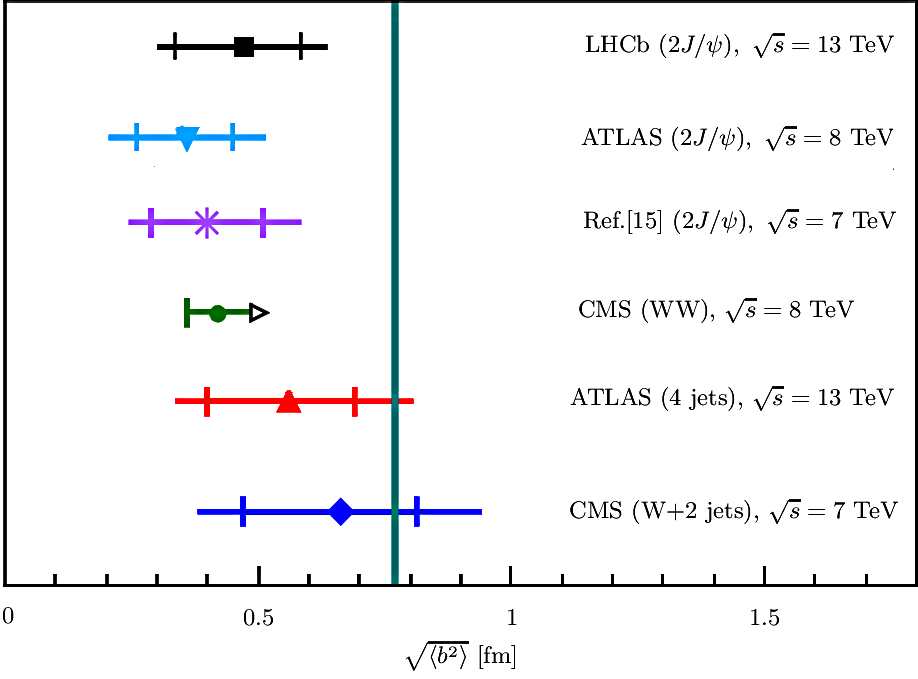}}
\caption{Left: Ratio $r_b(x_1,x_2,b_\perp,Q^2)$ quantifying the impact of relativistic correlations on digluon distributions~\cite{Rinaldi:2018bsf}. This quantity would be equal to 1 if parton correlations were absent.  Black and yellow curves shown the results of different models of double parton distribution functions for $Q^2 = m^2_H$ (full and dashed) and $Q^2 = 4m^2_c$ (dot-dashed and dotted).
Right: Range of allowed transverse partonic distances obtained from the extracted mean values of $\sigmaeff$. [Figure adapted from~\cite{Rinaldi:2018slz}.]}
\label{fig:dPDFcorrel}
\end{figure}

The pocket formula (\ce{eq:pocketDPS}) provides the baseline purely geometric DPS cross section expected in the absence of any parton correlations. Obviously, longitudinal, transverse, and spin correlations, among others, are present at the parton level, and are expected to modify the $\sigmaeff$ value extracted from the ratio of squared SPS over DPS cross sections~\cite{Rinaldi:2013vpa,Rinaldi:2014ddl,Rinaldi:2016jvu}. The role of these effects in digluon distributions, fundamental in  gluon initiated processes such as those relevant for quarkonium production, has been covered in~\cite{Rinaldi:2018bsf} in the covariant relativistic Light-Front (LF) approach adopted to calculate the dPDFs. In this case, rotations between the canonical and the LF spin induce model-independent correlations that prevent a factorisation between the $(x_1,x_2)$--$b_\perp$ dependence (\cf{fig:dPDFcorrel}, Left), where $b_\perp$ is the transverse partonic distance. By properly considering general features of moments of dPDFs, the following relationship has been derived for $\sigmaeff$~\cite{Rinaldi:2018slz,Rinaldi:2018bsf}:
\begin{align}
 \dfrac{\sigmaeff}{3 \pi} \leq \langle b_\perp^2 \rangle \leq 
\dfrac{\sigmaeff}{ \pi}~.
\end{align}
The right panel of Fig.~\ref{fig:dPDFcorrel} shows how data on $\sigmaeff$ can thereby constrain the mean transverse distance between two partons. 

\begin{figure}[h!]
\centering
\includegraphics[width=0.30\textwidth,draft=false]{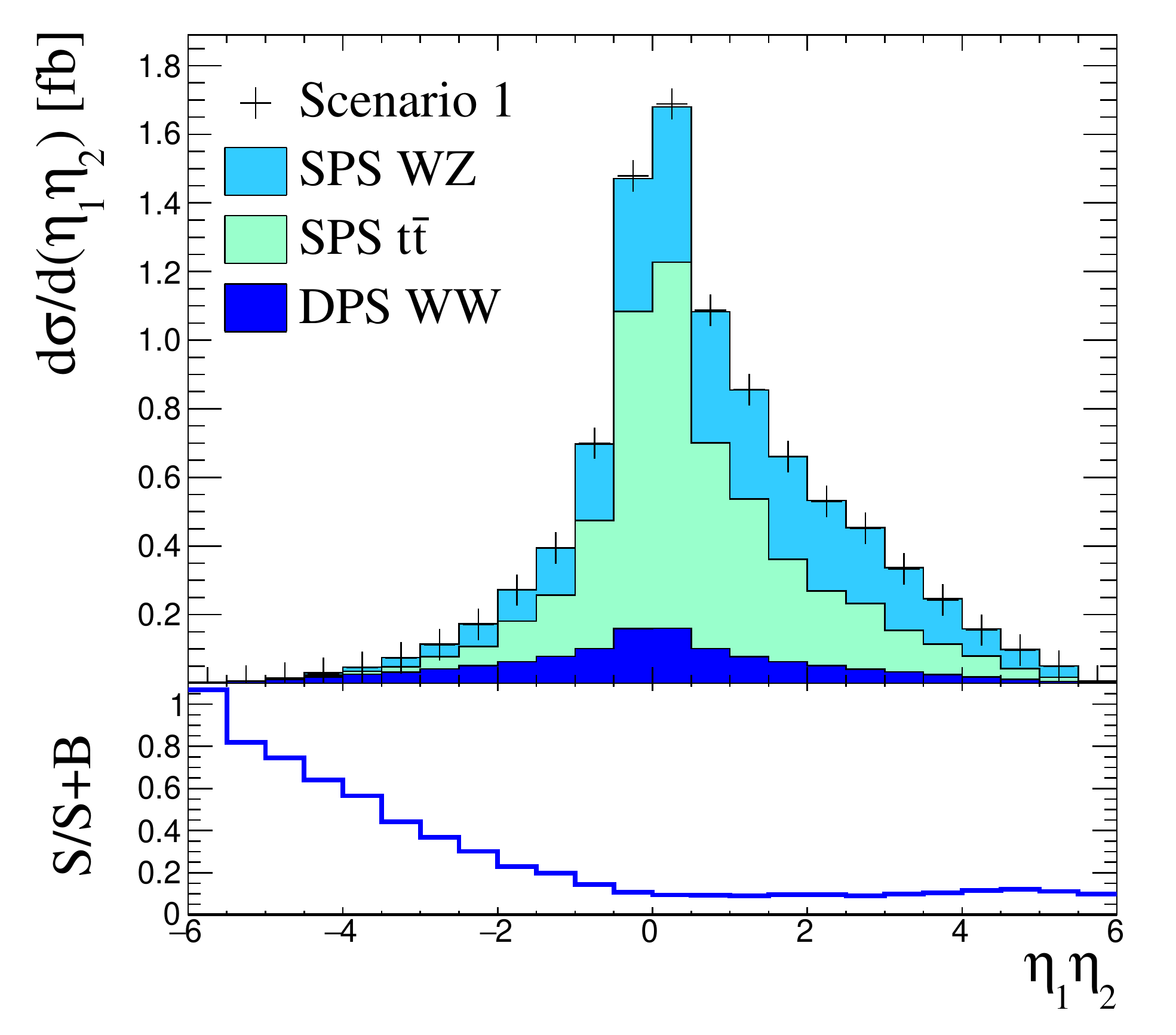}
\includegraphics[width=0.30\textwidth,draft=false]{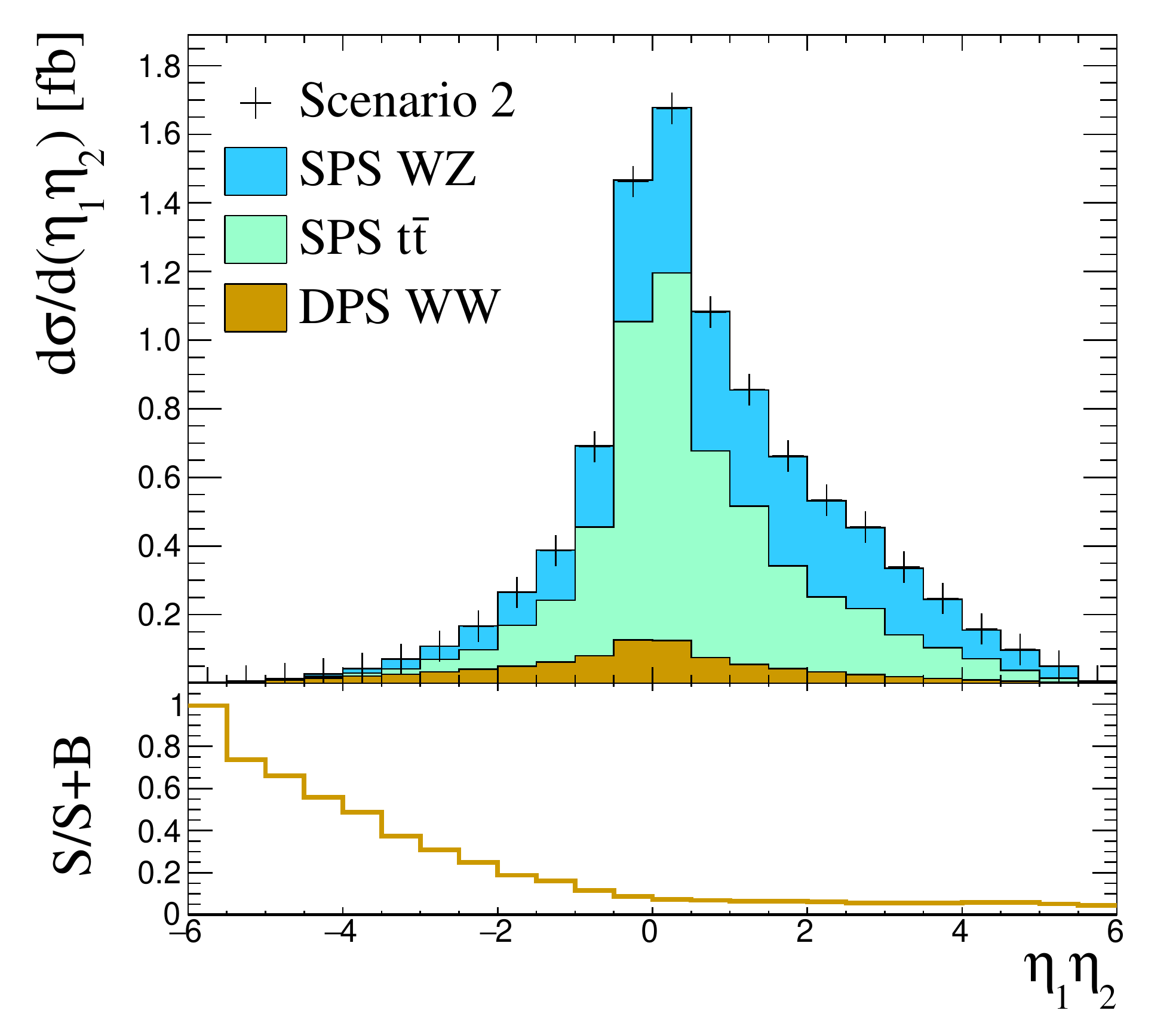}
\includegraphics[width=0.36\textwidth,draft=false]{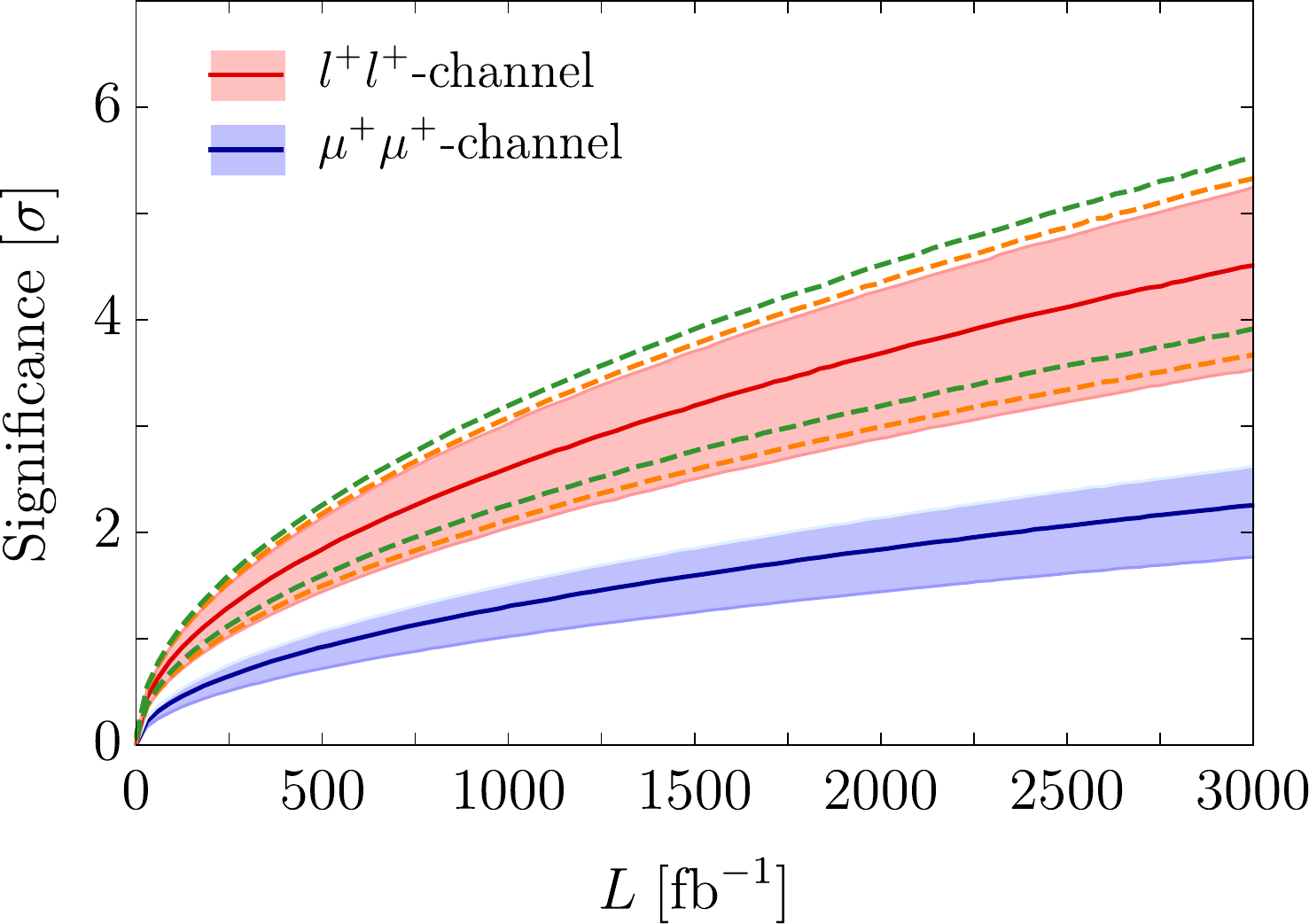}
\caption{Results of template fits in Scenario 1 (correlated DPS with uncorrelated extraction, Left) and Scenario 2 (uncorrelated DPS with correlated extraction, centre) for the product of pseudorapidity densities of same-sign leptons from the decay of DPS $W+W$ production. Right: Estimate of the significance (in standard deviations) of an assumed asymmetry of 0.11  for a signal cross section of 0.29~fb. Blue line/band corresponds to $\mu^+\mu^+$ only, while the red line/band includes all positively charged combinations of $e^+$ and $\mu^+$. Dashed curves show the sensitivity of the central red curve to changes in the asymmetry of $\pm$20\% (orange dashed curves) and the magnitude of the DPS cross section by a factor of $3/2$ or $3/4$ (green dashed curves). [Plots are taken from~\cite{Cotogno:2020iio}].
\label{fig:spin_in_dps}}
\end{figure}

Since correlations between the spin of partons have direct consequences on the angular distribution of the particles produced in the final state, it has been proposed to study various asymmetries in DPS processes to extract information on correlated quantum properties of two partons inside a proton. A calculation of double same-sign $W$-boson production cross sections has recently demonstrated that spin correlations can have large effects on the distribution of particles, and that the HL-LHC phase (if not before) opens up the possibility to measure them~\cite{Cotogno:2020iio,Cotogno:2018mfv}. A promising variable for spin correlation measurements is the asymmetry between the DPS cross section for the case when the leptons from the $W$-boson decay go towards the same or opposite hemispheres. The rightmost panel of Fig.~\ref{fig:spin_in_dps} shows the estimated significance of a possible observation of such an asymmetry as a function of the integrated luminosity collected. Even a {\it null measurement}, \ie a precisely measured zero asymmetry, would be interesting, as it would severely constrain the spin correlations inside the proton. 

\cf{fig:spin_in_dps} (Left and Centre) shows two template fits to a combination of DPS signal and backgrounds. In  Scenario 1 (Fig.~\ref{fig:spin_in_dps}, Left) partons are assumed to be correlated, but the extraction assumes uncorrelated DPS. In Scenario 2 (Fig.~\ref{fig:spin_in_dps}, centre) the roles are reversed.  %
Assuming an underlying effective cross section of $\sigmaeff = 15$~mb, the values for the fiducial cross sections and associated $\sigmaeff$ derived after analyses of the angular distributions in the two scenarios are: $\sigmadps = 0.59$~fb, $\sigmaeff = 12.2$~mb, and 
$\sigmadps = 0.44$~fb, $\sigmaeff = 16.4$~mb, respectively. As one can see, the 30\% span of the DPS production cross section and the corresponding variation of $\sigmaeff$, found in this simple treatment, illustrate the danger of neglecting correlations in DPS measurements in general, and of using correlation-sensitive variables in template fits in particular.  %

Quantum number correlations will also be present in DPS involving one or more quarkonium states. The size of these effects are largely unknown, and the lessons learned from double-$W$ production can be directly applied to DPS quarkonium production. The smaller momentum fractions probed tend to decrease the relevance of correlations, but the corresponding lower energy scale tends to increase their effects. If difficulties in isolating the DPS contribution in quarkonium production can be overcome, the large DPS cross sections provide unique possibilities to study interparton correlations.

\subsection{DPS studies with $\Q$}
\label{sec:DPSonia}

\subsubsection{Current status}

Thanks to their large production yields in hadronic collisions, multiple measurements exist now of the cross sections for the production of two quarkonium states, or a quarkonium plus another high-$\pT$ or heavy particle, in proton-(anti)proton collisions at the LHC and Tevatron. 
The corresponding SPS studies, as tools to understand the quarkonium production mechanism itself, are discussed in Sec.~\ref{sec:oniumassociate}. The measurements can be generally categorised as diquarkonium processes: $\jpsi+\jpsi$~\cite{Aaij:2011yc,Abazov:2014qba,Khachatryan:2014iia,Aaboud:2016fzt,Aaij:2016bqq}, $\jpsi+\Upsilon$~\cite{Abazov:2015fbl}, and $\Upsilon+\Upsilon$~\cite{Khachatryan:2016ydm,Sirunyan:2020txn}, quarkonium in association with a vector boson: $\jpsi+W^{\pm}$~\cite{Aad:2014rua,Aaboud:2019wfr} and $\jpsi+Z$~\cite{Aad:2014kba}, or with an open heavy-flavour hadron: $\jpsi$+open-charm hadron~\cite{Aaij:2012dz}, $\Upsilon$+open-charm hadron~\cite{Aaij:2015wpa}. All these processes have recently been reviewed in~\cite{Lansberg:2019adr}.

The standard DPS measurements proceed as follows. Since DPS are by nature more kinematically uncorrelated than single scattering processes, the DPS contribution preferentially populates the regions with larger azimuthal and rapidity separations between the two produced objects, compared to the SPS production mechanisms. The $y$ and $\phi$ differential cross sections measured in data are compared to the expectations of SPS models, and any excess with respect to the SPS predictions is attributed to DPS contributions. 
An example of a differential production cross section in bins of rapidity difference (of two $\jpsi$ mesons), is shown in~\cf{fig:DPSexp}\,(Left). This plot also illustrates a typical misconception of theoretical uncertainties in which the shapes of the extremal curves of an uncertainty band are assumed to give the shape of all the possible theory curves. Theory uncertainty bands rather indicate where one expects to find curves computed at a higher precision; these may not follow the same shape. In the present case, such an approximation artificially underestimates the SPS uncertainty and overestimates the
discriminating power of the $| \Delta y|$ spectrum between the DPS and SPS contributions. 

\begin{figure}[h!]
\centering
\includegraphics[width=0.475\textwidth]{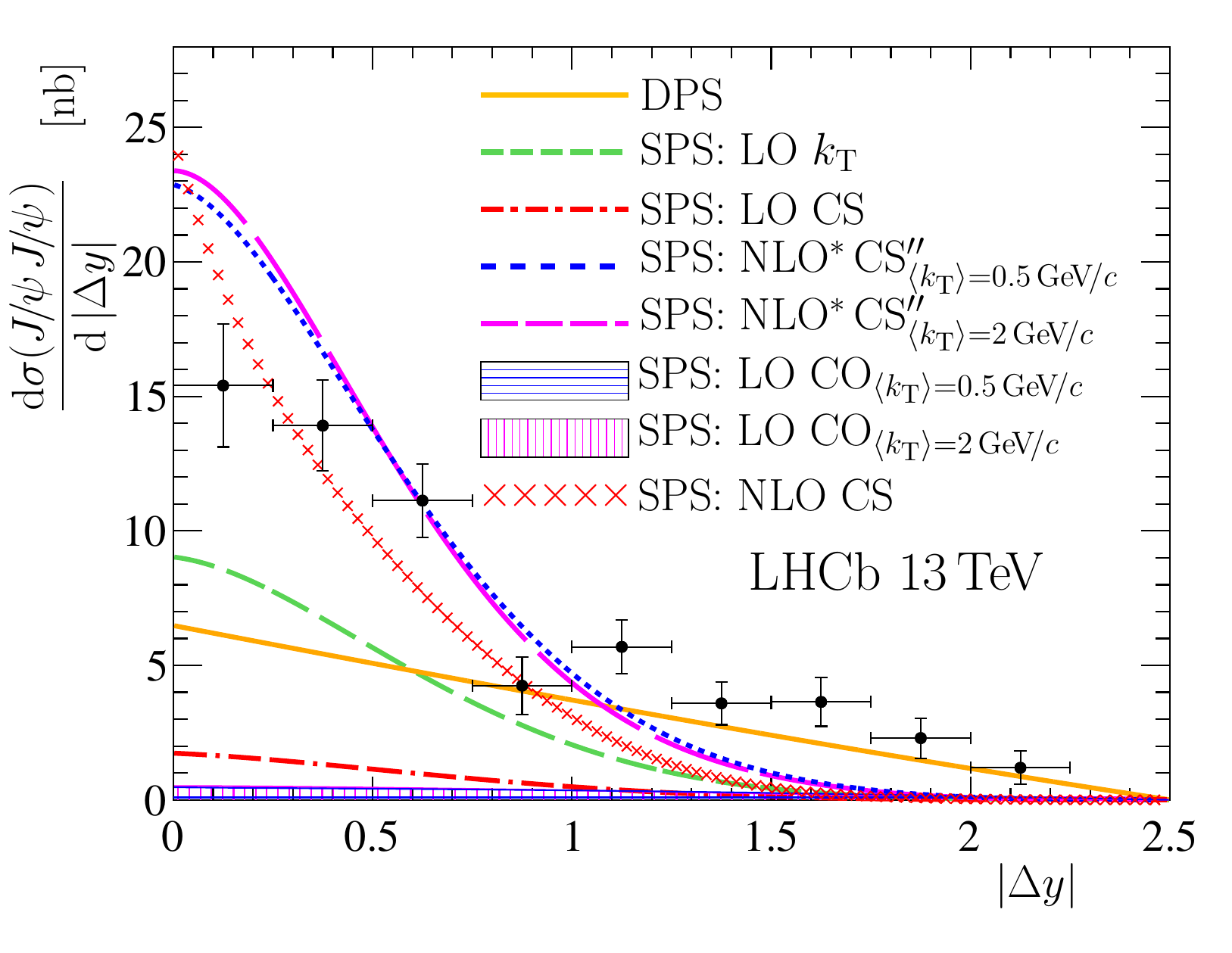}
\includegraphics[width=0.515\textwidth]{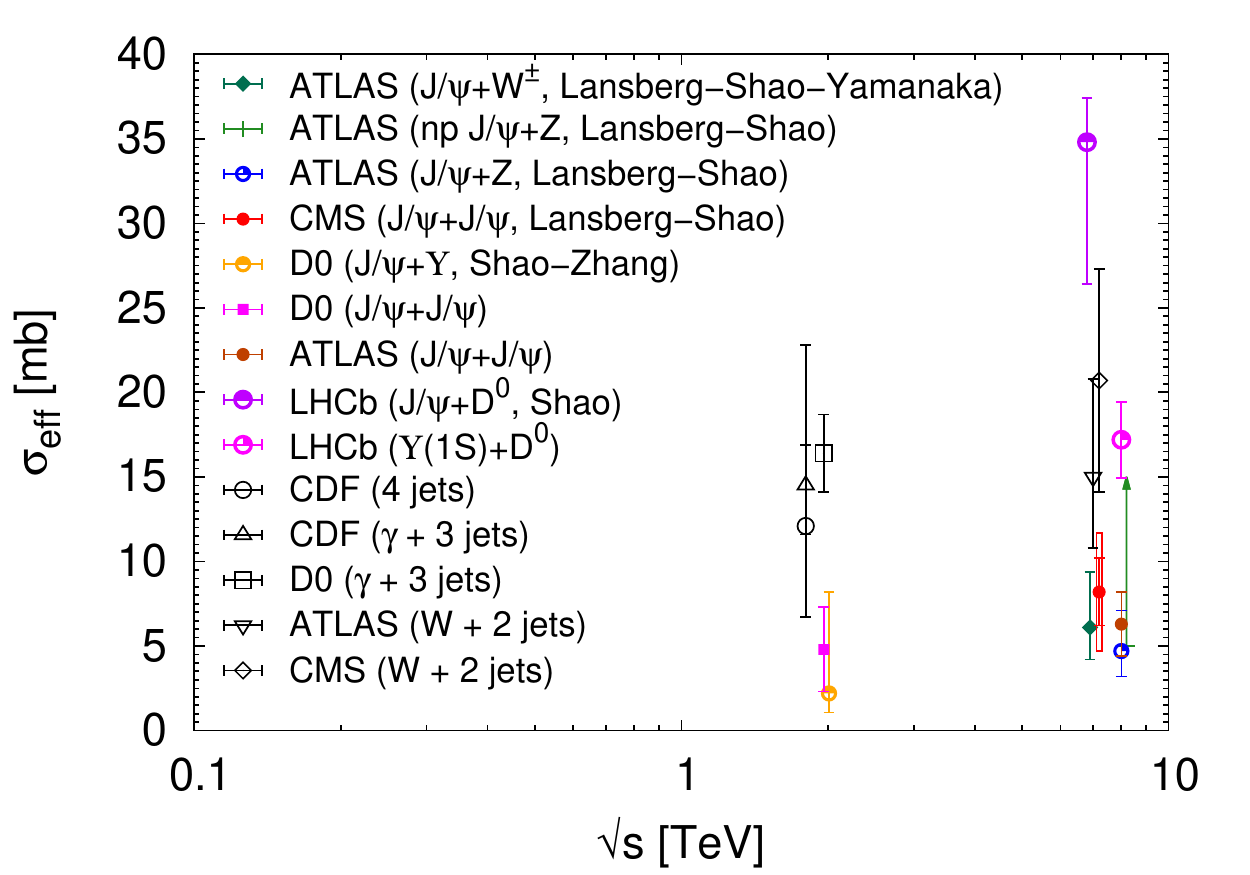}
\caption{Left: Differential $\jpsi$-pair production cross section as a function of rapidity difference~$\left| \Delta y \right|$ of two $\jpsi$ mesons measured by LHCb in \pp collisions at $\sqrts=13$ TeV, compared to SPS (various models, without uncertainties) and DPS predictions. 
Right: Comparison of $\sigmaeff$ values extracted with different processes in \pp and \ppbar\ collisions. [Left plot is from~\cite{Aaij:2016bqq}, and right plot is from~\cite{Shao:2020lqk}.]
\label{fig:DPSexp}}
\end{figure}

 From the derived value of $\sigmadps$, one can then usually derive the associated effective cross section taking its ratio to the product of corresponding SPS cross sections as per \ce{eq:pocketDPS}, $\sigmaeff \propto (\sigma^{H_1}_{\rm SPS}\sigma^{H_2}_{\rm SPS})/\sigma^{H_1+H_2}_{\rm DPS}$. Smaller values of $\sigmaeff$ correspond to larger DPS cross sections. \cf{fig:DPSexp}\,(Right) summarises the current status of $\sigmaeff$ extractions, based on the DPS pocket formula, with different final states~\cite{Abe:1993rv,Abe:1997xk,Abazov:2009gc,Aad:2013bjm,Chatrchyan:2013xxa,Abazov:2014qba,Lansberg:2014swa,Aaboud:2016fzt,Aaij:2012dz,Aaij:2015wpa,Shao:2016wor,Lansberg:2016rcx,Lansberg:2016muq,Lansberg:2017chq,Shao:2020kgj}. Values of $\sigmaeff\approx 2$--30~mb have been derived, though with large errors, with a simple (unweighted) average giving $\sigmaeff \approx 15$~mb. %
 This summary plot indicates that $\sigmaeff$ is smaller when derived from measurements of quarkonium processes in the ATLAS and CMS experiments, which typically cover central rapidities and require $\jpsi$ mesons with relatively large $\pT$ in order to ensure the decayed muons can reach the muon chambers. On the contrary, the other (forward) quarkonium-based extractions lead to larger $\sigmaeff$ values, indicating smaller DPS contributions. 
 
 Such differences can be interpreted as indicative of the non-universality of $\sigmaeff$ when measured in different kinematic ranges (LHCb vs ATLAS/CMS), (\eg\ due to the different relative weight of gluon vs.\ quark initial states), of non-universal parton correlations, and/or attributed to poorly controlled subtractions of SPS contributions. One typical example for the last point is the process of production of a $\jpsi$ meson in association with a $D^0$ meson discussed in~\cite{Shao:2020kgj}. This more recent and more refined analysis with improved SPS calculations yields a factor of two larger $\sigmaeff$ value than the one presented in the original LHCb paper~\cite{Aaij:2012dz}, %
 where the SPS contribution was assumed to be negligible.
 Similar caution is necessary when using associated production of \ups plus open-flavour to extract DPS cross sections, as done by LHCb Collaboration with the $\Upsilon+D$ final state~\cite{Aaij:2015wpa}. As shown in~\cite{Karpishkov:2019vyt}, taking into account NRQCD CO processes, feed-down decays, and $g\to D$ fragmentation contributions, %
one can obtain SPS cross section values of the same order as observed in the data. Moreover, the initial-state radiation effects, considered in~\cite{Karpishkov:2019vyt} within the HE factorisation, lead to kinematic distributions very similar to those observed in the experiment (with the notable exception of the $\Delta\phi$ distribution, which has a hard-to-explain enhancement towards $\Delta\phi\to 0$ in the data). Therefore, the conclusion that $\Upsilon+D$ production is dominated by DPS is significantly weakened.

\begin{figure}[h!]
\centering
\includegraphics[width=0.475\textwidth]{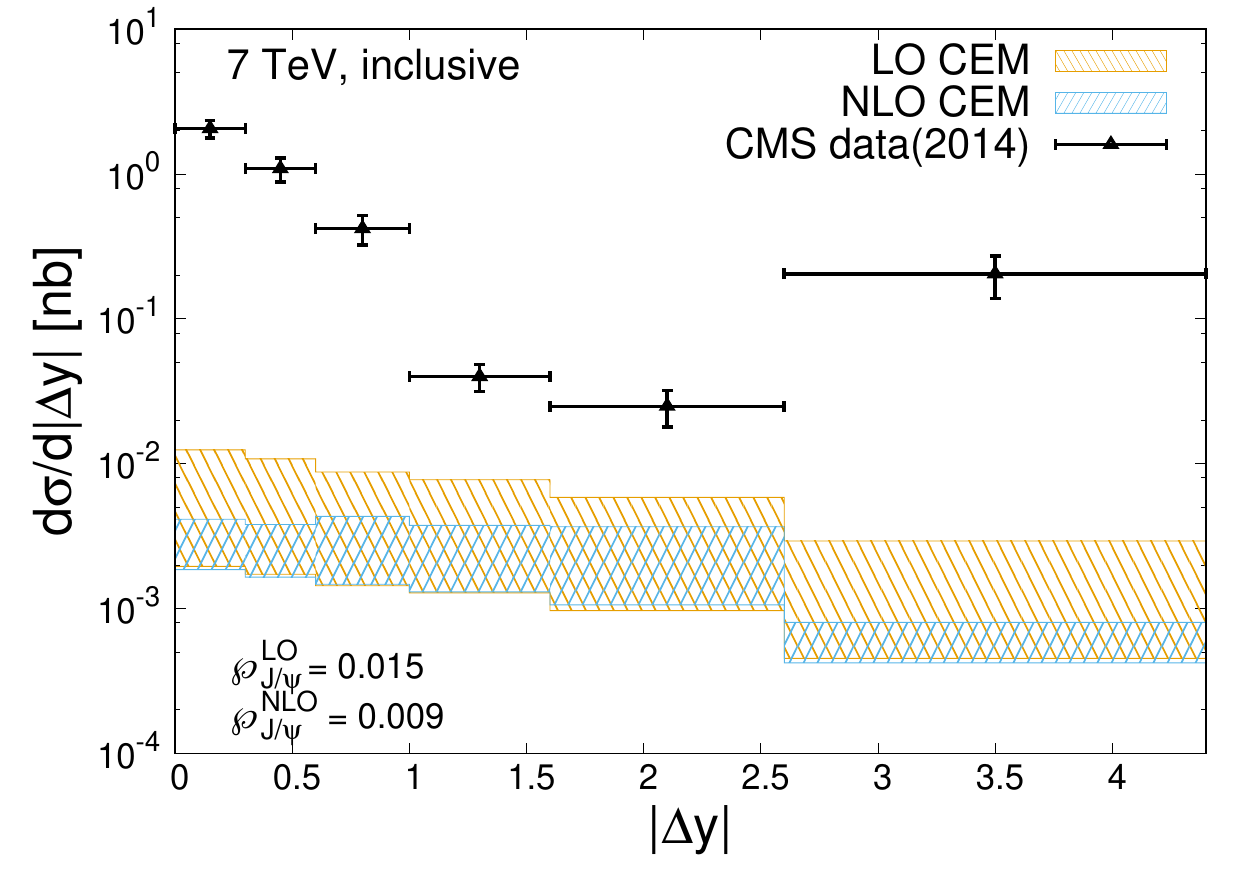}
\includegraphics[width=0.51\textwidth]{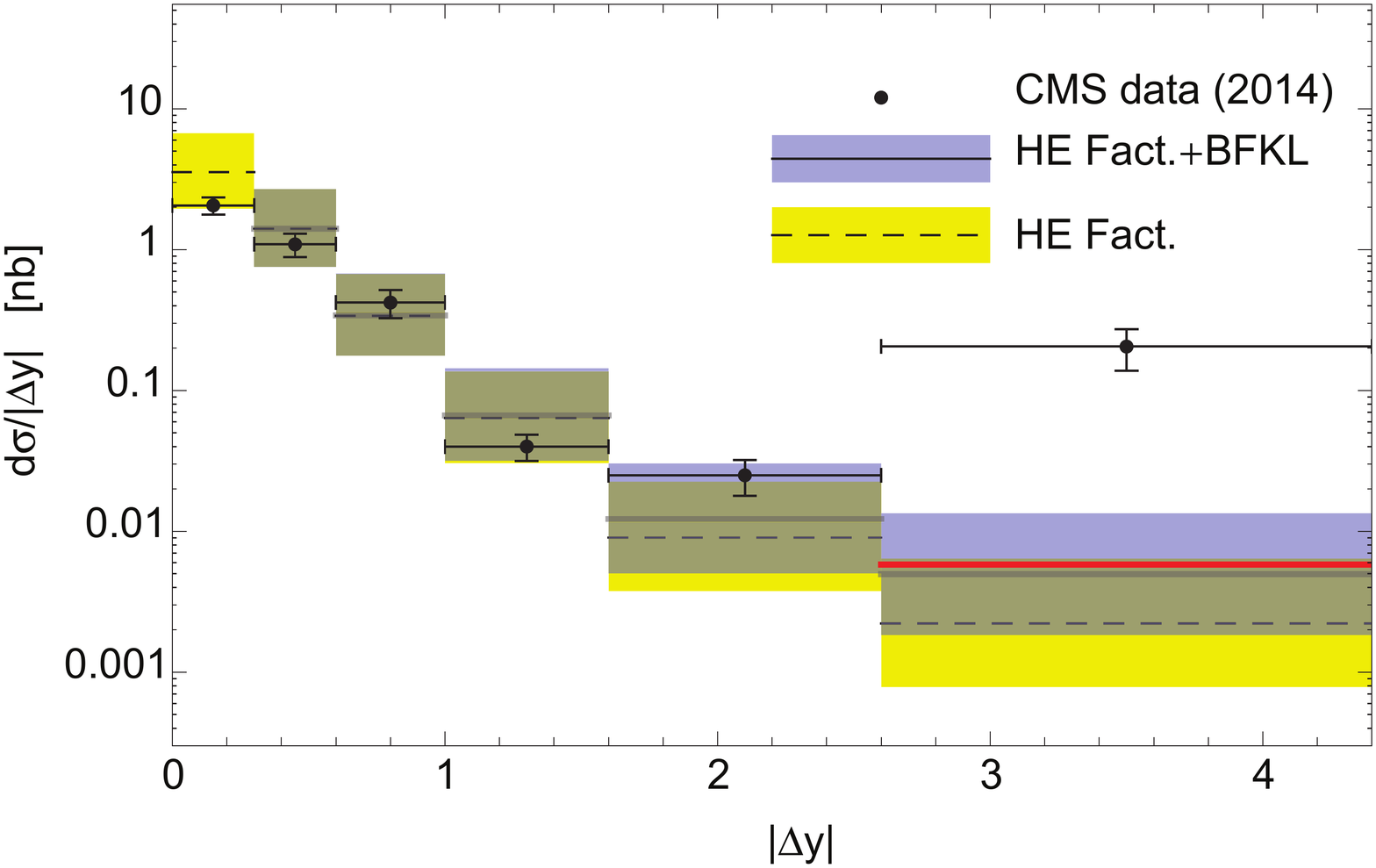}
  \caption{Differential $\jpsi$-pair production cross section as a function of the rapidity difference between the two $\jpsi$ mesons, $\left| \Delta y \right|$, measured by the CMS experiment in \pp collisions at $\sqrts=7$ TeV~\cite{Khachatryan:2014iia}, compared to the predictions of the LO and NLO CEM (Left) and of the HE factorisation (Right). [Left plot is from~\cite{Lansberg:2020rft}, and right plot adapted from~\cite{He:2019qqr}].}
  \label{fig:diJpsi_DY_CMS} 
\end{figure}

The importance of appropriately controlling the SPS production mechanism before attempting to extract any DPS cross section is further illustrated in Fig.~\ref{fig:diJpsi_DY_CMS} for double-\jpsi\ production. In NRQCD factorisation, the total cross section of (direct) $\jpsi$ pair production is dominated by double CS ${}^3S_1^{[1]}$ contribution as confirmed both in the collinear factorisation up to NLO accuracy~\cite{Lansberg:2013qka,Sun:2014gca,Lansberg:2014swa,Lansberg:2019fgm} and in HE factorisation~\cite{He:2019qqr} discussed in~Section~\ref{sec:HEfactorisation}. On the other hand, in the CEM with collinear factorisation at LO and NLO accuracy, such a contribution is absent, thereby leading to an underestimation of the SPS cross sections in the whole $\Delta y$ range~\cite{Lansberg:2020rft} (Fig.~\ref{fig:diJpsi_DY_CMS}, Left). If the CEM prediction was to be trusted, then practically the whole double-$\jpsi$ production cross section should be attributed to DPS, which would lead to a correspondingly reduced value of $\sigmaeff$ and would require unrealistically strong partonic correlations to describe the $\jpsi$ momenta distributions observed in data. 

In addition to the effects related to unknowns arising in the SPS cross section, further caution must be exercised with the assumptions on the DPS cross section itself. In particular, as discussed above, the shape of the DPS signal is unknown in the presence of parton correlations and can have an impact on the extractions of the DPS cross section.

In the next section, we discuss how upcoming measurements, in particular with the new opportunities opened up at the HL-LHC, can help to clarify the aforementioned experimental and theoretical issues and thereby improve our understanding of DPS processes with quarkonium-based studies.

\subsubsection{HL-LHC prospects}

A first step to exploit quarkonium DPS measurements in order to extract quantitative information on the hadronic wave functions (in particular, on the various underlying sources of partonic correlations) and their energy evolution, is to understand the wide span of $\sigmaeff$ extractions shown in Fig.~\ref{fig:DPSexp} (Right). As mentioned above, leading sources of confusion are the different techniques used in each measurement to determine and remove the contamination from SPS contributions in the DPS signal region.
{Key to the ability to impose more stringent cuts and to better probe different corners of phase space are, firstly 
the very large data samples, and secondly, the upgraded charged-particle tracking over a wide pseudorapidity, $|\eta| \lesssim 5$, (with muon acceptance extended by half a unit, up to $|\eta| \lesssim 3$) in the ATLAS and CMS detectors during the HL-LHC phase}. Both advantages will allow a better study of the azimuthal and rapidity separations 
between quarkonium states simultaneously produced in SPS and DPS processes. The following concrete experimental proposals are suggested for DPS studies at HL-LHC:
\begin{itemize}
\item %
In order to better extract the DPS signal from the data, rather than simple standard cut-based analyses used so far, more advanced multi-variate analyses of the relevant quarkonium pair kinematic variables ($y_{ij}$, $\phi_{ij}$, ${\pT}_{ij}$,...) should be carried out. SPS predictions with the highest order of accuracy (ideally, at least, NLO plus resummation and/or parton showering) should be used only, and effects of variations in the shape of the DPS cross section should be explicitly investigated. The theoretical uncertainty associated with the SPS cross sections should be properly propagated into any experimentally extracted DPS cross section and $\sigmaeff$ value.

\item Final states with $\psip$ mesons, free of feed-down contributions in constrast to the $\jpsi$ mesons commonly studied so far, should be considered. 
High-statistics measurements of the $\Delta\phi$-differential distribution of $Q\bar Q$-pair production at $|\Delta y|\gtrsim2.5$, where all SPS models tend to fail, should be performed.
The $\jpsi + \psip$ and $\jpsi + \chi_{c}$ final states are of particular interest as they can provide new ways to differentiate the SPS and DPS contributions
since their feed-down fractions to $\jpsi$ pairs are significantly different when they are produced by SPS and DPS~\cite{Lansberg:2014swa,Lansberg:2019adr}.
\item The production of charmonium (\eg $\psip$ as an essentially feed-down-free state) plus a $B$-meson (or a non-prompt \jpsi) is an interesting candidate for future DPS studies, since the leading $v$ contribution from the CSM is suppressed at LO~\cite{Lansberg:2019adr} and the $B$ fragmentation function is better controlled than that of the $D$. 
\item Beyond the first study of the associated production of a $\jpsi$ with a charmed meson carried out by LHCb~\cite{Aaij:2012dz}, it will be instructive to perform precise comparisons of the $\pT$ spectra associated with different charm hadrons and those produced alone. 
With more precise measurements, it will be possible to confirm the hint of a slight difference in the \pT\ spectrum of the $\jpsi$ produced alone or with a charmed hadron. Such a confirmation would go against the DPS dominance, thus along the lines of~\cite{Shao:2020kgj}. Obviously, this could be complemented by the extraction of the DPS yields using data from control regions and DPS MC simulations.
\item The unique feature of the ALICE detector for quarkonium studies is a forward muon system, covering $2.5<\eta<4$, combined with a central-barrel tracking/PID system ($|\eta|<0.9$) to achieve pseudorapidity differences $|\Delta\eta|$ up to 4.9, exceeding the capabilities of ATLAS and CMS. %
With an expected \pp-collision data set corresponding to $\sim0.2$ fb$^{-1}$, measurements of $D$ or $B$ mesons in the central region, associated with a quarkonium in the forward region, become possible with very low limits on $\pT$ for both objects. Compared to ATLAS and CMS, the relatively low integrated luminosity of ALICE  will be compensated by the much smaller pileup probability (and lower $\pT$). %
\item During Run-2, the LHCb experiment collected around 6~fb$^{-1}$ of \pp collisions at $\sqrts=13$ TeV which, for double-$\jpsi$ production, translates into data samples 20 times larger than in previous DPS measurements. This data sample remains to be analysed and will make it possible to study doubly differential production cross sections, \eg\ in two-dimensional bins of $\jpsi$-pair transverse momentum and invariant mass, especially probing the momentum distribution of linearly polarised gluons inside unpolarised protons~\cite{Lansberg:2017dzg,Scarpa:2019fol} (Section \ref{sec:psi-psi-TMDs}). Using the data sets expected at the HL-LHC, one can carry out a similar programme of measurements for rarer processes, such as those, for example, involving $\psip$, $\Upsilon$ or even $\eta_c$~\cite{Lansberg:2013qka}.
\item In all the above cases, quarkonium polarisation measurements can be instrumental in disentangling DPS from SPS. If the former are dominant, the quarkonium polarisation should be identical in both single and associated production.
\item \jpsi-pair production could also be studied at the FT-LHC at $\sqrts=115$~GeV with large enough yields to look for possible DPS contributions~\cite{Lansberg:2015lva}. This would provide a possibly unique measurement of $\sigmaeff$ in this energy range.
\end{itemize}

On the theoretical side, the following developments, among others, are needed to fully exploit the experimental data made available:
\begin{itemize}
\item The theoretical SPS cross sections (``subtracted'' from the experimental data in order to identify the DPS contributions) need to include the largest number of perturbative corrections possible, both for FO and resummed logarithms terms. Additionally, and particularly relevant for quarkonium production, efforts to significantly reduce the model dependence will be crucial to isolate the DPS cross section. Predictions with limited theoretical accuracy should be avoided as they {consequently degrade the DPS cross-section extraction}. %
\item Progress towards full-NLO corrections for the DPS cross sections for double-quarkonium production, including pQCD-induced partonic correlations, computed via \ce{eq:dps_xsec}, must be made.    %
\item {Studies should be undertaken of} the impact of perturbative and non-perturbative effects on gluon-gluon double parton distribution functions calculated within phenomenological models, such as constituent quark models.
\item A consistent treatment of heavy-quark-mass thresholds and {the} evaluation of their numerical effect in DPS cross sections are required.
\item Cross-section calculations should be performed for double quarkonium production including explicitly the effects of parton correlations of (i) kinematic (momentum fractions, transverse separation), (ii) quantum (flavour, spin, colour, fermion number), and (iii) mixed (involving interplay between the two) origins.
\item Explicit studies of the $x$-dependence of the effective cross section $\sigmaeff$, and identification of experimental observables sensitive to such an evolution, are required.
\end{itemize}

\subsection{TPS studies with $\Q$ in \pp collisions}

As discussed in the previous Section, the wide span of $\sigmaeff$ extractions based on the DPS pocket formula for double-quarkonium measurements (Fig.~\ref{fig:DPSexp}, right) calls for alternative studies that can shed light on the origin of the ranges of derived values. In~\cite{dEnterria:2016ids}, it was pointed out for the first time that the study of triple parton scatterings (TPS) can further help to independently improve our understanding of the transverse proton profile and estimate the impact of parton correlations. The pocket formula for triple parton scattering reads, based on \ce{eq:pocketNPS}, 
\begin{equation} 
\sigma^{hh' \to H_1+H_2+H_3}_{\rm TPS} =  \left(\frac{\mathpzc{m}}{3!}\right)\, \frac{\sigma^{hh' \to H_1}_{\rm SPS} \cdot
\sigma^{hh' \to H_2}_{\rm SPS} \cdot \sigma^{hh' \to H_3}_{\rm SPS}}{\sigmaefftps^2}, \;\;\;\mbox{with}\;\;\;
\sigmaefftps^2=\left[ \int d^2b \,T^3(b)\right]^{-1}\,.
\label{eq:sigmaTPS}
\end{equation}
In this purely geometric approach, it was demonstrated that for a wide range of proton transverse profiles (encoded in the cube of the overlap function $T^3(b)$), the effective triple and double effective cross sections are actually proportional and very similar numerically~\cite{dEnterria:2016ids}:
\begin{equation}
\sigmaefftps = k\times\sigmaeff, \; {\rm with}\;\; k = 0.82\pm 0.11\,.%
\label{eq:TPS_DPS_factor}
\end{equation}
Therefore, from the $\sigmaefftps$ values extracted from the data, one can derive independent values of $\sigmaeff$. However, since TPS cross sections depend on the cube of the corresponding SPS cross sections, a triple hard process $\pp\to H_i+H_i+H_i$, with SPS cross sections 
$\sigma_{\rm SPS}^{\pp\to H_i}\approx \rm 1\;\mu b$, %
has a very small TPS cross section %
$\sigma_{\rm TPS}^{\pp \to H_i+H_i+H_i}\approx 1$~fb,
and perturbative processes with large enough SPS cross sections are needed in order to have visible number of events. The very large yields of quarkonium expected at the HL-LHC allow one to carry out triple parton scattering (TPS) studies for the first time~\cite{dEnterria:2016ids}.

As TPS is a priori of subleading power with respect to single and double parton scattering, its theoretical investigation is challenging. As one goes to high scale $Q$, TPS contributions will rapidly diminish as $\Lambda^4_{\rm QCD}/Q^4$ compared to SPS.  On the other hand, if one goes to few-GeV-scale observables, usually the theoretical predictions are plagued with very large intrinsic theoretical uncertainties. In addition, extracting TPS contributions requires an accurate control, not only of the SPS but also, of the SPS+DPS contributions, as sources of the same final states. These three facts may reduce the eventual potential of TPS studies at the LHC. %
Production modes that have been studied in the literature are the triple-$\jpsi$~\cite{Shao:2019qob} and triple $D\bar{D}$-mesons~\cite{Maciula:2017meb} production, while other processes, like $\jpsi+$two same-sign open charm, and $\jpsi+\jpsi$ plus open charm production~\cite{Lansberg:2014swa}, are also worth pursuing. %
We will focus here on the triple-$\jpsi$ production process.

\begin{figure}[h!]
\centering
\subfloat[$P_{T,{\rm min}}$]{
\includegraphics[width=0.40\textwidth,draft=false]{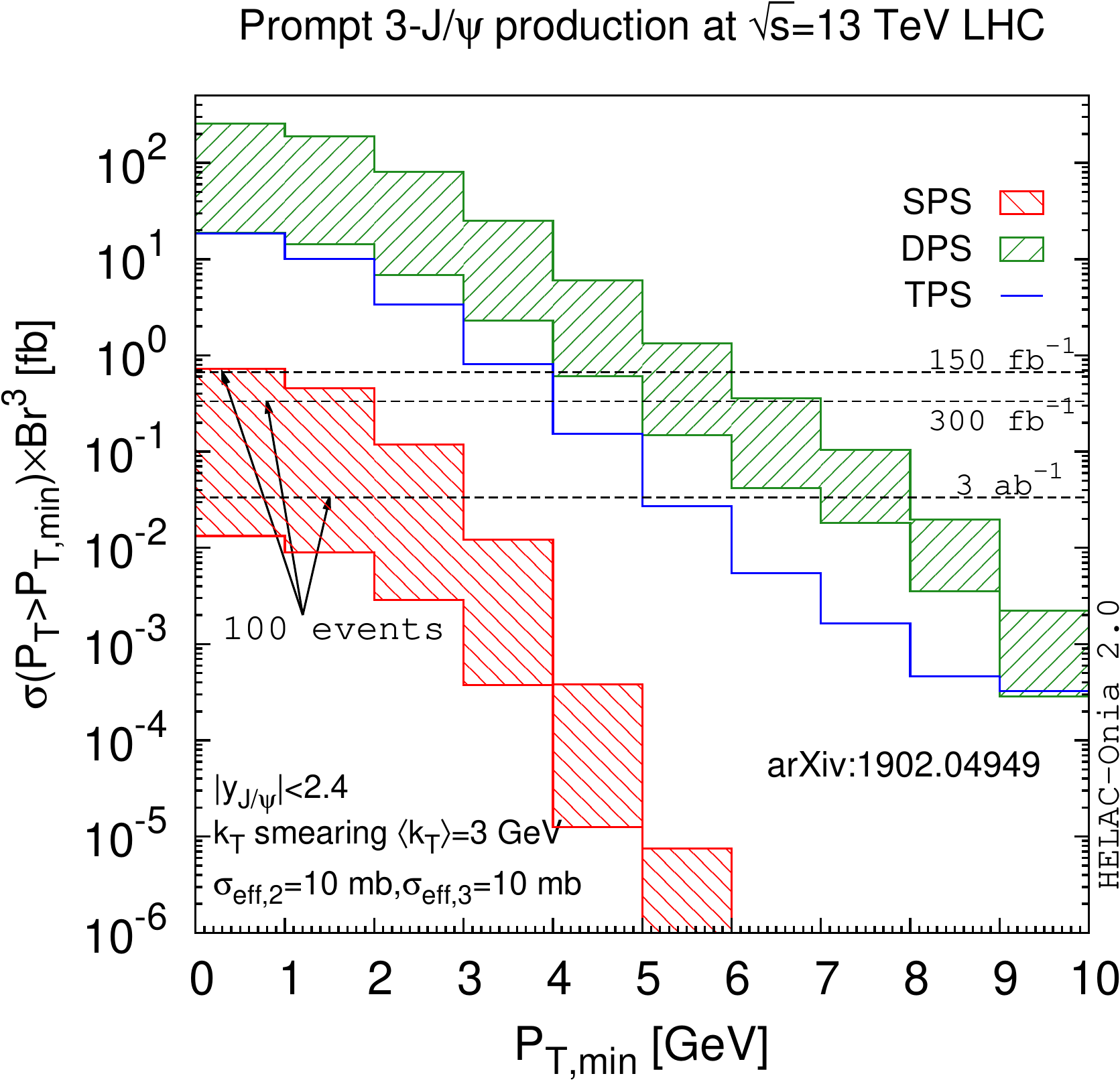}\label{fig:d3psiplota}}
\hspace{0.5cm}
\subfloat[$|\Delta y|_{{\rm min}}$]{\includegraphics[width=0.40\textwidth,draft=false]{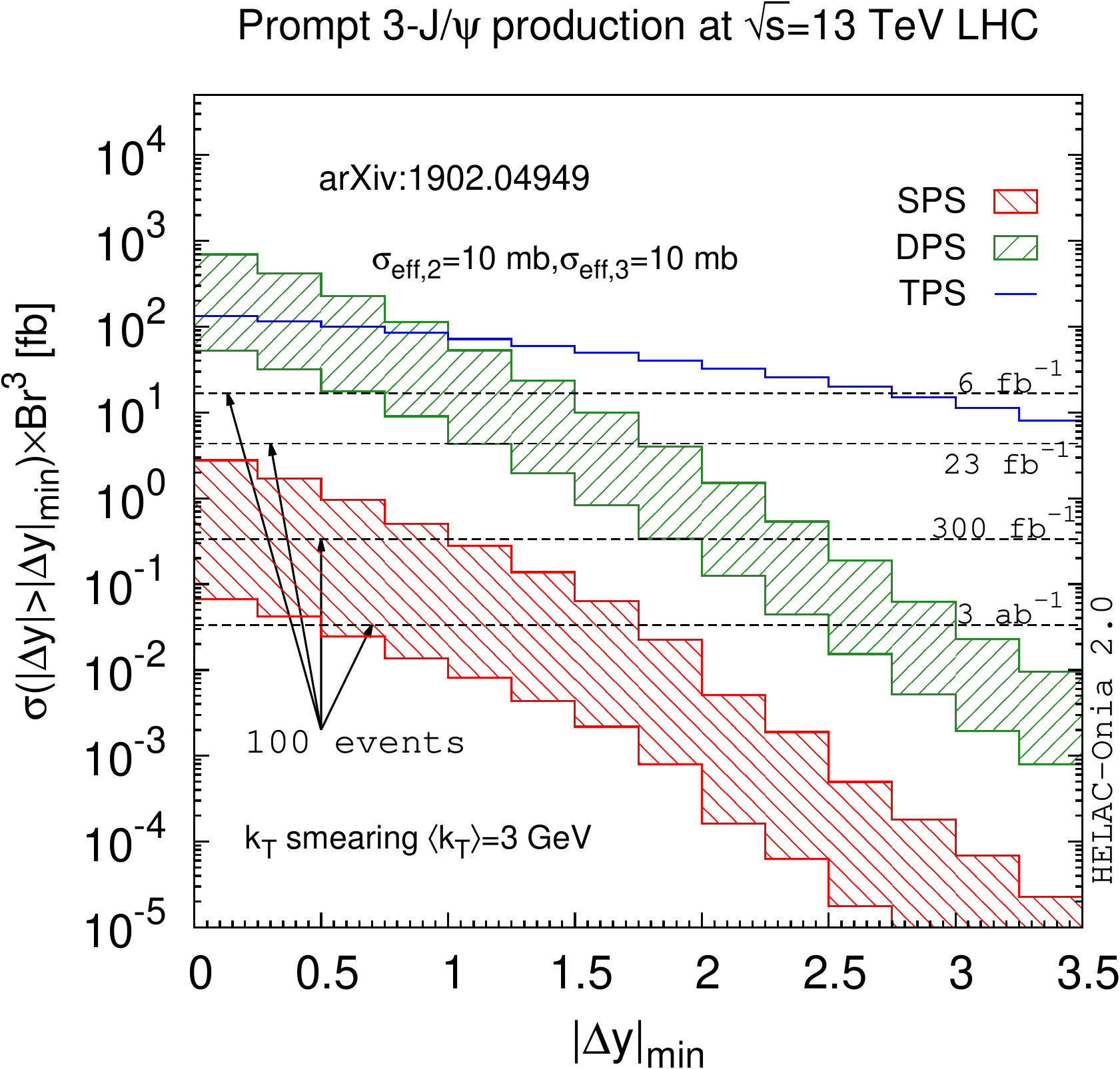}\label{fig:d3psiplotb}}
\caption{Cumulative cross section of the dependence of the triple-prompt $\jpsi$ production ($\sigma(pp\rightarrow 3\jpsi)\times{\rm BR}^3(\jpsi\rightarrow \mu^+\mu^-)$, in fb) on the minimal transverse momentum cut $\pT^{\jpsi}>P_{\rm T, min}$ (Left) and of the minimal rapidity gap cut $|\Delta y(\jpsi,\jpsi)|>|\Delta y|_{\rm min}$ (Right) among three $\jpsi$'s in \pp collisions at $\sqrts=13$ TeV.}
\end{figure}

A complete study of triple-prompt $\jpsi$ production in \pp collisions at 13 TeV has been carried out in~\cite{Shao:2019qob}, by computing SPS, DPS, and TPS contributions simultaneously for the first time based on the event generator {\sc\small HELAC-Onia}~\cite{Shao:2012iz,Shao:2015vga}. The study shows that the process receives a suppressed SPS contribution with respect to the DPS and TPS ones. Thus, it becomes a golden channel for the first-ever observation of TPS processes, and to provide new valuable insights into double-quarkonium production by comparing the value of $\sigmaeff$ obtained from the DPS contribution measured directly in the process to that derived from the TPS yields via \ce{eq:TPS_DPS_factor}. The cumulative cross section $\sigma(pp\rightarrow 3\jpsi)\times{\rm BR}^3(\jpsi\rightarrow \mu^+\mu^-)$ after imposing the $\pT^{\jpsi}>P_{T,{\rm min}}$ cut  and the rapidity gap cut $|\Delta y(\jpsi,\jpsi)|>|\Delta y|_{\rm min}$ on each $\jpsi$ pair can be found in Figs.~\ref{fig:d3psiplota} and~\ref{fig:d3psiplotb}, respectively. By assuming $100\%$ event-reconstruction efficiency, the horizontal lines in the two plots indicate the cross sections at which 100 events are collected for several integrated luminosities. In particular, with the nominal HL-LHC luminosity of 3~ab$^{-1}$, 100 events are anticipated with $\pT^{\jpsi}>7$ GeV. Moreover, Fig.~\ref{fig:d3psiplotb} shows that the minimal rapidity gap cut between $\jpsi$ pairs can be used to improve the purity of the TPS signal. Such a study was carried out by assuming zero correlation between the partonic scatterings following \ce{eq:sigmaTPS} above. The measurement of this novel process with HL-LHC data should definitely clarify whether such a simple geometric hypothesis is justified.

\subsection{DPS and TPS studies with $\Q$ in \pA\ collisions}

\pA and \AaAa collisions also provide new handles on improving our understanding of DPS, and in general NPS, processes. DPS~\cite{Strikman:2001gz,Cattaruzza:2004qb,dEnterria:2012jam} and TPS~\cite{dEnterria:2016yhy} are significantly enhanced in \pA collisions compared to \pp\ collisions thanks to the (much) larger transverse parton density of nuclei compared to protons. As discussed in the \pp case, final states with quarkonia benefit from large production yields that have lead to the first measurements of DPS processes, and to future more detailed analyses, as discussed below.

In the case of DPS, the cross section receives contributions from interactions where the two partons of the nucleus belong to the same nucleon ($\sigmaDPSone$), and two different nucleons ($\sigmaDPStwo$). The pocket formula for the DPS cross section of particles $H_1,H_2$ in \pA collisions can be written as a function of the elementary proton-nucleon ($pN$) SPS cross sections to produce $H_1$ and $H_2$ separately as~\cite{dEnterria:2012jam}
\begin{eqnarray} 
\sigma_{\rm DPS}^{\pA\to H_1 + H_2} = \left(\frac{\mathpzc{m}}{2}\right) \frac{\sigma_{\rm SPS}^{\pN \to H_1} \cdot \sigma_{\rm SPS}^{\pN \to H_2}}{\sigmaeffdpspA}\,,
\label{eq:sigmapA_DPS}
\end{eqnarray}%
where the effective DPS \pA\ cross section in the denominator, $\sigmaeffdpspA$, depends on the standard $\sigmaeff$ parameter measured in \pp collisions, \ce{eq:pocketDPS}, and on a pure geometric quantity, $T_{\AaAa}(0)$, that is directly derivable from the well-known nuclear transverse profile via a Glauber model~\cite{dEnterria:2020dwq}. 
The overall expected DPS enhancement in \pA compared to \pp collisions is $\sigmaeffdps/\sigmaeffdpspA \approx [A+A^{4/3}/\pi]$ which in the case of \pPb\ amounts to a factor of $\sim$600 relative to \pp, \ie\ a factor of $[1+A^{1/3}/\pi]\approx$~3 higher than the naive expectation assuming the same $A$-scaling of the single parton cross sections~\cite{dEnterria:2012jam}. The relative weights of the two DPS contributions are $\sigmaDPSone:\sigmaDPStwo = 0.7 : 0.3$ (for small mass number $A$), and $0.33 : 0.66$ (for large $A$)~\cite{dEnterria:2012jam}.
One can thus exploit such large expected DPS signals over the SPS backgrounds in \pA\ collisions to study double parton scatterings in detail and, in particular, to extract the value of $\sigmaeffdps$ independently of measurements in \pp collisions.
In addition, recent studies that incorporate impact-parameter-dependent nPDF effects~\cite{Shao:2020acd}, have pointed out that the study of DPS processes in heavy-ion collisions provide useful information on the (unknown) spatial-dependence of nuclear parton densities. 

In the case of triple parton scatterings, a similar formula to \ce{eq:sigmapA_DPS} has been derived~\cite{dEnterria:2016ids}, that includes now three types of contributions from  interactions where the three partons of the nucleus belong to the same nucleon ($\sigmaDPSone$), two ($\sigmaDPStwo$) and three different nucleons ($\sigmaDPSthree$). For \pPb\ collisions, the
three TPS terms are $\sigmaTPSone:\sigmaTPStwo:\sigmaTPSthree = 1 : 4.54 : 3.56$, and their sum amounts to 9.1, namely the TPS cross sections are nine times larger than the naive expectation based on an {$A$ scaling} of the corresponding proton-nucleon TPS cross sections. Generic pocket formulas exist that allow the determination of the cross sections for any combination of three final-state particles, including quarkonium states in \pA\ collisions~\cite{dEnterria:2017yhd}. Using NNLO predictions for single heavy-quark production, the authors of~\cite{dEnterria:2016ids} have shown that three $D\bar{D}$-pairs are produced from separate parton interactions in about 10\% of the \pPb\ events at the LHC. The study of TPS in \pA\ scattering at the HL-LHC will provide novel experimental and theoretical handles to understand double and triple parton scatterings, constrain the parton transverse profile of the proton, and clarify the role of partonic correlations in the proton and ion wave functions.

\subsubsection{Current status}
The first-ever experimental study of DPS in \pA\ collision has been carried out by  LHCb, measuring the like-sign $D+D$ ($D^0+D^0$, $D^0+D^+$ and $D^++D^+$) and $\jpsi + D$ production in \pPb\ collisions at $\sqrtsnn = 5.02$~TeV~\cite{Aaij:2020smi}. 
The azimuthal angle between the two charm hadrons in a pair, $\Delta\phi$, is measured to be flat, independent of a cut on the charm transverse momentum for $DD$ pairs, while that of $D\bar{D}$ pairs tends to peak at $\Delta\phi\approx 0$ for higher charm $\pT$. The ratio of cross sections between $DD$ and $D\bar{D}$ pairs is shown in \cf{fig:DPS_LHCb} (Left), with a magnitude of about $0.3$, while the measurement in \pp collisions is about 0.1~\cite{Aaij:2012dz}. %
The forward-backward ratio ($R_\mathrm{FB}$) quantifying the production at positive rapidities over that at negative rapidities in the common range $2.7<|y(D)|<3.7$, is measured for $D\bar{D}$ pairs to be $R_\mathrm{FB}(D\bar{D})=0.61\,\pm0.04\,(\mathrm{stat})\,\pm0.12\,(\mathrm{syst})$, which is consistent with that of inclusive charm production $R_\mathrm{FB}(D)$~\cite{LHCb-CONF-2019-004}, but that of $DD$ pairs is $R_\mathrm{FB}(DD)=0.40\,\pm0.05\,(\mathrm{stat})\,\pm0.10\,(\mathrm{syst})\approx R_\mathrm{FB}^2(D\bar{D})$. Since the forward-backward ratio being lower than unity is explained by the modification of nuclear PDF, the value for like-sign $DD$ pairs is consistent with two pairs of partons participating in the hard scattering. These observations support a significant contribution of DPS in the like-sign $DD$ production while the opposite-sign $D\bar{D}$ production has a large component of SPS, namely the inclusive production of a single charm quark pair.

\begin{figure}[h!]
\centering
\includegraphics[width=0.45\textwidth]{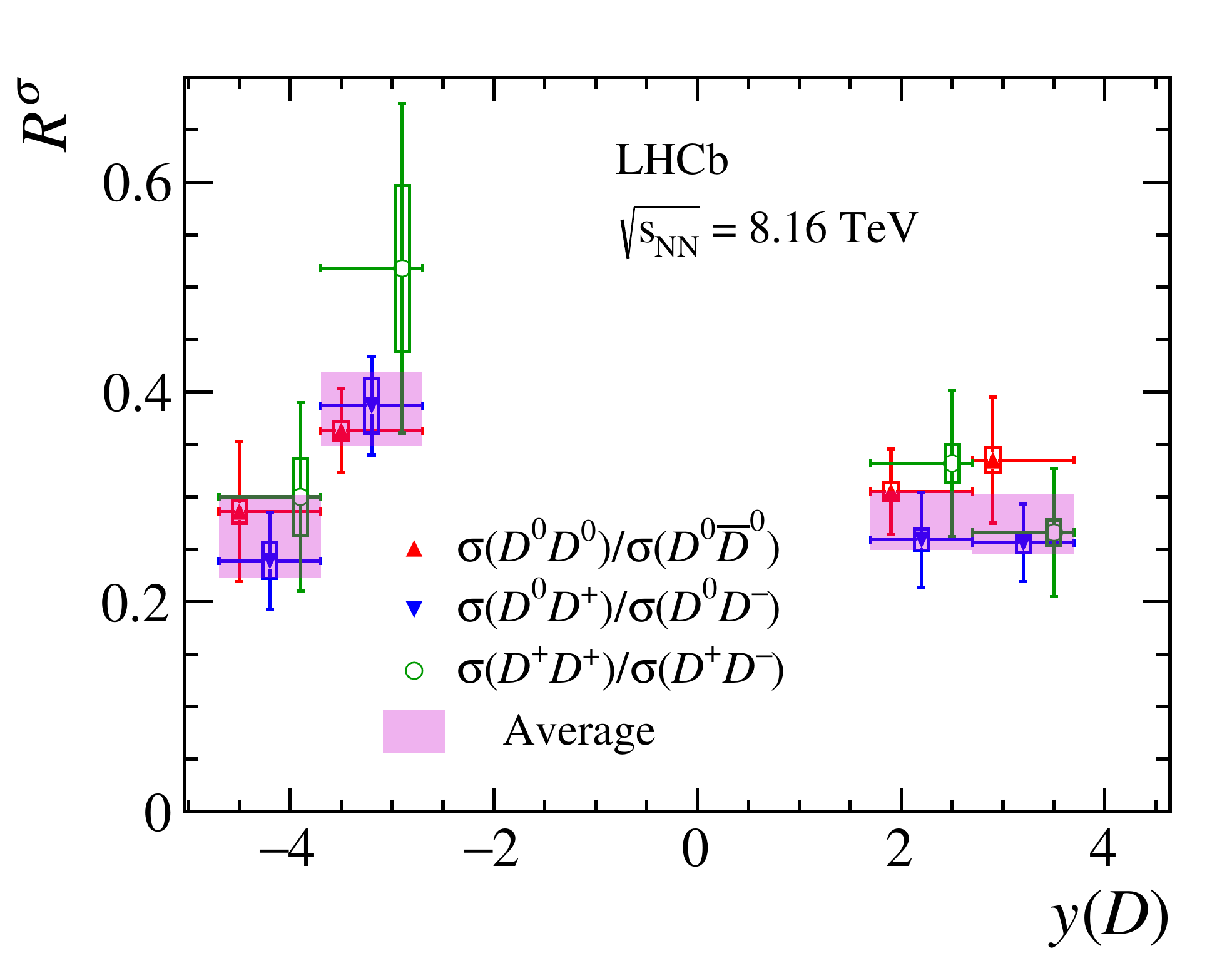}
\hspace{0.2cm}
\includegraphics[width=0.45\textwidth]{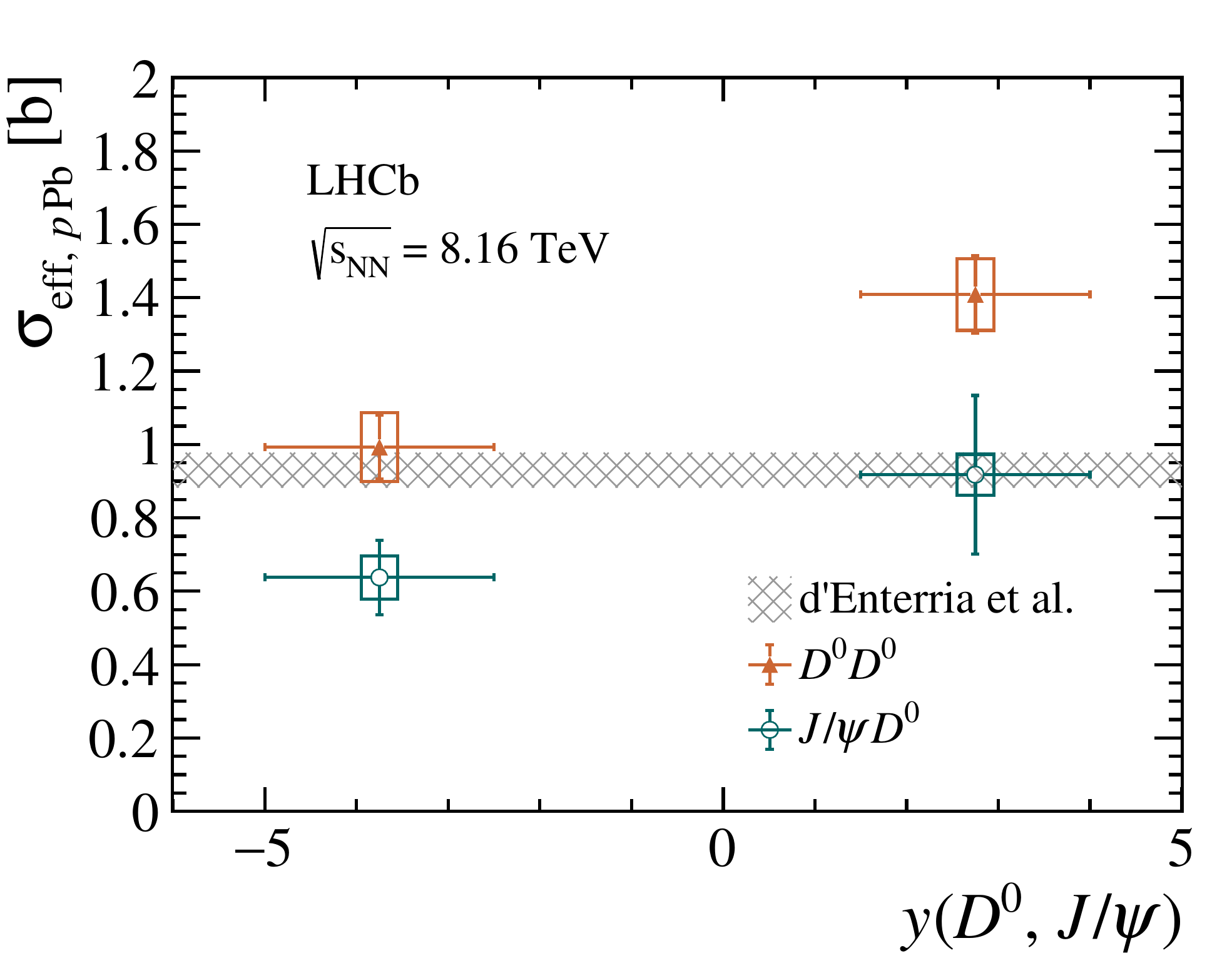}
  \caption{Left: Ratios of cross sections between like-sign and opposite-sign open charm pairs for different rapidity regions of charm hadrons~\cite{Aaij:2020smi}. The weighted average of ratios for different pairs is shown as a shaded magenta box. Right: $\sigmaeff$ parameter derived using $D^0+D^0$ and $\jpsi+D^0$ production for both negative and positive rapidities. The shaded area corresponds to the prediction from~\cite{dEnterria:2012jam} scaled by $A^2$, which predicts around a factor of three relative enhancement for DPS production compared to a naive scaling from \pp collisions. Vertical bars (boxes) are statistical (systematic) uncertainties. 
  \label{fig:DPS_LHCb}}
\end{figure}

The $\sigmaeffpA$ parameter is obtained using $D^0+D^0$ and $\jpsi+D^0$ production assuming solely DPS contribution (\cf{fig:DPS_LHCb}, Right). The LHCb derivation of $\sigmaeffpA$  results in a value that is (arbitrarily) normalised to be $A^2 = 208^2$ times larger than that defined in \ce{eq:sigmapA_DPS}. The theoretical prediction, shown as the grey band in the plot, amounts to $\sigmaeffpA \approx 1$~b, %
and is supported by the data. This result confirms the predicted factor of three enhancement for DPS compared to a simple $A$ scaling~\cite{Strikman:2001gz,Cattaruzza:2004qb,dEnterria:2012jam}. Looking in more detail, the positive-rapidity data exhibit a higher $\sigmaeffpA$ value compared to the negative rapidity one, which implies the necessity of considering the impact-parameter dependence in nPDFs~\cite{Shao:2020acd} (see below). The $\sigmaeffpA$ parameter measured for $\jpsi + D^0$  production hints at smaller values than that derived from $D^0+D^0$ production, and the same behaviour was observed in \pp data~\cite{Aaij:2012dz}. This is suggestive of a non-negligible contribution of SPS in $\jpsi + D^0$ production~\cite{Shao:2020kgj} which is not subtracted in the LHCb analysis. Due to limited statistics, the kinematic correlation between $\jpsi$ and $D^0$, \eg the $\Delta\phi$ distribution, does not provide yet enough information to identify the SPS component.

The LHCb observation of non-identical $\frac{\sigma^2_{p{\rm Pb}\to D^0}}{2\sigma_{p{\rm Pb}\to D^0+D^0}}$ values in the forward and backward regions indicates the presence of more effects beyond the expected geometrical DPS enhancement in \pp compared to \pA collisions. Assuming the same test function of the transverse spatial dependence $G(x)\propto x^a$ in the nuclear PDF modifications suggested in~\cite{Shao:2020acd}, \cf{fig:D0D0nPDF} shows that the LHCb data supports exponent values $a>1.5$, which however suffer from the uncertainty of $\sigmaeffpp$. A smaller $\sigmaeffpp$ value in fact requires stronger impact-parameter dependence (\ie larger $a$). The nuclear modification factors of single inclusive $D^0$ production in the two rapidity intervals are from independent measurements of the single inclusive process. This example corroborates the conclusion of~\cite{Shao:2020acd} that DPS in \pA collisions can be used to probe the impact-parameter-dependent nPDFs.  %

\begin{figure}[htbp]
\centering
\includegraphics[width=0.44\textwidth]{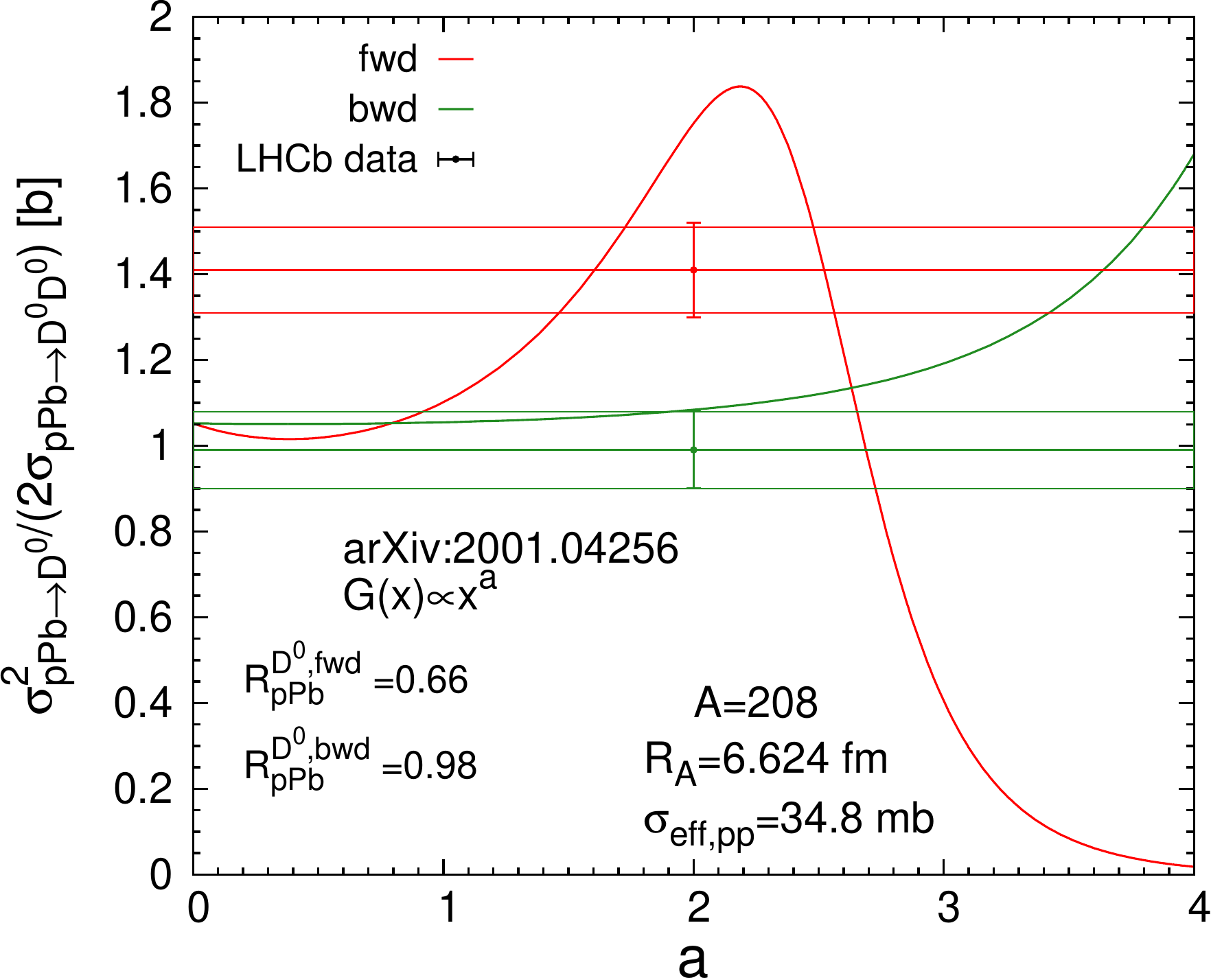}
\hspace{0.2cm}
\includegraphics[width=0.44\textwidth]{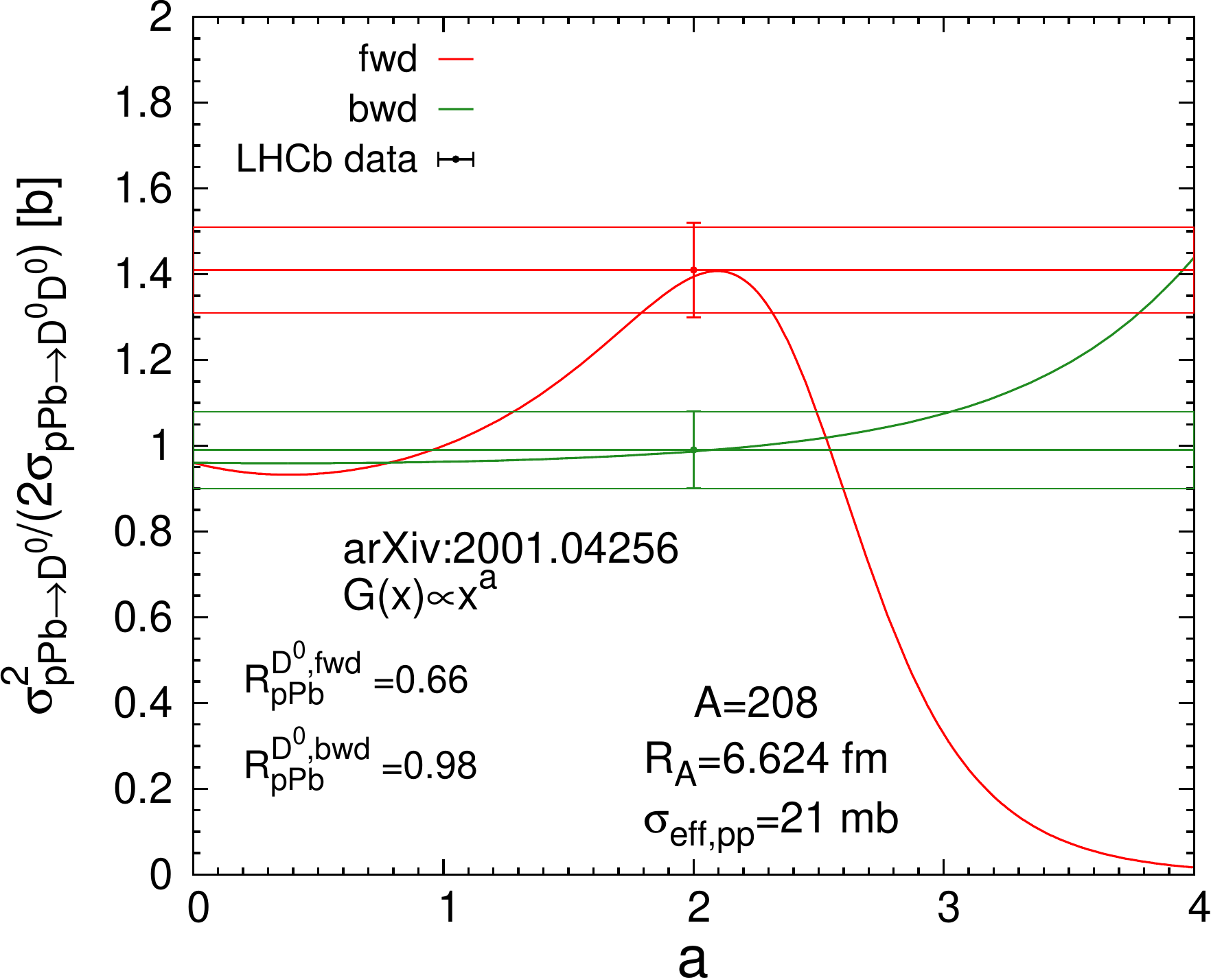}
\caption{Comparison of the ratio $\sigma^2_{p{\rm Pb}\to D^0}/(2\sigma_{p{\rm Pb}\to D^0+D^0})$ between the impact-parameter-dependent DPS calculation~\cite{Shao:2020acd} and the LHCb data~\cite{Aaij:2020smi} in both forward ($1.5<y(D^0)<4.0$) and backward ($-5.0<y(D^0)<-2.5$) rapidity intervals. Two different $\sigmaeffpp$ values are shown in the left and right plots respectively.}
\label{fig:D0D0nPDF} 
\end{figure}

\subsubsection{HL-LHC prospects}

At the HL-LHC, with the size of \pPb\ data samples increased by about a factor of ten compared to Run-2, one can exploit the large expected DPS signals over the SPS backgrounds in quarkonium final states as a means to scrutinise double and triple parton scatterings and, in particular in the purely geometric picture neglecting parton correlations, to extract the value of the effective DPS cross section $\sigmaeff$ independently of (and complementarily to) measurements in \pp collisions. First off, measurements in fine bins of final-state kinematics can be obtained for $\jpsi + D^0$  pairs in order to understand the possible difference of the $\sigmaeffpA$ parameter derived from $\jpsi+D^0$ and $D^0+D^0$ data, and shed light on the varying values at negative and positive rapidities (Fig.~\ref{fig:DPS_LHCb}). %

Table~\ref{tab:3} collects the expected DPS cross sections for the combined production of quarkonia ($\jpsi, \Upsilon$) and/or electroweak bosons ($W,\,Z$) in \pPb\ collisions at the nominal LHC energy of $\sqrtsnn = 8.8$~TeV. The individual SPS \pN cross sections have been derived in~\cite{dEnterria:2014lwk} at NLO accuracy with the colour evaporation model (CEM)~\cite{Vogt:2012vr} for quarkonia, and with \mcfm\ for the electroweak bosons, using the CT10~\cite{Lai:2010vv} proton and EPS09~\cite{Eskola:2009uj} nPDFs. The EPS09 nPDF does not include any impact-parameter dependence of nuclear effects, \ie it ignores the effects discussed in Fig.~\ref{fig:D0D0nPDF}. The DPS cross sections are estimated with the factorised expression for \pA\ collisions, \ce{eq:sigmapA_DPS} with $\sigmaeffdpspA = 22.5$~$\mu$b. The visible DPS yields (${\NDPS}_{p\mathrm{Pb}}$ values quoted) are estimated taking into account the relevant di-lepton decay branching fractions BR($\jpsi,\,\Upsilon,\,W,\,Z) = 6\%$, 2.5\%, 11\%, 3.4\%, plus simplified acceptance and efficiency losses. For \jpsi, the following value $\left({\cal A\times E}\right)_{\jpsi}\approx$~0.01 was assumed over merely one unit of rapidity at $|y|=0$, and $|y|=2$, corresponding to ATLAS/CMS central, and ALICE/LHCb forward, acceptances. For $\Upsilon$ and $W,Z$, the following values $\left({\cal A\times E}\right)_{\Upsilon}\approx 0.2$ and $\left({\cal A\times E}\right)_{W,Z}\approx 0.5$ were assumed over $|y|<2.5$. The quoted numbers were evaluated for an integrated luminosity amounting to ${\cal L}_{\rm int}=1$~pb$^{-1}$ . The quoted ${\NDPS}_{p\mathrm{Pb}}$ values are conservative for two reasons. First, ATLAS/CMS may ultimately integrate about ${\cal L}_{\rm int}=2$~pb$^{-1}$ \pPb\ collisions (although ALICE/LHCb should record half this value, see Table~\ref{tab:yrlumis})~\cite{Citron:2018lsq}. Second, for final states with $\jpsi$, the expected number of visible events can be easily multiplied by a factor of 3--5, taking into account the full rapidity acceptance (enlarged after Run-2, in some cases) of the ALICE/LHCb and ATLAS/CMS detectors. All listed processes are therefore in principle observable in the LHC proton-lead runs. Rarer DPS processes like $W+Z$ and $Z+Z$ have much lower cross sections and will require much higher integrated luminosities at the HL-LHC and/or \cm\ energies such as those reachable at the CERN Future Circular Collider~\cite{Dainese:2016gch,Benedikt:2018csr}.

\renewcommand{\arraystretch}{1.3}
\begin{table}[h!]
\centering
\caption{\label{tab:3}Estimated production cross sections at $\sqrtsnn = 8.8$~TeV for SPS quarkonia and electroweak bosons in \pN collisions, and for DPS double-$\jpsi$, $\jpsi+\Upsilon$, $\jpsi+W$, $\jpsi+Z$,  double-$\Upsilon$, $\Upsilon+W$, $\Upsilon+Z$, and same-sign $W+W$, in \pPb.  DPS cross sections are obtained via \ce{eq:sigmapA_DPS} for $\sigmaeffdpspA = 22.5$~$\mu$b (uncertainties, not quoted, are of the order of 30\%), and the associated yields for 1~pb$^{-1}$ integrated luminosity, after di-lepton decays and acceptance+efficiency losses~\protect\cite{dEnterria:2014lwk,Snigirev:2018arx}. %
We note that the $\jpsi$ yields quoted are only {\it per unit of rapidity} at mid- or forward-$y$.}
\vspace{0.25cm}
\begin{tabular}{lcccccccc}\hline
\pPb, $\sqrtsnn=8.8$ TeV & \multicolumn{4}{c}{final states}\\\hline
& $\jpsi+\jpsi$ \hspace{0.5cm}& $\jpsi+\Upsilon$ \hspace{0.5cm}& $\jpsi+W$ \hspace{0.5cm}& $\jpsi+Z$ \hspace{0.5cm} \\\hline
$\sigSPS^{\pN\to a},\,\sigSPS^{\pN\to b}$  & 45~$\mu$b ($\times2$) & 45~$\mu$b, 2.6~$\mu$b & 45~$\mu$b, 60~nb & 45~$\mu$b, 35~nb \\
 $\sigDPS^{p\mathrm{Pb}}$             & 45 $\mu$b &  5.2 $\mu$b & 120~nb & 70~nb \\
 $\NDPS^{p\mathrm{Pb}}$ (1 pb$^{-1}$)\hspace{0.5cm}& $\sim$65 & $\sim$60 & $\sim$15 & $\sim$3 \\\hline
 & $\Upsilon+\Upsilon$ & $\Upsilon+W$ & $\Upsilon+Z$ & ss\,$W+W$ \\\hline
$\sigSPS^{{\pN}\to a},\,\sigSPS^{\pN\to b}$  & 2.6~$\mu$b ($\times2$) & 2.6~$\mu$b, 60~nb & 2.6~$\mu$b, 35~nb & 60~nb ($\times2$) \\
 $\sigDPS^{p\mathrm{Pb}}$             & 150~nb & 7~nb & 4~nb & 150 pb \\
 $\NDPS^{p\mathrm{Pb}}$ (1 pb$^{-1}$)&  $\sim$15 & $\sim$8 & $\sim$1.5 & $\sim$4 \\\hline
\end{tabular}
\end{table}

\section{Summary}

Quarkonium measurements at the LHC are not only motivated by the intrinsic goal of advancing our understanding of their underlying production mechanisms, which are still not fully understood today, but also by the broad and unique  opportunities  they offer to perform a wide range of studies. In this document, we have reviewed the prospects for quarkonium studies in the upcoming high-luminosity phases of the LHC, with proton-proton (\pp), proton-nucleus (\pA), and nucleus-nucleus (\AaAa) collisions. Among the research topics highlighted are:   
opportunities in multi-quark spectroscopy; in new probes of the proton parton distributions including transverse-momentum dependent and spin effects; in sensitive observables for the study of double parton scattering interactions; in nuclear PDFs or other nuclear effects; and in studies to determine the properties of the quark-gluon plasma.

Section~\ref{sec:pp} surveyed the prospects for measurements of %
quarkonium production in \pp\ collisions in the coming years and into the HL-LHC era. The motivations for future measurements of the quarkonium \pT spectra and polarisation were discussed, as well as the possibilities for more detailed characterisations of the properties of quarkonium-production events, in particular for \jpsi and \ups. Such investigations can be undertaken through the study of (i) the hadronic activity accompanying scatterings where quarkonia are produced, (ii) the formation of quarkonia within high-energy jets, and (iii) their associated production alongside highly energetic objects such as jets, vector bosons, or other quarkonium states. We have highlighted where such measurements can provide new insights into broader fields such as in the search for new physics phenomena and in the study of multi-parton interactions, and where such measurements have already shown significant promise. The potential for the study of $C$-even quarkonia as well as multi-quark and molecular states was presented. Opportunities for HL-LHC quarkonium data to provide constraints on proton PDFs in the low-$x$ and low-scale regime were also outlined.

Section~\ref{sec:excl_diff} addressed diffractive and, mainly, exclusive photoproduction of quarkonia in hadron-hadron collisions. After a short description of selected experimental results and the discussion of open points in experiment and theory, this section focused on measurements possible at the HL-LHC, both in the collider and fixed-target modes, that either have not yet been performed or that have not yet been sufficiently exploited. In particular, the study of forward \jpsi production in combination with a backward jet and the study of exclusive single-quarkonium and quarkonium-pair production, which provide access to the multi-dimensional nucleon and nucleus partonic structure, have been discussed. Here, the advantage of \pA collisions in the collider data-taking mode, and the need for high integrated luminosities, have been highlighted in order to fully exploit the potential of exclusive measurements.

Section~\ref{sec:spin} focused on studies of the transverse-momentum-dependent and spin dynamics in quarkonium production in \pp\ collisions. Having first reviewed the two main frameworks that account for transverse-momentum-dependent effects, \ie\ TMD factorisation and HE factorisation, a discussion followed on their applicability to quarkonium production along with potential challenges, open issues, and opportunities. In particular, the discussion covered those quarkonium-production processes that can be used to study the impact of factorisation-breaking effects and the region of applicability of these frameworks. Single transverse-spin asymmetries, believed to be generated in quarkonium production by the gluon Sivers effect, that arises from the correlation between the proton spin and the gluon motion, were also addressed. Three approaches that can account for this correlation have been discussed, as well as a selection of experimental projections for the HL-LHC in unpolarised collisions, and in polarised collisions in the LHC FT mode.

Section~\ref{sec:pa} focused on inclusive quarkonium studies in \pA\ collisions at the LHC. First, a survey of the different phenomena at play was given. This was followed by an overview of the current status of the use of quarkonium data to constrain nPDFs in the collider and FT modes, and of low-$x$ parton saturation calculations applied to quarkonium production. Second, experimental observables used to compare quarkonium production in \pA\ and \pp\ collisions were discussed. The section concluded with a discussion of the status and prospects for the understanding of flow-like phenomena observed in \pA\ collisions, as well as of the experimental and theoretical status of quarkonium-hadronisation modifications in \pA compared to \pp collisions.

Section~\ref{sec:aa} focused on quarkonium production in \AaAa collisions. The main physics phenomena at play in quarkonium physics in heavy-ion collisions were introduced, as well as the theoretical state-of-the-art and experimental prospects for the HL-LHC. Recent theory developments were been discussed, including the semi-classical transport in open quantum systems, a density-operator model, and an advanced effective-field-theory model. A selection of opportunities offered by the HL-LHC was presented, including the investigation of the collision-energy dependence of various observables through comparisons of FT and collider data, and prospects for studies of the $X(3872)$ state and for measurements of the \jpsi polarisation.

Section~\ref{sec:dps} discussed the current theoretical and experimental status of the physics of double and triple parton scatterings (DPS and TPS) in \pp and \pA collisions, with an emphasis on the role of measurements of the production of multiple quarkonia, or quarkonia plus electroweak gauge boson, as a means to clarify the multiple open issues in the field. Detailed theoretical perspectives and experimental prospects of relevance for the HL-LHC operation, including expected number of events for various DPS and TPS final states with quarkonia, were provided.

Overall, this document reviewed how the HL-LHC will, on the one hand, help to understand quarkonium production better and, on the other, help to advance the use of quarkonia as tools for multiple aspects of QCD physics.

\section*{Acknowledgements}
This project has received funding from the European Union’s Horizon 2020 research and innovation programme under the grant agreement No.824093 (STRONG-2020). 
This project has also received funding from the French ANR under the grant ANR-20-CE31-0015 (``PrecisOnium'').
This work was also partly supported by the French CNRS via the IN2P3 project GLUE@NLO, via the Franco-Chinese LIA FCPPL (Quarkonium4AFTER), via the IEA No.205210 (``GlueGraph") and ``Excitonium'', by the Paris-Saclay U. via the P2I Department and by the P2IO Labex via the Gluodynamics project. 
D.Y.A.V. and P.B.G. acknowledge the support of the "R\'egion Pays de la Loire" under the contract No. 2015-08473.
M.A.O.’s work was partly supported by the ERC grant 637019 “MathAm”.
The work of B.D. has been supported by the ERC Starting Grant 715049 ``QCDforfuture''.
C.V.H.\ has received funding from the European Union's Horizon 2020 research and innovation programme under the Marie Sklodowska--Curie grant agreement No 792684.
The work of F.G.C.\ has been supported by the Italian MIUR under the FARE program (code n.\ R16XKPHL3N, 3DGLUE), of U.D. and C.P. by Fondazione di Sardegna under the projects “Quarkonium at LHC energies”, No. F71I17000160002 (University of Cagliari) and  ''Proton tomography at the LHC”, No. F72F20000220007 (University of Cagliari). 
D.P.\ is supported by the Science and Technology Facilities Council under grants ST/M005437/1 and  ST/N000374/1.
The work of S.B. has been supported by the National Science Foundation under Contract No.\ PHY-1516088. The work of XY is supported by the U.S. Department of Energy, Office of Science, Office of Nuclear Physics grant DE-SC0011090.
J.-W.Q and K.W. are supported by Jefferson Science Associates, LLC under U.S.\ DOE Contract No.DE-AC05-06OR23177. This work is also supported within the framework of the TMD Topical Collaboration.
The work of V.K. was supported in part by the Shota Rustaveli National Science Foundation of Georgia (SRNSFG) under grant FR17-184.
M.G.E.\ is supported by the Spanish MICINN grant PID2019-106080GB-C21.
E.G.F.\ is supported by Ministerio de Ciencia e Innovaci\'on of Spain under project FPA2017-83814-P; Unidad de Excelencia Mar\'ia de Maetzu under project MDM-2016-0692; and Xunta de Galicia (Conseller\'ia de Educaci\'on) and FEDER. 
J. He is partly supported by NSFC (No. 11775227).
E. Chapon is supported by MoST (No. 2018YFA0403901) and NSFC (No. 11875275, 12061141003) and partially by CCEPP (China).
The work of M.N. has been supported by the Ministry of Education and Science of Russia via the State assignment to educational and research institutions under the project FSSS-2020-0014.
S.B.\ would like to thank Andreas Metz for insightful discussions, and Jian Zhou for the collaboration. 
F.G.C.\ thanks Alessandro Bacchetta, Alessandro Papa, and Michael Fucilla for fruitful conversations.

\bibliographystyle{utphys}
\bibliography{references}

\end{document}